\documentstyle[epsf,epic,eepic]{article}

     \def\tenrm{\rm}

     \def\sR{\scriptscriptstyle R}
     
     \def\sD{\scriptscriptstyle D}
     \def\ps{\hbox{\raisebox{.2ex}{$p$}}\mbox{\hspace{-0.9ex}}/}
     \def\ks{k\mbox{\hspace{-0.95ex}}/}
     
     \def\nab{\nabla\mbox{\hspace{-1.5ex}}/}
     
     \def\11{\mbox{\boldmath $1$}}
     \def\xx{\mbox{\boldmath $x$}}
     \def\yy{\mbox{\boldmath $y$}}
     \def\tr{\mbox{tr}}
     \def\Det{\mbox{Det}}
     \def\lsim{\mbox{\hspace{1ex}}
               \hbox{\raisebox{.4ex}{$<$}}\mbox{\hspace{-1.7ex}}
              {\lower.8ex\hbox{$\sim$}}\mbox{\hspace{1.1ex}}}
     \def\ssim{\mbox{\hspace{1ex}}
               \hbox{\raisebox{.4ex}{$>$}}\mbox{\hspace{-1.7ex}}
              {\lower.8ex \hbox{$\sim$}}\mbox{\hspace{1.1ex}}}

     \def\beq{\begin{equation}}
     \def\eeq{\end{equation}}
     \def\bce{\begin{center}}
     \def\ece{\end{center}}
     \def\bea{\begin{eqnarray}}
     \def\eea{\end{eqnarray}}
     \def\ben{\begin{enumerate}}
     \def\een{\end{enumerate}}

     \def\nn{\nonumber}

     \def\wt{\widetilde}

     \def\brr{\begin{array}}
     \def\err{\end{array}}
     
     \def\nabar{\not\!\nabla}
     \def\nabar2{{\not\!\nabla}^2}

     \def\ltextindent#1{\hbox to\hangindent{#1\hss}\ignorespaces}

     \newcommand{\VT}{V^{T}}
     
     \newcommand{\VTR}{V^{TR}}
     \newcommand{\betacr}{\beta_{\mbox{cr}}}
     \newcommand{\alphacr}{\alpha_{\mbox{cr}}}
     
     \newcommand{\omegacr}{\omega_{\mbox{cr}}}
     \newcommand{\sumomeg}{\sum_{n=-\infty}^{\infty}}
     \newcommand{\intsig}{\int_{0}^{\sigma}\!\!\!ds}
     \newcommand{\diracsh}[1]{{\not\hspace{0.1em}}#1}
\begin{document}
\vspace*{15mm}

\begin{center}
{\Large \bf 
Dynamical Symmetry Breaking in Curved Spacetime}\\
{\large -- Four-Fermion Interactions --}\\[8mm]
Tomohiro Inagaki
\footnote{e-mail : inagaki@ipc.hiroshima-u.ac.jp},\\
{\it Information Processing Center, Hiroshima University,
Higashi-Hiroshima, Hiroshima 739}\\[3mm]
Taizo Muta
\footnote{e-mail:muta@sci.hiroshima-u.ac.jp},\\
{\it Department of Physics, Hiroshima University,
Higashi-Hiroshima, Hiroshima 739}\\[3mm]
Sergei D. Odintsov
\footnote{e-mail:sergei@ecm.ub.es, odintsov@quantum.univalle.edu.co},
{\it Departamento de Fisica, Universidad del Valle,
A. A. 25360 Cali, Colombia\\
and\\
Dept. Math. and Physics, Tomsk Pedagogical University,
634041 Tomsk, Russia}
\\[3mm]
\end{center}

\begin{abstract}
   This review deals with the theory of
four-fermion interactions in curved spacetime.
Starting with the $D$-dimensional Minkowski spacetime
($2 \leq D \leq 4$) the effective potential in the 
leading order of $1/N$-expansion is calculated
and the phase structure of the theory is investigated.
Using the same technique the effective potential
for composite operator $\bar{\psi}\psi$ 
in four-fermion models is calculated under the following 
circumstances:

a) $D$-dimensional weakly curved spacetime (in linear
curvature approximation),

b) $D$-dimensional de Sitter and anti-de Sitter universe,

c) $D$-dimensional Einstein universe.

\noindent
The phase structure of the theory is investigated analytically
as well as numerically.
Curvature induced phase transitions are discussed where
fermion masses are dynamically generated.

As an extension of four-fermion models we consider the gauged
Nambu-Jona-Lasinio (NJL) model, higher derivative
NJL model and supersymmetric NJL model
in weakly curved spacetime where the effective
potential is analytically evaluated.
The phase structure of the models is again analyzed and 
the condition for the chiral symmetry breaking in the 
gauged NJL model is given in an analytical form.

Finally the influence of two external effects
(non-zero temperature and gravitational field,
non-trivial topology and gravitational field as well as
magnetic and gravitational field) to the phase structure
of four-fermion models is analyzed.
The possibility of curvature and temperature-induced
or curvature- and topology-induced
phase transitions is discussed.
It is also argued that the chiral symmetry broken
by a weak magnetic field may be restored due to
the presence of gravitational field.
Some applications of four-fermion models in quantum gravity 
are also briefly investigated.
\end{abstract}

\begin{center}
{\bf This paper was published in Prog. Theor. Phys. Suppl. 127 (1997) 93.}
\end{center}

\newpage

\tableofcontents

\newpage

\section{Introduction}

An idea of dynamical symmetry breaking (DSB) was introduced
by Nambu and Jona-Lasinio in quantum field theory 
in Ref. \cite{NJL} 
(for modern reviews see, e.g., Ref. \cite{DSB}).
The original Nambu-Jona-Lasinio (NJL) model and its
generalizations (all of which are somehow related to the
theory of superconductivity a la Bardeen, Cooper and Schriffer
\cite{BCS}) are extremely useful toy models in the study of
composite states in various circumstances. To discuss the dynamical
symmetry breaking nonperturbative treatments are essential. One of
the simplest nonperturbative techniques is the $1/N$-expansion
scheme which works well in the four-fermion models.

In the simplest version of the NJL model \cite{NJL}
(or four-fermion model) the chiral symmetry of the
theory is spontaneously broken according to the emergence of
a vacuum condensate of the composite field of fermions
$\psi$ : $\langle \bar{\psi}\psi\rangle\neq 0$.
At the same time the non-vanishing fermion mass is generated.
In the spectrum one finds a mode of the composite scalar field : 
$\sigma\sim\bar{\psi}{\psi}$, which may correspond to the Higgs
field in the theory with the elementary Higgs scalar field.
Clearly the model may be regarded as a composite Higgs model
based on the mechanism of dynamical symmetry breaking.

Four-fermion models (gauged four-fermion models) have been 
often considered as an effective low energy theory for
quantum chromodynamics 
(QCD, see Ref. \cite{Taizo} for an introduction).
The chiral symmetry of such a theory is broken by 
the creation of quark-anti-quark condensates and
quark masses are dynamically generated. Such a theory is extremely
useful in dealing with non-perturvative phenomena of QCD.

Moreover the (gauged) four-fermion models may be well applied to
the description of the dynamical symmetry breaking in standard
model (SM). At the electroweak scale the dynamical symmetry
breaking may be caused by the fundamental fermion condensate
as in the technicolor model.\cite{TC} The similar mechanism could
work in the symmetry breaking at the GUT era
in early universe.

The description of elementary particle theories
(QCD, SM, GUT, $\cdots$) in terms of the effective four-fermion
model is particularly convenient when the theories are considered
under some external conditions (finite temperature, finite density,
non-trivial topology, external gauge and gravitational fields, etc.).
As an extreme external condition one immediately considers the early
universe where the gravity is strong and the temperature is high.
Since the dynamical symmetry breaking mechanism replaces the
spontaneous symmetry breaking caused by the elementary Higgs scalar
field, it is natural for us to employ the composite field $\sigma$
as an inflaton in the inflationary universe.\cite{Inf}
Through renormalization group arguments \cite{BOS}
one finds the necessity of
the scalar-gravitational non-minimal coupling $\xi$ \cite{BOS}
whose value is governed by the renormalization group running.
The value of $\xi$ varies in different theories and is crucial
to have a successful inflation.

One realizes an importance of studying the phase structure
of the four-fermion model in curved background and finds the cosmological
motivation to deal with this model.

The main purpose of the present article is to present what is
and is not established in the study of the four-fermion models
in curved spacetime in full detail.
We do not consider in the present article any cosmological applications.
Instead we give a review of the phase structure of the different
kinds of the four-fermion model with different gravitational backgrounds.
Hence the classical external gravity is considered here as a probe
of quantum four-fermion models
and is regarded as a source of the curvature-induced phase transition
in these models.

The paper is organized as following.
In \S 2 fundamental properties of four-fermion models
are briefly reviewed.
To investigate the phase structure of the four-fermion theory
the effective potential is introduced in curved spacetime.
We adopt the $1/N$ expansion method as a nonperturbative
approach and calculate the effective potential
in the leading order of the $1/N$ expansion.
As an instructive rehearsal we work in the Minkowski spacetime and
evaluate the effective potential analytically as well as numerically.
On the basis of the effective potential we discuss the phase 
structure of the four-fermion model in the Minkowski spacetime.

In \S 3 the weakly curved spacetime is considered.
We keep only terms independent of and linear in the curvature.
Evaluating the effective potential in the leading order
of $1/N$ expansion the phase structure of the theory is clarified
and the curvature-induced phase transition is discussed
in arbitrary dimensions ($2 \leq D \leq 4$).

In \S 4 we consider the spacetimes with the constant 
curvature (de Sitter background, anti-de Sitter
background and Einstein universe background).
In these spacetimes we can evaluate the effective
potential of the four-fermion model without
making any approximation with regard to the curvature.
With the aim of justifying the validity of the weak curvature
approximation we investigate the phase structure in the constant
curvature spacetime. We then compare the result with the one in
the weakly curved spacetime.

Section 5 is devoted to the study of the gauged NJL model,
supersymmetric NJL model and 
higher derivative four-fermion model in the weakly curved 
spacetime. 
We evaluate the effective potential in the gauged NJL model 
with finite cut-off and let the cut-off tend to infinity. 
In order to do so we use the renormalization group method
and appeal to the equivalence with the gauge Higgs-Yukawa model.
The analytical form of the condition for the chiral symmetry breaking 
is found. 
For the four-fermion model with higher derivative terms we also study
the effective potential and the phase structure of the model.
The phase structure of supersymmetric NJL model is
also discussed in detail.

In \S 6 the combined effect of curvature and temperature
is discussed in curved spacetime.
To introduce temperature we suppose the thermal equilibrium
and we restrict ourselves to the Einstein universe background.
The effective potential is evaluated at finite
temperature in spacetimes with positive or negative curvature.
For sufficiently high temperature
we show that the broken chiral symmetry is restored
and the symmetric phase can be realized
even in the spacetime with negative curvature.

In \S 7 we discuss the influence of the non-trivial topology
to the chiral symmetry breaking in the weakly curved spacetime. 
We start with the calculation of the effective 
potential in the four-fermion model in the flat torus-compactified 
spacetime. 
We find the same quantity as in the weakly curved spacetime 
with the torus-compactified dimension. 
The possibility of curvature and topology-induced phase
transition is explicitly shown.

Section 8 is devoted to the investigation of chiral symmetry
breaking in the weakly curved spacetime under the influence of
the external magnetic field. 
The effective potential is found and its phase structure 
is studied. 
The chiral symmetry is shown to be broken in the NJL model 
under the external magnetic field. However, the inclusion of
the external gravitational field with positive curvature acts
in opposite direction and it may restore chiral symmetry.

In \S 9 some applications of four-fermion models in quantum
gravity are presented. In particulary, we discuss semiclassical
approach in 2D Gross-Neveu-dilaton model.
The conformal factor dynamics based on four-fermion theory
is studied in 4D quantum gravity.
The effective potential for composite gravitino in $N=1$
supergravity on de Sitter background is also calculated.

In \S 10 we give an outlook and prospects for
future researches.

Our notation is basically in conformity with the $(-,-,-)$
convention in the book by Misner, Thorne and Wheeler.\cite{MTW}

\section{Four-fermion models in Minkowski spacetime}

The four-fermion model is one of the prototype models of
the dynamical symmetry breaking. 
In the present section we briefly review the fundamental properties
of the four-fermion models.
A four-fermion model and its effective potential are 
introduced in curved spacetime.
We adopt the $1/N$ expansion as a non-perturbative
approach to investigate the dynamical symmetry breaking
in the model. 
The phase structure of the theory is discussed in Minkowski
spacetime as an instructive example of our method.

\subsection{Gross-Neveu type model}

Here we consider the Gross-Neveu type model in 
curved spacetime 
and discuss the symmetry and the symmetry breaking
in arbitrary dimensions ($2\leq D\leq 4$).
It is defined by the action
\begin{equation}
     S = \int\! \sqrt{-g} d^{\sD}x \left[
     \sum^{N}_{k=1}\bar{\psi}_{k}i\gamma^{\mu}\nabla_{\mu}\psi_{k}
     +\frac{\lambda_0}{2N}
      \left(\sum^{N}_{k=1}\bar{\psi}_{k}\psi_{k}\right)^{2}
     \right]\, ,
\label{ac:gn}
\end{equation}
where index $k$ represents the flavors of the fermion field
$\psi$, $N$ is the number of fermion species, $g$ the determinant
of the metric tensor $g_{\mu\nu}$,
$\gamma^{\mu}$ the Dirac matrix in curved
spacetime and $\nabla_{\mu} \psi$ the covariant derivative of the
fermion field $\psi$.
In two spacetime dimensions the theory is nothing but the
Gross-Neveu model.\cite{GN}
For simplicity, we neglect the flavor index below. 

The action (\ref{ac:gn}) is invariant under the discrete 
transformation
\begin{equation}
\left\{
\begin{array}{l}
\bar{\psi}\psi \longrightarrow -\bar{\psi}\psi\, ,\\
\bar{\psi}\gamma_{\mu}\psi \longrightarrow \bar{\psi}\gamma_{\mu}\psi\, .
\end{array}
\right.
\end{equation}
In two or four spacetime dimensions this transformation is realized
by the discrete chiral transformation
\begin{equation}
     \psi \longrightarrow \gamma_{5}\psi\, .
\label{t:dischi}
\end{equation}
Below we call this $Z^{2}$ symmetry the discrete 
chiral symmetry.

The discrete chiral symmetry prohibits the fermion mass term.
If the composite operator constructed by the fermion and anti-fermion
develops the non-vanishing vacuum expectation value 
\begin{equation}
     \langle \bar{\psi}\psi\rangle \neq 0\, ,
\end{equation}
fermion mass term appears in the four-fermion interaction
term and the chiral symmetry is broken down dynamically.

We consider the theory with $N$ flavors of the fermion fields.
Thus the theory also has $SU(N)$ flavor symmetry
\begin{equation}
     \psi \longrightarrow e^{i\sum_{a} g_{a}T^{a}}\psi\, ,
\end{equation}
where $T^{a}$ are generators of the $SU(N)$ symmetry.
Under the circumstance of the global $SU(N)$ flavor symmetry
we may work in a scheme of the $1/N$ expansion.

For practical calculations in four-fermion theories 
it is more convenient to introduce auxiliary 
field $\sigma$.\cite{GN}
The generating functional is given by
\begin{equation}
    Z
   = \int[d\tilde{\psi}][d\tilde{\bar{\psi}}]e^{iS}\, ,
\label{def:gfunc0}
\end{equation}
where we choose the path-integral measure in terms of the field
variables
\begin{equation}
\left\{
\begin{array}{l}
     \tilde{\psi}=\sqrt[4]{-g}\psi\, ,\\
     \tilde{\bar{\psi}}=\sqrt[4]{-g}\bar{\psi}\, ,
\end{array}
\right.
\end{equation}
so that the covariance under the general coordinate
transformation \cite{Fj} is guaranteed.

We consider the Gaussian integral 
\begin{equation}
     \mbox{C}=\int [d\sigma'] \exp i\!\int\! \sqrt{-g}d^{\sD}x
     \left(
     -\frac{N}{2\lambda_{0}}{\sigma'}^{2}
     \right)\, .
\label{gauss}
\end{equation}
Obviously C is a constant.
The path integral measure is invariant
under the redefinition of the parameter $\sigma'$
\begin{equation}
     \sigma' \longrightarrow \sigma=\sigma' 
     -\frac{\lambda_{0}}{N}\bar{\psi}\psi\, .
\end{equation}
After the redefinition, Eq. (\ref{gauss}) becomes
\begin{equation}
     \mbox{C}=\int [d\sigma] \exp i\!\int\! \sqrt{-g} d^{\sD}x
     \left[
     -\frac{N}{2\lambda_{0}}
     \left(\sigma +\frac{\lambda_{0}}{N}\bar{\psi}\psi\right)^{2}
     \right]\, .
\label{gauss:tran}
\end{equation}
As Eq. (\ref{gauss:tran}) is a constant, we are free to insert
it in the right-hand side of Eq. (\ref{def:gfunc0}) without
any change. 
We obtain
\begin{equation}
    Z
   = \frac{1}{\mbox{C}}\int[d\tilde{\psi}][d\tilde{\bar{\psi}}][d\sigma]
       e^{iS_{y}}\, ,
\label{aux:gfunc0}
\end{equation}
where action $S_{y}$ is given by
\begin{equation}
     S_{y} = \!\int\! \sqrt{-g}d^{\sD}x \left(\bar{\psi}i\gamma_{\mu}
      \nabla^{\mu}\psi
     -\frac{N}{2\lambda_0}\sigma^{2}-\bar{\psi}\sigma\psi
     \right)\, .
\label{ac:yukawa}
\end{equation}
The four-fermion interaction term 
in Eq. (\ref{ac:gn}) has been canceled out by 
introducing the Gaussian integral (\ref{gauss:tran}).

The action $S_{y}$ is equivalent to the original action 
(\ref{ac:gn}).\cite{HMC}
If the non-vanishing vacuum expectation value is assigned to
the auxiliary field $\sigma$
\begin{equation}
     \langle \sigma \rangle =m \neq 0\, ,
\end{equation}
there appears a mass term for the fermion field $\psi$
and the discrete chiral symmetry (the $Z_{2}$ symmetry)
is eventually broken.
\vglue 2ex
$\psi$ propagator:
\begin{figure}
    \begin{minipage}{.53\linewidth}
     \epsfxsize = 50mm
     \centerline{\epsfbox{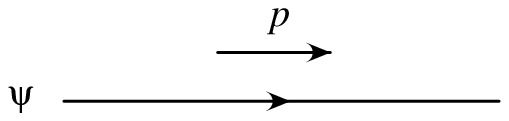}}
    \end{minipage}
\hfill
    \begin{minipage}{.46\linewidth}
    \begin{center}
        \vglue 4ex
        $\displaystyle\frac{i}{\ps}\, ,$
    \end{center}
    \end{minipage}
\end{figure}
\vglue 4ex
$\sigma$ propagator:
\begin{figure}
    \begin{minipage}{.53\linewidth}
     \epsfxsize = 50mm
     \centerline{\epsfbox{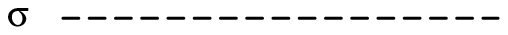}}
    \end{minipage}
\hfill
    \begin{minipage}{.46\linewidth}
    \begin{center}
        $\displaystyle -i\ \frac{\lambda_{0}}{N}\, ,$
    \end{center}
    \end{minipage}
\end{figure}
\vglue 4ex
$\bar{\psi}\psi\sigma$ vertex:
\begin{figure}
    \begin{minipage}{.53\linewidth}
     \epsfxsize = 50mm
     \centerline{\epsfbox{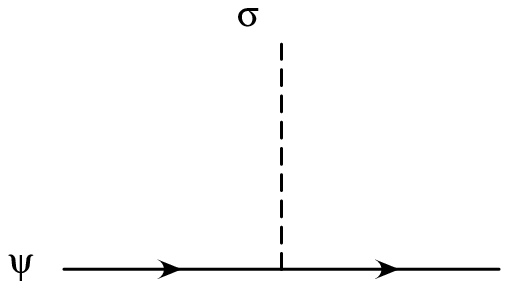}}
    \end{minipage}
\hfill
    \begin{minipage}{.46\linewidth}
    \begin{center}
        $i\, ,$
    \end{center}
    \end{minipage}
\end{figure}
\vglue 4ex
$\psi$ loop:
\begin{figure}
    \begin{minipage}{.53\linewidth}
     \epsfxsize = 50mm
     \centerline{\epsfbox{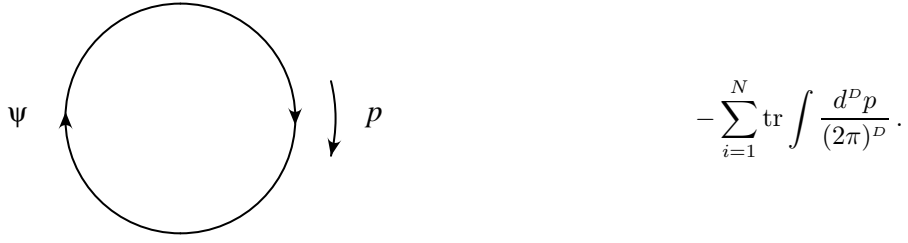}}
    \end{minipage}
\hfill
    \begin{minipage}{.46\linewidth}
    \begin{center}
        $\displaystyle{-\sum^{N}_{i=1}
        \mbox{tr}\int \frac{d^{\sD}p}{(2\pi)^{\sD}}}\, .$
    \end{center}
    \end{minipage}
\caption{Feynman rules of Gross-Neveu type model
in Minkowski spacetime.}
\label{feyn:rules}
\end{figure}

In Fig. 1 we present the Feynman rules for the action 
(\ref{ac:yukawa}) in Minkowski spacetime.
These Feynman rules are useful to make quantum calculations
in the Gross-Neveu type model.
The Feynman rule for the $\sigma$ propagator contains the 
factor $1/N$. Thus the radiative corrections including the 
$\sigma$ propagator are suppressed at the large $N$ limit.
Internal $\sigma$ lines have no contributions
in the leading order of the $1/N$ expansion.
Therefore we easily calculate the radiative correction in
the Gross-Neveu type model 
at the leading order of the $1/N$ expansion.
For example the effective potential in Gross-Neveu 
type model is calculated from the two-point function of massive
free fermions (see \S 2.2).

\subsection{Effective potential}

We would like to find a ground state of the system described by 
the four-fermion model.
For this purpose we evaluate an effective potential for the field 
$\sigma$.
The ground state is determined by observing the minimum of the
effective potential.
If the auxiliary field $\sigma$ develops the non-vanishing
vacuum expectation value, the ground state is no longer invariant 
under the discrete chiral transformation and then
the chiral symmetry is broken down dynamically.
In the present subsection we briefly review how to calculate the 
effective potential in general situation, i.e., curved spacetime.
The case of the Minkowski space is included as a special case.

We start with the generating functional
\begin{equation}
     Z[J]=\int [d\tilde{\psi}][d\tilde{\bar{\psi}}][d\sigma]\
          \mbox{exp}\left(i S_{y} 
           +iN\int \sqrt{-g}d^{\sD}x \sigma(x) J(x)\right)\, ,
\label{z0:gn}
\end{equation}
where $J(x)$ is the source function for the field $\sigma(x)$.
Performing the integration over the fermion fields  $\psi$ and
$\bar{\psi}$, we get
\begin{equation}
     Z[J]=\int [d\sigma]\ \mbox{exp}\
          iN\left[\int \sqrt{-g}d^{\sD}x
          \left(
                 -\frac{\sigma^{2}}{2\lambda_{0}}+\sigma J
          \right)
         -i\mbox{ln Det}(i\gamma^{\mu}\nabla_{\mu}-\sigma)
          \right]\, .
\label{z02:gn}
\end{equation}
We divide the field $\sigma$ into two parts
\begin{equation}
     \sigma=\sigma_{0}+\tilde{\sigma}\, ,
\end{equation}
where $\sigma_{0}$ is a classical background which
satisfies the classical equation of motion and $\tilde{\sigma}$
is a quantum fluctuation.
In terms of $\sigma_{0}$ and $\tilde{\sigma}$ Eq. (\ref{z02:gn}) 
is rewritten as
\begin{eqnarray}
     Z[J]&=&\mbox{exp}\
          iN\left[\int \sqrt{-g}d^{\sD}x
          \left(
                 -\frac{\sigma_{0}^{2}}{2\lambda_{0}}+\sigma_{0} J
          \right)
                 -i\mbox{ln Det}(i\gamma^{\mu}\nabla_{\mu}-\sigma_{0})
          \right] \nonumber \\
         &&\times \int [d\tilde{\sigma}]\ \mbox{exp}\
          iN\left[\int \sqrt{-g}d^{\sD}x
          \left(
                 -\frac{2\sigma_{0}\tilde{\sigma}+\tilde{\sigma}^{2}}
                       {2\lambda_{0}}+\tilde{\sigma} J
          \right)
          \right.\nonumber \\
          &&\left.-i\mbox{ln Det}
          \frac{\tilde{\sigma}}{(i\gamma^{\mu}\nabla_{\mu}-\sigma_{0})}
          \right]\, .
\end{eqnarray}
As may be seen through Fig. 1 radiative corrections in $\tilde{\sigma}$ 
are higher order corrections in the $1/N$ expansion.
In the leading order of the $1/N$ expansion the generating functional
$W[J]$ for connected Green functions is given by
\begin{eqnarray}
     N W[J] &=& -i\mbox{ln}Z[J]\nonumber \\
            &=& \int \sqrt{-g}d^{\sD}x
              \left(
                 -\frac{\sigma_{0}^{2}}{2\lambda_{0}}+\sigma_{0} J
              \right)
                 -i\mbox{ln Det}(i\gamma^{\mu}\nabla_{\mu}-\sigma_{0})
              \nonumber \\
              && +O\left(\frac{1}{N}\right)\, .
\label{w0:gn}
\end{eqnarray}
We pull out an obvious factor $N$ in defining the generating
functional $W[J]$.

The effective action $\Gamma[\sigma_{c}]$ is defined to be the
Legendre transform of $W[J]$
\begin{equation}
     \Gamma[\sigma_{c}]=W[J]-\int \sqrt{-g}d^{\sD}x\ \sigma_{c}(x)J(x)\, ,
\label{def:gamma}
\end{equation}
where the new variable $\sigma_{c}$ is given by
\begin{equation}
     \sigma_{c}=\frac{\delta W[J]}{\delta J}=\sigma_{0}
               +O\left(\frac{1}{N}\right)\, ,
\label{def:sigmac}
\end{equation}
which corresponds to the vacuum expectation value of $\sigma$
in the presence of the source $J$.
We substitute Eqs. (\ref{w0:gn}) and (\ref{def:sigmac}) into 
Eq. (\ref{def:gamma}) to get
\begin{eqnarray}
     \Gamma[J] = -\int \sqrt{-g}d^{\sD}x
                 \frac{\sigma_{c}^{2}}{2\lambda_{0}}
                 -i\mbox{ln Det}(i\gamma^{\mu}\nabla_{\mu}-\sigma_{c})
              +O\left(\frac{1}{N}\right)\, .
\label{w0:gamma}
\end{eqnarray}

The effective potential $V(\sigma)$ (with $N$ factored out)
is defined by
\begin{equation}
    V(\sigma)=-\frac{\Gamma(\sigma)}{\int\sqrt{-g}d^{\sD}x}\, ,
\label{def:v0}
\end{equation}
where we put $\sigma_{c}(x)=\sigma$, a constant
independent of the coordinate $x$.
The effective potential gives the energy density under 
the constant background $\sigma$. 
The vacuum state should minimize the effective potential.
Thus the effective potential is useful to determine the vacuum
state in a static and homogeneous universe where
the assumption $\sigma_{c}(x)=\sigma$, a constant,
is adopted.
According to the Schwinger proper time method \cite{SP}
the second term of the right-hand side of
Eq. (\ref{w0:gamma}) for constant $\sigma$ reads
\begin{eqnarray}
     &&\mbox{ln Det}(i\gamma^{\mu}\nabla_{\mu}-\sigma) \nonumber \\
     &&=\mbox{tr} \int d^{\sD}x \mbox{ln}(i\gamma^{\mu}\nabla_{\mu}-\sigma)\\
     &&=-\mbox{tr} \int d^{\sD}x \sqrt{-g}\int_{0}^{\sigma}ds S(x,x;s)
        +\mbox{const}\, ,
     \nonumber
\label{lndet}
\end{eqnarray}
where const is a constant number which is independent
of the field $\sigma$ and $S(x,x;s)$ is the spinor two-point
function which satisfies the Dirac equation
\begin{equation}
     (i\gamma^{\mu}\nabla_{\mu}-s)S(x,y;s)
     =\frac{1}{\sqrt{-g}}\delta^{\sD}(x,y)\, .
\end{equation}
Thus the effective potential (\ref{def:v0}) reads
\begin{equation}
     V(\sigma ) = \frac{1}{2\lambda_0}\sigma^{2}
                  -i\mbox{tr} \int_{0}^{\sigma}ds S(x,x;s)
                  +O\left(\frac{1}{N}\right)\, .
\label{v:gn}
\end{equation}
It should be noted that the effective potential is normalized so that
$V(0)=0$.
From Eq. (\ref{v:gn}) we recognize that
the effective potential is described by
the two-point function $S(x,x;s)$ of the free 
fermion with mass $s$.

We turn our attention to the theory governed by the action
\begin{eqnarray}
     S&=&\int\! \sqrt{-g} d^{\sD}x \biggl\{
     \sum^{N}_{k=1}\bar{\psi}_{k}i\gamma^{\mu}\nabla_{\mu}\psi_{k}
      \nonumber \\
     &&+\frac{\lambda_0}{2N}
      \biggl[
      \left(\sum^{N}_{k=1}\bar{\psi}_{k}\psi_{k}\right)^{2}
     +\left(\sum^{N}_{k=1}\bar{\psi}_{k}i\gamma_{5}\psi_{k}\right)^{2}
     \biggr]\biggr\}\, .
\label{ac:njl}
\end{eqnarray}
In four dimensions the action (\ref{ac:njl}) describes the 
NJL (Nambu-Jona-Lasinio) model.\cite{NJL}
The action is invariant under the chiral $U(1)$
transformation in even dimensions,
\begin{equation}
     \psi \longrightarrow e^{i\theta\gamma_{5}}\psi\, .
\end{equation}
The chiral $U(1)$ symmetry prevents the action from
having mass terms. Using the auxiliary field method
and applying the similar method mentioned above,
we obtain the effective potential for the NJL model:
\begin{equation}
     V(\sigma ) = \frac{1}{2\lambda_0}{\sigma'}^{2}
                  -i\mbox{tr} \int_{0}^{\sigma'}ds S(x,x;s)
                  +O\left(\frac{1}{N}\right)\, ,
\label{v:njl}
\end{equation}
\begin{equation}
     \sigma'=\sqrt{\sigma^{2}+\pi^{2}}\, .
\end{equation}
We need, corresponding to the two four-fermion interaction
terms in the action (\ref{ac:njl}), two kinds of auxiliary
fields $\sigma$ and $\pi$.
If $\sigma'$ develops the non-vanishing vacuum
expectation value the chiral $U(1)$ symmetry is broken down
and then Nambu-Goldstone (NG) mode appears \cite{NG}
(i.e., $\pi$ becomes massless\footnote{
In Gross-Neveu type model the symmetry under the
transformation (\ref{t:dischi})
is not a continuous symmetry.
Thus no NG boson appears.
Our analysis is valid for Gross-Neveu type model
in two dimensions.
}).
In two dimensions it is well-known that the $\pi$ loop
has an infrared divergence and we cannot neglect the
next-to-leading order terms of the $1/N$ 
expansion.\cite{2DNJL}
Therefore our present analysis is not sufficient
for dealing with the action  (\ref{ac:njl})
in two dimensions.

The effective potentials (\ref{v:gn}) and (\ref{v:njl})
have the same form in the leading order of the $1/N$
expansion.
Thus, using the effective potential (\ref{v:gn}),
we can discuss the phase structure of both models
except in two dimensions.

\subsection{Effective potential in Minkowski spacetime}

An instructive example to study the phase structure
of the four-fermion model may be found in an
infinite volume flat
spacetime (i.e., Minkowski spacetime).
In the present subsection we calculate the effective potential 
$V(\sigma)$ and show the phase structure of the Gross-Neveu
type model in Minkowski spacetime.

The two-point function $S(x,x;s)$
in Minkowski spacetime is given by
\begin{equation}
S(x,x;s)=\int\frac{d^{\sD}k}{(2\pi)^{\sD}}\frac{1}{\ks -s}\, , 
\label{tpf:mink}
\end{equation}
where $\sqrt{-g}=1$.
Inserting Eq. (\ref{tpf:mink}) into Eq. (\ref{v:gn}) and performing 
the integration, we obtain the effective potential in Minkowski spacetime,
\begin{equation}
     V_{0}(\sigma) = \frac{1}{2\lambda_0}\sigma^{2}
                 -\frac{\tr\11}{(4\pi)^{\sD/2}D}
                  \Gamma \left( 1-\frac{D}{2} \right)\sigma^{\sD} \, ,
\label{v:nonren}
\end{equation}
where $\tr\11$ is the trace of the unit Dirac matrix,
the suffix $0$ for $V_0(\sigma)$ is introduced to keep
the memory that $R=0$ and $L\rightarrow\infty$ with $R$
the spacetime curvature and $L$ the size of a space
or time component.

The effective potential (\ref{v:nonren}) is divergent in two
and four dimensions. It happens to be finite in three dimensions 
in the leading order of the $1/N$ expansion.
As is well-known, the four-fermion model is renormalizable in
the two-dimensional Minkowski spacetime.  Therefore the potential
(\ref{v:nonren}) is made finite at $D=2$ by the usual
renormalization procedure. 
For $D=3$, four-fermion model is known to be renormalizable
in the sense of the $1/N$ expansion.\cite{3DFF}
The effective potential (\ref{v:nonren})
happens to be finite in three dimensions in the leading order
of the $1/N$ expansion.\footnote{If we perform the momentum
integration by using the
cut-off regularization, linear divergence appears in three
dimensions. The divergence is dropped in Eq. (\ref{v:nonren})
by the analytic continuation of $\Gamma$ function.
After renormalization both regularization methods give 
the same results.}
In four dimensions the four-fermion 
model is not renormalizable and the finite effective potential 
cannot be defined. In the present section we regard the effective 
potential for $D=4-\epsilon$ with $\epsilon$ sufficiently 
small positive as a regularization of the one in four dimensions
and consider the theory for $D=4-\epsilon$ as a low energy
effective theory stemming from more fundamental theories.

We introduce the renormalization procedure by imposing
the renormalization condition,
\begin{equation}
     \left. 
     \frac{\partial^{2}V_{0}(\sigma)}{\partial \sigma^{2}}
     \right|_{\sigma = \mu}
     =\frac{\mu^{\sD-2}}{\lambda}\, ,
\label{cond:ren}
\end{equation}
where $\mu$ is the renormalization scale.  From
the renormalization condition (\ref{cond:ren}) 
we get the renormalized
coupling $\lambda$,
\begin{equation}
     \frac{1}{\lambda_0}=\frac{\mu^{\sD-2}}{\lambda}
                       +\frac{\tr\11}{(4\pi)^{\sD/2}}(D-1)
                        \Gamma \left( 1-\frac{D}{2} \right)
                        \mu^{\sD-2}\, .
\label{eqn:ren}
\end{equation}
Replacing the bare coupling constant $\lambda_{0}$ with the renormalized
one $\lambda$ we obtain the renormalized effective potential
\begin{eqnarray}
     V_{0}(\sigma)
     & = &  \displaystyle \frac{1}{2\lambda}\sigma^{2}\mu^{\sD-2}
            +\frac{\tr \11}{2(4\pi)^{\sD/2}}(D-1)
             \Gamma \left( 1-\frac{D}{2}\right)
             \sigma^{2}\mu^{\sD-2}
\nonumber \\
     &   &  -\frac{\tr \11}{(4\pi)^{\sD/2}D}
             \Gamma \left( 1-\frac{D}{2} \right)\sigma^{\sD}\, .
\label{v:ren}
\end{eqnarray}
In Minkowski spacetime the renormalized effective potential
$V_{0}(\sigma)$ is no longer divergent in the whole range 
of the spacetime dimensions considered here, i.e. for $2 \leq D < 4$. 

We consider the two-, three- and four-dimensional limit
of the effective potential (\ref{v:ren}).
Taking the two dimensional limit, $D\rightarrow 2$, we get
\begin{equation}
       \frac{V_{0}^{D=2}(\sigma)}{\mu^{2}}=
                   \frac{1}{2 \lambda}\left(\frac{\sigma}{\mu}\right)^{2}
                  +\frac{\tr\11}{8 \pi}\left[-3+\ln\left(\frac{\sigma}{\mu}
                   \right)^{2}
                   \right]\left(\frac{\sigma}{\mu}\right)^{2}\, .
\label{v:2d}
\end{equation}
Taking the three dimensional limit $D\rightarrow 3$, we find
\begin{equation}
     \frac{V_{0}^{D=3}(\sigma)}{\mu^{3}}=
                   \frac{1}{2 \lambda}\left(\frac{\sigma}{\mu}\right)^{2}
                  -\frac{\tr\11}{4\pi}
                   \left[\left(\frac{\sigma}{\mu}\right)^{2}
                  -\frac{1}{3}\left(\frac{\sigma}{\mu}\right)^{3}\right]
     \, .
\label{v:3d}
\end{equation}
Equations (\ref{v:2d}) and (\ref{v:3d}) are not divergent for
finite $\sigma$ confirming that
the divergent part in the effective potential
(\ref{v:nonren}) is removed by the renormalization procedure
for $2 \leq D < 4$.

If we take the four-dimensional limit $D\rightarrow 4$, the effective
potential (\ref{v:ren}) reads
\begin{eqnarray}
     \frac{V_{0}^{D=4}(\sigma)}{\mu^{\sD}}&=&
                   \frac{1}{2 \lambda}\left(\frac{\sigma}{\mu}\right)^{2}
     -\frac{\tr \11}{4(4 \pi)^{2}}
                   \left\{6\left(C_{\mbox{div}}-\frac{2}{3}\right)
                   \left(\frac{\sigma}{\mu}\right)^{2}
     \right.\nonumber \\
     &&\left.      -\left[C_{\mbox{div}}+\frac{1}{2}
                   -\ln\left(\frac{\sigma}{\mu}\right)^{2}
                   \right]\left(\frac{\sigma}{\mu}\right)^{4}\right\}\, ,
\label{v:4d}
\end{eqnarray}
where we express the divergent part by
\begin{equation}
     C_{\mbox{div}}=\frac{2}{4-D}-\gamma+\ln 4\pi +1\, .
\label{div:c}
\end{equation}

On the other hand we can calculate the effective potential
for $D=4$ by using the cut-off regularization.
Inserting Eq. (\ref{tpf:mink}) into Eq. (\ref{v:gn}) at $D=4$ 
we obtain the effective potential for $D=4$, 
\begin{equation}
     V_{0}^{D=4}(\sigma) = \frac{1}{2\lambda_0}\sigma^{2}
                 -i\tr \int^{\sigma}_{0}ds
                 \int\frac{d^{4}k}{(2\pi)^{4}}\frac{1}{\ks -s}\, .
\label{v:nonren:cut}
\end{equation}
We cut off the higher momentum region of the divergent integral
in Eq. (\ref{v:nonren:cut}) and get
\begin{equation}
     V_{0}^{D=4}(\sigma) = \frac{1}{2\lambda_0}\sigma^{2}
                 -\frac{\tr\11}{4(4\pi)^{2}}
                 \left[
                 2\sigma^{2}\Lambda^{2}-\frac{1}{2}\sigma^{4}
                 -\sigma^{4}
                 \ln\left(\frac{\Lambda}{\sigma}\right)^{2}
                 \right]
                 +O\left(\frac{\sigma^{2}}{\Lambda^{2}}\right)\, ,
\label{v:nonren:cut2}
\end{equation}
where $\Lambda$ is a cut-off scale of the divergent integral.
We apply the same renormalization condition\footnote{
Since the four-fermion model is not renormalizable in four dimensions,
we do not apply the renormalization in four dimensions.
Note here that the direct comparison of the dimensional
reguralization with the cut-off reguralization is possible only 
after renormalizing the coupling constant $\lambda$ under the common
renormalization condition.
} 
as shown 
in Eq. (\ref{cond:ren}) and obtain the renormalized coupling $\lambda$:
\begin{equation}
    \frac{1}{\lambda_{0}}=\frac{\mu^{2}}{\lambda}
    +\frac{\tr\11}{(4\pi)^{2}}
    \left[
    \Lambda^{2}+2\mu^{2}-3\mu^{2}
    \ln\left(\frac{\Lambda}{\mu}\right)^{2}
    \right]\, .
\label{ren:cut}
\end{equation}
Substituting Eq. (\ref{ren:cut}) into Eq. (\ref{v:nonren:cut2})
we obtain the renormalized effective potential for $D=4$
\begin{eqnarray}
     \frac{V_{0}^{D=4}(\sigma)}{\mu^{4}}&=&
     \frac{1}{2 \lambda}\left(\frac{\sigma}{\mu}\right)^{2}
     -\frac{\tr \11}{4(4 \pi)^{2}}
                   \left\{6\left[\ln\left(\frac{\Lambda}{\mu}\right)^{2}
                          -\frac{2}{3}\right]
                   \left(\frac{\sigma}{\mu}\right)^{2}
                   \right.\nonumber \\
                   &&\left.-\left[\frac{1}{2}
                   -\ln\left(\frac{\sigma}{\Lambda}\right)^{2}
                   \right]\left(\frac{\sigma}{\mu}\right)^{4}\right\}\, .
\label{v:d4:cut}
\end{eqnarray}
We find that there is a correspondence \cite{IKM} between 
Eq. (\ref{v:4d}) and Eq. (\ref{v:d4:cut}) if we 
make a replacement
\begin{equation}
     C_{\mbox{div}} 
     \leftrightarrow \ln\left(\frac{\Lambda}{\mu}\right)^{2}\, .
\label{corr:flat}
\end{equation}
This correspondence rule is to be modified 
in curved spacetime (see \S 3.2).

\subsection{Phase structure in Minkowski spacetime}

In terms of the effective potential $V_{0}(\sigma)$ we can argue
the phase structure of the four-fermion model.
Using Eq. (\ref{v:ren}) the effective potential is calculated 
numerically.
In Fig. 2 we present the typical behavior of the effective potential
evaluated at $D=2.5$.
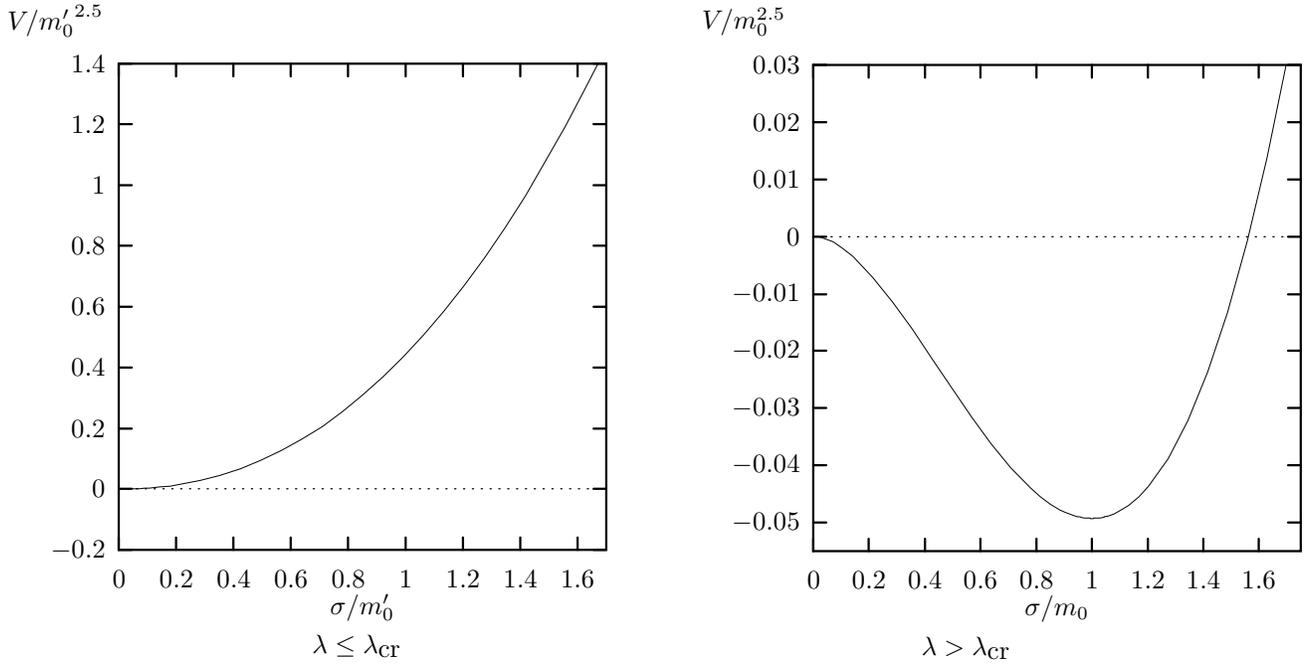
\begin{figure}
\vglue 2ex
    \begin{minipage}{.49\linewidth}
    \begin{center}
\hspace*{-4em}
\setlength{\unitlength}{0.240900pt}
\begin{picture}(1049,900)(0,0)
\tenrm
\thicklines \path(220,113)(240,113)
\thicklines \path(985,113)(965,113)
\put(198,113){\makebox(0,0)[r]{$-0.2$}}
\thicklines \path(220,209)(240,209)
\thicklines \path(985,209)(965,209)
\put(198,209){\makebox(0,0)[r]{0}}
\thicklines \path(220,304)(240,304)
\thicklines \path(985,304)(965,304)
\put(198,304){\makebox(0,0)[r]{0.2}}
\thicklines \path(220,400)(240,400)
\thicklines \path(985,400)(965,400)
\put(198,400){\makebox(0,0)[r]{0.4}}
\thicklines \path(220,495)(240,495)
\thicklines \path(985,495)(965,495)
\put(198,495){\makebox(0,0)[r]{0.6}}
\thicklines \path(220,591)(240,591)
\thicklines \path(985,591)(965,591)
\put(198,591){\makebox(0,0)[r]{0.8}}
\thicklines \path(220,686)(240,686)
\thicklines \path(985,686)(965,686)
\put(198,686){\makebox(0,0)[r]{1}}
\thicklines \path(220,782)(240,782)
\thicklines \path(985,782)(965,782)
\put(198,782){\makebox(0,0)[r]{1.2}}
\thicklines \path(220,877)(240,877)
\thicklines \path(985,877)(965,877)
\put(198,877){\makebox(0,0)[r]{1.4}}
\thicklines \path(220,113)(220,133)
\thicklines \path(220,877)(220,857)
\put(220,68){\makebox(0,0){0}}
\thicklines \path(310,113)(310,133)
\thicklines \path(310,877)(310,857)
\put(310,68){\makebox(0,0){0.2}}
\thicklines \path(400,113)(400,133)
\thicklines \path(400,877)(400,857)
\put(400,68){\makebox(0,0){0.4}}
\thicklines \path(490,113)(490,133)
\thicklines \path(490,877)(490,857)
\put(490,68){\makebox(0,0){0.6}}
\thicklines \path(580,113)(580,133)
\thicklines \path(580,877)(580,857)
\put(580,68){\makebox(0,0){0.8}}
\thicklines \path(670,113)(670,133)
\thicklines \path(670,877)(670,857)
\put(670,68){\makebox(0,0){1}}
\thicklines \path(760,113)(760,133)
\thicklines \path(760,877)(760,857)
\put(760,68){\makebox(0,0){1.2}}
\thicklines \path(850,113)(850,133)
\thicklines \path(850,877)(850,857)
\put(850,68){\makebox(0,0){1.4}}
\thicklines \path(940,113)(940,133)
\thicklines \path(940,877)(940,857)
\put(940,68){\makebox(0,0){1.6}}
\thicklines \path(220,113)(985,113)(985,877)(220,877)(220,113)
\put(45,945){\makebox(0,0)[l]{\shortstack{$V/{m'_{0}}^{2.5}$}}}
\put(602,23){\makebox(0,0){$\sigma/m'_{0}$}}
\thinlines \path(220,209)(220,209)(223,209)(224,209)(225,209)(226,209)(228,209)(230,209)(232,209)(236,209)(240,209)(244,209)(252,209)(260,210)(268,210)(284,212)(300,213)(316,216)(347,222)(379,230)(411,241)(443,254)(475,269)(507,287)(539,307)(571,330)(602,356)(634,385)(666,416)(698,451)(730,488)(762,529)(794,572)(826,619)(858,669)(889,723)(921,779)(953,840)(972,877)
\dottedline{14}(220,209)(985,209)
\end{picture}

        $\lambda \leq \lambda_{\mbox{cr}}$
    \end{center}
    \end{minipage}
\hfill
    \begin{minipage}{.49\linewidth}
    \begin{center}
\setlength{\unitlength}{0.240900pt}
\begin{picture}(1049,900)(0,0)
\tenrm
\thicklines \path(220,158)(240,158)
\thicklines \path(985,158)(965,158)
\put(198,158){\makebox(0,0)[r]{$-0.05$}}
\thicklines \path(220,248)(240,248)
\thicklines \path(985,248)(965,248)
\put(198,248){\makebox(0,0)[r]{$-0.04$}}
\thicklines \path(220,338)(240,338)
\thicklines \path(985,338)(965,338)
\put(198,338){\makebox(0,0)[r]{$-0.03$}}
\thicklines \path(220,428)(240,428)
\thicklines \path(985,428)(965,428)
\put(198,428){\makebox(0,0)[r]{$-0.02$}}
\thicklines \path(220,517)(240,517)
\thicklines \path(985,517)(965,517)
\put(198,517){\makebox(0,0)[r]{$-0.01$}}
\thicklines \path(220,607)(240,607)
\thicklines \path(985,607)(965,607)
\put(198,607){\makebox(0,0)[r]{0}}
\thicklines \path(220,697)(240,697)
\thicklines \path(985,697)(965,697)
\put(198,697){\makebox(0,0)[r]{0.01}}
\thicklines \path(220,787)(240,787)
\thicklines \path(985,787)(965,787)
\put(198,787){\makebox(0,0)[r]{0.02}}
\thicklines \path(220,877)(240,877)
\thicklines \path(985,877)(965,877)
\put(198,877){\makebox(0,0)[r]{0.03}}
\thicklines \path(220,113)(220,133)
\thicklines \path(220,877)(220,857)
\put(220,68){\makebox(0,0){0}}
\thicklines \path(307,113)(307,133)
\thicklines \path(307,877)(307,857)
\put(307,68){\makebox(0,0){0.2}}
\thicklines \path(395,113)(395,133)
\thicklines \path(395,877)(395,857)
\put(395,68){\makebox(0,0){0.4}}
\thicklines \path(482,113)(482,133)
\thicklines \path(482,877)(482,857)
\put(482,68){\makebox(0,0){0.6}}
\thicklines \path(570,113)(570,133)
\thicklines \path(570,877)(570,857)
\put(570,68){\makebox(0,0){0.8}}
\thicklines \path(657,113)(657,133)
\thicklines \path(657,877)(657,857)
\put(657,68){\makebox(0,0){1}}
\thicklines \path(745,113)(745,133)
\thicklines \path(745,877)(745,857)
\put(745,68){\makebox(0,0){1.2}}
\thicklines \path(832,113)(832,133)
\thicklines \path(832,877)(832,857)
\put(832,68){\makebox(0,0){1.4}}
\thicklines \path(919,113)(919,133)
\thicklines \path(919,877)(919,857)
\put(919,68){\makebox(0,0){1.6}}
\thicklines \path(220,113)(985,113)(985,877)(220,877)(220,113)
\put(45,945){\makebox(0,0)[l]{\shortstack{$V/m_{0}^{2.5}$}}}
\put(602,23){\makebox(0,0){$\sigma/m_{0}$}}
\thinlines \path(220,607)(220,607)(223,607)(224,607)(226,607)(228,607)(232,606)(235,605)(243,602)(251,599)(266,589)(282,576)(313,544)(344,505)(375,462)(406,416)(437,370)(468,324)(499,282)(530,244)(561,212)(576,199)(592,187)(607,178)(615,174)(623,171)(630,169)(634,167)(638,167)(642,166)(646,165)(648,165)(650,165)(652,165)(653,165)(654,165)(654,165)(655,164)(656,164)(657,164)(658,164)(659,164)(660,165)(661,165)(662,165)(663,165)(665,165)(667,165)(669,165)(673,166)(677,167)
\thinlines \path(677,167)(684,169)(692,172)(700,176)(715,185)(731,199)(746,215)(777,259)(808,319)(839,395)(870,488)(901,600)(932,731)(962,877)
\dottedline{14}(220,607)(985,607)
\end{picture}

        $\lambda > \lambda_{\mbox{cr}}$
    \end{center}
    \end{minipage}
\caption{Behavior of the effective potential in Minkowski spacetime
         for $D=2.5$.}
\end{figure}
In drawing Fig. 2 we introduce, for convenience, 
the new variables $m_{0}$
which will be defined by Eq. (\ref{def:mass}) and
$m'_{0}$ defined by 
\begin{equation}
   m'_{0}=
   \mu\left[
       -\frac{(4\pi)^{\sD/2}}{\tr\11\Gamma(1-D/2)}\frac{1}{\lambda}
       -D+1
   \right]^{1/(D-2)}\, .
\label{m0p}
\end{equation}
As is shown in Fig. 2 the shape of the effective potential 
is of a single and a double well for sufficiently small 
and large coupling constant $\lambda$ respectively. 
Hence we expect that the effective potential
changes its shape
at a critical value of the coupling constant $\lambda_{\mbox{cr}}$. 
We can see in Fig. 2 that the ground state is invariant 
under the discrete chiral transformation for a sufficiently 
small coupling constant. 
On the other hand for $\lambda > \lambda_{\mbox{cr}}$
the minimum of the effective potential
is located at non-vanishing $\sigma$.
In the latter case the auxiliary field $\sigma$ develops
the non-vanishing vacuum expectation value
and the discrete chiral symmetry is broken down dynamically.
As the result the fermion acquires the dynamical mass.

Next we calculate the dynamically generated fermion mass
for $\lambda > \lambda_{\mbox{cr}}$.
It is equal to the vacuum expectation value of $\sigma$
which is obtained by observing the minimum of the effective potential.
The necessary condition for the minimum is given by the gap
equation:
\begin{equation}
     \left. 
     \frac{\partial V_{0}(\sigma)}{\partial \sigma}
     \right|_{\sigma = m_{0}} =m_{0}f(m_{0},\lambda)=0\, ,
\label{dmass}
\end{equation}
where $f(m_{0},\lambda)$ is given by
\begin{eqnarray}
     f(m_{0},\lambda)&=&
     \frac{1}{\lambda}\mu^{D-2}+\frac{\tr \11}{(4\pi)^{D/2}}
     (D-1)\Gamma\left(1-\frac{D}{2}\right)\mu^{D-2}\nonumber \\
     &&-\frac{\tr \11}{(4\pi)^{D/2}}
     \Gamma\left(1-\frac{D}{2}\right)
     m_{0}^{D-2}.
\end{eqnarray}
The dynamical fermion mass is obtained by the non-trivial
solution of the gap equation (\ref{dmass}).
Differentiating the effective potential (\ref{v:ren})
and solving the equation $f(m_{0},\lambda)=0$
for $m_{0}$, we find the dynamical
fermion mass:
\begin{equation}
   m_{0}=\mu
   \left[
       \frac{(4\pi)^{\sD/2}}{\tr\11\Gamma(1-D/2)}\frac{1}{\lambda}
       +D-1
   \right]^{1/(D-2)}\, .
\label{def:mass}
\end{equation}
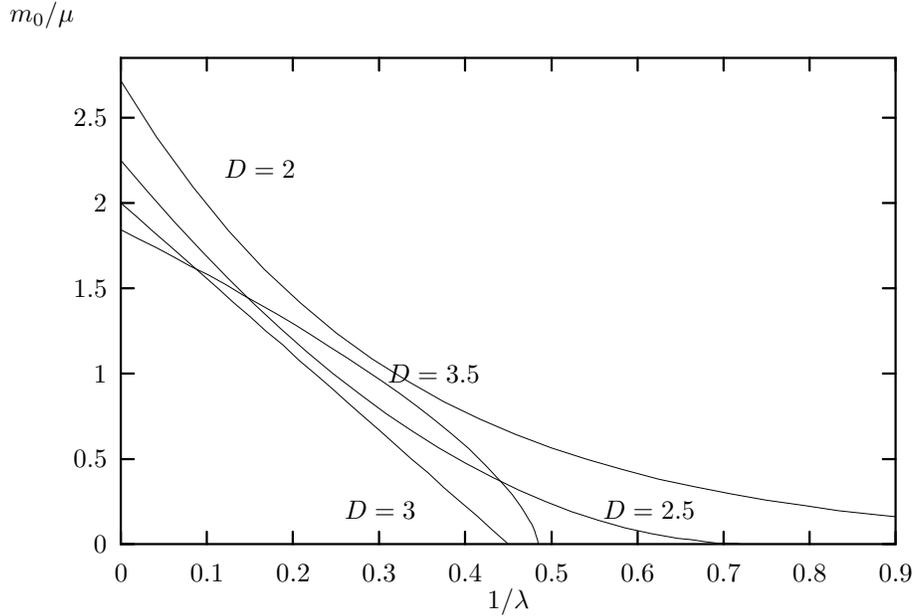
\begin{figure}
\vglue 2ex
    \begin{center}
\setlength{\unitlength}{0.240900pt}
\begin{picture}(1500,900)(0,0)
\tenrm
\thicklines \path(220,113)(240,113)
\thicklines \path(1436,113)(1416,113)
\put(198,113){\makebox(0,0)[r]{0}}
\thicklines \path(220,247)(240,247)
\thicklines \path(1436,247)(1416,247)
\put(198,247){\makebox(0,0)[r]{0.5}}
\thicklines \path(220,381)(240,381)
\thicklines \path(1436,381)(1416,381)
\put(198,381){\makebox(0,0)[r]{1}}
\thicklines \path(220,515)(240,515)
\thicklines \path(1436,515)(1416,515)
\put(198,515){\makebox(0,0)[r]{1.5}}
\thicklines \path(220,649)(240,649)
\thicklines \path(1436,649)(1416,649)
\put(198,649){\makebox(0,0)[r]{2}}
\thicklines \path(220,783)(240,783)
\thicklines \path(1436,783)(1416,783)
\put(198,783){\makebox(0,0)[r]{2.5}}
\thicklines \path(220,113)(220,133)
\thicklines \path(220,877)(220,857)
\put(220,68){\makebox(0,0){0}}
\thicklines \path(355,113)(355,133)
\thicklines \path(355,877)(355,857)
\put(355,68){\makebox(0,0){0.1}}
\thicklines \path(490,113)(490,133)
\thicklines \path(490,877)(490,857)
\put(490,68){\makebox(0,0){0.2}}
\thicklines \path(625,113)(625,133)
\thicklines \path(625,877)(625,857)
\put(625,68){\makebox(0,0){0.3}}
\thicklines \path(760,113)(760,133)
\thicklines \path(760,877)(760,857)
\put(760,68){\makebox(0,0){0.4}}
\thicklines \path(896,113)(896,133)
\thicklines \path(896,877)(896,857)
\put(896,68){\makebox(0,0){0.5}}
\thicklines \path(1031,113)(1031,133)
\thicklines \path(1031,877)(1031,857)
\put(1031,68){\makebox(0,0){0.6}}
\thicklines \path(1166,113)(1166,133)
\thicklines \path(1166,877)(1166,857)
\put(1166,68){\makebox(0,0){0.7}}
\thicklines \path(1301,113)(1301,133)
\thicklines \path(1301,877)(1301,857)
\put(1301,68){\makebox(0,0){0.8}}
\thicklines \path(1436,113)(1436,133)
\thicklines \path(1436,877)(1436,857)
\put(1436,68){\makebox(0,0){0.9}}
\thicklines \path(220,113)(1436,113)(1436,877)(220,877)(220,113)
\put(45,945){\makebox(0,0)[l]{\shortstack{$m_{0}/\mu$}}}
\put(828,23){\makebox(0,0){$1/\lambda$}}
\put(382,703){\makebox(0,0)[l]{$D=2$}}
\put(977,167){\makebox(0,0)[l]{$D=2.5$}}
\put(571,167){\makebox(0,0)[l]{$D=3$}}
\put(639,381){\makebox(0,0)[l]{$D=3.5$}}
\thinlines \path(220,841)(220,841)(276,752)(333,674)(389,605)(445,545)(502,492)(558,445)(614,404)(670,369)(727,337)(783,310)(839,286)(896,264)(952,246)(1008,230)(1064,215)(1121,203)(1177,192)(1233,182)(1290,174)(1346,166)(1402,160)(1436,156)
\thinlines \path(220,716)(220,716)(262,667)(303,620)(345,575)(387,532)(428,491)(470,452)(511,416)(553,381)(595,349)(636,318)(678,290)(719,264)(761,240)(803,218)(844,198)(886,180)(927,164)(969,151)(1011,139)(1052,130)(1094,122)(1115,120)(1135,117)(1156,115)(1167,115)(1177,114)(1187,114)(1193,113)(1198,113)(1203,113)(1206,113)(1208,113)(1211,113)(1212,113)(1213,113)(1215,113)(1216,113)(1219,113)
\thinlines \path(220,649)(220,649)(245,627)(271,604)(296,582)(321,560)(347,537)(372,515)(397,493)(423,470)(448,448)(474,426)(499,403)(524,381)(550,359)(575,336)(600,314)(626,292)(651,269)(676,247)(702,225)(727,202)(752,180)(778,158)(803,135)(828,113)
\thinlines \path(220,607)(220,607)(247,593)(275,579)(302,565)(329,550)(357,536)(384,521)(411,505)(438,490)(466,474)(493,458)(520,441)(547,424)(575,406)(602,388)(629,370)(657,350)(684,330)(711,309)(738,287)(766,263)(793,236)(820,207)(834,191)(848,172)(854,162)(861,150)(865,144)(868,137)(870,132)(871,128)(872,125)(873,122)(874,119)(875,113)
\end{picture}

    \end{center}
\caption{Dynamically generated fermion mass for $D=2,2.5,3,3.5$.}
\end{figure}

In Fig. 3 we plot the dynamical fermion mass $m_{0}$ as a function
of the coupling constant $\lambda$ for $D=2,2.5,3,3.5$.
The dynamical fermion mass decreases as $1/\lambda$ increases,
so that the dynamical fermion mass disappears at a critical
value of $\lambda$.
Since no mass gap is observed at the critical point
in Fig. 3, the phase transition at the critical
coupling $\lambda_{\mbox{cr}}$ is of the second order.
In two dimensions the dynamical fermion mass
decreases exponentially as $1/\lambda$ increases.
Thus the critical coupling constant $\lambda_{\mbox{cr}}=0$ 
for $D=2$.

For second order phase transition
the critical coupling constant $\lambda_{\mbox{cr}}$ is obtained by
taking the limit $m_{0}\rightarrow 0$:
\begin{equation}
   \lim_{m_{0}\rightarrow 0} f(m_{0},\lambda_{\mbox{cr}})=0\, ,
\label{crp:sec}
\end{equation}
where $f(m_{0},\lambda_{\mbox{cr}})$ is defined in
Eq. (\ref{dmass}).
Solving Eq. (\ref{crp:sec}) we easily find the critical coupling
constant
\begin{equation}
     \lambda_{cr} = \frac{(4\pi)^{\sD/2}}{\tr\11}
                        \left[
                        (1-D)\Gamma \left( 1-\frac{D}{2} \right)
                        \right]^{-1}\, .
\label{cr:l:d}
\end{equation}
For $\lambda > \lambda_{\mbox{cr}}$ the dynamical fermion mass is
generated and the discrete chiral symmetry is broken down.
In Fig. 4 we plot the critical coupling constant $\lambda_{\mbox{cr}}$
as a function of the spacetime dimension $D$.
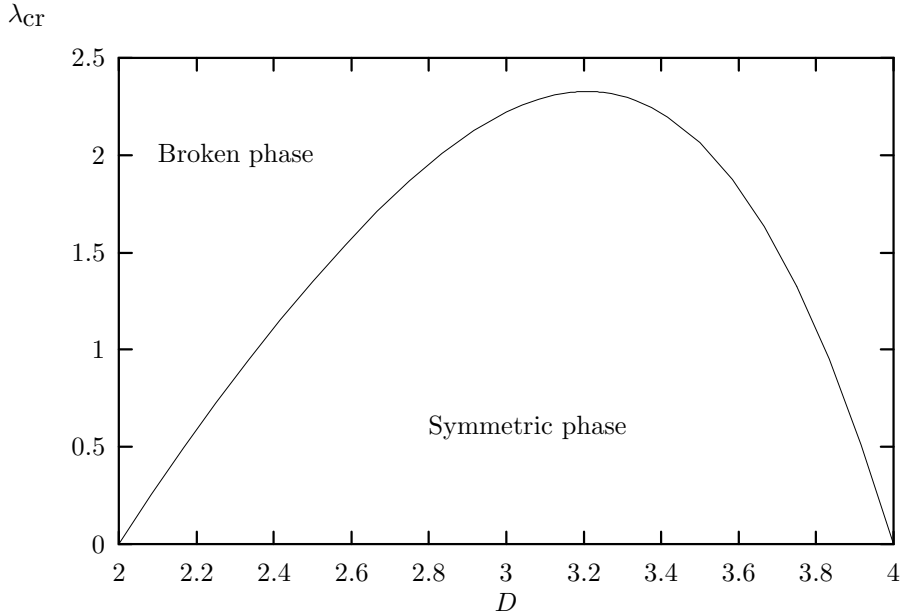
\begin{figure}
\vglue 2ex
    \begin{center}
\setlength{\unitlength}{0.240900pt}
\begin{picture}(1500,900)(0,0)
\tenrm
\thicklines \path(220,113)(240,113)
\thicklines \path(1436,113)(1416,113)
\put(198,113){\makebox(0,0)[r]{0}}
\thicklines \path(220,266)(240,266)
\thicklines \path(1436,266)(1416,266)
\put(198,266){\makebox(0,0)[r]{0.5}}
\thicklines \path(220,419)(240,419)
\thicklines \path(1436,419)(1416,419)
\put(198,419){\makebox(0,0)[r]{1}}
\thicklines \path(220,571)(240,571)
\thicklines \path(1436,571)(1416,571)
\put(198,571){\makebox(0,0)[r]{1.5}}
\thicklines \path(220,724)(240,724)
\thicklines \path(1436,724)(1416,724)
\put(198,724){\makebox(0,0)[r]{2}}
\thicklines \path(220,877)(240,877)
\thicklines \path(1436,877)(1416,877)
\put(198,877){\makebox(0,0)[r]{2.5}}
\thicklines \path(220,113)(220,133)
\thicklines \path(220,877)(220,857)
\put(220,68){\makebox(0,0){2}}
\thicklines \path(342,113)(342,133)
\thicklines \path(342,877)(342,857)
\put(342,68){\makebox(0,0){2.2}}
\thicklines \path(463,113)(463,133)
\thicklines \path(463,877)(463,857)
\put(463,68){\makebox(0,0){2.4}}
\thicklines \path(585,113)(585,133)
\thicklines \path(585,877)(585,857)
\put(585,68){\makebox(0,0){2.6}}
\thicklines \path(706,113)(706,133)
\thicklines \path(706,877)(706,857)
\put(706,68){\makebox(0,0){2.8}}
\thicklines \path(828,113)(828,133)
\thicklines \path(828,877)(828,857)
\put(828,68){\makebox(0,0){3}}
\thicklines \path(950,113)(950,133)
\thicklines \path(950,877)(950,857)
\put(950,68){\makebox(0,0){3.2}}
\thicklines \path(1071,113)(1071,133)
\thicklines \path(1071,877)(1071,857)
\put(1071,68){\makebox(0,0){3.4}}
\thicklines \path(1193,113)(1193,133)
\thicklines \path(1193,877)(1193,857)
\put(1193,68){\makebox(0,0){3.6}}
\thicklines \path(1314,113)(1314,133)
\thicklines \path(1314,877)(1314,857)
\put(1314,68){\makebox(0,0){3.8}}
\thicklines \path(1436,113)(1436,133)
\thicklines \path(1436,877)(1436,857)
\put(1436,68){\makebox(0,0){4}}
\thicklines \path(220,113)(1436,113)(1436,877)(220,877)(220,113)
\put(45,945){\makebox(0,0)[l]{\shortstack{$\lambda_{\mbox{cr}}$}}}
\put(828,23){\makebox(0,0){$D$}}
\put(706,296){\makebox(0,0)[l]{Symmetric phase}}
\put(281,724){\makebox(0,0)[l]{Broken phase}}
\thinlines \path(220,113)(220,113)(271,191)(321,264)(372,335)(423,402)(473,466)(524,526)(575,583)(625,636)(676,684)(727,727)(777,763)(828,792)(853,803)(879,812)(904,819)(917,821)(923,822)(929,823)(936,823)(939,824)(942,824)(944,824)(945,824)(947,824)(948,824)(950,824)(952,824)(953,824)(955,824)(956,824)(958,824)(959,824)(961,824)(964,824)(967,824)(971,824)(974,823)(980,823)(986,822)(993,821)(1005,818)(1018,815)(1031,810)(1056,799)(1081,784)(1132,744)(1183,686)(1233,612)(1284,518)
\thinlines \path(1284,518)(1335,404)(1385,269)(1436,113)
\end{picture}

    \end{center}
\caption{Critical coupling constant as a function of the spacetime
         dimension $D$.}
\end{figure}

For some specific values of $D$ the solutions $m_{0}$
and $\lambda_{\mbox{cr}}$ simplify:
\begin{equation}
\begin{array}{lll}
     m_{0}=\mu e^{1-2\pi/(\tr\11\lambda)}, & 
     \lambda_{\mbox{cr}}=0\, ; & D=2,\\
     \displaystyle m_{0}=\mu \left(2-\frac{4\pi}{\tr\11\lambda}\right), 
     & \displaystyle \lambda_{\mbox{cr}}=\frac{2\pi}{\tr\11}\, ; & D=3.
\end{array}
\end{equation}
Taking the four-dimensional limit, $D\rightarrow 4$, we find
\begin{equation}
     \left[
     C_{\mbox{div}}-\ln\left(\frac{m_{0}}{\mu}\right)^{2}
     \right]\left(\frac{m_{0}}{\mu}\right)^{2}
     =\frac{(4\pi)^{2}}{\tr\11\lambda}
     +3\left(C_{\mbox{div}}-\frac{2}{3}\right)\, ,
\end{equation}
\begin{equation}
     \frac{1}{\lambda_{\mbox{cr}}}=\frac{3\tr\11}{(4\pi)^2}
     \left(C_{\mbox{div}}-\frac{2}{3}\right)\, ,
\end{equation}
where the divergent part $C_{\mbox{div}}$ is defined 
in Eq. (\ref{div:c}).

\subsection{Renormalization group analysis in Minkowski spacetime}

We analyze the $\beta$-function for the four-fermion
coupling $\lambda$.
It has been often pointed out that the fermions acquire 
the dynamical mass for a larger coupling than the ultraviolet 
stable fixed point.\cite{GN,RGA}

The renormalization group $\beta$-function for the four-fermion coupling
is defined by
\begin{equation}
     \beta(\lambda)=\left.
     \mu \frac{\partial \lambda}{\partial \mu}
     \right|_{\lambda_0}\, .
\label{def:beta}
\end{equation}
Taking into account Eq. (\ref{eqn:ren}) with Eq. (\ref{cr:l:d})
we obtain
\begin{equation}
     \beta(\lambda)=\frac{D-2}{\lambda_{\mbox{cr}}}\lambda
     (\lambda_{\mbox{cr}}-\lambda)\, ,
\label{beta:gn}
\end{equation}
in the leading order of the $1/N$ expansion.
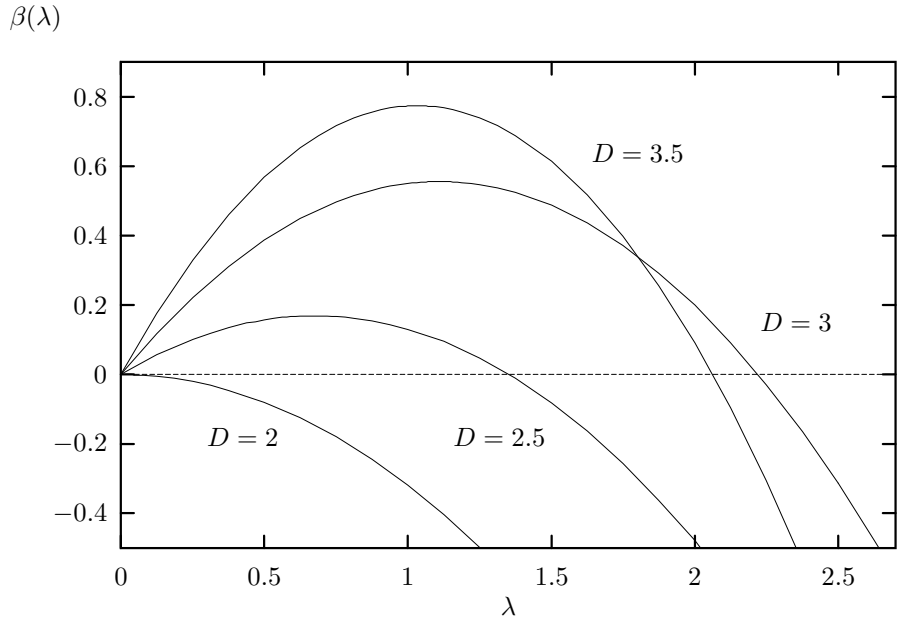
\begin{figure}
\vglue 2ex
    \begin{center}
\setlength{\unitlength}{0.240900pt}
\begin{picture}(1500,900)(0,0)
\tenrm
\thicklines \path(220,168)(240,168)
\thicklines \path(1436,168)(1416,168)
\put(198,168){\makebox(0,0)[r]{$-0.4$}}
\thicklines \path(220,277)(240,277)
\thicklines \path(1436,277)(1416,277)
\put(198,277){\makebox(0,0)[r]{$-0.2$}}
\thicklines \path(220,386)(240,386)
\thicklines \path(1436,386)(1416,386)
\put(198,386){\makebox(0,0)[r]{0}}
\thicklines \path(220,495)(240,495)
\thicklines \path(1436,495)(1416,495)
\put(198,495){\makebox(0,0)[r]{0.2}}
\thicklines \path(220,604)(240,604)
\thicklines \path(1436,604)(1416,604)
\put(198,604){\makebox(0,0)[r]{0.4}}
\thicklines \path(220,713)(240,713)
\thicklines \path(1436,713)(1416,713)
\put(198,713){\makebox(0,0)[r]{0.6}}
\thicklines \path(220,822)(240,822)
\thicklines \path(1436,822)(1416,822)
\put(198,822){\makebox(0,0)[r]{0.8}}
\thicklines \path(220,113)(220,133)
\thicklines \path(220,877)(220,857)
\put(220,68){\makebox(0,0){0}}
\thicklines \path(445,113)(445,133)
\thicklines \path(445,877)(445,857)
\put(445,68){\makebox(0,0){0.5}}
\thicklines \path(670,113)(670,133)
\thicklines \path(670,877)(670,857)
\put(670,68){\makebox(0,0){1}}
\thicklines \path(896,113)(896,133)
\thicklines \path(896,877)(896,857)
\put(896,68){\makebox(0,0){1.5}}
\thicklines \path(1121,113)(1121,133)
\thicklines \path(1121,877)(1121,857)
\put(1121,68){\makebox(0,0){2}}
\thicklines \path(1346,113)(1346,133)
\thicklines \path(1346,877)(1346,857)
\put(1346,68){\makebox(0,0){2.5}}
\thicklines \path(220,113)(1436,113)(1436,877)(220,877)(220,113)
\put(45,945){\makebox(0,0)[l]{\shortstack{$\beta(\lambda)$}}}
\put(828,23){\makebox(0,0){$\lambda$}}
\put(355,288){\makebox(0,0)[l]{$D=2$}}
\put(742,288){\makebox(0,0)[l]{$D=2.5$}}
\put(1224,468){\makebox(0,0)[l]{$D=3$}}
\put(959,735){\makebox(0,0)[l]{$D=3.5$}}
\thinlines \path(220,386)(220,386)(224,386)(225,386)(227,386)(231,386)(234,386)(238,386)(241,385)(248,385)(255,385)(262,384)(276,383)(290,382)(304,380)(333,375)(361,369)(389,361)(445,342)(501,318)(558,288)(614,253)(670,212)(727,166)(783,114)(784,113)
\thinlines \path(220,386)(220,386)(276,417)(333,441)(361,451)(389,460)(417,467)(431,469)(445,472)(459,474)(473,476)(480,476)(487,477)(494,477)(501,478)(505,478)(509,478)(512,478)(514,478)(516,478)(517,478)(519,478)(521,478)(523,478)(524,478)(526,478)(528,478)(530,478)(531,478)(533,478)(537,478)(540,478)(544,478)(551,477)(558,477)(572,476)(586,474)(614,470)(642,464)(670,457)(727,438)(783,412)(839,380)(896,341)(952,297)(1008,246)(1064,188)(1121,125)(1130,113)
\thinlines \path(220,386)(220,386)(276,450)(333,507)(389,556)(445,597)(501,631)(558,657)(586,667)(614,675)(628,679)(642,682)(656,684)(670,686)(677,687)(684,687)(691,688)(699,688)(702,689)(706,689)(709,689)(711,689)(713,689)(714,689)(716,689)(718,689)(720,689)(721,689)(723,689)(725,689)(727,689)(728,689)(730,689)(734,689)(737,689)(741,688)(748,688)(755,687)(769,686)(783,684)(811,679)(839,672)(896,652)(952,624)(1008,589)(1064,545)(1121,495)(1177,436)(1233,370)(1290,296)(1346,215)
\thinlines \path(1346,215)(1402,126)(1410,113)
\thinlines \path(220,386)(220,386)(276,482)(333,566)(389,637)(445,696)(501,742)(530,761)(558,777)(586,789)(600,794)(614,798)(628,802)(642,805)(649,806)(656,807)(663,807)(667,807)(670,808)(674,808)(676,808)(677,808)(679,808)(681,808)(683,808)(684,808)(686,808)(688,808)(690,808)(691,808)(695,808)(699,808)(702,807)(706,807)(713,807)(720,806)(727,805)(741,802)(755,798)(783,789)(811,777)(839,761)(896,721)(952,668)(1008,603)(1064,526)(1121,436)(1177,334)(1233,219)(1280,113)
\thinlines \drawline[-50](220,386)(220,386)(232,386)(245,386)(257,386)(269,386)(281,386)(294,386)(306,386)(318,386)(331,386)(343,386)(355,386)(367,386)(380,386)(392,386)(404,386)(417,386)(429,386)(441,386)(453,386)(466,386)(478,386)(490,386)(503,386)(515,386)(527,386)(539,386)(552,386)(564,386)(576,386)(588,386)(601,386)(613,386)(625,386)(638,386)(650,386)(662,386)(674,386)(687,386)(699,386)(711,386)(724,386)(736,386)(748,386)(760,386)(773,386)(785,386)(797,386)(810,386)(822,386)
\thinlines \drawline[-50](822,386)(834,386)(846,386)(859,386)(871,386)(883,386)(896,386)(908,386)(920,386)(932,386)(945,386)(957,386)(969,386)(982,386)(994,386)(1006,386)(1018,386)(1031,386)(1043,386)(1055,386)(1068,386)(1080,386)(1092,386)(1104,386)(1117,386)(1129,386)(1141,386)(1153,386)(1166,386)(1178,386)(1190,386)(1203,386)(1215,386)(1227,386)(1239,386)(1252,386)(1264,386)(1276,386)(1289,386)(1301,386)(1313,386)(1325,386)(1338,386)(1350,386)(1362,386)(1375,386)(1387,386)(1399,386)(1411,386)(1424,386)(1436,386)
\end{picture}

    \end{center}
\caption{Behavior of the $\beta$-function for $D=2,2.5,3,3.5$.}
\end{figure}

In Fig. 5 we show the behavior of the $\beta$-functions.
As is obviously observed in Fig. 5 the theory is asymptotically
free for $D=2$ and the theory for $2 < D < 4$ has a
nontrivial ultraviolet stable fixed point at
\begin{equation}
     \lambda=\lambda_{\mbox{cr}}\, .
\label{fixp:ultra}
\end{equation}
It simplifies for some special values of $D$:
\begin{equation}
\begin{array}{lll}
     \displaystyle
     \beta(\lambda)=-\frac{\tr\11}{2\pi}\lambda^{2}\, \, \,
     ; & D=2,\\[2mm]
     \displaystyle \beta(\lambda)
     =-\lambda\left(\frac{\tr\11}{2\pi}\lambda-1\right)\, ; & D=3.
\end{array}
\end{equation}
Taking the four-dimensional limit, $D\rightarrow 4$, we find
\begin{equation}
     \beta(\lambda)=-2\lambda\left[
     \frac{3\tr\11}{(4\pi)^{2}}
     \left(C_{\mbox{div}}-\frac{5}{3}\right)\lambda -1
     \right]\,\, ;\, D\rightarrow 4\, ,
\end{equation}
where $C_{\mbox{div}}$ is defined in Eq. (\ref{div:c}).
For a larger coupling than the ultraviolet stable
fixed point a tachyon pole appears in the $\sigma$
propagator.
Then the classical vacuum state becomes unstable.
The generation of the dynamical fermion mass helps recovering
the stability of the vacuum.

The basic properties of the four-fermion models
have been briefly reviewed.
We calculated the effective potential and discussed
the phase structure of the four-fermion model 
(Gross-Neveu type model) in Minkowski spacetime.
In the leading order of the $1/N$ expansion the effective
potential is given by the integration of the spinor
two-point function $S(x,x;s)$ over mass variable $s$.
Evaluating the effective potential in the leading
order of the $1/N$ expansion it was shown that the 
discrete chiral symmetry is broken down dynamically
for a sufficiently large coupling constant 
$\lambda > \lambda_{\mbox{cr}}$.
Dynamically generated fermion mass $m_{0}$ and the critical
point $\lambda_{\mbox{cr}}$ was obtained analytically by solving the
gap equation (\ref{dmass}).
Since no mass gap appears at the critical point,
the phase transition from the broken phase to the symmetric
phase is of the second order.
Applying the renormalization group analysis it was shown that
the critical coupling constant $\lambda_{\mbox{cr}}$ corresponds to
the ultraviolet stable fixed point.

In the present section we basically used the dimensional 
reguralization method to investigate the phase structure
for $2 \leq D < 4$ and considered the theory for
$D=4-\epsilon$ with $\epsilon$ sufficiently small positive
as a regularized version of the one in four dimensions.
As was shown the effective potential can be calculated
also by cut-off regularization.
In this case the same result is obtained with recourse to
the relation between $\epsilon$ and $\Lambda$ (\ref{corr:flat}).

In the following sections we will apply the similar method
to the case of the curved spacetime.

\section{Four-fermion models in weakly curved spacetime}

At an early stage of the universe it is generally assumed
that broken symmetries are restored and the
physical phenomena are
described by a  more fundamental theory with a higher
symmetry.
The spontaneous symmetry breaking may be induced under the 
influence of
the strong curvature, finite volume, non-trivial topology
and so on.

Here we discuss the curvature induced phase transition in 
four-fermion models. 
For this purpose the effective potential\footnote{Since 
the assumption $\sigma_{c}(x)=\sigma$
(constant) in (\ref{def:v0}) is not always acceptable 
in general curved spacetimes, it is necessary to evaluate 
the effective action instead of the effective potential to 
study the phase structure.
Here we restrict ourselves to the static homogeneous 
spacetime where the assumption is acceptable.}
of the four-fermion
models need to be calculated in a general curved spacetime.
In the present section we consider the theory in weakly 
curved spacetime.
We shall confine ourselves to the case
that the spacetime is curved very moderately and shall
neglect terms involving derivatives of the metric
higher than second order.

There are many studies of the phase transition
in weakly curved spacetime.
Spontaneous symmetry breaking of
the massless scalar theory with $\lambda \phi^{4}$ 
interaction has been studied in four dimensions 
(see for example Refs. \cite{scalar,BO2} and the references
there in).
In the scalar theory the continuous $U(1)$ symmetry 
is restored for a sufficiently large positive curvature.
The pioneering works in the four-fermion model have
been performed by Itoyama,\cite{Ito} and Buchbinder
and Kirillova\cite{BK} in two dimensions.
In Refs. \cite{WC1}$\sim$\cite{WC2} four-fermion models have been
studied in three, four and arbitrary dimensions
($3 \leq D < 4$) respectively.
In these literatures it is found that the possibility of the
curvature-induced phase transition in four-fermion models.

In this section we mainly follow Refs. \cite{WC1} and
\cite{WC2}
and discuss the
phase structure of the Gross-Neveu type model in weakly
curved spacetime by using the similar method explained
in the previous section.

\subsection{Riemann normal coordinate expansion of $S(x,y;s)$}

As is shown in the previous section the effective potential
of the Gross-Neveu type model is described by the spinor
two-point function $S(x,x;s)$ in the leading order of the $1/N$
expansion.
Thus we start with the analysis of the two-point function 
$S(x,x;s)$ in weakly curved spacetime.

For this purpose it is convenient to introduce
the Riemann normal coordinate system
which is a coordinate system with the affine connection
${\Gamma^{\alpha}}_{\mu\nu}$ which vanishes at least locally 
(i.e., locally inertial frame).
At the origin $x_{0}$ of the coordinates the metric tensor
$g_{\mu\nu}$ and
the affine connection ${\Gamma^{\alpha}}_{\mu\nu}$ have the
property
\begin{eqnarray}
     g_{\mu\nu}(x_{0})&=&\eta_{\mu\nu}\, ,\nonumber \\
     {\Gamma^{\alpha}}_{\mu\nu}(x_{0})&=&0\, ,
\end{eqnarray}
where $\eta_{\mu\nu}$ is diag$(1,-1,-1,-1)$.
Near the origin $x_{0}$ the metric tensor is expanded to be \cite{RNC}
\begin{eqnarray}
     g_{\mu\nu}(y)&=&\eta_{\mu\nu}+\frac{1}{3}R^{0}_{\mu\alpha\nu\beta}
     (y-x_{0})^{\alpha}(y-x_{0})^{\beta}+O(R_{;\mu},R^{2})\, ,\nonumber
\\
     g(y)&=&-1-\frac{1}{3}R^{0}_{\alpha\beta}
     (y-x_{0})^{\alpha}(y-x_{0})^{\beta}+O(R_{;\mu},R^{2})\, ,
\label{def:rnce}
\end{eqnarray}
where index $0$ for $R^{0}_{\mu\alpha\nu\beta}$ and $R^{0}_{\alpha\beta}$
designates the tensor at the origin $x_{0}$.

By the use of the Riemann normal coordinate expansion 
(\ref{def:rnce}) we calculate
the spinor two-point function $S(x,x;s)$ in the
approximation of slowly varying curvature where we neglect
any terms involving derivatives of the metric
higher than second order.
According to the method developed by Parker and
Toms\cite{PT}
the two-point function $S(x,x;s)$ is expanded asymptotically
around $R^{0}=0$.
We introduce the bispinor function $G(x_{0},y;s)$ defined by
\begin{equation}
     (i \gamma_{\mu}\nabla^{\mu}+s)G(x_{0},y;s)=S(x_{0},y;s)\, .
\label{def:g}
\end{equation}
It satisfies the following equation :
\begin{equation}
     \left(-\nabla^{\mu}\nabla_{\mu}-\frac{R}{4}-s^{2}\right)
     G(x_{0},y;s)
     =\frac{1}{\sqrt{-g}}\delta^{\sD}(x_{0},y)\, .
\label{eq:g}
\end{equation}

Using the Riemann normal coordinate expansion (\ref{def:rnce})
we expand the first term on the left-hand
side of Eq. (\ref{eq:g}) and find
\begin{equation}
\begin{array}{rcl}
     \displaystyle
\sqrt{-g}\nabla^{\mu}\nabla_{\mu}G(x_{0},y;s)&=& \displaystyle
     \left[\eta^{\mu\nu}\partial_{\mu}\partial_{\nu}
     +\frac{1}{6}{R^{0}}_{\alpha\beta}(y-x_{0})^{\alpha}(y-x_{0})^{\beta}
     \eta^{\mu\nu}\partial_{\mu}\partial_{\nu}\right.
\\[5mm]
     && \displaystyle
-\frac{1}{3}{{{{R^{0}}^{\mu}}_{\alpha}}^{\nu}}_{\beta}
        (y-x_{0})^{\alpha}(y-x_{0})^{\beta}\partial_{\mu}\partial_{\nu}
\\[5mm]
     && \displaystyle
-\frac{2}{3}{{R^{0}}^{\mu}}_{\alpha}(y-x_{0})^{\alpha}\partial_{\mu}
\\[5mm]
    && \displaystyle \left.
+\frac{1}{4}{{R^{0}}^{\mu}}_{\alpha a b}
       \sigma^{ab}(y-x_{0})^{\alpha}\partial_{\mu}+\cdots
     \right]G(x_{0},y;s)\, ,
\end{array}\label{eq:rne}
\end{equation}
where
\begin{equation}
     \sigma^{ab}=\frac{1}{2}[\gamma^{a},\gamma^{b}]\, ,
\end{equation}
and Latin indices $a$ and $b$ are vierbein indices.
Here we keep only terms independent of the curvature $R$ and
terms linear in $R$.
Inserting the Eq. (\ref{eq:rne}) into Eq. (\ref{eq:g}) and
performing the following Fourier transformation,
\begin{equation}
     G(x_{0},y;s)=\int
\frac{d^{\sD}p}{(2\pi)^{\sD}}e^{-ip(x_{0}-y)}\tilde{G}(p,y;s)\, ,
\label{fou:g}
\end{equation}
we find that Eq. (\ref{eq:g}) reduces to
\begin{equation}
\begin{array}{l}
     \displaystyle
\left[\eta^{\mu\nu}p_{\mu}p_{\nu}-\frac{1}{4}R^{0}-s^{2}
     -\frac{2}{3}{{R^{0}}^{\mu}}_{\alpha}p_{\mu}
     \frac{\partial}{\partial p_{\alpha}}
     -\frac{1}{6}{R^{0}}^{\alpha\beta}\eta^{\mu\nu}p_{\mu}p_{\nu}
     \frac{\partial}{\partial p_{\alpha}}\frac{\partial}{\partial p_{\beta}}
     \right.
\\[5mm]
     \displaystyle
     +\frac{1}{3}{{{{R^{0}}^{\mu}}_{\alpha}}^{\nu}}_{\beta}p_{\mu}p_{\nu}
     \frac{\partial}{\partial
p_{\alpha}}\frac{\partial}{\partial p_{\beta}}
     +\frac{1}{6}{R^{0}}^{\alpha\beta}s^{2}
     \frac{\partial}{\partial
p_{\alpha}}\frac{\partial}{\partial p_{\beta}}
\\[5mm]
     \displaystyle \left.
     -\frac{1}{4}{{R^{0}}^{\mu}}_{\alpha ab}\sigma^{ab}p_{\mu}
     \frac{\partial}{\partial p_{\alpha}}
     \right]\tilde{G}(p,y;s)=1\, .
\end{array}
\label{eq:tildeg}
\end{equation}
From Eq. (\ref{eq:tildeg}) $\tilde{G}(p,y)$ is found to be
\begin{equation}
     \tilde{G}(p,y;s)=\frac{1}{p^{2}-s^{2}}
     -\frac{1}{12}\frac{R^{0}}{(p^{2}-s^{2})^{2}}
     +\frac{2}{3}\frac{{R^{0}}^{\mu\nu}p_{\mu}p_{\nu}}{(p^{2}-s^{2})^{3}}
     +\mbox{O}(R^{0}_{;\mu},(R^{0})^{2})\, .
\label{exp:tildeg}
\end{equation}
Inserting Eqs. (\ref{fou:g}) and (\ref{exp:tildeg}) into
Eq. (\ref{def:g})
the two-point function $S(x_{0},y;s)$ in the weak
curvature
expansion is obtained:\cite{WC2}
\begin{equation}
\begin{array}{rcl}
     \displaystyle S(x_{0},y;s)&=&\displaystyle (i
\gamma_{\mu}\nabla^{\mu}+s)
     \int
\frac{d^{\sD}p}{(2\pi)^{\sD}}e^{-ip(x_{0}-y)}\tilde{G}(p,y)
\\[5mm]
     &=&\displaystyle \int
\frac{d^{\sD}p}{(2\pi)^{\sD}}e^{-ip(x_{0}-y)}\left[
        \frac{\gamma^{a}p_{a}+s}{p^{2}-s^{2}}
        -\frac{1}{12}R^{0}\frac{\gamma^{a}p_{a}+s}{(p^{2}-s^{2})^{2}}
     \right.
\\[5mm]
&&\displaystyle\left.+\frac{2}{3}{R^{0}}^{\mu\nu}p_{\mu}p_{\nu}
     \frac{\gamma^{a}p_{a}+s}{(p^{2}-s^{2})^{3}}
     +\frac{1}{4}\gamma^{a}\sigma^{cd}{R^{0}}_{cda\mu}p^{\mu}
     \frac{1}{(p^{2}-s^{2})^{2}}
     \right]
\\[5mm]
     && \displaystyle
     +\mbox{O}(R^{0}_{;\mu},(R^{0})^{2})\, .
\end{array}
\label{exp:s}
\end{equation}
It may be easily checked that in
the limit $R^{0}\rightarrow 0$ the spinor two-point function
in Minkowski space (\ref{tpf:mink}) is reproduced.

\subsection{Effective potential in weakly curved spacetime}

We would like to discuss the curvature induced phase transition
in weakly curved spacetime.
For this purpose we evaluate the effective potential.
In the Gross-Neveu type model the effective potential
is given by Eq. (\ref{v:gn}) in the leading order of the
$1/N$ expansion.
Substituting Eq. (\ref{exp:s}) to Eq. (\ref{v:gn})
we obtain the effective potential in weakly curved
spacetime up to terms linear in $R$
\begin{equation}
\begin{array}{rcl}
\displaystyle
     V(\sigma)&=&\displaystyle
     \frac{1}{2\lambda_{0}}\sigma^{2}
     -i\tr\11 \int^{\sigma}_{0}ds
     \frac{d^{\sD}p}{(2\pi)^{\sD}}\left[
        \frac{s}{p^{2}-s^{2}}
        -\frac{1}{12}R\frac{s}{(p^{2}-s^{2})^{2}}\right.
\\[5mm]
     && \displaystyle\left.
     +\frac{2}{3}{R}^{\mu\nu}p_{\mu}p_{\nu}
     \frac{s}{(p^{2}-s^{2})^{3}}\right]\, .
\end{array}
\label{v:wcurv}
\end{equation}
Here we neglect the index $0$ of $R^{0}$ for simplicity.
In order to regularize the divergent integral in
the second term of the right-hand side
of Eq. (\ref{v:wcurv}) we apply the dimensional regularization.
Integrating over $p$ and $s$ in arbitrary spacetime dimensions 
we obtain the effective potential (\ref{v:wcurv}),
\begin{equation}
     V(\sigma)=V_{0}(\sigma)+V_{\sR}(\sigma)+O(R_{;\mu},R^{2})\, ,
\label{v:nonren:r}
\end{equation}
where $V_{0}(\sigma)$ is the effective potential at $R=0$
given in Eq. (\ref{v:nonren})
and $V_{\sR}(\sigma)$ is the effective potential linear in
$R$,\cite{WC1,WC2}
\begin{equation}
     V_{\sR}(\sigma)=-\frac{\tr \11}{(4\pi)^{\sD/2}}
     \frac{R}{24}\Gamma\left(1-\frac{D}{2}\right)\sigma^{\sD-2}\, .
\label{vr:nonren:r}
\end{equation}

As is mentioned in \S 2.3 the effective potential
$V_{0}(\sigma)$ is divergent in two and four dimensions.
We apply the same renormalization condition as the one in
Eq. (\ref{cond:ren}) and obtain the renormalized effective 
potential as in the $R=0$ case.
Replacing the bare coupling constant $\lambda_{0}$ with the 
renormalized one $\lambda$ in Eq. (\ref{eqn:ren})
we obtain
the renormalized effective potential,
\begin{equation}
\begin{array}{rcl}
     \displaystyle V(\sigma)
     & = &  \displaystyle
\frac{1}{2\lambda}\sigma^{2}\mu^{\sD-2}
            +\frac{\tr \11}{2(4\pi)^{\sD/2}}(D-1)
             \Gamma \left( 1-\frac{D}{2}\right)
             \sigma^{2}\mu^{\sD-2}
\\[5mm]
     &   &  \displaystyle -\frac{\tr \11}{(4\pi)^{\sD/2}D}
             \Gamma \left( 1-\frac{D}{2} \right)\sigma^{\sD}
            -\frac{\tr \11}{(4\pi)^{\sD/2}}
            \frac{R}{24}\Gamma\left(1-\frac{D}{2}\right)\sigma^{\sD-2}
\, .
\end{array}
\label{v:ren:w}
\end{equation}

Before the discussion of the phase structure
we consider the two-, three- and four-dimensional
limits of the effective potential (\ref{v:ren:w}).
Taking the two dimensional limit, $D\rightarrow 2$, we get
\begin{equation}
\begin{array}{rcl}
     \displaystyle \frac{V^{D=2}(\sigma)}{\mu^{\sD}}&=&
\displaystyle
                   \frac{1}{2
\lambda}\left(\frac{\sigma}{\mu}\right)^{2}
                  +\frac{\tr\11}{8
\pi}\left[-3+\ln\left(\frac{\sigma}{\mu}
                   \right)^{2}
                   \right]\left(\frac{\sigma}{\mu}\right)^{2}
\\[5mm]
     \displaystyle & &  \displaystyle
      -\frac{\tr\11}{96 \pi}\frac{R}{\mu^{2}}
                        \left[\frac{2}{2-D}-\gamma+\ln 4\pi
                        -\ln
\left(\frac{\sigma}{\mu}\right)^{2}
                        \right]\, .
\end{array}
\label{v:2d:w}
\end{equation}
It is different from the expression obtained in
Ref. \cite{BK}.\footnote{Differentiating
Eq. (\ref{v:2d:w}) we find
\begin{equation}
     \frac{\partial V^{D=2}}{\partial \sigma}
     =\sigma\left\{
     \frac{1}{\lambda}+\frac{\tr\11}{4\pi}
     \left[ -2+\ln\left(\frac{\sigma}{\mu}\right)^{2}
     +\frac{R}{12\sigma^{2}}\right]
     \right\}\, .
\end{equation}
On the other hand Eq. (4.3) in Ref. \cite{BK} reads
\begin{equation}
     \frac{\partial V^{D=2}}{\partial \sigma}
     =\sigma\left\{
     \frac{1}{\lambda}+\frac{\tr\11}{4\pi}
     \left[ -2+\ln\left(\frac{\sigma}{\mu}\right)^{2}
     +\frac{R}{4\sigma^{2}}\right]
     \right\}\, ,
\end{equation}
in the weak curvature limit where we adopt the 
notation in the present paper.
The common renormalization condition is adopted
in the above two cases.
As is mentioned in \S 2 the four-fermion theory
is renormalizable. The theory is independent
of the regularization method.
Thus the above two results should agree with each other.
The solution of the equation for the evolution operator 
in Ref. \cite{BK} seems not to be the general solution
but a special solution.}
The effective potential $V_{\sR}(\sigma)$ 
(i.e. the third term on the right-hand side in the Eq. (\ref{v:2d:w}))
is also divergent in two dimensions.
For $D=2$ the divergence comes from the lower momentum 
region of the $p$-integral in Eq. (\ref{v:wcurv}).
To show it we cut off the higher and lower momentum regions
of the divergent integral in Eq. (\ref{v:wcurv}) 
and find
\begin{eqnarray}
     V^{D=2}(\sigma)&=&\frac{1}{2\lambda_{0}}\sigma^{2}
     +\frac{\tr\11}{8\pi}\left[1+
      \ln\left(\frac{\Lambda_{UV}}{\sigma}\right)^{2}
      \right] \nonumber \\
     &&-\frac{\tr\11 R}{48\pi}\left[\frac{1}{2}
      \ln\left(\frac{\Lambda_{IR}}{\sigma}\right)^{2}
     +\frac{\Lambda_{UV}^{2}}{\Lambda_{UV}^{2}+\sigma^{2}}
     +\frac{\Lambda_{IR}^{2}}{\Lambda_{IR}^{2}+\sigma^{2}}
     \right]
     \nonumber \\
     && +O\left(
     \frac{\sigma^{2}}{\Lambda_{UV}^{2}},
     \frac{\Lambda_{IR}^{2}}{\sigma^{2}}\right)\, ,
\label{v:2d:cut:wc}
\end{eqnarray}
where $\Lambda_{UV}$ is the ultraviolet cut-off
scale and $\Lambda_{IR}$ is the infrared cut-off
scale of the divergent integral.
The ultraviolet divergence is cancelled out 
after renormalization and the terms
independent of $R$ reduce to Eq. (\ref{v:2d}).
As is seen in Eq. (\ref{v:2d:cut:wc}) the infrared 
divergence appears in the terms linear in $R$ for $D=2$.
Because of the infrared divergence the symmetry restoration 
always occurs for any positive values of the spacetime
curvature $R$ whereas only a broken phase is realized
for negative $R$.

Taking the three-dimensional limit $D\rightarrow 3$ we find
\cite{EOS}
\begin{equation}
     \frac{V^{D=3}(\sigma)}{\mu^{3}}=
                   \frac{1}{2
\lambda}\left(\frac{\sigma}{\mu}\right)^{2}
                  -\frac{\tr\11}{4\pi}
                   \left[\left(\frac{\sigma}{\mu}\right)^{2}
                  -\frac{1}{3}\left(\frac{\sigma}{\mu}\right)^{3}\right]
     +\frac{\tr\11}{96 \pi}\frac{R}{\mu^{2}}
                        \frac{\sigma}{\mu}\, .
\label{v:3d:w}
\end{equation}

If we take the four-dimensional limit $D\rightarrow 4$, the
effective potential (\ref{v:ren:w}) reduces to
\begin{equation}
\begin{array}{rcl}
     \displaystyle \frac{V^{D=4}(\sigma)}{\mu^{\sD}}&=&
     \displaystyle
           \frac{1}{2\lambda}\left(\frac{\sigma}{\mu}\right)^{2}
     -\frac{\tr \11}{4(4 \pi)^{2}}
                   \left\{6\left(C_{\mbox{div}}-\frac{2}{3}\right)
                   \left(\frac{\sigma}{\mu}\right)^{2}
\right.
\\[5mm]
     && \displaystyle\left.
                   -\left[C_{\mbox{div}}+\frac{1}{2}
                   -\ln\left(\frac{\sigma}{\mu}\right)^{2}
                   \right]\left(\frac{\sigma}{\mu}\right)^{4}\right\}
\\[5mm]
     \displaystyle & &  \displaystyle
                   +\frac{\tr \11}{4(4\pi)^{2}}\frac{1}{6}\frac{R}{\mu^{2}}
                        \left[C_{\mbox{div}}
                        -\ln\left(\frac{\sigma}{\mu}\right)^{2}
                        \right]
                        \left(\frac{\sigma}{\mu}\right)^{2}\, ,
\end{array}
\label{v:4d:w}
\end{equation}
where $C_{\mbox{div}}$ is defined in Eq. (\ref{div:c}).

To see the relationship between $C_{\mbox{div}}$ and a cut-off 
parameter
we calculate the effective potential for $D=4$
by using the cut-off regularization.
Performing the integration in Eq. (\ref{v:wcurv})
with the cut-off $\Lambda$ in the higher momentum region
one gets \cite{WC1}
\begin{equation}
     V^{D=4}(\sigma)=V_{0}^{D=4}(\sigma)
     +\frac{\tr\11}{(4\pi)^{2}}\frac{R}{6}
     \left[\ln\left(\frac{\Lambda}{\sigma}\right)^{2}-1
     \right]\sigma^{2}+O\left(\frac{\sigma^{2}}{\Lambda^{2}}\right)\, ,
\label{v:4d:cut:w}
\end{equation}
where $V_{0}(\sigma)$ is given by Eq. (\ref{v:nonren:cut2}). 
Applying the renormalization condition given by Eq. (\ref{cond:ren})
$V_{0}(\sigma)$ reduces to (\ref{v:d4:cut}).
After the renormalization
we find that there is a correspondence between two results
(\ref{v:4d:w}) and (\ref{v:4d:cut:w}), if we make a
replacement \cite{WC2}
\begin{equation}
     C_{\mbox{div}}+\frac{R}{6}\frac{1}{6\mu^{2}-\sigma^{2}}
     \leftrightarrow \ln \frac{\Lambda^{2}}{\mu^{2}}\, .
\label{corr:wc}
\end{equation}
We recognize that the previous correspondence
rule (\ref{corr:flat}) is now modified by the term
including curvature in Eq. (\ref{corr:wc}).

Now we have succeeded to expand the effective potential 
asymptotically
in terms of the spacetime curvature $R$ by using
the Riemann normal coordinate.
Starting from the effective potential (\ref{v:ren:w})
we investigate the phase structure of the theory 
in weakly curved spacetime in the following subsection.

\subsection{Curvature induced phase transition}

The vacuum expectation value is determined by observing
the minimum of the effective potential.
In weakly curved spacetime the effective potential
is modified as is mentioned above. Thus the vacuum expectation
value of $\sigma$ will be changed by the curvature effect.
We expect that the phase transition takes place 
by varying the spacetime curvature $R$.

\subsubsection{Phase structure for $\lambda > \lambda_{\mbox{cr}}$}

In Minkowski spacetime,
as is shown in \S 2.3, the system is in broken phase
for $\lambda > \lambda_{\mbox{cr}}$.
\begin{figure}
\setlength{\unitlength}{0.240900pt}
\begin{picture}(1500,900)(0,0)
\tenrm
\thinlines \drawline[-50](220,113)(220,877)
\thicklines \path(220,113)(240,113)
\thicklines \path(1436,113)(1416,113)
\put(198,113){\makebox(0,0)[r]{$-0.08$}}
\thicklines \path(220,226)(240,226)
\thicklines \path(1436,226)(1416,226)
\put(198,226){\makebox(0,0)[r]{$-0.06$}}
\thicklines \path(220,339)(240,339)
\thicklines \path(1436,339)(1416,339)
\put(198,339){\makebox(0,0)[r]{$-0.04$}}
\thicklines \path(220,453)(240,453)
\thicklines \path(1436,453)(1416,453)
\put(198,453){\makebox(0,0)[r]{$-0.02$}}
\thicklines \path(220,566)(240,566)
\thicklines \path(1436,566)(1416,566)
\put(198,566){\makebox(0,0)[r]{0}}
\thicklines \path(220,679)(240,679)
\thicklines \path(1436,679)(1416,679)
\put(198,679){\makebox(0,0)[r]{0.02}}
\thicklines \path(220,792)(240,792)
\thicklines \path(1436,792)(1416,792)
\put(198,792){\makebox(0,0)[r]{0.04}}
\thicklines \path(220,113)(220,133)
\thicklines \path(220,877)(220,857)
\put(220,68){\makebox(0,0){0}}
\thicklines \path(363,113)(363,133)
\thicklines \path(363,877)(363,857)
\put(363,68){\makebox(0,0){0.2}}
\thicklines \path(506,113)(506,133)
\thicklines \path(506,877)(506,857)
\put(506,68){\makebox(0,0){0.4}}
\thicklines \path(649,113)(649,133)
\thicklines \path(649,877)(649,857)
\put(649,68){\makebox(0,0){0.6}}
\thicklines \path(792,113)(792,133)
\thicklines \path(792,877)(792,857)
\put(792,68){\makebox(0,0){0.8}}
\thicklines \path(935,113)(935,133)
\thicklines \path(935,877)(935,857)
\put(935,68){\makebox(0,0){1}}
\thicklines \path(1078,113)(1078,133)
\thicklines \path(1078,877)(1078,857)
\put(1078,68){\makebox(0,0){1.2}}
\thicklines \path(1221,113)(1221,133)
\thicklines \path(1221,877)(1221,857)
\put(1221,68){\makebox(0,0){1.4}}
\thicklines \path(1364,113)(1364,133)
\thicklines \path(1364,877)(1364,857)
\put(1364,68){\makebox(0,0){1.6}}
\thicklines \path(220,113)(1436,113)(1436,877)(220,877)(220,113)
\put(45,945){\makebox(0,0)[l]{\shortstack{$V/m_{0}^{2.5}$}}}
\put(828,23){\makebox(0,0){$\sigma/m_{0}$}}
\put(721,764){\makebox(0,0)[l]{$R=3R_{\mbox{cr}}/2$}}
\put(792,622){\makebox(0,0)[l]{$R=R_{\mbox{cr}}$}}
\put(792,481){\makebox(0,0)[l]{$R=R_{\mbox{cr}}/2$}}
\put(864,339){\makebox(0,0)[l]{$R=0$}}
\put(849,209){\makebox(0,0)[l]{$R=-R_{\mbox{cr}}/2$}}
\thinlines \path(220,566)(220,566)(222,559)(223,556)(226,552)(233,546)(245,537)(271,522)(321,492)(372,460)(423,425)(473,389)(524,352)(575,315)(625,280)(676,246)(727,216)(777,190)(828,169)(853,160)(879,153)(891,150)(904,148)(917,146)(929,144)(942,143)(948,142)(955,142)(958,142)(959,142)(961,142)(963,142)(964,142)(966,142)(967,142)(969,142)(971,142)(972,142)(974,142)(975,142)(977,142)(980,142)(983,142)(986,142)(993,142)(999,143)(1005,143)(1018,145)(1031,147)(1056,153)(1081,161)
\thinlines \path(1081,161)(1107,172)(1132,184)(1183,218)(1233,261)(1284,316)(1335,382)(1385,461)(1436,552)
\thinlines \path(220,566)(220,566)(222,566)(224,566)(226,566)(227,566)(229,566)(231,565)(235,565)(242,565)(250,564)(265,561)(280,558)(309,550)(339,540)(399,513)(458,482)(518,449)(578,414)(637,381)(697,351)(756,325)(786,314)(816,305)(846,297)(861,294)(876,292)(891,290)(905,288)(913,288)(917,287)(920,287)(924,287)(926,287)(928,287)(930,287)(932,287)(933,287)(935,287)(937,287)(939,287)(941,287)(943,287)(945,287)(946,287)(950,287)(954,287)(958,288)(965,288)(973,289)(980,290)
\thinlines \path(980,290)(995,292)(1010,295)(1025,298)(1055,308)(1084,320)(1114,336)(1174,377)(1233,432)(1293,502)(1353,589)(1412,693)(1436,742)
\thinlines \path(220,566)(220,566)(222,573)(224,576)(227,580)(231,583)(235,586)(242,590)(250,593)(257,595)(265,597)(272,599)(280,600)(283,600)(287,600)(289,601)(291,601)(293,601)(295,601)(296,601)(298,601)(300,601)(302,601)(304,601)(306,601)(308,601)(309,601)(311,601)(313,601)(317,601)(324,600)(332,599)(339,598)(354,596)(369,593)(399,585)(458,565)(518,541)(578,516)(637,491)(697,468)(756,449)(786,442)(816,436)(831,433)(846,431)(853,431)(861,430)(868,429)(876,429)(879,429)
\thinlines \path(879,429)(883,429)(885,429)(887,429)(889,429)(891,429)(892,429)(894,429)(896,429)(898,429)(900,429)(902,429)(905,429)(907,429)(909,429)(913,429)(920,429)(928,430)(935,430)(950,432)(965,435)(995,441)(1025,451)(1055,463)(1114,496)(1174,543)(1233,603)(1293,678)(1353,770)(1412,877)
\thinlines \path(220,566)(220,566)(222,580)(224,586)(227,595)(231,601)(235,607)(250,622)(257,628)(265,633)(280,641)(295,647)(309,652)(317,653)(324,655)(332,656)(339,657)(343,657)(347,658)(350,658)(354,658)(356,658)(358,658)(360,658)(362,658)(363,658)(365,658)(367,658)(369,658)(371,658)(373,658)(375,658)(376,658)(380,658)(384,658)(391,658)(399,657)(414,656)(429,653)(458,648)(518,634)(578,617)(637,601)(697,585)(727,579)(756,574)(771,571)(786,569)(801,568)(816,567)(824,566)
\thinlines \path(824,566)(831,566)(835,566)(837,566)(838,566)(840,566)(842,566)(844,566)(846,566)(848,566)(850,566)(851,566)(853,566)(855,566)(857,566)(861,566)(865,566)(868,566)(876,567)(883,567)(891,568)(905,569)(920,571)(935,574)(965,581)(995,591)(1025,603)(1055,618)(1114,657)(1174,708)(1233,774)(1293,854)(1307,877)
\thinlines \path(220,566)(220,566)(222,588)(224,597)(227,610)(231,619)(235,627)(250,652)(265,669)(280,683)(295,694)(309,702)(339,716)(354,720)(369,724)(384,727)(399,729)(406,730)(414,730)(417,731)(421,731)(425,731)(429,731)(432,731)(434,731)(436,731)(438,731)(440,731)(442,731)(444,731)(445,731)(447,731)(449,731)(451,731)(455,731)(458,731)(466,731)(473,731)(488,730)(503,728)(518,727)(578,719)(637,710)(667,706)(697,703)(712,701)(727,700)(742,699)(756,698)(764,698)(771,697)
\thinlines \path(771,697)(775,697)(777,697)(779,697)(781,697)(783,697)(784,697)(786,697)(788,697)(790,697)(792,697)(794,697)(796,697)(797,697)(801,697)(805,697)(809,698)(816,698)(824,698)(831,699)(846,700)(861,702)(876,704)(905,710)(935,718)(965,728)(995,740)(1055,773)(1114,817)(1174,874)(1176,877)
\dottedline{14}(220,566)(220,566)(1436,566)
\end{picture}

                \vglue 1ex
                \hspace*{18em}\mbox{$(a) D=2.5$}
                \vglue 7ex
\setlength{\unitlength}{0.240900pt}
\begin{picture}(1500,900)(0,0)
\tenrm
\dottedline{14}(220,566)(1436,566)
\thicklines \path(220,113)(240,113)
\thicklines \path(1436,113)(1416,113)
\put(198,113){\makebox(0,0)[r]{$-0.08$}}
\thicklines \path(220,226)(240,226)
\thicklines \path(1436,226)(1416,226)
\put(198,226){\makebox(0,0)[r]{$-0.06$}}
\thicklines \path(220,339)(240,339)
\thicklines \path(1436,339)(1416,339)
\put(198,339){\makebox(0,0)[r]{$-0.04$}}
\thicklines \path(220,453)(240,453)
\thicklines \path(1436,453)(1416,453)
\put(198,453){\makebox(0,0)[r]{$-0.02$}}
\thicklines \path(220,566)(240,566)
\thicklines \path(1436,566)(1416,566)
\put(198,566){\makebox(0,0)[r]{0}}
\thicklines \path(220,679)(240,679)
\thicklines \path(1436,679)(1416,679)
\put(198,679){\makebox(0,0)[r]{0.02}}
\thicklines \path(220,792)(240,792)
\thicklines \path(1436,792)(1416,792)
\put(198,792){\makebox(0,0)[r]{0.04}}
\thicklines \path(220,113)(220,133)
\thicklines \path(220,877)(220,857)
\put(220,68){\makebox(0,0){0}}
\thicklines \path(363,113)(363,133)
\thicklines \path(363,877)(363,857)
\put(363,68){\makebox(0,0){0.2}}
\thicklines \path(506,113)(506,133)
\thicklines \path(506,877)(506,857)
\put(506,68){\makebox(0,0){0.4}}
\thicklines \path(649,113)(649,133)
\thicklines \path(649,877)(649,857)
\put(649,68){\makebox(0,0){0.6}}
\thicklines \path(792,113)(792,133)
\thicklines \path(792,877)(792,857)
\put(792,68){\makebox(0,0){0.8}}
\thicklines \path(935,113)(935,133)
\thicklines \path(935,877)(935,857)
\put(935,68){\makebox(0,0){1}}
\thicklines \path(1078,113)(1078,133)
\thicklines \path(1078,877)(1078,857)
\put(1078,68){\makebox(0,0){1.2}}
\thicklines \path(1221,113)(1221,133)
\thicklines \path(1221,877)(1221,857)
\put(1221,68){\makebox(0,0){1.4}}
\thicklines \path(1364,113)(1364,133)
\thicklines \path(1364,877)(1364,857)
\put(1364,68){\makebox(0,0){1.6}}
\thicklines \path(220,113)(1436,113)(1436,877)(220,877)(220,113)
\put(45,945){\makebox(0,0)[l]{\shortstack{$V/m_{0}^{3.5}$}}}
\put(828,23){\makebox(0,0){$\sigma/m_{0}$}}
\put(435,679){\makebox(0,0)[l]{$R=3R_{\mbox{cr}}/2$}}
\put(635,611){\makebox(0,0)[l]{$R=R_{\mbox{cr}}$}}
\put(663,520){\makebox(0,0)[l]{$R=R_{\mbox{cr}}/2$}}
\put(864,396){\makebox(0,0)[l]{$R=0$}}
\put(900,238){\makebox(0,0)[l]{$R=-R_{\mbox{cr}}/2$}}
\thinlines \path(220,566)(220,566)(222,566)(223,566)(226,566)(230,565)(233,565)(239,565)(245,564)(258,562)(271,560)(296,554)(321,547)(372,527)(423,502)(473,472)(524,439)(575,403)(625,366)(676,328)(727,291)(777,256)(828,224)(879,197)(929,176)(955,169)(967,166)(980,163)(993,161)(999,160)(1005,160)(1008,159)(1012,159)(1015,159)(1018,159)(1020,159)(1021,159)(1023,159)(1024,159)(1026,158)(1028,158)(1029,158)(1031,158)(1032,159)(1034,159)(1035,159)(1037,159)(1040,159)(1043,159)(1050,159)(1056,160)
\thinlines \path(1056,160)(1069,162)(1081,165)(1094,168)(1107,172)(1132,183)(1157,197)(1183,215)(1233,262)(1284,327)(1335,411)(1385,515)(1436,642)
\thinlines \path(220,566)(220,566)(222,566)(224,566)(226,566)(227,566)(229,566)(231,566)(235,566)(239,565)(242,565)(250,565)(265,564)(280,562)(309,557)(339,551)(399,534)(458,511)(518,485)(578,456)(637,427)(697,398)(756,372)(786,360)(816,350)(846,342)(861,339)(876,336)(891,334)(905,332)(913,331)(917,331)(920,331)(924,331)(928,331)(930,331)(932,331)(933,331)(935,331)(937,331)(939,331)(941,331)(943,331)(945,331)(946,331)(950,331)(954,331)(958,331)(965,332)(973,333)(980,334)
\thinlines \path(980,334)(995,337)(1010,340)(1025,345)(1055,357)(1084,373)(1114,393)(1174,448)(1233,526)(1293,627)(1353,756)(1398,877)
\thinlines \path(220,566)(220,566)(227,566)(235,566)(239,566)(242,566)(244,566)(246,566)(248,566)(250,566)(252,566)(254,566)(255,566)(257,566)(259,566)(261,566)(265,566)(268,566)(272,566)(280,566)(287,565)(295,565)(309,564)(324,563)(339,562)(369,558)(399,553)(458,542)(518,527)(578,512)(637,496)(697,483)(727,477)(756,473)(771,472)(779,471)(786,470)(794,470)(797,470)(801,470)(805,469)(809,469)(810,469)(812,469)(814,469)(816,469)(818,469)(820,469)(822,469)(824,469)(825,469)
\thinlines \path(825,469)(827,469)(831,469)(835,469)(838,470)(846,470)(853,471)(861,471)(876,473)(891,475)(905,479)(935,487)(965,498)(995,513)(1025,531)(1055,554)(1114,611)(1174,689)(1233,789)(1275,877)
\thinlines \path(220,566)(220,566)(280,569)(295,570)(309,571)(324,572)(339,572)(347,573)(354,573)(362,573)(369,573)(373,573)(375,573)(376,573)(378,573)(380,573)(382,573)(384,573)(386,573)(388,573)(390,573)(391,573)(393,573)(395,573)(399,573)(403,573)(406,573)(414,573)(429,572)(444,572)(458,572)(518,569)(548,568)(578,567)(593,566)(600,566)(607,566)(611,566)(615,566)(619,566)(622,566)(624,566)(626,566)(628,566)(630,566)(632,566)(634,566)(635,566)(637,566)(639,566)(641,566)
\thinlines \path(641,566)(645,566)(648,566)(652,566)(660,566)(667,566)(675,567)(682,567)(697,568)(712,569)(727,570)(756,575)(786,580)(816,588)(846,598)(876,610)(935,643)(995,689)(1055,750)(1114,830)(1142,877)
\thinlines \path(220,566)(220,566)(280,573)(339,583)(399,592)(458,602)(518,611)(578,622)(637,635)(697,653)(756,676)(816,707)(876,747)(935,799)(995,865)(1004,877)
\end{picture}

                \vglue 1ex
                \hspace*{18em}\mbox{$(b) D=3.5$}
                \vglue 1ex
\caption{Behavior of the effective potential is shown at $D=2.5$
         and $D=3.5$ for fixed $\lambda$ $( > \lambda_{\mbox{cr}})$
         with the varying curvature where
         $R_{\mbox{cr}}=6(D-2)(D(4-D)/4)^{(4-\sD)/(\sD-2)}m_{0}^{2}>0$.}
\label{[3fig:pot25}
\end{figure}
Here in the present subsection we fix the coupling 
constant $\lambda$
larger than the critical value $\lambda_{\mbox{cr}}$ and see
whether the chiral symmetry is restored by the curvature effect.
To study the phase structure in curved spacetime we
evaluate the effective
potential (\ref{v:ren:w}) numerically with the varying curvature.
In Fig. 6 we present the typical behavior of the effective
potential (\ref{v:ren:w}) for several values of the curvature 
in the case $D=2.5$ and $D=3.5$.
Here we adopted the formula $\tr\11=2^{\sD/2}$.
We find that the chiral symmetry is restored as $R$ is
increased with $\lambda$ fixed.
As can be seen in Fig. 6 the phase transition induced by
curvature effects is of the first order.
On the other hand no phase transition takes place and
the vacuum expectation value of the
auxiliary field simply increases for the negative curvature
as its absolute value increases.
We observe the similar behavior of the effective potential
in the spacetime dimensions, $2 < D < 4$.

In the leading order of the $1/N$ expansion
the dynamical fermion mass is equal to the vacuum
expectation value of the auxiliary field $\langle\sigma\rangle$.
The dynamical fermion mass is given through
the minimum of the effective potential.
The necessary condition for the minimum is given by the gap
equation :
\begin{equation}
     \left.\frac{\partial V(\sigma)}{\partial
\sigma}\right|_{\sigma=m}
     =0\, .
\label{gap:w}
\end{equation}
The non-trivial solution of the gap equation corresponds to
the dynamical mass of the fermion.
Inserting the Eqs. (\ref{v:ren:w}) and (\ref{def:mass}) into
Eq. (\ref{gap:w})
the non-trivial solution of the gap equation is expressed as
\begin{equation}
     m_{0}^{\sD-2}-m^{\sD-2}
     +\frac{R}{12}\left(1-\frac{D}{2}\right)m^{\sD-4}=0\, .
\label{nontri:w}
\end{equation}
In Fig. 7 we plot the non-trivial solution of the gap
equation as a function of the spacetime dimension $D$.
\begin{figure}
\setlength{\unitlength}{0.240900pt}
\begin{picture}(1500,900)(0,0)
\tenrm
\thinlines \dottedline{14}(506,113)(506,877)
\thicklines \path(220,113)(240,113)
\thicklines \path(1436,113)(1416,113)
\put(198,113){\makebox(0,0)[r]{0}}
\thicklines \path(220,240)(240,240)
\thicklines \path(1436,240)(1416,240)
\put(198,240){\makebox(0,0)[r]{0.2}}
\thicklines \path(220,368)(240,368)
\thicklines \path(1436,368)(1416,368)
\put(198,368){\makebox(0,0)[r]{0.4}}
\thicklines \path(220,495)(240,495)
\thicklines \path(1436,495)(1416,495)
\put(198,495){\makebox(0,0)[r]{0.6}}
\thicklines \path(220,622)(240,622)
\thicklines \path(1436,622)(1416,622)
\put(198,622){\makebox(0,0)[r]{0.8}}
\thicklines \path(220,750)(240,750)
\thicklines \path(1436,750)(1416,750)
\put(198,750){\makebox(0,0)[r]{1}}
\thicklines \path(220,877)(240,877)
\thicklines \path(1436,877)(1416,877)
\put(198,877){\makebox(0,0)[r]{1.2}}
\thicklines \path(220,113)(220,133)
\thicklines \path(220,877)(220,857)
\put(220,68){\makebox(0,0){$-4$}}
\thicklines \path(363,113)(363,133)
\thicklines \path(363,877)(363,857)
\put(363,68){\makebox(0,0){$-2$}}
\thicklines \path(506,113)(506,133)
\thicklines \path(506,877)(506,857)
\put(506,68){\makebox(0,0){0}}
\thicklines \path(649,113)(649,133)
\thicklines \path(649,877)(649,857)
\put(649,68){\makebox(0,0){2}}
\thicklines \path(792,113)(792,133)
\thicklines \path(792,877)(792,857)
\put(792,68){\makebox(0,0){4}}
\thicklines \path(935,113)(935,133)
\thicklines \path(935,877)(935,857)
\put(935,68){\makebox(0,0){6}}
\thicklines \path(1078,113)(1078,133)
\thicklines \path(1078,877)(1078,857)
\put(1078,68){\makebox(0,0){8}}
\thicklines \path(1221,113)(1221,133)
\thicklines \path(1221,877)(1221,857)
\put(1221,68){\makebox(0,0){10}}
\thicklines \path(1364,113)(1364,133)
\thicklines \path(1364,877)(1364,857)
\put(1364,68){\makebox(0,0){12}}
\thicklines \path(220,113)(1436,113)(1436,877)(220,877)(220,113)
\put(45,945){\makebox(0,0)[l]{\shortstack{$m / m_{0}$}}}
\put(828,23){\makebox(0,0){$R/m_{0}^2$}}
\put(1150,495){\makebox(0,0)[l]{$D=4$}}
\put(649,495){\makebox(0,0)[l]{$D=2$}}
\thinlines \path(506,114)(506,114)(506,115)(506,116)(506,117)(507,118)(507,120)(508,123)(509,126)(512,133)(516,140)(536,166)(562,193)(592,219)(623,246)(655,272)(686,299)(716,325)(743,352)(767,378)(787,405)(796,418)(804,431)(810,445)(815,458)(817,464)(819,471)(820,478)(821,484)(821,488)(821,489)(822,491)(822,493)(822,494)(822,495)(822,496)(822,497)(822,498)(822,498)(822,499)(822,500)(822,501)(822,502)(822,503)(822,504)(822,506)(822,508)(821,511)(821,514)(820,518)(819,524)
\thinlines \path(819,524)(818,531)(815,537)(810,551)(803,564)(784,591)(757,617)(723,644)(682,670)(632,697)(573,723)(506,750)(506,750)(414,782)(308,813)(220,837)
\thinlines \path(506,113)(506,113)(506,114)(507,115)(507,116)(509,120)(515,126)(529,140)(565,166)(604,193)(644,219)(684,246)(721,272)(755,299)(785,325)(812,352)(833,378)(842,392)(850,405)(857,418)(862,431)(864,438)(865,445)(867,451)(867,455)(867,458)(868,461)(868,463)(868,464)(868,466)(868,467)(868,468)(868,469)(868,469)(868,470)(868,471)(868,472)(868,473)(868,474)(868,474)(868,476)(868,478)(868,479)(868,481)(867,484)(867,488)(867,491)(865,498)(864,504)(861,511)(856,524)
\thinlines \path(856,524)(849,537)(830,564)(805,591)(773,617)(734,644)(688,670)(634,697)(574,723)(506,750)(492,755)(398,787)(293,819)(220,839)
\thinlines \path(506,113)(506,113)(575,140)(637,166)(694,193)(745,219)(789,246)(828,272)(861,299)(888,325)(908,352)(917,365)(923,378)(929,392)(931,398)(932,405)(934,411)(934,415)(935,418)(935,421)(935,423)(935,425)(935,426)(935,427)(935,428)(935,429)(935,430)(935,431)(935,431)(935,432)(935,433)(935,434)(935,435)(935,435)(935,436)(935,438)(935,440)(935,441)(935,445)(934,448)(934,451)(932,458)(931,464)(929,471)(923,484)(917,498)(908,511)(888,537)(861,564)(828,591)(789,617)
\thinlines \path(789,617)(745,644)(694,670)(637,697)(575,723)(506,750)(477,760)(382,793)(279,825)(220,842)
\thinlines \path(506,113)(506,113)(547,114)(565,115)(589,116)(623,120)(649,123)(671,126)(738,140)(788,153)(829,166)(893,193)(942,219)(979,246)(1007,272)(1027,299)(1034,312)(1040,325)(1042,332)(1044,338)(1045,345)(1046,352)(1046,355)(1047,358)(1047,360)(1047,361)(1047,362)(1047,363)(1047,363)(1047,364)(1047,365)(1047,366)(1047,367)(1047,368)(1047,368)(1047,369)(1047,370)(1047,372)(1047,373)(1046,375)(1046,378)(1046,382)(1045,385)(1044,392)(1042,398)(1041,405)(1036,418)(1029,431)(1013,458)(991,484)(964,511)
\thinlines \path(964,511)(932,537)(895,564)(853,591)(807,617)(756,644)(700,670)(640,697)(575,723)(506,750)(462,766)(368,798)(266,831)(220,845)
\thinlines \path(1364,113)(1364,113)(1364,115)(1364,115)(1364,116)(1364,117)(1364,118)(1364,120)(1364,121)(1364,123)(1364,126)(1364,130)(1364,133)(1363,140)(1362,146)(1361,153)(1359,166)(1355,179)(1351,193)(1341,219)(1327,246)(1311,272)(1291,299)(1269,325)(1244,352)(1215,378)(1184,405)(1150,431)(1113,458)(1072,484)(1029,511)(983,537)(934,564)(882,591)(827,617)(768,644)(707,670)(643,697)(576,723)(506,750)(448,771)(354,804)(255,837)(220,848)
\end{picture}

\vglue 1ex
\caption{Solutions of the gap equation for $D=2,2.5,3,3.5,4.$}
\vglue 7ex
\setlength{\unitlength}{0.240900pt}
\begin{picture}(1500,900)(0,0)
\tenrm
\thicklines \path(220,113)(240,113)
\thicklines \path(1436,113)(1416,113)
\put(198,113){\makebox(0,0)[r]{0}}
\thicklines \path(220,240)(240,240)
\thicklines \path(1436,240)(1416,240)
\put(198,240){\makebox(0,0)[r]{0.2}}
\thicklines \path(220,368)(240,368)
\thicklines \path(1436,368)(1416,368)
\put(198,368){\makebox(0,0)[r]{0.4}}
\thicklines \path(220,495)(240,495)
\thicklines \path(1436,495)(1416,495)
\put(198,495){\makebox(0,0)[r]{0.6}}
\thicklines \path(220,622)(240,622)
\thicklines \path(1436,622)(1416,622)
\put(198,622){\makebox(0,0)[r]{0.8}}
\thicklines \path(220,750)(240,750)
\thicklines \path(1436,750)(1416,750)
\put(198,750){\makebox(0,0)[r]{1}}
\thicklines \path(220,877)(240,877)
\thicklines \path(1436,877)(1416,877)
\put(198,877){\makebox(0,0)[r]{1.2}}
\thicklines \path(220,113)(220,133)
\thicklines \path(220,877)(220,857)
\put(220,68){\makebox(0,0){$-4$}}
\thicklines \path(363,113)(363,133)
\thicklines \path(363,877)(363,857)
\put(363,68){\makebox(0,0){$-2$}}
\thicklines \path(506,113)(506,133)
\thicklines \path(506,877)(506,857)
\put(506,68){\makebox(0,0){0}}
\thicklines \path(649,113)(649,133)
\thicklines \path(649,877)(649,857)
\put(649,68){\makebox(0,0){2}}
\thicklines \path(792,113)(792,133)
\thicklines \path(792,877)(792,857)
\put(792,68){\makebox(0,0){4}}
\thicklines \path(935,113)(935,133)
\thicklines \path(935,877)(935,857)
\put(935,68){\makebox(0,0){6}}
\thicklines \path(1078,113)(1078,133)
\thicklines \path(1078,877)(1078,857)
\put(1078,68){\makebox(0,0){8}}
\thicklines \path(1221,113)(1221,133)
\thicklines \path(1221,877)(1221,857)
\put(1221,68){\makebox(0,0){10}}
\thicklines \path(1364,113)(1364,133)
\thicklines \path(1364,877)(1364,857)
\put(1364,68){\makebox(0,0){12}}
\thicklines \path(220,113)(1436,113)(1436,877)(220,877)(220,113)
\put(45,945){\makebox(0,0)[l]{\shortstack{$m / m_{0}$}}}
\put(828,23){\makebox(0,0){$R/m_{0}^2$}}
\put(1150,495){\makebox(0,0)[l]{$D=4$}}
\put(1021,368){\makebox(0,0)[l]{$D=3.5$}}
\put(849,368){\makebox(0,0)[l]{$D=3$}}
\put(510,368){\makebox(0,0)[l]{$D=2.5$}}
\put(327,368){\makebox(0,0)[l]{$D=2$}}
\thinlines \path(506,750)(506,750)(506,750)(414,782)(308,813)(220,837)
\thinlines \path(683,673)(683,673)(634,697)(574,723)(506,750)(492,755)(398,787)(293,819)(220,839)
\thinlines \path(828,591)(828,591)(789,617)(745,644)(694,670)(637,697)(575,723)(506,750)(477,760)(382,793)(279,825)(220,842)
\thinlines \path(995,480)(995,480)(991,484)(964,511)(932,537)(895,564)(853,591)(807,617)(756,644)(700,670)(640,697)(575,723)(506,750)(462,766)(368,798)(266,831)(220,845)
\thinlines \path(1364,113)(1364,113)(1364,115)(1364,115)(1364,116)(1364,117)(1364,118)(1364,120)(1364,121)(1364,123)(1364,126)(1364,130)(1364,133)(1363,140)(1362,146)(1361,153)(1359,166)(1355,179)(1351,193)(1341,219)(1327,246)(1311,272)(1291,299)(1269,325)(1244,352)(1215,378)(1184,405)(1150,431)(1113,458)(1072,484)(1029,511)(983,537)(934,564)(882,591)(827,617)(768,644)(707,670)(643,697)(576,723)(506,750)(448,771)(354,804)(255,837)(220,848)
\dottedline{14}(506,750)(506,113)
\dottedline{14}(683,673)(683,113)
\dottedline{14}(828,591)(828,113)
\dottedline{14}(995,480)(995,113)
\end{picture}

\vglue 1ex
\caption{Dynamical fermion mass as a function of the
spacetime curvature.}
\label{fig:gap:w}
\end{figure}

Since the phase transition induced by curvature effects
is of the first-order, two independent solutions appear for
a small positive curvature.
The larger solution corresponds to the local minimum
and the smaller solution represents the first extremum
of the effective potential.

At the critical point the two local minima of
the effective potential acquire the
same value which is zero.
Thus the critical point is obtained by the solution of
Eq. (\ref{nontri:w}) supplemented by the condition
\begin{equation}
     V(m)=V(0)=0\, .
\label{cri1st:w}
\end{equation}
We solve Eqs. (\ref{nontri:w}) and (\ref{cri1st:w})
and obtain the critical curvature,\cite{WC2}
\begin{equation}
     R_{\mbox{cr}}
     =6(D-2)\left(\frac{D(4-D)}{4}\right)^{(4-D)/(D-2)}m_{0}^{2}\, ,
\label{cr:r:w}
\end{equation}
and the mass gap at the critical point,\cite{WC2}
\begin{equation}
     m_{\mbox{cr}}=\left(\frac{D(4-D)}{4}\right)^{1/(\sD-2)}m_{0}\, .
\end{equation}

If the spacetime curvature $R$ is less than the critical
value $R_{\mbox{cr}}$,
the vacuum expectation value of the auxiliary field
$\langle\sigma\rangle$
is given by the larger solution of Eq. (\ref{nontri:w}).
In the case that the spacetime curvature $R$ is larger than the critical
value $R_{\mbox{cr}}$,
the chiral symmetry is restored and the vacuum expectation
value of the auxiliary field $\langle\sigma\rangle$ disappears.
The behavior of the dynamical fermion mass as a function
of $R$ is given in Fig. 8.
We observe that the mass gap appears at $R=R_{\mbox{cr}}$ and
disappears in the four-dimensional limit.
For some special values of $D$ Eq. (\ref{cr:r:w}) simplifies:
\begin{equation}
\begin{array}{ll}
     R_{\mbox{cr}}=0\, ; & D=2\, ,
\\[5mm]
     \displaystyle R_{\mbox{cr}}=\frac{9}{2}m_{0}^{2}\, ; & D=3\, ,
\\[5mm]
     R_{\mbox{cr}}=12 m_{0}^{2}\, ; & D=4\, .
\end{array}
\label{rc234:wc}
\end{equation}
\begin{figure}
\vspace*{2ex}
\setlength{\unitlength}{0.240900pt}
\begin{picture}(1500,900)(0,0)
\tenrm
\thicklines \path(220,113)(240,113)
\thicklines \path(1436,113)(1416,113)
\put(198,113){\makebox(0,0)[r]{0}}
\thicklines \path(220,235)(240,235)
\thicklines \path(1436,235)(1416,235)
\put(198,235){\makebox(0,0)[r]{2}}
\thicklines \path(220,357)(240,357)
\thicklines \path(1436,357)(1416,357)
\put(198,357){\makebox(0,0)[r]{4}}
\thicklines \path(220,480)(240,480)
\thicklines \path(1436,480)(1416,480)
\put(198,480){\makebox(0,0)[r]{6}}
\thicklines \path(220,602)(240,602)
\thicklines \path(1436,602)(1416,602)
\put(198,602){\makebox(0,0)[r]{8}}
\thicklines \path(220,724)(240,724)
\thicklines \path(1436,724)(1416,724)
\put(198,724){\makebox(0,0)[r]{10}}
\thicklines \path(220,846)(240,846)
\thicklines \path(1436,846)(1416,846)
\put(198,846){\makebox(0,0)[r]{12}}
\thicklines \path(220,113)(220,133)
\thicklines \path(220,877)(220,857)
\put(220,68){\makebox(0,0){2}}
\thicklines \path(342,113)(342,133)
\thicklines \path(342,877)(342,857)
\put(342,68){\makebox(0,0){2.2}}
\thicklines \path(463,113)(463,133)
\thicklines \path(463,877)(463,857)
\put(463,68){\makebox(0,0){2.4}}
\thicklines \path(585,113)(585,133)
\thicklines \path(585,877)(585,857)
\put(585,68){\makebox(0,0){2.6}}
\thicklines \path(706,113)(706,133)
\thicklines \path(706,877)(706,857)
\put(706,68){\makebox(0,0){2.8}}
\thicklines \path(828,113)(828,133)
\thicklines \path(828,877)(828,857)
\put(828,68){\makebox(0,0){3}}
\thicklines \path(950,113)(950,133)
\thicklines \path(950,877)(950,857)
\put(950,68){\makebox(0,0){3.2}}
\thicklines \path(1071,113)(1071,133)
\thicklines \path(1071,877)(1071,857)
\put(1071,68){\makebox(0,0){3.4}}
\thicklines \path(1193,113)(1193,133)
\thicklines \path(1193,877)(1193,857)
\put(1193,68){\makebox(0,0){3.6}}
\thicklines \path(1314,113)(1314,133)
\thicklines \path(1314,877)(1314,857)
\put(1314,68){\makebox(0,0){3.8}}
\thicklines \path(1436,113)(1436,133)
\thicklines \path(1436,877)(1436,857)
\put(1436,68){\makebox(0,0){4}}
\thicklines \path(220,113)(1436,113)(1436,877)(220,877)(220,113)
\put(45,945){\makebox(0,0)[l]{\shortstack{$R_{\mbox{cr}}/m_{0}^{2}$}}}
\put(828,23){\makebox(0,0){$D$}}
\put(342,755){\makebox(0,0)[l]{Symmetric phase}}
\put(919,235){\makebox(0,0)[l]{Broken phase}}
\thinlines \path(220,113)(220,113)(221,113)(271,143)(321,170)(372,195)(422,219)(473,242)(523,264)(573,285)(624,306)(674,326)(724,346)(775,366)(825,387)(876,408)(926,429)(976,452)(1027,476)(1077,501)(1128,528)(1178,558)(1228,592)(1279,630)(1329,676)(1354,704)(1380,735)(1405,774)(1430,826)(1436,846)
\end{picture}

\caption{Critical curvature as a function of dimension $D$
in weakly curved spacetime.}
\label{fig:crcurall}
\end{figure}
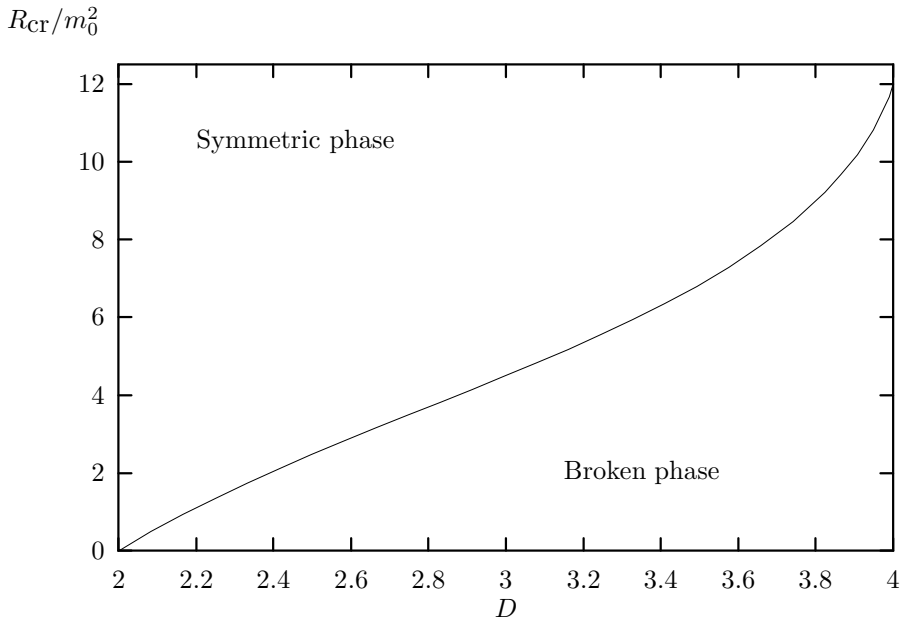
In Fig. 9 we show the critical curvature $R_{\mbox{cr}}$ as a
function of the spacetime dimensions $D$.
In two dimensions the weak curvature expansion shows that
the broken chiral symmetry is restored by the effect of an
infrared divergence for any positive curvature.
The critical curvature $R^{D=2}_{\mbox{cr}}$ is thus zero.

\subsubsection{Phase structure for $\lambda \leq \lambda_{\mbox{cr}}$}

For $\lambda \leq \lambda_{\mbox{cr}}$ 
the system is in symmetric phase
in Minkowski spacetime.
In the present subsection we investigate
whether the chiral symmetry is 
broken down by the curvature effect for
$\lambda \leq \lambda_{\mbox{cr}}$.
We introduce, for convenience, the scale $m'_{0}$ defined
in Eq. (\ref{m0p}) and calculate the effective potential
numerically.
In Fig. 10 the behavior of the effective potential
(\ref{v:ren:w}) is presented with the varying
curvature at $D=2.5$ and $D=3.5$.
\begin{figure}
\setlength{\unitlength}{0.240900pt}
\begin{picture}(1500,900)(0,0)
\tenrm
\thinlines \dottedline{14}(220,368)(1436,368)
\thicklines \path(220,113)(240,113)
\thicklines \path(1436,113)(1416,113)
\put(198,113){\makebox(0,0)[r]{$-0.02$}}
\thicklines \path(220,240)(240,240)
\thicklines \path(1436,240)(1416,240)
\put(198,240){\makebox(0,0)[r]{$-0.01$}}
\thicklines \path(220,368)(240,368)
\thicklines \path(1436,368)(1416,368)
\put(198,368){\makebox(0,0)[r]{0}}
\thicklines \path(220,495)(240,495)
\thicklines \path(1436,495)(1416,495)
\put(198,495){\makebox(0,0)[r]{0.01}}
\thicklines \path(220,622)(240,622)
\thicklines \path(1436,622)(1416,622)
\put(198,622){\makebox(0,0)[r]{0.02}}
\thicklines \path(220,750)(240,750)
\thicklines \path(1436,750)(1416,750)
\put(198,750){\makebox(0,0)[r]{0.03}}
\thicklines \path(220,877)(240,877)
\thicklines \path(1436,877)(1416,877)
\put(198,877){\makebox(0,0)[r]{0.04}}
\thicklines \path(220,113)(220,133)
\thicklines \path(220,877)(220,857)
\put(220,68){\makebox(0,0){0}}
\thicklines \path(423,113)(423,133)
\thicklines \path(423,877)(423,857)
\put(423,68){\makebox(0,0){0.05}}
\thicklines \path(625,113)(625,133)
\thicklines \path(625,877)(625,857)
\put(625,68){\makebox(0,0){0.1}}
\thicklines \path(828,113)(828,133)
\thicklines \path(828,877)(828,857)
\put(828,68){\makebox(0,0){0.15}}
\thicklines \path(1031,113)(1031,133)
\thicklines \path(1031,877)(1031,857)
\put(1031,68){\makebox(0,0){0.2}}
\thicklines \path(1233,113)(1233,133)
\thicklines \path(1233,877)(1233,857)
\put(1233,68){\makebox(0,0){0.25}}
\thicklines \path(1436,113)(1436,133)
\thicklines \path(1436,877)(1436,857)
\put(1436,68){\makebox(0,0){0.3}}
\thicklines \path(220,113)(1436,113)(1436,877)(220,877)(220,113)
\put(45,945){\makebox(0,0)[l]{\shortstack{$V/{m'}_{0}^{2.5}$}}}
\put(828,23){\makebox(0,0){$\sigma/m'_{0}$}}
\put(504,597){\makebox(0,0)[l]{$R=K/2$}}
\put(706,495){\makebox(0,0)[l]{$R=0$}}
\put(787,419){\makebox(0,0)[l]{$R=-K/2$}}
\put(1112,240){\makebox(0,0)[l]{$R=-K$}}
\thinlines \path(220,368)(220,368)(222,368)(223,368)(225,368)(226,368)(228,368)(230,368)(233,368)(236,368)(239,368)(245,368)(252,368)(258,368)(271,368)(283,369)(296,369)(321,370)(347,371)(372,373)(423,377)(473,382)(524,389)(575,397)(625,407)(676,418)(727,431)(777,445)(828,460)(879,477)(929,496)(980,516)(1031,538)(1081,562)(1132,587)(1183,614)(1233,642)(1284,672)(1335,704)(1385,738)(1436,774)
\thinlines \path(220,368)(220,368)(222,374)(223,377)(226,380)(233,386)(245,393)(271,404)(321,421)(372,435)(423,449)(473,463)(524,478)(575,493)(625,509)(676,526)(727,545)(777,564)(828,585)(879,607)(929,631)(980,656)(1031,683)(1081,711)(1132,740)(1183,771)(1233,804)(1284,838)(1335,874)(1339,877)
\thinlines \path(220,368)(220,368)(222,361)(223,359)(226,355)(233,350)(239,346)(245,342)(271,332)(296,325)(321,319)(347,314)(372,310)(397,307)(423,305)(435,304)(448,303)(461,302)(473,302)(486,301)(492,301)(499,301)(502,301)(505,301)(508,301)(511,301)(513,301)(515,301)(516,301)(518,301)(519,301)(521,301)(522,301)(524,301)(526,301)(527,301)(530,301)(533,301)(537,301)(543,301)(549,301)(562,301)(575,302)(600,303)(625,305)(676,310)(727,316)(777,325)(828,335)(879,347)(929,361)
\thinlines \path(929,361)(980,376)(1031,394)(1081,413)(1132,433)(1183,456)(1233,481)(1284,507)(1335,535)(1385,565)(1436,597)
\thinlines \path(220,368)(220,368)(222,355)(223,350)(226,342)(233,332)(245,317)(258,305)(271,296)(321,268)(372,248)(423,232)(473,221)(499,216)(524,212)(549,209)(575,206)(600,204)(625,203)(638,202)(644,202)(651,202)(657,202)(663,201)(666,201)(670,201)(673,201)(674,201)(676,201)(678,201)(679,201)(681,201)(682,201)(684,201)(686,201)(687,201)(689,201)(690,201)(692,201)(695,201)(701,201)(708,202)(714,202)(727,202)(739,203)(752,203)(777,205)(803,207)(828,210)(879,217)(929,226)
\thinlines \path(929,226)(980,236)(1031,249)(1081,264)(1132,280)(1183,299)(1233,319)(1284,341)(1335,366)(1385,392)(1436,420)
\end{picture}

                \vglue 1ex
                \hspace*{15em}\mbox{$(a) D=2.5$}
                \vglue 7ex
\setlength{\unitlength}{0.240900pt}
\begin{picture}(1500,900)(0,0)
\tenrm
\thinlines \dottedline{14}(220,368)(1436,368)
\thicklines \path(220,113)(240,113)
\thicklines \path(1436,113)(1416,113)
\put(198,113){\makebox(0,0)[r]{$-0.0015$}}
\thicklines \path(220,198)(240,198)
\thicklines \path(1436,198)(1416,198)
\put(198,198){\makebox(0,0)[r]{$-0.001$}}
\thicklines \path(220,283)(240,283)
\thicklines \path(1436,283)(1416,283)
\put(198,283){\makebox(0,0)[r]{$-0.0005$}}
\thicklines \path(220,368)(240,368)
\thicklines \path(1436,368)(1416,368)
\put(198,368){\makebox(0,0)[r]{0}}
\thicklines \path(220,453)(240,453)
\thicklines \path(1436,453)(1416,453)
\put(198,453){\makebox(0,0)[r]{0.0005}}
\thicklines \path(220,537)(240,537)
\thicklines \path(1436,537)(1416,537)
\put(198,537){\makebox(0,0)[r]{0.001}}
\thicklines \path(220,622)(240,622)
\thicklines \path(1436,622)(1416,622)
\put(198,622){\makebox(0,0)[r]{0.0015}}
\thicklines \path(220,707)(240,707)
\thicklines \path(1436,707)(1416,707)
\put(198,707){\makebox(0,0)[r]{0.002}}
\thicklines \path(220,792)(240,792)
\thicklines \path(1436,792)(1416,792)
\put(198,792){\makebox(0,0)[r]{0.0025}}
\thicklines \path(220,877)(240,877)
\thicklines \path(1436,877)(1416,877)
\put(198,877){\makebox(0,0)[r]{0.003}}
\thicklines \path(220,113)(220,133)
\thicklines \path(220,877)(220,857)
\put(220,68){\makebox(0,0){0}}
\thicklines \path(423,113)(423,133)
\thicklines \path(423,877)(423,857)
\put(423,68){\makebox(0,0){0.05}}
\thicklines \path(625,113)(625,133)
\thicklines \path(625,877)(625,857)
\put(625,68){\makebox(0,0){0.1}}
\thicklines \path(828,113)(828,133)
\thicklines \path(828,877)(828,857)
\put(828,68){\makebox(0,0){0.15}}
\thicklines \path(1031,113)(1031,133)
\thicklines \path(1031,877)(1031,857)
\put(1031,68){\makebox(0,0){0.2}}
\thicklines \path(1233,113)(1233,133)
\thicklines \path(1233,877)(1233,857)
\put(1233,68){\makebox(0,0){0.25}}
\thicklines \path(1436,113)(1436,133)
\thicklines \path(1436,877)(1436,857)
\put(1436,68){\makebox(0,0){0.3}}
\thicklines \path(220,113)(1436,113)(1436,877)(220,877)(220,113)
\put(45,945){\makebox(0,0)[l]{\shortstack{$V/{m'}_{0}^{3.5}$}}}
\put(828,23){\makebox(0,0){$\sigma/m'_{0}$}}
\put(321,622){\makebox(0,0)[l]{$R=K/2$}}
\put(747,622){\makebox(0,0)[l]{$R=0$}}
\put(1071,622){\makebox(0,0)[l]{$R=-K/2$}}
\put(787,266){\makebox(0,0)[l]{$R=-K$}}
\thinlines \path(220,368)(220,368)(222,368)(223,368)(225,368)(226,368)(228,368)(230,368)(233,368)(236,368)(239,368)(245,368)(252,369)(258,369)(271,370)(283,372)(296,373)(321,378)(347,384)(372,391)(423,409)(473,433)(524,461)(575,496)(625,535)(676,580)(727,631)(777,688)(828,750)(879,819)(919,877)
\thinlines \path(220,368)(220,368)(222,368)(223,368)(226,368)(230,368)(233,369)(245,371)(258,373)(271,377)(296,385)(321,396)(372,425)(423,461)(473,506)(524,558)(575,617)(625,683)(676,757)(727,838)(749,877)
\thinlines \path(220,368)(220,368)(271,364)(321,359)(334,359)(347,358)(359,357)(372,357)(378,357)(382,357)(385,357)(388,357)(391,357)(393,356)(394,356)(396,356)(397,356)(399,356)(400,356)(402,356)(404,356)(405,356)(407,356)(410,356)(412,357)(413,357)(416,357)(423,357)(429,357)(435,357)(448,358)(461,358)(473,359)(499,362)(524,365)(575,374)(625,387)(676,404)(727,424)(777,449)(828,478)(879,512)(929,550)(980,593)(1031,641)(1081,693)(1132,752)(1183,815)(1228,877)
\thinlines \path(220,368)(220,368)(222,368)(223,367)(226,367)(233,366)(245,364)(271,357)(321,341)(372,323)(423,304)(473,286)(524,269)(575,253)(625,239)(676,227)(727,217)(752,213)(777,210)(790,209)(803,208)(815,207)(828,206)(834,206)(841,205)(847,205)(853,205)(856,205)(860,205)(863,205)(864,205)(866,205)(868,205)(869,205)(871,205)(872,205)(874,205)(876,205)(877,205)(879,205)(880,205)(882,205)(885,205)(888,205)(891,205)(898,205)(904,205)(917,206)(929,207)(955,209)(980,212)
\thinlines \path(980,212)(1005,216)(1031,222)(1081,235)(1132,252)(1183,273)(1233,298)(1284,329)(1335,364)(1385,403)(1436,448)
\end{picture}

                \vglue 1ex
                \hspace*{15em}\mbox{$(b) D=3.5$}
                \vglue 1ex
\caption{The behavior of the effective potential is shown at
         $D=2.5$ and $D=3.5$ for fixed $\lambda$ 
         $( \leq \lambda_{\mbox{cr}})$
         with the varying curvature where
         $K=6(D-2)(D(4-D)/4)^{(4-\sD)/(\sD-2)}{m'}_{0}^{2}>0$.}
\label{fig:potsc}
\end{figure}
We observe the second order phase transition as the
curvature decreases.
The chiral symmetry is always broken down for the negative
curvature even if the coupling constant $\lambda$
is very small.
Thus the critical curvature is given by $R_{\mbox{cr}}=0$ in the
whole range of the spacetime dimensions $D$
considered here, $2 \leq D < 4$.

In the case with $\lambda \leq \lambda_{\mbox{cr}}$ Eq. (\ref{nontri:w}) 
is modified to be
\begin{equation}
     -{m'}_{0}^{\sD-2}-m^{\sD-2}
     +\frac{R}{12}\left(1-\frac{D}{2}\right)m^{\sD-4}=0\, .
\label{nontri2:w}
\end{equation}
The dynamical mass of the fermion is given by the solution
of the Eq. (\ref{nontri2:w}).
In Fig. 11 we present the dynamical mass of the fermion as a
function of the spacetime dimensions $D$.
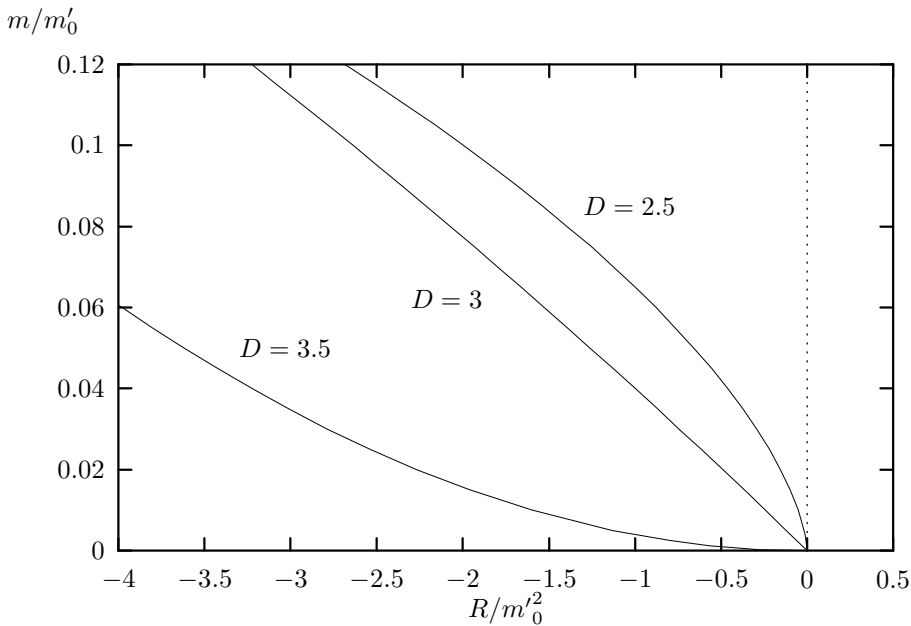
\begin{figure}
\vspace*{2ex}
\setlength{\unitlength}{0.240900pt}
\begin{picture}(1500,900)(0,0)
\tenrm
\thinlines \dottedline{14}(1301,113)(1301,877)
\thicklines \path(220,113)(240,113)
\thicklines \path(1436,113)(1416,113)
\put(198,113){\makebox(0,0)[r]{0}}
\thicklines \path(220,240)(240,240)
\thicklines \path(1436,240)(1416,240)
\put(198,240){\makebox(0,0)[r]{0.02}}
\thicklines \path(220,368)(240,368)
\thicklines \path(1436,368)(1416,368)
\put(198,368){\makebox(0,0)[r]{0.04}}
\thicklines \path(220,495)(240,495)
\thicklines \path(1436,495)(1416,495)
\put(198,495){\makebox(0,0)[r]{0.06}}
\thicklines \path(220,622)(240,622)
\thicklines \path(1436,622)(1416,622)
\put(198,622){\makebox(0,0)[r]{0.08}}
\thicklines \path(220,750)(240,750)
\thicklines \path(1436,750)(1416,750)
\put(198,750){\makebox(0,0)[r]{0.1}}
\thicklines \path(220,877)(240,877)
\thicklines \path(1436,877)(1416,877)
\put(198,877){\makebox(0,0)[r]{0.12}}
\thicklines \path(220,113)(220,133)
\thicklines \path(220,877)(220,857)
\put(220,68){\makebox(0,0){$-4$}}
\thicklines \path(355,113)(355,133)
\thicklines \path(355,877)(355,857)
\put(355,68){\makebox(0,0){$-3.5$}}
\thicklines \path(490,113)(490,133)
\thicklines \path(490,877)(490,857)
\put(490,68){\makebox(0,0){$-3$}}
\thicklines \path(625,113)(625,133)
\thicklines \path(625,877)(625,857)
\put(625,68){\makebox(0,0){$-2.5$}}
\thicklines \path(760,113)(760,133)
\thicklines \path(760,877)(760,857)
\put(760,68){\makebox(0,0){$-2$}}
\thicklines \path(896,113)(896,133)
\thicklines \path(896,877)(896,857)
\put(896,68){\makebox(0,0){$-1.5$}}
\thicklines \path(1031,113)(1031,133)
\thicklines \path(1031,877)(1031,857)
\put(1031,68){\makebox(0,0){$-1$}}
\thicklines \path(1166,113)(1166,133)
\thicklines \path(1166,877)(1166,857)
\put(1166,68){\makebox(0,0){$-0.5$}}
\thicklines \path(1301,113)(1301,133)
\thicklines \path(1301,877)(1301,857)
\put(1301,68){\makebox(0,0){0}}
\thicklines \path(1436,113)(1436,133)
\thicklines \path(1436,877)(1436,857)
\put(1436,68){\makebox(0,0){0.5}}
\thicklines \path(220,113)(1436,113)(1436,877)(220,877)(220,113)
\put(45,945){\makebox(0,0)[l]{\shortstack{$m / m'_{0}$}}}
\put(828,23){\makebox(0,0){$R/{m'}_{0}^2$}}
\put(950,654){\makebox(0,0)[l]{$D=2.5$}}
\put(679,508){\makebox(0,0)[l]{$D=3$}}
\put(409,431){\makebox(0,0)[l]{$D=3.5$}}
\thinlines \path(1301,113)(1301,113)(1301,114)(1301,115)(1301,117)(1301,119)(1300,121)(1299,129)(1298,137)(1296,145)(1287,177)(1274,209)(1259,240)(1242,272)(1222,304)(1200,336)(1176,368)(1151,400)(1123,431)(1094,463)(1064,495)(1031,527)(997,559)(962,591)(924,622)(886,654)(846,686)(804,718)(761,750)(717,782)(671,813)(624,845)(575,877)
\thinlines \path(1301,113)(1301,113)(1268,145)(1235,177)(1202,209)(1169,240)(1135,272)(1100,304)(1066,336)(1031,368)(996,400)(960,431)(925,463)(888,495)(852,527)(815,559)(778,591)(741,622)(703,654)(665,686)(626,718)(588,750)(548,782)(509,813)(469,845)(429,877)
\thinlines \path(1301,113)(1301,113)(1247,114)(1224,115)(1193,117)(1148,121)(1085,129)(995,145)(868,177)(770,209)(688,240)(615,272)(548,304)(487,336)(429,368)(375,400)(323,431)(274,463)(226,495)(220,499)
\end{picture}

\caption{Solutions of the gap equation for fixed $\lambda$
         smaller than $\lambda_{\mbox{cr}}$ at $D=2.5, 3, 3.5$.}
\label{fig:solsc}
\end{figure}
As is expected for the second order phase transition,
the dynamical mass of the fermion smoothly disappears at
$R=R_{\mbox{cr}}=0$.

\subsubsection{Phase structure in four dimensions}

We are interested in applying the results to the
critical phenomena at the early stage of the universe.
Thus we focus on the study of
the phase structure in four dimensions
here.

In four dimensions the four-fermion models are not
renormalizable. In Minkowski spacetime the correspondence
(\ref{corr:flat}) does not depend on $\sigma$.
Thus both the dimensional and cut-off regularization
describe the equivalent theory. 
The differences are removed by the finite renormalization. 
Therefore it is not necessary to investigate the theories
defined by the different regularizations in Minkowski
spacetime.

In curved spacetime the correspondence
(\ref{corr:wc}) depends on $\sigma$ and the curvature $R$.
Thus above results in weakly curved spacetime
may depend on the regularization methods.
We regard the theory for $D=4-\epsilon$
with $\epsilon$ sufficiently small positive as a
regularization of the one in four dimensions.
In the previous subsections we found that the broken chiral 
symmetry was restored
for a sufficiently large positive curvature $R > R_{\mbox{cr}}$
and the phase transition induced by curvature effects
was of the first order for a positive finite $\epsilon$.

To see whether the phase structure depends on the
regularization method
we define the theory by the cut-off
regularization.\cite{WC1}
In this case the effective potential is given by
Eqs. (\ref{v:nonren:cut2}) and (\ref{v:4d:cut:w}) and
the gap equation defined by Eq. (\ref{gap:w}) 
reads\footnote{In the present subsection we do not
adopt the renormalization procedure.
It is not necessary in four dimensions because
the theory is not renormalizable.
}
\begin{equation}
     \frac{1}{\lambda_{0}}-\frac{\tr\11}{(4\pi)^{2}}
                   \left[\Lambda^{2}
                   -m^{2}
                   \ln
     \left(\frac{\Lambda^{2}}{m^{2}}\right)
                   \right]
     -\frac{\tr\11}{192\pi^{2}}R
                   \left[
                   -\ln
     \left(\frac{\Lambda^{2}}{\sigma_{0}^{2}}\right)
                   +2
                   \right]
              =  0 \: .
\label{gap:cut:w4}
\end{equation}
The dynamical fermion mass is obtained by 
analyzing the gap equation (\ref{gap:cut:w4}).
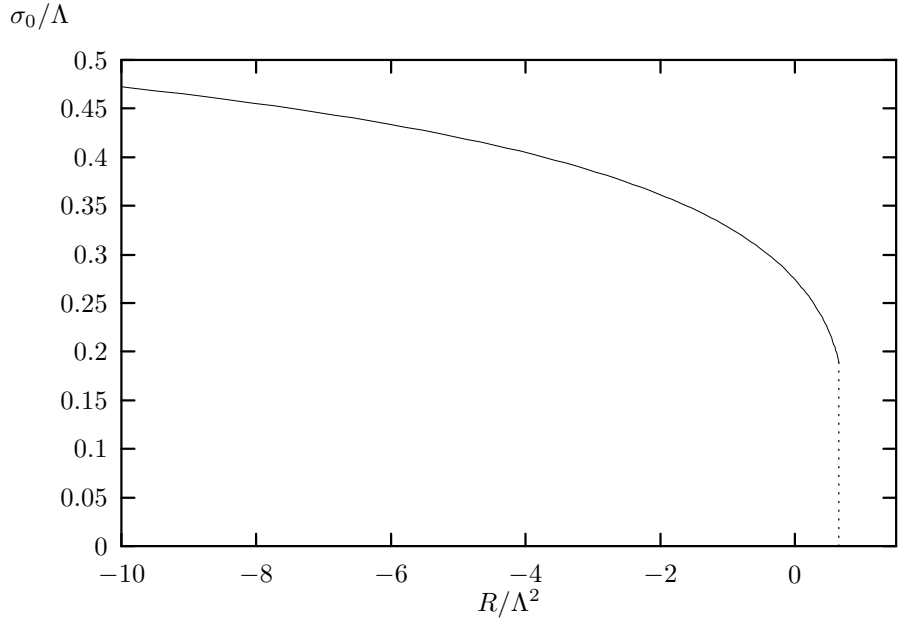
\begin{figure}
\vspace*{2ex}
\setlength{\unitlength}{0.240900pt}
\begin{picture}(1500,900)(0,0)
\tenrm
\thicklines \path(220,113)(240,113)
\thicklines \path(1436,113)(1416,113)
\put(198,113){\makebox(0,0)[r]{0}}
\thicklines \path(220,189)(240,189)
\thicklines \path(1436,189)(1416,189)
\put(198,189){\makebox(0,0)[r]{0.05}}
\thicklines \path(220,266)(240,266)
\thicklines \path(1436,266)(1416,266)
\put(198,266){\makebox(0,0)[r]{0.1}}
\thicklines \path(220,342)(240,342)
\thicklines \path(1436,342)(1416,342)
\put(198,342){\makebox(0,0)[r]{0.15}}
\thicklines \path(220,419)(240,419)
\thicklines \path(1436,419)(1416,419)
\put(198,419){\makebox(0,0)[r]{0.2}}
\thicklines \path(220,495)(240,495)
\thicklines \path(1436,495)(1416,495)
\put(198,495){\makebox(0,0)[r]{0.25}}
\thicklines \path(220,571)(240,571)
\thicklines \path(1436,571)(1416,571)
\put(198,571){\makebox(0,0)[r]{0.3}}
\thicklines \path(220,648)(240,648)
\thicklines \path(1436,648)(1416,648)
\put(198,648){\makebox(0,0)[r]{0.35}}
\thicklines \path(220,724)(240,724)
\thicklines \path(1436,724)(1416,724)
\put(198,724){\makebox(0,0)[r]{0.4}}
\thicklines \path(220,801)(240,801)
\thicklines \path(1436,801)(1416,801)
\put(198,801){\makebox(0,0)[r]{0.45}}
\thicklines \path(220,877)(240,877)
\thicklines \path(1436,877)(1416,877)
\put(198,877){\makebox(0,0)[r]{0.5}}
\thicklines \path(220,113)(220,133)
\thicklines \path(220,877)(220,857)
\put(220,68){\makebox(0,0){$-10$}}
\thicklines \path(431,113)(431,133)
\thicklines \path(431,877)(431,857)
\put(431,68){\makebox(0,0){$-8$}}
\thicklines \path(643,113)(643,133)
\thicklines \path(643,877)(643,857)
\put(643,68){\makebox(0,0){$-6$}}
\thicklines \path(854,113)(854,133)
\thicklines \path(854,877)(854,857)
\put(854,68){\makebox(0,0){$-4$}}
\thicklines \path(1066,113)(1066,133)
\thicklines \path(1066,877)(1066,857)
\put(1066,68){\makebox(0,0){$-2$}}
\thicklines \path(1277,113)(1277,133)
\thicklines \path(1277,877)(1277,857)
\put(1277,68){\makebox(0,0){0}}
\thicklines \path(220,113)(1436,113)(1436,877)(220,877)(220,113)
\put(45,945){\makebox(0,0)[l]{\shortstack{$\sigma_{0}/\Lambda$}}}
\put(828,23){\makebox(0,0){$R/\Lambda^{2}$}}
\thinlines \path(220,835)(235,833)(276,828)(316,824)(354,819)(391,814)(426,809)(460,805)(493,800)(524,795)(555,790)(584,786)(612,781)(640,776)(666,771)(691,767)(716,762)(739,757)(762,752)(784,748)(805,743)(826,738)(846,734)(865,729)(883,724)(901,719)(918,715)(935,710)(951,705)(966,700)(981,696)(996,691)(1010,686)(1023,681)(1036,677)(1049,672)(1061,667)(1073,662)(1084,658)(1095,653)(1106,648)(1116,644)(1126,639)(1136,634)(1145,629)(1154,625)(1162,620)(1171,615)(1179,610)(1186,606)(1194,601)
\thinlines \path(1194,601)(1201,596)(1208,591)(1215,587)(1221,582)(1228,577)(1234,572)(1240,568)(1245,563)(1251,558)(1256,554)(1261,549)(1266,544)(1270,539)(1275,535)(1279,530)(1283,525)(1287,520)(1291,516)(1295,511)(1299,506)(1302,501)(1305,497)(1308,492)(1311,487)(1314,482)(1317,478)(1320,473)(1322,468)(1324,464)(1327,459)(1329,454)(1331,449)(1333,445)(1335,440)(1336,435)(1338,430)(1340,426)(1341,421)(1343,416)(1344,411)(1345,407)(1346,402)
\dottedline{14}(1346,402)(1346,113)
\end{picture}

\vglue 1ex
\caption{The behavior of the dynamical fermion mass where
$\lambda_{0}=1.25\lambda_{\mbox{cr}}$ 
and $R_{\mbox{cr}}=0.656\Lambda^{2}$.}
\end{figure}
\begin{figure}
\vspace*{2ex}
\setlength{\unitlength}{0.240900pt}
\begin{picture}(1500,900)(0,0)
\tenrm
\thicklines \path(220,113)(240,113)
\thicklines \path(1436,113)(1416,113)
\put(198,113){\makebox(0,0)[r]{$-10$}}
\thicklines \path(220,240)(240,240)
\thicklines \path(1436,240)(1416,240)
\put(198,240){\makebox(0,0)[r]{$-5$}}
\thicklines \path(220,368)(240,368)
\thicklines \path(1436,368)(1416,368)
\put(198,368){\makebox(0,0)[r]{0}}
\thicklines \path(220,495)(240,495)
\thicklines \path(1436,495)(1416,495)
\put(198,495){\makebox(0,0)[r]{5}}
\thicklines \path(220,622)(240,622)
\thicklines \path(1436,622)(1416,622)
\put(198,622){\makebox(0,0)[r]{10}}
\thicklines \path(220,750)(240,750)
\thicklines \path(1436,750)(1416,750)
\put(198,750){\makebox(0,0)[r]{15}}
\thicklines \path(220,877)(240,877)
\thicklines \path(1436,877)(1416,877)
\put(198,877){\makebox(0,0)[r]{20}}
\thicklines \path(220,113)(220,133)
\thicklines \path(220,877)(220,857)
\put(220,68){\makebox(0,0){0}}
\thicklines \path(423,113)(423,133)
\thicklines \path(423,877)(423,857)
\put(423,68){\makebox(0,0){0.2}}
\thicklines \path(625,113)(625,133)
\thicklines \path(625,877)(625,857)
\put(625,68){\makebox(0,0){0.4}}
\thicklines \path(828,113)(828,133)
\thicklines \path(828,877)(828,857)
\put(828,68){\makebox(0,0){0.6}}
\thicklines \path(1031,113)(1031,133)
\thicklines \path(1031,877)(1031,857)
\put(1031,68){\makebox(0,0){0.8}}
\thicklines \path(1233,113)(1233,133)
\thicklines \path(1233,877)(1233,857)
\put(1233,68){\makebox(0,0){1}}
\thicklines \path(1436,113)(1436,133)
\thicklines \path(1436,877)(1436,857)
\put(1436,68){\makebox(0,0){1.2}}
\thicklines \path(220,113)(1436,113)(1436,877)(220,877)(220,113)
\put(45,945){\makebox(0,0)[l]{\shortstack{$R_{\mbox{cr}}/\Lambda^{2}$}}}
\put(828,23){\makebox(0,0){$\lambda^{R}_{\mbox{cr}}/\lambda_{\mbox{cr}}$}}
\put(808,673){\makebox(0,0)[l]{Symmetric phase}}
\put(321,368){\makebox(0,0)[l]{Broken phase}}
\thinlines \path(473,877)(473,874)(474,869)(475,865)(476,862)(477,856)(478,852)(479,849)(480,845)(481,842)(482,836)(483,832)(484,828)(485,825)(487,821)(488,817)(489,814)(490,810)(491,806)(492,802)(493,799)(494,797)(495,793)(496,789)(497,785)(498,781)(499,780)(500,776)(501,772)(502,770)(503,766)(504,762)(505,760)(506,756)(507,754)(508,750)(509,748)(510,744)(511,742)(512,738)(513,736)(514,734)(515,730)(516,728)(517,726)(518,722)(519,720)(520,718)(521,715)(522,711)(523,709)
\thinlines \path(523,709)(524,707)(525,705)(526,702)(527,700)(528,698)(529,694)(530,691)(531,689)(532,687)(533,684)(534,682)(535,680)(536,677)(537,675)(538,675)(539,672)(540,670)(541,667)(542,665)(543,663)(544,660)(545,660)(546,657)(547,655)(548,652)(549,652)(550,649)(551,646)(552,644)(553,643)(554,641)(555,638)(556,637)(557,635)(558,634)(559,631)(560,631)(561,628)(563,625)(564,625)(565,622)(566,621)(567,618)(568,618)(569,617)(570,614)(571,613)(572,610)(573,610)(574,607)
\thinlines \path(574,607)(575,606)(576,605)(577,602)(578,601)(579,600)(580,597)(581,596)(582,595)(583,595)(584,591)(585,590)(586,590)(587,589)(588,588)(589,584)(590,583)(591,582)(592,581)(593,580)(594,579)(595,576)(596,575)(597,574)(598,573)(599,571)(600,570)(601,569)(602,568)(603,567)(604,565)(605,564)(606,563)(607,562)(608,560)(609,559)(610,558)(611,557)(612,555)(613,554)(614,553)(615,551)(616,552)(617,551)(618,549)(619,548)(620,546)(621,545)(622,543)(623,544)(624,543)
\thinlines \path(624,543)(625,541)(626,540)(627,538)(628,539)(629,537)(630,536)(631,534)(632,535)(633,533)(634,532)(635,530)(636,531)(637,529)(639,527)(640,528)(641,526)(642,524)(643,525)(644,523)(645,524)(646,522)(647,520)(648,521)(649,519)(650,520)(651,518)(652,515)(653,517)(654,514)(655,515)(656,513)(733,461)(734,462)(735,461)(736,460)(737,460)(738,459)(739,458)(740,459)(741,458)(742,457)(743,456)(744,457)(745,456)(746,455)(747,455)(748,455)(749,454)(750,453)(751,453)
\thinlines \path(751,453)(752,452)(753,452)(754,452)(755,451)(756,451)(757,450)(758,450)(759,449)(760,449)(761,449)(762,448)(763,448)(764,447)(765,447)(766,446)(767,446)(768,445)(769,445)(770,445)(771,444)(772,444)(773,443)(774,443)(775,442)(776,442)(777,442)(778,441)(779,441)(780,441)(781,440)(782,440)(783,440)(784,439)(785,438)(786,438)(787,438)(788,437)(789,437)(791,437)(792,436)(793,436)(794,435)(795,435)(796,435)(797,435)(798,434)(799,434)(800,433)(801,433)(802,433)
\thinlines \path(802,433)(803,433)(804,432)(805,432)(806,431)(807,431)(808,431)(809,430)(810,430)(811,429)(812,429)(813,429)(814,429)(815,428)(816,428)(817,428)(818,427)(819,427)(820,427)(821,426)(822,426)(823,426)(824,425)(825,425)(826,425)(827,425)(828,424)(829,424)(830,424)(831,423)(832,423)(833,423)(834,422)(835,422)(836,422)(837,421)(838,421)(839,421)(840,421)(841,420)(842,420)(843,420)(844,420)(845,419)(846,419)(847,419)(848,418)(849,418)(850,418)(851,418)(852,417)
\thinlines \path(852,417)(853,417)(854,417)(855,417)(856,416)(857,416)(858,416)(859,416)(860,415)(861,415)(862,415)(863,414)(864,414)(865,414)(867,414)(868,413)(869,413)(870,413)(871,413)(872,412)(873,412)(874,412)(875,412)(876,411)(877,411)(878,411)(879,411)(880,410)(881,410)(882,410)(883,410)(884,410)(885,409)(886,409)(887,409)(888,409)(889,408)(890,408)(891,408)(892,408)(893,407)(894,407)(895,407)(896,407)(897,407)(898,406)(899,406)(900,406)(901,406)(902,405)(903,405)
\thinlines \path(903,405)(904,405)(905,405)(906,405)(907,404)(908,404)(909,404)(910,404)(911,404)(912,403)(913,403)(914,403)(915,403)(916,403)(917,402)(918,402)(919,402)(920,402)(921,402)(922,401)(923,401)(924,401)(925,401)(926,401)(927,400)(928,400)(929,400)(930,400)(931,400)(932,399)(933,399)(934,399)(935,399)(936,399)(937,398)(938,398)(939,398)(940,398)(941,398)(943,398)(944,397)(945,397)(946,397)(947,397)(948,397)(949,397)(950,396)(951,396)(952,396)(953,396)(954,396)
\thinlines \path(954,396)(955,395)(956,395)(957,395)(958,395)(959,395)(960,395)(961,394)(962,394)(963,394)(964,394)(965,394)(966,394)(967,393)(968,393)(969,393)(970,393)(971,393)(972,393)(973,392)(974,392)(975,392)(976,392)(977,392)(978,392)(979,392)(980,391)(981,391)(982,391)(983,391)(984,391)(985,391)(986,391)(987,390)(988,390)(989,390)(990,390)(991,390)(992,390)(993,389)(994,389)(995,389)(996,389)(997,389)(998,389)(999,389)(1000,388)(1001,388)(1002,388)(1003,388)(1004,388)
\thinlines \path(1004,388)(1005,388)(1006,388)(1007,387)(1008,387)(1009,387)(1010,387)(1011,387)(1012,387)(1013,387)(1014,386)(1015,386)(1016,386)(1017,386)(1019,386)(1020,386)(1021,386)(1022,386)(1023,385)(1024,385)(1025,385)(1026,385)(1027,385)(1028,385)(1029,385)(1030,385)(1031,384)(1032,384)(1033,384)(1034,384)(1035,384)(1036,384)(1037,384)(1038,384)(1039,383)(1040,383)(1041,383)(1042,383)(1043,383)(1044,383)(1045,383)(1046,383)(1047,382)(1048,382)(1049,382)(1050,382)(1051,382)(1052,382)(1053,382)(1054,382)(1055,381)
\thinlines \path(1055,381)(1056,381)(1057,381)(1058,381)(1059,381)(1060,381)(1061,381)(1062,381)(1063,381)(1064,380)(1065,380)(1066,380)(1067,380)(1068,380)(1069,380)(1070,380)(1071,380)(1072,380)(1073,379)(1074,379)(1075,379)(1076,379)(1077,379)(1078,379)(1079,379)(1080,379)(1081,379)(1082,379)(1083,378)(1084,378)(1085,378)(1086,378)(1087,378)(1088,378)(1089,378)(1090,378)(1091,378)(1092,378)(1093,377)(1095,377)(1096,377)(1097,377)(1098,377)(1099,377)(1100,377)(1101,377)(1102,377)(1103,377)(1104,376)(1105,376)(1106,376)
\thinlines \path(1106,376)(1107,376)(1108,376)(1109,376)(1110,376)(1111,376)(1112,376)(1113,376)(1114,376)(1115,375)(1116,375)(1117,375)(1118,375)(1119,375)(1120,375)(1121,375)(1122,375)(1123,375)(1124,375)(1125,375)(1126,374)(1127,374)(1128,374)(1129,374)(1130,374)(1131,374)(1132,374)(1133,374)(1134,374)(1135,374)(1136,374)(1137,374)(1138,373)(1139,373)(1140,373)(1141,373)(1142,373)(1143,373)(1144,373)(1145,373)(1146,373)(1147,373)(1148,373)(1149,373)(1150,373)(1151,372)(1152,372)(1153,372)(1154,372)(1155,372)(1156,372)
\thinlines \path(1156,372)(1157,372)(1158,372)(1159,372)(1160,372)(1161,372)(1162,372)(1163,372)(1164,371)(1165,371)(1166,371)(1167,371)(1168,371)(1169,371)(1171,371)(1172,371)(1173,371)(1174,371)(1175,371)(1176,371)(1177,371)(1178,371)(1179,371)(1180,370)(1181,370)(1182,370)(1183,370)(1184,370)(1185,370)(1186,370)(1187,370)(1188,370)(1189,370)(1190,370)(1191,370)(1192,370)(1193,370)(1194,370)(1195,370)(1196,369)(1197,369)(1198,369)(1199,369)(1200,369)(1201,369)(1202,369)(1203,369)(1204,369)(1205,369)(1206,369)(1207,369)
\thinlines \path(1207,369)(1208,369)(1209,369)(1210,369)(1211,369)(1212,369)(1213,369)(1214,368)(1215,368)(1216,368)(1217,368)(1218,368)(1219,368)(1220,368)(1221,368)(1222,368)(1223,368)(1224,368)(1225,368)(1226,368)(1227,368)(1228,368)(1229,368)(1230,368)(1231,368)(1232,368)(1233,368)(1436,368)
\end{picture}

\vglue 1ex
\caption{The phase diagram in $\lambda_{0}$-$R$ plane
for the four-fermion model defined by the cut-off regularization.}
\label{fig:gap:w4}
\end{figure}
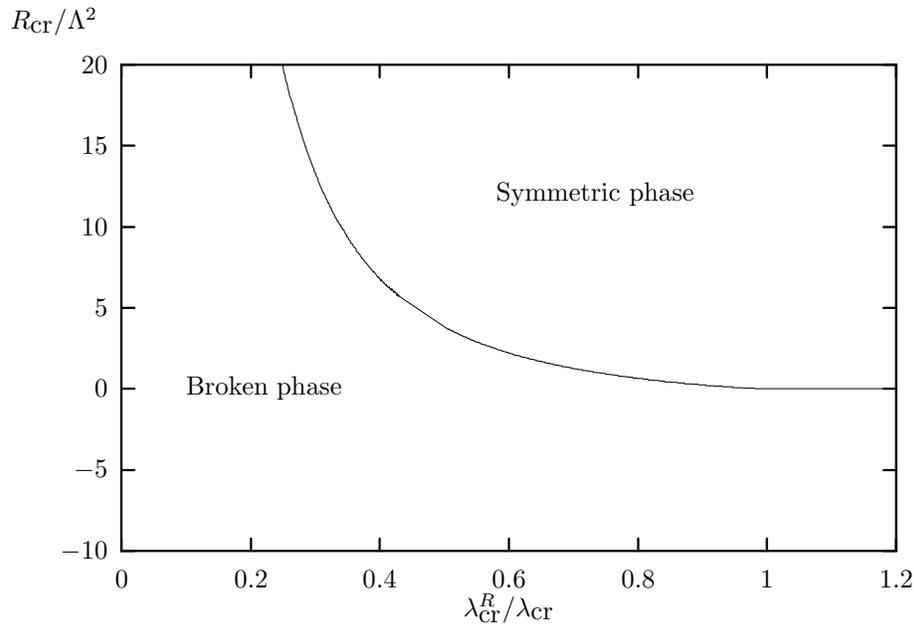
In Fig. 12 the dynamical mass of the fermion is plotted as
a function of the curvature $R$ for fixed $\lambda_{0}$.
The coupling constant $\lambda_{0}$ is kept in the range
$\lambda_{0} > \lambda_{\mbox{cr}}=4\pi^{2}/\Lambda^{2}$.
The behavior of the dynamical mass as shown in Fig. 12 is
characteristic for the relatively small coupling constant,
$\lambda_{\mbox{cr}} < \lambda_{0} \lsim 2\lambda_{\mbox{cr}}$.
For larger coupling, $\lambda_{0} \ssim 2\lambda_{\mbox{cr}}$, the
behavior near the
critical point $R=R_{\mbox{cr}}$ is quite different from the one
in Fig. 12:
the curve representing the dynamical mass is bent upward
near $R=R_{\mbox{cr}}$.
At any rate there is observed the gap in the dynamical mass
at the critical
point $R=R_{\mbox{cr}}$ reflecting the nature of the first order
phase transition.
By the direct numerical analysis we find that there is no
gap at $R=R_{\mbox{cr}}=0$ if 
$\lambda_{0} \leq \lambda_{\mbox{cr}}$ 
and so the phase transition is of the second order.

It is possible to obtain the critical value of the curvature
$R$ and the coupling constant $\lambda_{0}$ by observing the behavior of
the effective potential.
The critical values $R_{\mbox{cr}}$ and $\lambda^{R}_{\mbox{cr}}$ 
constitute
a critical curve in the $R-\lambda_{0}$ plane as shown in Fig. 13.
It is found from Fig. 13 that for large positive $R$ the
chiral symmetry is restored even if $\lambda$ is kept 
in the region of the broken phase for $R = 0$.
On the other hand the chiral symmetry is always broken for
negative $R$ irrespective of the value of $\lambda$.
Thus the similar phase structure was found in the theory
defined by the cut-off regularization.

To summarize this section we have employed the weak
curvature expansion to evaluate the effective potential
in the leading order of the $1/N$ expansion.
We assumed that the spacetime is weakly curved and 
have investigated the phase structure.

Starting from the theory with broken chiral symmetry
for vanishing $R$ we calculated the effective potential
for finite $R$  in arbitrary dimensions
$2 \leq D <4$.
We found that the chiral symmetry is restored for 
a large positive curvature.\footnote{
Large positive curvature may spoil the validity
of the weak curvature expansion. This point will be
discussed in \S 4.}
The critical curvature dividing the symmetric phase
and broken phase is obtained analytically.
Starting from the theory with chiral symmetry
for vanishing $R$ it was found that the symmetry is 
broken down in the spacetime with the negative curvature.
Therefore only a broken phase occurs
for the negative $R$.\cite{WC1,WC2}

\section{Exact solutions in special spacetimes}

In the previous section the curvature induced 
phase transition was discussed by using the weak 
curvature approximation.
It is found that the broken chiral symmetry
is restored for a large positive curvature.
But the large positive curvature may spoil the
validity of the weak curvature approximation.
To study the phase transition in the large positive
curvature in more confident manner we need to 
avoid any approximation in dealing with the spacetime 
curvature.
In some specific spacetimes the effective
potential is calculable rigorously in the leading order 
of the $1/N$ expansion without using weak curvature
approximation because of the symmetry of the spacetime.

In the present section we try to find the effective
potential of the four-fermion model
without making any approximation in the spacetime curvature.
For this purpose we restrict ourselves to specific
spacetimes, i.e., maximally symmetric spacetime
($S^{D},H^{D}$) \cite{IMM,SACH,ELOS,b10} 
and Einstein universe 
($R\otimes S^{D-1}$) \cite{DA1,DA2,DA3,IIM} 
and calculate the effective
potential in the leading order of the $1/N$ expansion.
The phase structure of the four-fermion model is obtained
by using the same method developed in \S 2 and \S 3.
To justify the weak curvature approximation
we compare the result with the one obtained in \S 3.
It is shown that the weak curvature approximation
gives the exact result \cite{WC2,SACH,IIM} in the limit 
$D \rightarrow 4$ or $\Lambda\rightarrow\infty$.
We follow here Refs. \cite{IMM,IIM} and \cite{IIM2}.

\subsection{Spinor two-point function in $S^{D}$ and $H^{D}$}

As we have seen in \S 2 the effective potential for
the composite field in the leading order of the $1/N$
expansion for the Gross-Neveu type model 
is described by the two-point function $S(x,x;s)$ 
of the massive free fermion in curved spacetime.
Thus we start with the analysis of the two-point function
$S(x,y;s)$ in the maximally symmetric spacetime.

In the maximally symmetric spacetime $S^{D}$ and $H^{D}$ 
$(2 \leq D < 4)$
the exact expression of the two-point function is known
without making any approximation with respect to
the spacetime curvature.\cite{CR,ALL,AJ,CA}
We then restrict ourselves to the maximally symmetric
spacetime and closely follow the method developed by 
Camporesi.\cite{CA}
Here we consider the manifolds $S^{D}$ and $H^{D}$.
The manifold $S^{D}$ is defined by the metric
\begin{equation}
     ds^{2}=a^{2}(d\theta^{2}+\sin^{2}\theta d\Omega_{\sD-1})\, ,
\end{equation}
where $d\Omega_{\sD-1}$ is the metric on a unit sphere
$S^{\sD-1}$ while the manifold $H^{D}$ is defined by
\begin{equation}
     ds^{2}=a^{2}(d\theta^{2}+\sinh^{2}\theta d\Omega_{\sD-1})\, .
\end{equation}
The manifold $S^{D}$ and $H^{D}$ are constant curvature spacetimes
with positive and negative curvature
\begin{equation}
     R=\pm D(D-1)a^{-2}\, ,
\end{equation}
respectively ($2 \leq D < 4$).

The spinor two-point function $S(x,y;s)$ is defined by the Dirac
equation
\begin{equation}
     (\nab +s)S(x,y;s)=-\frac{1}{\sqrt{g}}\delta^{D}(x,y)\, ,
\label{eq:D}
\end{equation}
where $\delta^{D}(x,y)$ is the Dirac delta function in the 
maximally symmetric spacetime.
We introduce the bispinor function $G$ defined by
\begin{equation}
     (\nab -s)G(x,y;s)=S(x,y;s)\, .
\label{DfromG}
\end{equation}
According to Eq. (\ref{eq:D}) $G(x,y;s)$ satisfies the following
equation,
\begin{equation}
     (\nab\nab -s^{2})G(x,y;s)=-\frac{1}{\sqrt{g}}
     \delta^{D}(x,y)\, .
\label{eq:Gsp}
\end{equation}
On $S^{D}$ and $H^{D}$ we rewrite Eq. (\ref{eq:Gsp}) in the
following form
\begin{equation}
     \left(\Box_{D}-\frac{R}{4}-s^{2}\right)G(x,y;s)
     =-\frac{1}{\sqrt{g}}\delta^{D}(x,y)\, ,
\label{eq:G}
\end{equation}
where $\Box_{D}$ is the Laplacian on the maximally symmetric
spacetime.

The general form of the Green function $G(x,y;s)$ 
is written as\cite{CA}
\begin{equation}
     G(x,y;s)=U(x,y)g_{\sD}(l)\, ,
\label{ansatz}
\end{equation}
where $U$ is a matrix in the spinor indices,
$g_{\sD}$ is a scalar function depending only
on $l$, $l = a\theta$ which is
the geodesic distance between $x$ and $y$ on 
the maximally symmetric spacetime,
$n_{i}$ is a unit vector tangent to the geodesic
$n_{i}=\nabla_{i}l$.
Inserting Eq. (\ref{ansatz}) into Eq. (\ref{eq:G}) we get
\begin{equation}
     \left[U\Box_{\sD}g_{\sD}
     +2(\nabla_{j}U)\nabla^{j}g_{\sD}
     +(\Box_{\sD}U)g_{\sD}
     -\left(\frac{R}{4}+s^{2}\right)U
     g_{\sD}\right]=0\, ,
\label{eq:sn}
\end{equation}
where we restrict
ourselves to the region $l\neq 0$.
To evaluate Eq. (\ref{eq:sn}) we have to calculate the
covariant derivative of $U$ and $n_{i}$.
$U$ is the operator which makes parallel transport of the
spinor at
point $x$ along the geodesic to point $y$.
Thus the operator $U$ must satisfy the following parallel
transport
equations:\cite{ALL}
\begin{equation}
\left\{
\begin{array}{rcl}
     n^{i}\nabla_{i}U&=&0\, ,\\[4mm]
     U(x,x)&=&\11\, .
\end{array}
\right.
\label{eq:para}
\end{equation}
To evaluate the second derivative of $U$ we set
\begin{equation}
     \nabla_{i}U \equiv V_{i}U\, .
\end{equation}
From the integrability condition\cite{CR} on $V_{i}$,
\begin{equation}
     \nabla_{i}V_{j}-\nabla_{j}V_{i}-[V_{i},V_{j}]
     =\frac{R}{D(D-1)}\sigma_{ij}\, ,
\label{cond:int}
\end{equation}
and the parallel transport equation (\ref{eq:para}) we
easily find that
\begin{equation}
     V_{i}=-\frac{1}{a}\tan\left(\frac{l}{2a}\right)\sigma_{ij}n^{j}\, ;
     \mbox{on }S^{D},
\end{equation}
\begin{equation}
     V_{i}=\frac{1}{a}\tanh\left(\frac{l}{2a}\right)\sigma_{ij}n^{j}\, :
     \mbox{on }H^{D},
\end{equation}
where $\sigma_{ij}$ are the antisymmetric tensors
constructed by the Dirac gamma matrices,
$\sigma_{ij}=\frac{1}{4}[\gamma_{i},\gamma_{j}]$.
To find $V_{i}$ we have used the fact that the maximally
symmetric bitensors are represented as a sum of products 
of $n_{i}$ and
$g_{ij}$ with coefficients which are functions 
only of $l$.\cite{AJ}
After some calculations we get the Laplacian acting on $U$
\begin{equation}
\begin{array}{ll}
     \displaystyle
     \Box_{\sD}U=-\frac{D-1}{4a^{2}}
     \tan^{2}\left(\frac{l}{2a}\right)U\, ,& \mbox{on }S^{D},\\[4mm]
     \displaystyle
     \Box_{\sD}U=-\frac{D-1}{4a^{2}}
     \tanh^{2}\left(\frac{l}{2a}\right)U\, ,& \mbox{on }H^{D}.
\end{array}
\label{lap:U}
\end{equation}
The derivative of $n_{i}$ is also the
maximally symmetric
bitensor \cite{AJ} and found to be 
\begin{equation}
\begin{array}{rcll}
     \nabla_{i}n_{j}&=&\displaystyle
     \frac{1}{a}\cot\left(\frac{l}{a}\right)
     (g_{ij}-n_{i}n_{j})\, ,& \,\,\mbox{on }S^{D},\\[4mm]
     \nabla_{i}n_{j}&=&\displaystyle
     \frac{1}{a}\coth\left(\frac{l}{a}\right)
     (g_{ij}-n_{i}n_{j})\, ,& \,\,\mbox{on }H^{D}.
\end{array}
\label{der:n}
\end{equation}
Therefore Eq. (\ref{eq:sn}) reads
\begin{equation}
     \left({\partial_{l}}^{2}
          +\frac{D-1}{a}\cot\left(\frac{l}{a}\right)\partial_{l}
          -\frac{D-1}{4a^{2}}\tan^{2}\left(\frac{l}{2a}\right)
          -\frac{R}{4}-s^{2}
     \right)g_{\sD}=0\, , \mbox{on }S^{D}, 
\label{eq:sig:s}
\end{equation}
\begin{equation}
     \left({\partial_{l}}^{2}
          +\frac{D-1}{a}\coth\left(\frac{l}{a}\right)\partial_{l}
          -\frac{D-1}{4a^{2}}\tanh^{2}\left(\frac{l}{2a}\right)
          -\frac{R}{4}-s^{2}
     \right)g_{\sD}=0\, , \mbox{on }H^{D}.
\label{eq:sig:h}
\end{equation}
We define the functions $h_{SD}(l)$ and $h_{HD}(l)$
by $\displaystyle g_{\sD}(l)
=\cos\left(l/2a\right)h_{SD}(l)$ and
$\displaystyle g_{\sD}(l)
=\cosh\left(l/2a\right)h_{HD}(l)$
respectively and make a change of variable by
$\displaystyle z=\cos^{2}\left(l/2a\right)$
in Eq. (\ref{eq:sig:s})
and $\displaystyle z'=\cosh^{2}\left(l/2a\right)$
in Eq. (\ref{eq:sig:h}).
We then find that Eqs. (\ref{eq:sig:s}) and (\ref{eq:sig:h}) 
are rewritten in the forms of
hypergeometric differential equations:
\begin{equation}
     \left[z(1-z){\partial_{z}}^{2}
           +\left(\frac{D+2}{2}-(D+1)z\right)\partial_{z}
           -\frac{D^{2}}{4}-s^{2}a^{2}
     \right]h_{SD}(z)=0\, ,
\label{eq:HGsp:s}
\end{equation}
\begin{equation}
     \left[z'(1-z'){\partial_{z'}}^{2}
           +\left(\frac{D+2}{2}-(D+1)z'\right)\partial_{z'}
           -\frac{D^{2}}{4}+s^{2}a^{2}
     \right]h_{HD}(z')=0\, .
\label{eq:HGsp:h}
\end{equation}
Noting that the Green functions are regular at the point
$l=a\pi$ and fall off for $l\rightarrow\infty$
we write the solutions of Eqs. (\ref{eq:HGsp:s}) and
(\ref{eq:HGsp:h}) by the hypergeometric function,
\begin{equation}
     h_{SD}(z)
     =c_{SD}F\left(\frac{D}{2}+i s a, \frac{D}{2}-i s a,
             \frac{D+2}{2};z\right)\, ,
\label{eq:h:s}
\end{equation}
\begin{equation}
     h_{HD}(z')
     =c_{HD}(-z')^{-D/2-sa}F\left(\frac{D}{2}+s a, s a,
             2 s a+1;\frac{1}{z'}\right)\, .
\label{eq:h:h}
\end{equation}
As we remained in the region where $l\neq 0$ the
normalization
constants $c_{SD}$ and $c_{HD}$ are yet
undetermined.
To obtain $c_{SD}$ and $c_{HD}$ we consider
the singularity of $G(x,y;s)$ in the limit
$l \rightarrow 0$,
\begin{eqnarray}
     G&\longrightarrow& c_{SD}
     \frac{\displaystyle \Gamma\left(\frac{D+2}{2}\right)
                         \Gamma\left(\frac{D-2}{2}\right)}
          {\displaystyle \Gamma\left(\frac{D}{2}+i s
     a\right)
                         \Gamma\left(\frac{D}{2}-i s
     a\right)}
     \left(\frac{l}{2a}\right)^{2-\sD}\, ,\nonumber \\
     G&\longrightarrow& c_{HD}(-1)^{-D/2-sa}
     \frac{\displaystyle \Gamma\left(2sa+1\right)
                         \Gamma\left(\frac{D-2}{2}\right)}
          {\displaystyle \Gamma\left(\frac{D}{2}+ s a\right)
                         \Gamma\left(s a\right)}
     \left(\frac{l}{2a}\right)^{2-\sD}\, ,
\label{lim:g}
\end{eqnarray}
and compare them with the singularity of the Green function
in the flat spacetime.
This procedure is justified because the singularity on a
curved spacetime
background has the same structure as that in the flat
spacetime.
For $l \sim 0$ the Green function in the flat spacetime
behaves as
\begin{equation}
     G^{\mbox{flat}}(l)\sim\frac{1}{4\pi^{\sD/2}}
                        \Gamma\left(\frac{D-2}{2}\right)
                        l^{2-\sD}\, .
\label{fla:g}
\end{equation}
Comparing Eq. (\ref{lim:g}) with Eq. (\ref{fla:g}) the
over-all factors $c_{SD}$ and $c_{HD}$ are obtained:
\begin{eqnarray}
     c_{SD}&=&\frac{a^{2-\sD}}{(4\pi)^{\sD /2}}
     \frac{\displaystyle \Gamma\left(\frac{D}{2}+i s a\right)
                         \Gamma\left(\frac{D}{2}-i s a\right)}
          {\displaystyle \Gamma\left(\frac{D+2}{2}\right)}\, ,
     \nonumber \\
     c_{HD}&=&(-1)^{D/2+sa}\frac{a^{2-\sD}}{(4\pi)^{\sD /2}}
     \frac{\displaystyle \Gamma\left(\frac{D}{2}+s a\right)
                         \Gamma\left(sa \right)}
          {\displaystyle \Gamma\left( 2sa+1 \right)}\, .
\label{const}
\end{eqnarray}
Inserting the Eqs. (\ref{eq:h:s}), (\ref{eq:h:h}) and (\ref{const}) into
Eq. (\ref{ansatz})
we find on $S^{D}$
\begin{equation}
\begin{array}{rcl}
    \displaystyle G(x,y;s)&=&
    \displaystyle U(x,y)\frac{a^{2-\sD}}{(4\pi)^{\sD/2}}
     \frac{\displaystyle\Gamma\left(\frac{D}{2}+i s a\right)
                        \Gamma\left(\frac{D}{2}-i s
a\right)}
          {\displaystyle\Gamma\left(\frac{D+2}{2}\right)}\\[5mm]
    &&\displaystyle\times\cos\left(\frac{l}{2a}\right)
      F\left(\frac{D}{2}+i s a, \frac{D}{2}-i s a,
             \frac{D+2}{2};\cos^{2}\left(\frac{l}{2a}\right)
       \right) ,
\end{array}
\label{sol:g:s}
\end{equation}
and on $H^{D}$
\begin{equation}
\begin{array}{rcl}
    \displaystyle G(x,y;s)&=&
    \displaystyle U(x,y)\frac{a^{2-\sD}}{(4\pi)^{\sD/2}}
     \frac{\displaystyle\Gamma\left(\frac{D}{2}+s a\right)
                        \Gamma\left(sa\right)}
          {\displaystyle\Gamma\left(2sa+1\right)}\\[4mm]
    &&\displaystyle
      \times\left[\cosh\left(\frac{l}{2a}
       \right)\right]^{1-D-2sa}
      F\left(\frac{D}{2}+s a, s a,
             2sa+1;\cosh^{-2}\left(\frac{l}{2a}\right)
       \right) .
\end{array}
\label{sol:g:h}
\end{equation}
Thus the Green functions $G(x,y;s)$ on the maximally
symmetric spacetime are obtained.

The spinor two-point function $S(x,y;s)$ is derived from the
Green function $G(x,y;s)$.
From the Eq. (\ref{DfromG})
we get
\begin{eqnarray}
     S&=&(\gamma^{i}\nabla_{i}-s)Ug_{\sD} \nonumber \\
     &=&\left\{
\begin{array}{ll}
            \displaystyle
            \gamma_{i}n^{i}U\left(\partial_{l}
           -\frac{D-1}{2a}\tan\left(\frac{l}{2a}\right)\right)
           g_{\sD}-s U g_{\sD}\, ,& \mbox{on } S^{D}\, ,\\[2mm]
            \displaystyle
            \gamma_{i}n^{i}U\left(\partial_{l}
           +\frac{D-1}{2a}\tanh\left(\frac{l}{2a}\right)\right)
           g_{\sD}-s U g_{\sD}\, ,& \mbox{on } H^{D}\, .
\end{array}
     \right.
\label{eq:D:fin}
\end{eqnarray}
Substituting Eqs. (\ref{sol:g:s}) and (\ref{sol:g:h}) 
in Eq. (\ref{eq:D:fin}) the spinor
two-point function $S(x,y;s)$ is obtained \cite{CA}
\begin{equation}
\begin{array}{rcl}
    \displaystyle
S(x,y;s)&=&\displaystyle-\frac{a^{2-\sD}}{(4\pi)^{\sD/2}}
     \frac{\displaystyle\Gamma\left(\frac{D}{2}+i s a\right)
                        \Gamma\left(\frac{D}{2}-i s
a\right)}
          {\displaystyle\Gamma\left(\frac{D+2}{2}\right)}\\[5mm]
    &&\displaystyle
       \times\biggl[s U(x,y)
       \cos\left(\frac{l}{2a}\right)
      F\left(\frac{D}{2}+i s a, \frac{D}{2}-i s a,
             \frac{D+2}{2};\cos^{2}\left(\frac{l}{2a}\right)
       \right)\\[5mm]
    &&\displaystyle
       +\gamma_{i}n^{i}U(x,y)\frac{D}{2a}\sin\left(\frac{l}{2a}\right)
\\[5mm]
    &&\displaystyle\times
    F\left(\frac{D}{2}+i s a, \frac{D}{2}-i s a,
             \frac{D}{2};\cos^{2}\left(\frac{l}{2a}\right)
       \right)\biggr]\, ,
\end{array}
\label{stf:s}
\end{equation}
on $S^{D}$ and
\begin{equation}
\begin{array}{rcl}
    \displaystyle
S(x,y;s)&=&\displaystyle-\frac{a^{1-\sD}}{(4\pi)^{\sD/2}}
     \frac{\displaystyle\Gamma\left(\frac{D}{2}+s a\right)
                        \Gamma\left(sa+1\right)}
          {\displaystyle\Gamma\left(2sa+1\right)}
      \left[
      \cosh\left(\frac{l}{2a}\right)\right]^{1-D-2sa}\\[5mm]
    &&\displaystyle
       \times\biggl[U(x,y)
       \cosh\left(\frac{l}{2a}\right)
      F\left(\frac{D}{2}+s a, s a,
             2sa+1;\cosh^{-2}\left(\frac{l}{2a}\right)
       \right)\\[5mm]
    &&\displaystyle
       +\gamma_{i}n^{i}U(x,y)\sinh\left(\frac{l}{2a}\right)\\[5mm]
    &&\displaystyle \times
      F\left(\frac{D}{2}+s a, s a+1,
             2sa+1;\cosh^{-2}\left(\frac{l}{2a}\right)
       \right)\biggr]\, ,
\end{array}
\label{stf:h}
\end{equation}
on $H^{D}$.
According to the anticommutation relation of spinor fields
the two-point function (\ref{stf:s})
satisfies the antiperiodic
boundary condition $S(l)=-S(l +2 \pi n a)$
where $n$ is an arbitrary integer.

We succeeded to calculate the spinor two-point functions
without making any approximation in the spacetime
curvature. Using these functions we evaluate the exact expression
of the effective potential in curved spacetime in the following
subsections.

\subsection{de Sitter background ($S^{D}$)}

First we consider the four-fermion
model in de Sitter space \cite{IMM,SACH,ELOS}.
The $D$-dimensional de Sitter space is represented as a
hyperboloid,
\begin{equation}
     a^2={\xi_{0}}^{2}-{\xi_{1}}^{2}-\cdots-{\xi_{\sD}}^{2}\, ,
\label{def:desitter}
\end{equation}
embedded in the $(D+1)$-dimensional Minkowski space.
It is one of the maximally symmetric spacetime.
The de Sitter space is a constant curvature spacetime with
curvature
\begin{equation}
     R=D(D-1) a^{-2}\, .
\label{corr:ra}
\end{equation}

We consider the manifold $S^{\sD}$ as a Euclidean analog
of the $D$-dimensional de Sitter space.
As is mentioned in the previous subsection the spinor
two-point function $S(x,y;s)$ is known on the manifold 
$S^{D}$. $\tr S(x,x;s)$ is required in evaluating the effective
potential:
\begin{equation}
     \tr S(x,x;s)=\displaystyle -
     \frac{\tr\11 s a^{2-\sD}}{(4\pi)^{\sD/2}}
     \Gamma\left(1-\frac{D}{2}\right)
     \frac{\displaystyle \Gamma\left(\frac{D}{2}+i s
     a\right)
                         \Gamma\left(\frac{D}{2}-i s a\right)}
          {\displaystyle \Gamma\left(1+i s a\right)
                         \Gamma\left(1-i s a\right) }\, .
\label{trD:de}
\end{equation}
Performing the Wick rotation and inserting Eq. (\ref{trD:de})
into Eq. (\ref{v:gn}) our final
expression of the effective potential is obtained.\cite{IMM}
\begin{equation}
     V(\sigma)=\frac{1}{2\lambda_{0}}\sigma^{2}-
     \frac{\tr\11 a^{2-\sD}}{(4\pi)^{\sD/2}}
     \Gamma\left(1-\frac{D}{2}\right)\int^{\sigma}_{0}s ds
     \frac{\displaystyle \Gamma\left(\frac{D}{2}+i s a\right)
     \Gamma\left(\frac{D}{2}-i s a\right)}
     {\displaystyle \Gamma\left(1+i s a\right)
     \Gamma\left(1-i s a\right)}
     \, .
\label{v:nonren:de}
\end{equation}
Equation (\ref{v:nonren:de}) is an exact expression of the
effective potential for the model of four-fermion
interactions
in de Sitter space in the leading order of the $1/N$
expansion.
Substituting the renormalized coupling constant $\lambda$
defined
by Eq. (\ref{eqn:ren})
in the Eq. (\ref{v:nonren:de}) we find the renormalized
expression of the
effective potential $V(\sigma)$ in de Sitter space,
\begin{eqnarray}
     V(\sigma)&=&\frac{1}{2\lambda}\sigma^{2}\mu^{\sD-2}
     +\frac{\tr\11}{2(4\pi)^{\sD /2}}(D-1)\Gamma\left(
     1-\frac{D}{2}\right)\sigma^{2}\mu^{\sD-2}\nonumber \\
     &&-\frac{\tr\11 a^{2-\sD}}{(4\pi)^{\sD/2}}
     \Gamma\left(1-\frac{D}{2}\right)
     \int^{\sigma}_{0} s ds
     \frac{\displaystyle \Gamma\left(\frac{D}{2}+i s a\right)
     \Gamma\left(\frac{D}{2}-i s a\right)}
     {\displaystyle \Gamma\left(1+i s a\right)
     \Gamma\left(1-i s a\right)}\, .
\label{v:ren:de}
\end{eqnarray}
Note that Eq. (\ref{v:ren:de}) reduces to
\begin{eqnarray}
     \frac{V^{D=2}(\sigma)}{\mu^{2}}&=&
\left[\frac{1}{2\lambda}\right.
     -\left.\frac{\tr\11}{8\pi}\left(\ln(a\mu)^{2}+2\right)
\right]
     \left(\frac{\sigma}{\mu}\right)^{2}
\nonumber \\
     &&+\frac{\tr\11}{4\pi{\mu}^{2}} \int^{\sigma}_{0} ds s
     \left[ \psi(1 + isa)
       +\psi(1 - isa) \right]\, ,
\end{eqnarray}
in two dimensions.
Expanding Eq. (\ref{v:ren:de}) asymptotically around
$1/a=0$
(weak curvature expansion) the effective potential
(\ref{v:ren}) is reproduced.
\begin{figure}
\setlength{\unitlength}{0.240900pt}
\begin{picture}(1500,900)(0,0)
\tenrm
\thicklines \path(220,113)(240,113)
\thicklines \path(1436,113)(1416,113)
\put(198,113){\makebox(0,0)[r]{$-0.01$}}
\thicklines \path(220,240)(240,240)
\thicklines \path(1436,240)(1416,240)
\put(198,240){\makebox(0,0)[r]{0}}
\thicklines \path(220,368)(240,368)
\thicklines \path(1436,368)(1416,368)
\put(198,368){\makebox(0,0)[r]{0.01}}
\thicklines \path(220,495)(240,495)
\thicklines \path(1436,495)(1416,495)
\put(198,495){\makebox(0,0)[r]{0.02}}
\thicklines \path(220,622)(240,622)
\thicklines \path(1436,622)(1416,622)
\put(198,622){\makebox(0,0)[r]{0.03}}
\thicklines \path(220,750)(240,750)
\thicklines \path(1436,750)(1416,750)
\put(198,750){\makebox(0,0)[r]{0.04}}
\thicklines \path(220,877)(240,877)
\thicklines \path(1436,877)(1416,877)
\put(198,877){\makebox(0,0)[r]{0.05}}
\thicklines \path(220,113)(220,133)
\thicklines \path(220,877)(220,857)
\put(220,68){\makebox(0,0){0}}
\thicklines \path(423,113)(423,133)
\thicklines \path(423,877)(423,857)
\put(423,68){\makebox(0,0){0.05}}
\thicklines \path(625,113)(625,133)
\thicklines \path(625,877)(625,857)
\put(625,68){\makebox(0,0){0.1}}
\thicklines \path(828,113)(828,133)
\thicklines \path(828,877)(828,857)
\put(828,68){\makebox(0,0){0.15}}
\thicklines \path(1031,113)(1031,133)
\thicklines \path(1031,877)(1031,857)
\put(1031,68){\makebox(0,0){0.2}}
\thicklines \path(1233,113)(1233,133)
\thicklines \path(1233,877)(1233,857)
\put(1233,68){\makebox(0,0){0.25}}
\thicklines \path(1436,113)(1436,133)
\thicklines \path(1436,877)(1436,857)
\put(1436,68){\makebox(0,0){0.3}}
\thicklines \path(220,113)(1436,113)(1436,877)(220,877)(220,113)
\put(45,945){\makebox(0,0)[l]{\shortstack{$V/{m'}_{0}^{2.5}$}}}
\put(828,23){\makebox(0,0){$\sigma/{m'}_{0}$}}
\put(463,368){\makebox(0,0)[l]{$R=3K/2$}}
\put(1051,368){\makebox(0,0)[l]{$R=0$}}
\thinlines \path(220,240)(220,240)(222,240)(223,240)(225,240)(226,240)(228,240)(230,240)(233,240)(236,240)(239,240)(245,240)(252,241)(258,241)(271,241)(283,241)(296,242)(321,243)(347,244)(372,245)(423,250)(473,255)(524,262)(575,270)(625,280)(676,291)(727,303)(777,317)(828,333)(879,350)(929,369)(980,389)(1031,411)(1081,434)(1132,459)(1183,486)(1233,515)(1284,545)(1335,577)(1385,611)(1436,646)
\thinlines \path(220,240)(220,240)(222,240)(223,240)(225,240)(226,240)(228,240)(230,240)(233,240)(236,240)(239,240)(245,241)(252,241)(258,241)(271,241)(283,242)(296,242)(321,244)(347,246)(372,248)(423,255)(473,263)(524,273)(575,285)(625,298)(676,314)(727,331)(777,350)(828,371)(879,393)(929,418)(980,444)(1031,472)(1081,502)(1132,534)(1183,568)(1233,603)(1284,641)(1335,680)(1385,721)(1436,764)
\thinlines \path(220,240)(220,240)(222,240)(223,240)(225,240)(226,240)(228,240)(230,240)(233,240)(236,240)(239,240)(245,241)(252,241)(258,241)(271,241)(283,242)(296,243)(321,244)(347,246)(372,249)(423,256)(473,265)(524,276)(575,288)(625,303)(676,320)(727,338)(777,359)(828,382)(879,406)(929,433)(980,461)(1031,492)(1081,524)(1132,559)(1183,595)(1233,634)(1284,674)(1335,717)(1385,761)(1436,808)
\thinlines \path(220,240)(220,240)(222,240)(223,240)(225,240)(226,240)(228,240)(230,240)(233,240)(236,240)(239,240)(245,241)(252,241)(258,241)(271,241)(283,242)(296,243)(321,244)(347,247)(372,250)(423,257)(473,266)(524,278)(575,291)(625,306)(676,324)(727,344)(777,365)(828,389)(879,415)(929,443)(980,473)(1031,505)(1081,539)(1132,576)(1183,614)(1233,654)(1284,697)(1335,741)(1385,788)(1436,837)
\thinlines  \dottedline{14}(220,240)(220,240)(1436,240)
\end{picture}

                \vglue 1ex
                \hspace*{15em}\mbox{(a) $\lambda \leq
\lambda_{\mbox{cr}}$}
                \vglue 7ex
\setlength{\unitlength}{0.240900pt}
\begin{picture}(1500,900)(0,0)
\tenrm
\thicklines \path(220,113)(240,113)
\thicklines \path(1436,113)(1416,113)
\put(198,113){\makebox(0,0)[r]{$-0.06$}}
\thicklines \path(220,246)(240,246)
\thicklines \path(1436,246)(1416,246)
\put(198,246){\makebox(0,0)[r]{$-0.04$}}
\thicklines \path(220,379)(240,379)
\thicklines \path(1436,379)(1416,379)
\put(198,379){\makebox(0,0)[r]{$-0.02$}}
\thicklines \path(220,512)(240,512)
\thicklines \path(1436,512)(1416,512)
\put(198,512){\makebox(0,0)[r]{0}}
\thicklines \path(220,644)(240,644)
\thicklines \path(1436,644)(1416,644)
\put(198,644){\makebox(0,0)[r]{0.02}}
\thicklines \path(220,777)(240,777)
\thicklines \path(1436,777)(1416,777)
\put(198,777){\makebox(0,0)[r]{0.04}}
\thicklines \path(220,113)(220,133)
\thicklines \path(220,877)(220,857)
\put(220,68){\makebox(0,0){0}}
\thicklines \path(363,113)(363,133)
\thicklines \path(363,877)(363,857)
\put(363,68){\makebox(0,0){0.2}}
\thicklines \path(506,113)(506,133)
\thicklines \path(506,877)(506,857)
\put(506,68){\makebox(0,0){0.4}}
\thicklines \path(649,113)(649,133)
\thicklines \path(649,877)(649,857)
\put(649,68){\makebox(0,0){0.6}}
\thicklines \path(792,113)(792,133)
\thicklines \path(792,877)(792,857)
\put(792,68){\makebox(0,0){0.8}}
\thicklines \path(935,113)(935,133)
\thicklines \path(935,877)(935,857)
\put(935,68){\makebox(0,0){1}}
\thicklines \path(1078,113)(1078,133)
\thicklines \path(1078,877)(1078,857)
\put(1078,68){\makebox(0,0){1.2}}
\thicklines \path(1221,113)(1221,133)
\thicklines \path(1221,877)(1221,857)
\put(1221,68){\makebox(0,0){1.4}}
\thicklines \path(1364,113)(1364,133)
\thicklines \path(1364,877)(1364,857)
\put(1364,68){\makebox(0,0){1.6}}
\thicklines \path(220,113)(1436,113)(1436,877)(220,877)(220,113)
\put(45,945){\makebox(0,0)[l]{\shortstack{$V/m_{0}^{2.5}$}}}
\put(828,23){\makebox(0,0){$\sigma/m_{0}$}}
\put(613,744){\makebox(0,0)[l]{$R=3R_{\mbox{cr}}/2$}}
\put(864,578){\makebox(0,0)[l]{$R=R_{\mbox{cr}}$}}
\put(792,399){\makebox(0,0)[l]{$R=R_{\mbox{cr}}/2$}}
\put(864,259){\makebox(0,0)[l]{$R=0$}}
\thinlines \path(220,512)(220,512)(222,512)(223,512)(225,512)(226,511)(230,511)(233,511)(239,511)(245,510)(258,508)(271,505)(296,498)(321,489)(372,465)(423,436)(473,404)(524,370)(575,336)(625,303)(676,271)(727,243)(777,220)(803,210)(828,201)(853,194)(866,192)(879,189)(891,187)(898,186)(904,186)(910,185)(917,185)(920,185)(923,184)(926,184)(928,184)(929,184)(931,184)(933,184)(934,184)(936,184)(937,184)(939,184)(940,184)(942,184)(944,184)(945,184)(948,185)(952,185)(955,185)
\thinlines \path(955,185)(961,185)(967,186)(980,188)(993,190)(1005,192)(1031,200)(1056,209)(1081,222)(1132,254)(1183,298)(1233,355)(1284,424)(1335,506)(1385,603)(1436,715)
\thinlines \path(220,512)(220,512)(222,512)(223,512)(225,512)(226,512)(230,512)(233,512)(236,511)(239,511)(245,511)(252,511)(258,511)(271,510)(296,509)(321,506)(372,500)(423,492)(473,483)(524,473)(575,463)(625,455)(651,451)(676,448)(689,446)(701,445)(714,444)(720,444)(727,443)(733,443)(739,443)(746,443)(749,442)(750,442)(752,442)(754,442)(755,442)(757,442)(758,442)(760,442)(761,442)(763,442)(765,442)(766,442)(768,442)(771,442)(774,443)(777,443)(784,443)(790,443)(803,444)
\thinlines \path(803,444)(815,445)(828,447)(853,451)(879,456)(904,463)(929,472)(980,496)(1031,528)(1081,569)(1132,620)(1183,682)(1233,755)(1284,841)(1302,877)
\thinlines \path(220,512)(220,512)(222,512)(223,512)(225,512)(226,512)(228,512)(230,512)(231,512)(233,512)(234,512)(236,512)(237,512)(239,512)(241,512)(242,512)(244,512)(245,512)(248,512)(252,512)(255,512)(258,512)(261,512)(264,512)(267,512)(271,512)(277,512)(283,512)(290,512)(296,512)(302,512)(309,512)(315,512)(321,512)(334,512)(347,512)(359,512)(372,512)(385,512)(397,512)(410,512)(423,513)(448,513)(473,514)(499,515)(524,517)(549,519)(575,521)(600,524)(625,528)(676,538)
\thinlines \path(676,538)(727,551)(777,568)(828,591)(879,619)(929,654)(980,696)(1031,747)(1081,806)(1132,876)(1133,877)
\thinlines \path(220,512)(220,512)(222,512)(223,512)(225,512)(226,512)(230,512)(233,512)(236,512)(239,512)(245,512)(252,512)(258,512)(271,512)(283,513)(296,514)(321,515)(347,517)(372,520)(423,526)(473,535)(524,547)(575,562)(625,580)(676,602)(727,629)(777,661)(828,698)(879,742)(929,793)(980,851)(1000,877)
\thinlines  \dottedline{14}(220,512)(220,512)(1436,512)
\end{picture}

                \vglue 1ex
                \hspace*{15em}\mbox{(b) $\lambda >
\lambda_{\mbox{cr}}$}
                \vglue 1ex
\caption{Behavior of the effective potential is shown at
$D=2.5$
         for fixed $\lambda$ with the varying curvature
         where $K=8.23 {m'}_{0}^{2}$ and $R_{\mbox{cr}}=8.23
m_{0}^{2}$.}
\label{fig:epot:de}
\end{figure}
In Fig. 14 the behavior of the effective potential given by
Eq. (\ref{v:ren:de})
is illustrated in the case $D=2.5$ for several typical
values
of the curvature.
It is observed in Fig. 14 that, if $\lambda \leq
\lambda_{\mbox{cr}}$,
the theory is always in the symmetric phase as the curvature
changes
while, if   $\lambda > \lambda_{\mbox{cr}}$, the symmetry
restoration
takes place as the curvature exceeds its critical value.
This observation remains true if the spacetime dimension is
arbitrarily
changed.

To discuss the dynamical mass of the fermion
we study the minimum of the effective potential more
precisely.
A necessary condition for the minimum is given by
the gap equation (\ref{gap:w}).
Inserting Eq. (\ref{v:ren:de}) into Eq. (\ref{gap:w}) we find \cite{IMM}
\begin{eqnarray}
     &&\frac{1}{\lambda}
     +\tr\11\frac{D-1}{(4\pi)^{\sD /2}}\Gamma\left(
     1-\frac{D}{2}\right)\nonumber \\
     &&-\frac{\tr\11(a\mu)^{2-\sD}}{(4\pi)^{\sD/2}}
     \Gamma\left(1-\frac{D}{2}\right)
     \frac{\displaystyle \Gamma\left(\frac{D}{2}+i m a\right)
     \Gamma\left(\frac{D}{2}-i m a\right)}
     {\displaystyle \Gamma\left(1+i m a\right)
     \Gamma\left(1-i m a\right)}
     =0\, .
\label{eq:de:gap}
\end{eqnarray}
If the coupling constant $\lambda$ is no less than a
critical
value $\lambda_{cr}$, the gap equation allows a nontrivial
solution.
In Fig. 15 we present the solution of the gap equation (\ref{eq:de:gap}).
The dynamical fermion mass smoothly disappears as the
curvature increases.
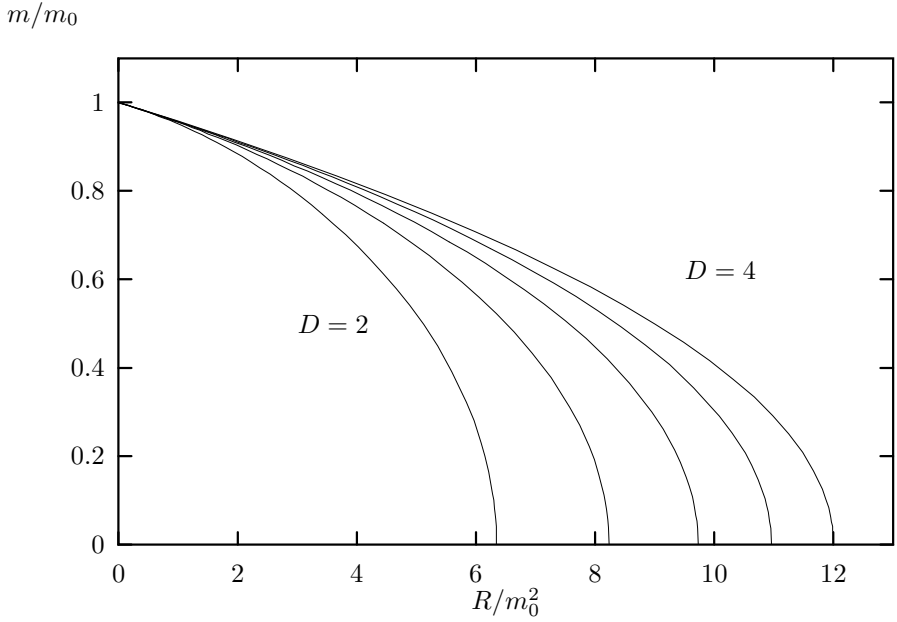
\begin{figure}
\vspace*{2ex}
\setlength{\unitlength}{0.240900pt}
\begin{picture}(1500,900)(0,0)
\tenrm
\thicklines \path(220,113)(240,113)
\thicklines \path(1436,113)(1416,113)
\put(198,113){\makebox(0,0)[r]{0}}
\thicklines \path(220,252)(240,252)
\thicklines \path(1436,252)(1416,252)
\put(198,252){\makebox(0,0)[r]{0.2}}
\thicklines \path(220,391)(240,391)
\thicklines \path(1436,391)(1416,391)
\put(198,391){\makebox(0,0)[r]{0.4}}
\thicklines \path(220,530)(240,530)
\thicklines \path(1436,530)(1416,530)
\put(198,530){\makebox(0,0)[r]{0.6}}
\thicklines \path(220,669)(240,669)
\thicklines \path(1436,669)(1416,669)
\put(198,669){\makebox(0,0)[r]{0.8}}
\thicklines \path(220,808)(240,808)
\thicklines \path(1436,808)(1416,808)
\put(198,808){\makebox(0,0)[r]{1}}
\thicklines \path(220,113)(220,133)
\thicklines \path(220,877)(220,857)
\put(220,68){\makebox(0,0){0}}
\thicklines \path(407,113)(407,133)
\thicklines \path(407,877)(407,857)
\put(407,68){\makebox(0,0){2}}
\thicklines \path(594,113)(594,133)
\thicklines \path(594,877)(594,857)
\put(594,68){\makebox(0,0){4}}
\thicklines \path(781,113)(781,133)
\thicklines \path(781,877)(781,857)
\put(781,68){\makebox(0,0){6}}
\thicklines \path(968,113)(968,133)
\thicklines \path(968,877)(968,857)
\put(968,68){\makebox(0,0){8}}
\thicklines \path(1155,113)(1155,133)
\thicklines \path(1155,877)(1155,857)
\put(1155,68){\makebox(0,0){10}}
\thicklines \path(1342,113)(1342,133)
\thicklines \path(1342,877)(1342,857)
\put(1342,68){\makebox(0,0){12}}
\thicklines \path(220,113)(1436,113)(1436,877)(220,877)(220,113)
\put(45,945){\makebox(0,0)[l]{\shortstack{$m / m_{0}$}}}
\put(828,23){\makebox(0,0){$R/m_{0}^2$}}
\put(501,460){\makebox(0,0)[l]{$D=2$}}
\put(1109,544){\makebox(0,0)[l]{$D=4$}}
\thinlines \path(220,808)(220,808)(220,808)(220,808)(220,808)(220,808)(220,808)(220,808)(220,808)(220,808)(220,808)(220,808)(220,807)(220,807)(221,807)(221,807)(222,807)(222,807)(224,806)(226,806)(229,805)(232,804)(236,803)(240,801)(245,800)(250,798)(256,796)(262,794)(268,792)(276,789)(283,786)(291,783)(300,780)(309,776)(319,772)(329,767)(340,762)(351,757)(362,751)(374,745)(387,738)(400,731)(414,723)(428,714)(442,705)(458,695)(473,684)(489,673)(506,660)(523,647)(540,632)
\thinlines \path(540,632)(558,616)(577,600)(596,581)(615,561)(635,539)(656,515)(677,489)(699,459)(721,425)(743,385)(766,337)(778,308)(789,273)(795,252)(801,227)(807,193)(809,183)(810,170)(811,162)(812,153)(813,142)(813,115)(813,113)
\thinlines \path(220,808)(220,808)(220,808)(220,808)(220,808)(220,808)(220,808)(220,808)(220,808)(220,808)(220,808)(220,807)(220,807)(221,807)(221,807)(221,807)(222,807)(223,807)(225,806)(228,805)(232,804)(236,803)(241,801)(246,799)(252,797)(259,795)(266,793)(274,790)(283,787)(292,784)(302,780)(313,777)(324,773)(336,768)(348,764)(361,758)(375,753)(390,747)(405,741)(420,734)(437,726)(454,719)(471,710)(490,701)(509,692)(528,681)(548,670)(569,659)(591,646)(613,632)(636,618)
\thinlines \path(636,618)(659,602)(683,585)(708,567)(733,547)(759,526)(786,502)(813,476)(841,447)(869,414)(899,376)(928,329)(944,301)(959,267)(967,247)(974,222)(982,191)(984,180)(986,168)(988,152)(989,140)(990,114)(990,113)
\thinlines \path(220,808)(220,808)(220,808)(220,808)(220,808)(220,808)(220,808)(220,808)(220,808)(220,808)(220,807)(220,807)(221,807)(221,807)(222,807)(222,807)(223,806)(226,806)(229,805)(234,803)(239,802)(244,800)(251,798)(258,796)(266,793)(275,790)(284,787)(294,784)(305,780)(317,776)(330,772)(343,767)(357,762)(372,757)(387,751)(403,745)(420,739)(438,732)(457,725)(476,717)(496,708)(517,700)(539,690)(561,680)(584,670)(608,658)(633,646)(658,633)(684,620)(711,605)(739,589)
\thinlines \path(739,589)(767,573)(796,555)(826,535)(857,514)(888,491)(921,465)(954,437)(987,405)(1022,367)(1057,322)(1075,295)(1093,262)(1102,242)(1111,219)(1116,205)(1121,188)(1123,178)(1125,166)(1126,159)(1128,150)(1129,139)(1130,113)
\thinlines \path(220,808)(220,808)(220,808)(220,808)(220,808)(220,808)(220,808)(220,808)(220,808)(220,807)(220,807)(220,807)(221,807)(221,807)(222,807)(223,807)(224,806)(227,805)(231,804)(235,803)(241,801)(247,799)(255,797)(263,794)(272,791)(282,788)(292,785)(304,781)(316,777)(329,772)(343,768)(358,763)(374,757)(391,752)(408,745)(427,739)(446,732)(466,725)(487,717)(509,709)(531,700)(555,691)(579,681)(604,671)(630,660)(657,648)(685,636)(714,623)(743,609)(773,594)(804,578)
\thinlines \path(804,578)(837,561)(869,543)(903,524)(938,503)(973,480)(1009,455)(1047,427)(1085,396)(1123,359)(1163,315)(1183,289)(1204,257)(1214,238)(1224,215)(1229,202)(1235,185)(1237,176)(1240,164)(1242,149)(1244,139)(1245,113)
\thinlines \path(1342,113)(1342,113)(1342,114)(1342,115)(1342,116)(1342,117)(1342,118)(1342,118)(1342,120)(1342,122)(1342,124)(1342,127)(1342,131)(1341,135)(1341,142)(1339,149)(1338,156)(1335,171)(1330,185)(1325,200)(1311,229)(1294,258)(1272,287)(1247,316)(1218,345)(1185,373)(1148,402)(1107,431)(1062,460)(1013,489)(961,518)(904,547)(844,576)(779,605)(711,634)(639,663)(563,692)(483,721)(399,750)(312,779)(220,808)
\end{picture}

\caption{Behavior of the dynamical fermion mass 
in de Sitter space as a
function of the
         curvature $R$ at $D=2.0,2.5,3.0,3.5,4$ where
$m_{0}$
         is the dynamical fermion mass in flat spacetime.}
\label{fig:mass:de}
\end{figure}

Taking the two-dimensional limit Eq. (\ref{eq:de:gap}) reduces to
\begin{equation}
     \frac{1}{\lambda}
     -\frac{\tr\11}{4\pi}\left[\ln (a^{2}\mu^{2})
     -\psi (1+i m a)-\psi(1-i m a)\right] = 0\, .
\label{two:gap}
\end{equation}
For three dimensions it reads \cite{ELOS}
\begin{equation}
     \frac{1}{\lambda}-\frac{2\tr\11}{4\pi}
     +\frac{\tr\11}{4\pi a\mu}
     \frac{\displaystyle \Gamma\left(\frac{3}{2}+i m
a\right)
     \Gamma\left(\frac{3}{2}-i m a\right)}
     {\displaystyle \Gamma\left(1+i m a\right)
     \Gamma\left(1-i m a\right)} =0\, ,
\label{three:gap}
\end{equation}
while it reduces to
\begin{eqnarray}
     &&\frac{1}{\lambda}
     -\frac{3\tr\11}{(4\pi)^{2}}
     \left(C_{\mbox{div}}-\frac{2}{3}\right)
     -\frac{2 \tr\11 a^{-2}}{(4\pi)^{2}\mu^{2}}
     \tr\11\nonumber \\
     &&+\frac{a^{-2}+m^{2}}{(4\pi)^{2}\mu^{2}}
     \left[C_{\mbox{div}}+\ln (a^{2}\mu^{2})
     -\psi(1+i m a)-\psi(1-i m a)\right]=0\, ,
\label{gap:4d}
\end{eqnarray}
in the limit of $D\rightarrow 4$, where the divergent part
is expressed
by $C_{\mbox{div}}$ in Eq. (\ref{div:c}).
Taking the four-dimensional limit, $D\rightarrow 4$,
the divergent parts of the gap equation (\ref{gap:4d})
is exactly equal to that given by Eq. (\ref{nontri:w}).
Hence the same critical behavior is observed for the
dynamically generated fermion mass in Figs. 8 and 13 at
$D=4$.

As is seen in Figs. 12 and 13 the phase transition is of the
second order.
Accordingly by setting $m=0$ in the gap equation (\ref{eq:de:gap})
we may derive the equation which determines
the critical radius $a_{\mbox{cr}}$,
\begin{equation}
     \frac{1}{\lambda}
     +\tr\11\frac{D-1}{(4\pi)^{\sD /2}}\Gamma\left(
     1-\frac{D}{2}\right)
     -\tr\11\frac{(a_{cr}\mu)^{2-\sD}}{(4\pi)^{\sD /2}}
     \Gamma^{2}\left(\frac{D}{2}\right)
     \Gamma\left(1-\frac{D}{2}\right)
     =0\, .
\label{eq:phase}
\end{equation}
Substituting Eqs. (\ref{def:mass}) and (\ref{cr:l:d})
in Eq. (\ref{eq:phase}) the critical radius $a_{\mbox{cr}}$
is found to be \cite{IMM}
\begin{equation}
     a_{\mbox{cr}}=\frac{1}{m_{0}}\left[\Gamma\left(\frac{D}{2}\right)
            \right]^{2/(\sD-2)}\, .
\label{r:cr}
\end{equation}
In Fig. 18(a) the critical radius $a_{\mbox{cr}}$ is plotted
as a function of the spacetime dimension $D$.
By using the critical radius (\ref{r:cr}) we find the
critical curvature:
\begin{equation}
     R_{\mbox{cr}}=D(D-1){a_{\mbox{cr}}}^{-2}\, .
\label{r:cr:de}
\end{equation}
For some special values of $D$ Eq. (\ref{r:cr:de}) simplifies
to:
\begin{equation}
\begin{array}{ll}
     {\displaystyle R_{\mbox{cr}}=2 e^{2\gamma} m_{0}^{2}}\,  &;
D=2\, ,\\[4mm]
     {\displaystyle R_{\mbox{cr}}=\frac{96}{\pi^{2}} m_{0}^{2}}\,
&; D=3\, ,\\[4mm]
     {\displaystyle R_{\mbox{cr}}=12 m_{0}^{2}}
     \,  &; D=4\, ,
\end{array}
\label{rc234:de}
\end{equation}
where $\gamma$ is the Euler constant.
In Fig. 18(b) the critical curvature $R_{\mbox{cr}}$ is presented as
a function of the spacetime dimension $D$.

\subsection{Anti-de Sitter background ($H^{D}$)}

Next we consider the anti-de Sitter space.
It is also the maximally
symmetric spacetime with negative curvature
\begin{equation}
     R=-D(D-1)a^{-2}\, .
\end{equation}
We consider the manifold $H^{D}$ as a Euclidean
analog of the $D$-dimensional anti-de Sitter space.

As is discussed in \S 4.1 the spinor two-point function
$S(x,x;s)$ is known in $H^{D}$ and $\tr S(x,x;s)$ is given
by
\begin{equation}
     \tr S(x,x;s)=-\frac{\tr\11 a^{1-D}}{(4\pi)^{D/2}}
     \frac{\displaystyle\Gamma\left(\frac{D}{2}+sa\right)
           \Gamma\left(1-\frac{D}{2}\right)}
          {\displaystyle\Gamma\left(1-\frac{D}{2}+sa\right)}\, .
\label{trD:hy}
\end{equation}
Performing the Wick rotation and inserting Eq. (\ref{trD:hy}) into
Eq. (\ref{v:gn}) the effective potential of the four-fermion
model reads
\begin{equation}
     V(\sigma)=\frac{1}{2\lambda_{0}}\sigma^{2}
     -\frac{\tr\11 a^{1-D}}{(4\pi)^{D/2}}
     \Gamma\left(1-\frac{D}{2}\right)
     \int^{\sigma}_{0}
     ds   \frac{\displaystyle\Gamma\left(\frac{D}{2}+sa\right)}
          {\displaystyle\Gamma\left(1-\frac{D}{2}+sa\right)}\, .
\label{v:nonren:hy}
\end{equation}
Thus we obtain the exact effective potential
(\ref{v:nonren:hy}) in anti-de Sitter space
in the leading order of the $1/N$ expansion.
Applying the same renormalization condition (\ref{cond:ren})
in Eq. (\ref{v:nonren:hy}) we find the renormalized
expression of the effective potential $V(\sigma)$ 
in anti-de Sitter space,
\begin{eqnarray}
     V(\sigma)&=&\frac{1}{2\lambda}\sigma^{2}\mu^{\sD-2}
     +\frac{\tr\11}{2(4\pi)^{\sD /2}}(D-1)\Gamma\left(
     1-\frac{D}{2}\right)\sigma^{2}\mu^{\sD-2}\nonumber \\
     &&-\frac{\tr\11 a^{1-D}}{(4\pi)^{D/2}}
     \Gamma\left(1-\frac{D}{2}\right)
     \int^{\sigma}_{0}
     ds   \frac{\displaystyle\Gamma\left(\frac{D}{2}+sa\right)}
          {\displaystyle\Gamma\left(1-\frac{D}{2}+sa\right)}\, .
\label{v:ren:hy}
\end{eqnarray}
Expanding Eq. (\ref{v:ren:hy}) asymptotically about $1/a=0$
(weak curvature expansion) the effective potential
(\ref{v:ren}) is reproduced.
To study the phase structure in 
anti-de Sitter space we evaluate
the effective potential (\ref{v:ren:hy}).

In order to illustrate phase structure of four-fermion model
in anti-de Sitter space we will consider two-dimensional case
only below.
That will be enough for our purposes because as we will see
negative curvature always supports chiral symmetry breaking.
Hence, there is no need to make the exposition of cases for all $D$
as it was above. Note that the fact that negative curvature supports
chiral symmetry breaking and only broken phase survives has been
already mentioned few times when we have discussed weak
curvature expansion.

In two dimensions the effective potential (\ref{v:ren:hy}) 
reduces to \cite{ELOS}
\begin{eqnarray}
     \frac{V^{D=2}(\sigma)}{\mu^{2}} &=&
     \left[\frac{1}{2\lambda}\right.
     +\left.\frac{\tr\11}{4\pi}\left(\ln\frac{1}{a\mu}-1\right)
     \right]
     \left(\frac{\sigma}{\mu}\right)^{2}\nonumber \\
     &&+\frac{\tr\11}{4\pi{\mu}^{2}} \int^{\sigma}_{0} ds s
     \left[ \psi(1 + sa)
       +\psi(sa) \right]\, .
\label{v:2d:hy}
\end{eqnarray}
Evaluating the effective potential (\ref{v:2d:hy}) 
numerically we easily see that only the broken phase 
is realized in two dimensions.
\begin{figure}
\setlength{\unitlength}{0.240900pt}
\begin{picture}(1500,900)(0,0)
\tenrm
\dottedline{14}(220,826)(1436,826)
\thicklines \path(220,113)(240,113)
\thicklines \path(1436,113)(1416,113)
\put(198,113){\makebox(0,0)[r]{$-0.14$}}
\thicklines \path(220,215)(240,215)
\thicklines \path(1436,215)(1416,215)
\put(198,215){\makebox(0,0)[r]{$-0.12$}}
\thicklines \path(220,317)(240,317)
\thicklines \path(1436,317)(1416,317)
\put(198,317){\makebox(0,0)[r]{$-0.1$}}
\thicklines \path(220,419)(240,419)
\thicklines \path(1436,419)(1416,419)
\put(198,419){\makebox(0,0)[r]{$-0.08$}}
\thicklines \path(220,520)(240,520)
\thicklines \path(1436,520)(1416,520)
\put(198,520){\makebox(0,0)[r]{$-0.06$}}
\thicklines \path(220,622)(240,622)
\thicklines \path(1436,622)(1416,622)
\put(198,622){\makebox(0,0)[r]{$-0.04$}}
\thicklines \path(220,724)(240,724)
\thicklines \path(1436,724)(1416,724)
\put(198,724){\makebox(0,0)[r]{$-0.02$}}
\thicklines \path(220,826)(240,826)
\thicklines \path(1436,826)(1416,826)
\put(198,826){\makebox(0,0)[r]{0}}
\thicklines \path(220,113)(220,133)
\thicklines \path(220,877)(220,857)
\put(220,68){\makebox(0,0){0}}
\thicklines \path(355,113)(355,133)
\thicklines \path(355,877)(355,857)
\put(355,68){\makebox(0,0){0.2}}
\thicklines \path(490,113)(490,133)
\thicklines \path(490,877)(490,857)
\put(490,68){\makebox(0,0){0.4}}
\thicklines \path(625,113)(625,133)
\thicklines \path(625,877)(625,857)
\put(625,68){\makebox(0,0){0.6}}
\thicklines \path(760,113)(760,133)
\thicklines \path(760,877)(760,857)
\put(760,68){\makebox(0,0){0.8}}
\thicklines \path(896,113)(896,133)
\thicklines \path(896,877)(896,857)
\put(896,68){\makebox(0,0){1}}
\thicklines \path(1031,113)(1031,133)
\thicklines \path(1031,877)(1031,857)
\put(1031,68){\makebox(0,0){1.2}}
\thicklines \path(1166,113)(1166,133)
\thicklines \path(1166,877)(1166,857)
\put(1166,68){\makebox(0,0){1.4}}
\thicklines \path(1301,113)(1301,133)
\thicklines \path(1301,877)(1301,857)
\put(1301,68){\makebox(0,0){1.6}}
\thicklines \path(1436,113)(1436,133)
\thicklines \path(1436,877)(1436,857)
\put(1436,68){\makebox(0,0){1.8}}
\thicklines \path(220,113)(1436,113)(1436,877)(220,877)(220,113)
\put(45,945){\makebox(0,0)[l]{\shortstack{$V/m_{0}^{2}$}}}
\put(828,23){\makebox(0,0){$\sigma/m_{0}$}}
\put(760,520){\makebox(0,0)[l]{$R=0$}}
\put(760,368){\makebox(0,0)[l]{$R=-2 m_{0}^{-2}$}}
\put(342,296){\makebox(0,0)[l]{$R=-4 m_{0}^{-2}$}}
\thinlines \path(220,826)(220,826)(223,826)(225,826)(226,826)(230,825)(233,825)(239,823)(245,822)(258,817)(271,812)(296,799)(321,782)(372,744)(423,702)(473,657)(524,613)(575,570)(625,531)(676,496)(727,467)(752,455)(777,444)(803,435)(815,432)(828,429)(841,426)(853,424)(860,423)(866,422)(872,422)(876,421)(879,421)(882,421)(885,421)(887,421)(888,421)(890,421)(891,421)(893,421)(895,421)(896,421)(898,421)(899,421)(901,421)(902,421)(904,421)(907,421)(910,421)(917,422)(923,422)
\thinlines \path(923,422)(929,423)(942,425)(955,427)(980,434)(1005,443)(1031,455)(1081,487)(1132,531)(1183,586)(1233,654)(1284,734)(1335,828)(1358,877)
\thinlines \path(220,826)(220,826)(271,763)(321,697)(372,629)(423,562)(473,496)(524,432)(575,373)(625,319)(676,270)(727,228)(777,194)(803,180)(828,168)(853,158)(866,154)(879,151)(891,148)(904,145)(910,145)(917,144)(923,143)(926,143)(929,143)(933,143)(936,143)(937,142)(939,142)(940,142)(942,142)(944,142)(945,142)(947,142)(948,142)(950,143)(951,143)(955,143)(958,143)(961,143)(967,144)(974,144)(980,145)(993,147)(1005,150)(1031,158)(1056,169)(1081,183)(1132,219)(1183,267)(1233,328)
\thinlines \path(1233,328)(1284,402)(1335,490)(1385,591)(1436,707)
\thinlines \path(220,826)(220,826)(271,779)(321,727)(372,671)(423,615)(473,558)(524,503)(575,452)(625,404)(676,362)(727,326)(777,297)(803,286)(828,276)(841,272)(853,269)(866,266)(879,264)(891,262)(898,262)(901,261)(904,261)(907,261)(910,261)(912,261)(913,261)(915,261)(917,261)(918,260)(920,260)(921,260)(923,260)(925,261)(926,261)(928,261)(929,261)(933,261)(936,261)(942,261)(948,262)(955,263)(967,265)(980,267)(1005,274)(1031,284)(1081,312)(1132,352)(1183,404)(1233,468)(1284,545)
\thinlines \path(1284,545)(1335,636)(1385,740)(1436,859)
\end{picture}

\caption{Behavior of the effective potential in $H^{2}$.}
\label{fig:epot:h2}
\end{figure}
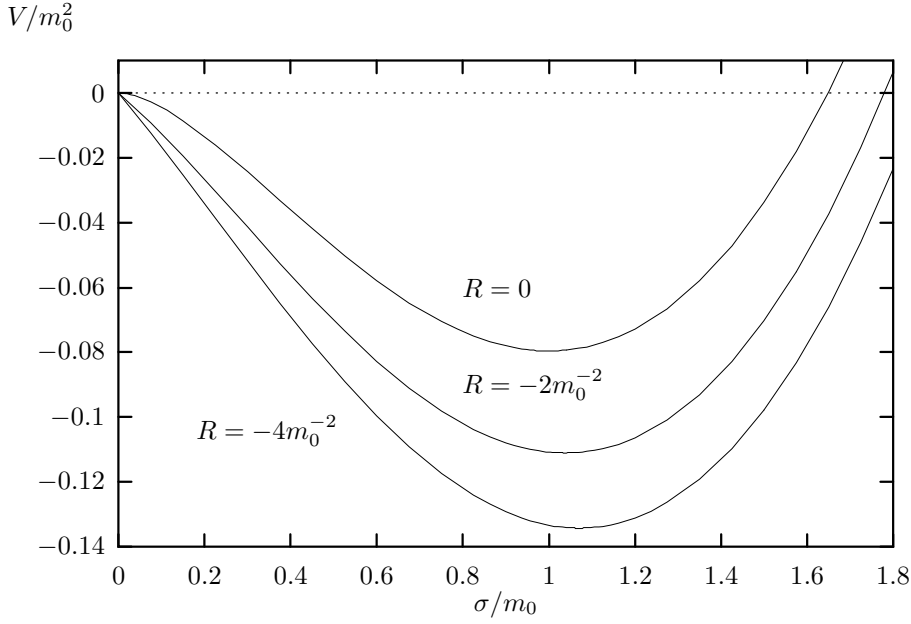
In Fig. 16 the behavior of the effective potential
given by Eq. (\ref{v:2d:hy}) is illustrated.

Differentiating (\ref{v:2d:hy}) with respect to $\sigma$
we get
\begin{equation}
     \frac{\partial V^{D=2}}{\partial\sigma}
     =\sigma\left\{\frac{1}{\lambda}
     -\frac{\tr\11}{4\pi}\left[\ln (a^{2}\mu^{2})
     -\psi (1+\sigma a)-\psi(\sigma a)\right]\right\}\, .
\label{two:gap:hy}
\end{equation}
A careful study of $\partial V/\partial\sigma$ 
shows that due to the presence of the last
term in (\ref{two:gap:hy}), $\sigma=0$ is never
stationary for any value of
$\lambda$ and finite $a$.
Owing to the fact that 
\begin{equation}
     \left.\frac{\partial V^{D=2}}{\partial\sigma}\right|_{\sigma=0}
     =-\frac{\tr\11}{4\pi a} < 0\, ,
\end{equation}
the chiral symmetry is always broken in two-dimensional
anti-de Sitter space.\cite{ELOS}

In arbitrary dimensions ($2 \leq D < 4$) we easily find
\begin{equation}
     \left.\frac{\partial V}{\partial\sigma}\right|_{\sigma=0}
     =-\frac{\tr\11}{(4\pi)^{\sD /2}}a^{1-D}\Gamma\left(\frac{D}{2}\right) 
     < 0\, .
\end{equation}
Thus the chiral symmetry is always broken down in
anti-de Sitter space (i.e., a negative
curvature spacetime) in arbitrary dimensions
$2 \leq D < 4$ as is mentioned in \S 3.

\subsection{Einstein universe background ($R \otimes S^{D-1}$)}

Another universe we can calculate the effective potential
without making any approximation in the spacetime curvature
is the static Einstein universe.\cite{IIM}
The $D$-dimensional Einstein universe is represented by the
metric
\begin{equation}
     ds^{2}=dr^{2}-a^{2}(d\theta^{2}+\sin^{2}\theta
d\Omega_{\sD-2})\, .
\end{equation}
It is a constant curvature spacetime with curvature
\begin{equation}
     R=(D-1)(D-2)\frac{1}{a^{2}}\, .
\label{rl:ein}
\end{equation}
Here we consider the manifold $R \otimes S^{D-1}$ as a
Euclidean analog
of the $D$-dimensional Einstein universe.
In two spacetime dimensions a Euclidean analog of the
Einstein universe is represented by a cylinder with radius $a$.

According to a similar analysis developed in \S 4.1
we are able to solve the Dirac equation in Einstein
universe.\cite{IIM2}
On $R \otimes S^{D-1}$ we rewrite Eq. (\ref{eq:Gsp}) in the
following form
\begin{equation}
     \left((\partial_{0})^{2}+\Box_{D-1}-\frac{R}{4}-s^{2}\right)G(x,y;s)
     =-\frac{1}{\sqrt{g}}\delta^{D}(x,y)\, ,
\label{eq:G:ein}
\end{equation}
where $\Box_{D-1}$ is the Laplacian on $S^{D-1}$.
Performing the Fourier transformation
\begin{equation}
     G(x,y;s)=\int \frac{d\omega}{2\pi} e^{-i\omega
(y-x)^{0}}\tilde{G}(\omega,\yy-\xx)\, ,
\label{fou}
\end{equation}
we rewrite Eq. (\ref{eq:G:ein}) in the form
\begin{equation}
     \left(\Box_{D-1}-\frac{R}{4}-(s^{2}+\omega^{2})\right)
     \tilde{G}(\omega,\yy-\xx)
     =-\frac{1}{\sqrt{g}}\delta^{D-1}(\yy)\, .
\label{eq:tildeG:ein}
\end{equation}
Equation (\ref{eq:tildeG:ein}) is of the same form as the one
for the spinor Green function with mass $\sqrt{s^{2}+\omega^{2}}$
on $S^{D-1}$.
Thus the explicit expression for the two-point function 
in $R\otimes S^{D-1}$ is given by
\begin{eqnarray}
     \mbox{tr} S(x,x;s)&=&\displaystyle -i
     \frac{\tr\11s a^{3-\sD}}{(4\pi)^{(\sD-1)/2}}
     \Gamma\left(\frac{3-D}{2}\right) \nonumber \\
     &&\times\int\frac{d\omega}{2\pi}
     \frac{\displaystyle
\Gamma\left(\frac{D-1}{2}+i\beta\right)
                         \Gamma\left(\frac{D-1}{2}-i\alpha\right)}
          {\displaystyle \Gamma\left(1+i\alpha\right)
                         \Gamma\left(1-i\alpha\right) }\, ,
\label{trS:ein}
\end{eqnarray}
where the parameter $\alpha$ is defined by
\begin{equation}
     \alpha = a \sqrt{s^{2}+\omega^{2}}\, .
\end{equation}

Inserting the two-point function (\ref{trS:ein}) into
Eq. (\ref{v:gn}) we obtain the effective potential for
the four-fermion model in Einstein universe in the leading
order of
the $1/N$ expansion.\cite{IIM}
\begin{eqnarray}
     V(\sigma)&=&\frac{1}{2\lambda_{0}}\sigma^{2}
     -\frac{\tr\11a^{3-\sD}}{(4\pi)^{(\sD-1)/2}}
     \Gamma\left(\frac{3-D}{2}\right)\nonumber \\
     &&\times\int^{\sigma}_{0}s ds
     \int\frac{d\omega}{2\pi}
     \frac{\displaystyle
     \Gamma\left(\frac{D-1}{2}+i\alpha\right)
                         \Gamma\left(\frac{D-1}{2}-i\alpha\right)}
          {\displaystyle \Gamma\left(1+i\alpha\right)
                         \Gamma\left(1-i\alpha\right) }\, .
\label{v:nonren:ein}
\end{eqnarray}
We apply the renormalization condition (\ref{cond:ren})
and obtain the renormalized effective potential by replacing
the coupling
constant $\lambda_{0}$ with the renormalized one $\lambda$
defined by Eq. (\ref{eqn:ren})
\begin{eqnarray}
     V(\sigma)&=&\frac{1}{2\lambda}\sigma^{2}\mu^{\sD-2}
     +\frac{\tr\11}{2(4\pi)^{\sD /2}}(D-1)\Gamma\left(
     1-\frac{D}{2}\right)\sigma^{2}\mu^{\sD-2}
     \nonumber\\
     &&-\frac{\tr\11a^{3-\sD}}{(4\pi)^{(\sD-1)/2}}
     \Gamma\left(\frac{3-D}{2}\right)
     \int^{\sigma}_{0}s ds
     \int\frac{d\omega}{2\pi}
     \frac{\displaystyle
     \Gamma\left(\frac{D-1}{2}+i\alpha\right)
                         \Gamma\left(\frac{D-1}{2}-i\alpha\right)}
          {\displaystyle \Gamma\left(1+i\alpha\right)
                         \Gamma\left(1-i\alpha\right) }\, .
\label{v:ren:ein}
\end{eqnarray}
Expanding Eq. (\ref{v:ren:ein}) asymptotically about
$1/a=0$ (weak curvature expansion) the effective potential
(\ref{v:ren}) is
reproduced again.
For convenience in a numerical calculation we rewrite
Eq. (\ref{v:ren:ein}) as
\begin{eqnarray}
  V(\sigma)&=&
  \frac{1}{2\lambda}\sigma^{2}\mu^{\sD-2}
  \nonumber\\
  &&+\frac{\tr\11}{(4\pi)^{(\sD-1)/2}}
  \Gamma\left(\frac{3-D}{2}\right)
  \int_{-\infty}^{\infty}\frac{d\omega}{2\pi}
  [\omega^{2}+(D-2)\mu^{2}]
  (\omega^{2}+\mu^{2})^{(\sD-5)/2}
  \sigma^{2}  \nonumber \\
  &&-   \frac{\tr\11 a^{3-\sD}}{(4\pi)^{(\sD-1)/2}}
  \Gamma\left(\frac{3-D}{2}\right)
  \int_{0}^{\sigma}\!\!\!s ds
  \, \int_{-\infty}^{\infty}\frac{d\omega}{2\pi}
     \frac{\displaystyle
     \Gamma\left(\frac{D-1}{2}+i\alpha\right)
                         \Gamma\left(\frac{D-1}{2}-i\alpha\right)}
          {\displaystyle \Gamma\left(1+i\alpha\right)
                         \Gamma\left(1-i\alpha\right) }\, .
\label{eq:v:ein}
\end{eqnarray}
By the use of the expression of the effective potential
(\ref{eq:v:ein})
we may study the behavior of the effective potential as a
function of
$\sigma$ through the numerical integrations.
 In the numerical integration in $\omega$ we introduced a
suitable
upper and lower bound and performed the numerical
integration between
these bounds.
By assuming the sufficiently large absolute value of these
bounds and
checking the stability of the integral under the change of
the bounds
we obtained the numerical value for $V(\sigma)$ for each
value of
$\sigma$ with $\lambda$ and $a$ kept fixed.

A behavior of the effective potential
is similar to that in de Sitter space illustrated in Fig. 14.
If $\lambda \leq \lambda_{\mbox{cr}}$,
the theory is always in the symmetric phase as the curvature
changes
while, if   $\lambda > \lambda_{\mbox{cr}}$, the symmetry
restoration
takes place as the curvature exceeds its critical value.
Only the second order phase transition occurs as the
curvature exceeds its critical value.\cite{IIM}

The dynamical fermion mass is given by the gap equation
(\ref{gap:w}):
\begin{eqnarray}
     &&\frac{1}{\lambda}\mu^{\sD-2}
     +\frac{\tr\11}{(4\pi)^{(\sD-1)/2}}
     \Gamma\left(\frac{3-D}{2}\right)
     \int\frac{d\omega}{2\pi}
     \Biggl\{ [\omega^{2}+(D-2)\mu^{2}](\omega^{2}+\mu^{2})^{(\sD-5)/2}
     \nonumber \\
     &&-a^{3-\sD}\frac{\displaystyle
\Gamma\left(\frac{D-1}{2}+i\alpha\right)
                         \Gamma\left(\frac{D-1}{2}-i\alpha\right)}
          {\displaystyle \Gamma\left(1+i\alpha\right)
                         \Gamma\left(1-i\alpha\right) }
     \Biggr\}=0\, .
\label{gap:ein}
\end{eqnarray}

Figure 17 represents the behavior of the dynamical fermion
mass which
is obtained by solving Eq. (\ref{gap:ein}) numerically for
$D=2.0,2,5,3.0$ and $3.5$.
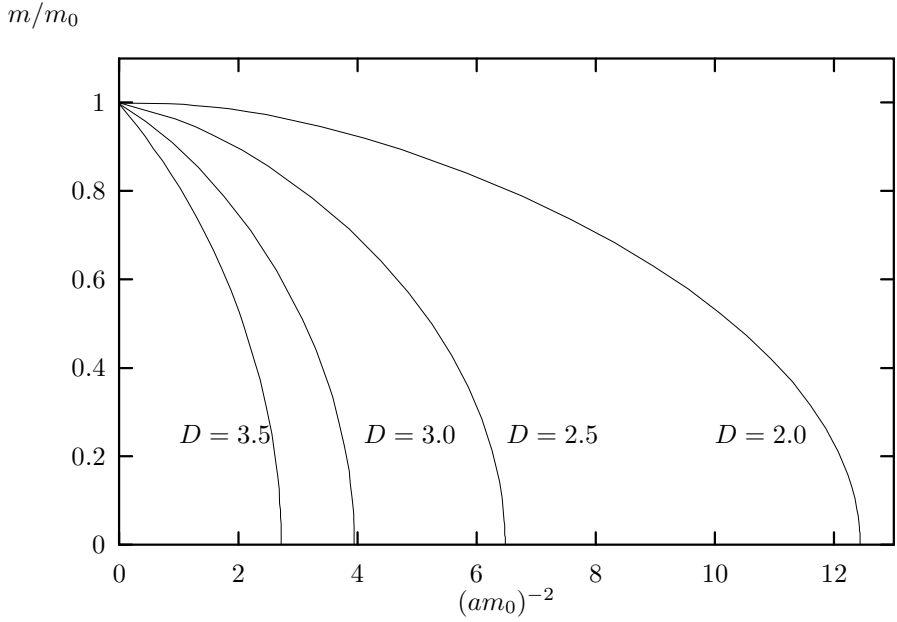
\begin{figure}
\setlength{\unitlength}{0.240900pt}
\begin{picture}(1500,900)(90,0)
\tenrm
\thicklines \path(220,113)(240,113)
\thicklines \path(1436,113)(1416,113)
\put(198,113){\makebox(0,0)[r]{0}}
\thicklines \path(220,252)(240,252)
\thicklines \path(1436,252)(1416,252)
\put(198,252){\makebox(0,0)[r]{0.2}}
\thicklines \path(220,391)(240,391)
\thicklines \path(1436,391)(1416,391)
\put(198,391){\makebox(0,0)[r]{0.4}}
\thicklines \path(220,530)(240,530)
\thicklines \path(1436,530)(1416,530)
\put(198,530){\makebox(0,0)[r]{0.6}}
\thicklines \path(220,669)(240,669)
\thicklines \path(1436,669)(1416,669)
\put(198,669){\makebox(0,0)[r]{0.8}}
\thicklines \path(220,808)(240,808)
\thicklines \path(1436,808)(1416,808)
\put(198,808){\makebox(0,0)[r]{1}}
\thicklines \path(220,113)(220,133)
\thicklines \path(220,877)(220,857)
\put(220,68){\makebox(0,0){0}}
\thicklines \path(407,113)(407,133)
\thicklines \path(407,877)(407,857)
\put(407,68){\makebox(0,0){2}}
\thicklines \path(594,113)(594,133)
\thicklines \path(594,877)(594,857)
\put(594,68){\makebox(0,0){4}}
\thicklines \path(781,113)(781,133)
\thicklines \path(781,877)(781,857)
\put(781,68){\makebox(0,0){6}}
\thicklines \path(968,113)(968,133)
\thicklines \path(968,877)(968,857)
\put(968,68){\makebox(0,0){8}}
\thicklines \path(1155,113)(1155,133)
\thicklines \path(1155,877)(1155,857)
\put(1155,68){\makebox(0,0){10}}
\thicklines \path(1342,113)(1342,133)
\thicklines \path(1342,877)(1342,857)
\put(1342,68){\makebox(0,0){12}}
\thicklines \path(220,113)(1436,113)(1436,877)(220,877)(220,113)
\put(45,945){\makebox(0,0)[l]{\shortstack{$m/m_{0}$}}}
\put(828,23){\makebox(0,0){$(a m_{0})^{-2}$}}
\put(1155,287){\makebox(0,0)[l]{$D=2.0$}}
\put(828,287){\makebox(0,0)[l]{$D=2.5$}}
\put(604,287){\makebox(0,0)[l]{$D=3.0$}}
\put(314,287){\makebox(0,0)[l]{$D=3.5$}}
\thinlines \path(1383,113)(1383,113)(1383,114)(1383,115)(1383,116)(1383,118)(1383,120)(1383,122)(1383,124)(1383,127)(1383,131)(1382,136)(1382,140)(1381,150)(1380,159)(1378,168)(1375,186)(1370,204)(1364,223)(1349,259)(1329,296)(1305,332)(1277,369)(1243,405)(1205,442)(1162,478)(1113,515)(1058,551)(997,588)(929,624)(852,661)(765,697)(663,734)(604,752)(536,770)(496,779)(451,789)(424,793)(393,798)(374,800)(351,802)(338,803)(321,805)(297,806)(221,807)
\thinlines \path(826,113)(826,113)(826,115)(826,116)(826,118)(826,119)(826,121)(826,122)(826,125)(825,128)(825,132)(825,138)(825,144)(824,150)(823,163)(822,175)(821,187)(817,212)(811,237)(805,262)(789,311)(768,361)(742,410)(711,460)(674,509)(631,559)(581,609)(523,658)(454,708)(413,733)(366,757)(339,770)(308,782)(221,807)
\thinlines \path(221,805)(221,805)(262,778)(303,745)(344,706)(385,660)(426,607)(467,544)(508,467)(529,420)(549,363)(555,345)(565,306)(574,267)(581,229)(583,209)(585,190)(587,171)(588,161)(588,152)(589,142)(589,137)(589,132)(589,127)(589,125)(589,123)(589,120)(589,118)(589,117)(589,115)(589,114)(589,113)
\thinlines \path(221,804)(221,804)(234,788)(248,771)(261,754)(274,735)(288,716)(301,696)(315,674)(328,652)(341,628)(355,602)(368,575)(381,546)(395,514)(408,479)(415,460)(441,373)(451,330)(460,287)(466,243)(469,222)(471,200)(472,178)(473,167)(473,156)(474,146)(474,140)(474,135)(474,129)(474,124)(474,121)(474,118)(474,116)(474,113)
\end{picture}

\caption{Behavior of the dynamical fermion mass 
in Einstein universe as a
function of
         $(a m_{0})^{-2}$ at $D=2.0,2.5,3.0,3.5$ where
$m_{0}$
         is the dynamical fermion mass in flat spacetime.}
\label{fig:mass:ein}
\end{figure}
In two dimensions Einstein universe becomes a flat
spacetime, $R=0$.
To see the mass dependence of the radius $a$ for $D=2$
we draw the dynamical fermion mass as a function of $(a
m_{0})^{-2}$
instead of the curvature $R$.

The critical radius is given by the massless limit of the
gap equation.
Taking the massless limit in Eq. (\ref{gap:ein}), we can
perform the
integration over $\omega$ analytically and find
\begin{eqnarray}
  &&\frac{1}{\lambda}+\frac{\tr\11}{(4\pi)^{\sD/2}}(D-1)
  \Gamma\left( 1-\frac{D}{2}\right)\nonumber \\
  &&-\frac{\tr\11(a \mu)^{2-\sD}}{\sqrt{\pi}(4\pi)^{\sD/2}}
  \Gamma\left(\frac{D-1}{2}\right)\Gamma\left(\frac{D}{2}\right)
  \Gamma\left(1-\frac{D}{2}\right)=0\, .
  \label{eq:ein:gap}
\end{eqnarray}
Taking into account Eq. (\ref{def:mass}) with
Eq. (\ref{cr:l:d}) we rewrite Eq. (\ref{eq:ein:gap}) in the
following form \cite{IIM}
\begin{equation}
  a_{\mbox{cr}}=\frac{1}{m_{0}}\left[\frac{1}{\sqrt{\pi}}
  \Gamma\left(\frac{D-1}{2}\right)\Gamma\left(\frac{D}{2}\right)
  \right]^{1/(D-2)}\, .
  \label{eqn:critgap}
\end{equation}
Solving Eq. (\ref{eqn:critgap}) the critical radius is
obtained
in an arbitrary dimension.
In Fig. 18(a) we show the dependence of the critical radius
to dimension $D$.
\begin{figure}
\setlength{\unitlength}{0.240900pt}
\begin{picture}(1500,900)(0,0)
\tenrm
\thicklines \path(220,113)(240,113)
\thicklines \path(1436,113)(1416,113)
\put(198,113){\makebox(0,0)[r]{0}}
\thicklines \path(220,252)(240,252)
\thicklines \path(1436,252)(1416,252)
\put(198,252){\makebox(0,0)[r]{0.2}}
\thicklines \path(220,391)(240,391)
\thicklines \path(1436,391)(1416,391)
\put(198,391){\makebox(0,0)[r]{0.4}}
\thicklines \path(220,530)(240,530)
\thicklines \path(1436,530)(1416,530)
\put(198,530){\makebox(0,0)[r]{0.6}}
\thicklines \path(220,669)(240,669)
\thicklines \path(1436,669)(1416,669)
\put(198,669){\makebox(0,0)[r]{0.8}}
\thicklines \path(220,808)(240,808)
\thicklines \path(1436,808)(1416,808)
\put(198,808){\makebox(0,0)[r]{1}}
\thicklines \path(220,113)(220,133)
\thicklines \path(220,877)(220,857)
\put(220,68){\makebox(0,0){2}}
\thicklines \path(342,113)(342,133)
\thicklines \path(342,877)(342,857)
\put(342,68){\makebox(0,0){2.2}}
\thicklines \path(463,113)(463,133)
\thicklines \path(463,877)(463,857)
\put(463,68){\makebox(0,0){2.4}}
\thicklines \path(585,113)(585,133)
\thicklines \path(585,877)(585,857)
\put(585,68){\makebox(0,0){2.6}}
\thicklines \path(706,113)(706,133)
\thicklines \path(706,877)(706,857)
\put(706,68){\makebox(0,0){2.8}}
\thicklines \path(828,113)(828,133)
\thicklines \path(828,877)(828,857)
\put(828,68){\makebox(0,0){3}}
\thicklines \path(950,113)(950,133)
\thicklines \path(950,877)(950,857)
\put(950,68){\makebox(0,0){3.2}}
\thicklines \path(1071,113)(1071,133)
\thicklines \path(1071,877)(1071,857)
\put(1071,68){\makebox(0,0){3.4}}
\thicklines \path(1193,113)(1193,133)
\thicklines \path(1193,877)(1193,857)
\put(1193,68){\makebox(0,0){3.6}}
\thicklines \path(1314,113)(1314,133)
\thicklines \path(1314,877)(1314,857)
\put(1314,68){\makebox(0,0){3.8}}
\thicklines \path(1436,113)(1436,133)
\thicklines \path(1436,877)(1436,857)
\put(1436,68){\makebox(0,0){4}}
\thicklines \path(220,113)(1436,113)(1436,877)(220,877)(220,113)
\put(45,945){\makebox(0,0)[l]{\shortstack{$a_{\mbox{cr}}m_{0}$}}}
\put(828,23){\makebox(0,0){$D$}}
\put(706,252){\makebox(0,0)[l]{Symmetric phase}}
\put(342,738){\makebox(0,0)[l]{Broken phase}}
\thinlines \path(222,503)(222,503)(223,504)(226,505)(233,506)(245,510)(271,516)(321,530)(372,543)(423,556)(473,569)(524,582)(575,595)(625,608)(676,620)(727,633)(777,646)(828,658)(879,671)(929,684)(980,696)(1031,709)(1081,721)(1132,734)(1183,746)(1233,758)(1284,771)(1335,783)(1385,795)(1436,808)
\thinlines \dashline[3]{8}(220,308)(220,308)(222,308)(223,309)(226,310)(233,311)(245,315)(271,321)(321,334)(372,347)(423,360)(473,373)(524,386)(575,398)(625,411)(676,423)(727,436)(777,448)(828,460)(879,472)(929,485)(980,497)(1031,509)(1081,521)(1132,533)(1183,545)(1233,557)(1284,569)(1335,580)(1385,592)(1436,604)
\end{picture}

     \vglue 1ex
     \hspace*{18ex}\mbox{(a) Critical radius $a_{\mbox{cr}}$ as a function
of dimension $D$.}
     \vglue 7ex
\setlength{\unitlength}{0.240900pt}
\begin{picture}(1500,900)(0,0)
\tenrm
\thicklines \path(220,113)(240,113)
\thicklines \path(1436,113)(1416,113)
\put(198,113){\makebox(0,0)[r]{0}}
\thicklines \path(220,235)(240,235)
\thicklines \path(1436,235)(1416,235)
\put(198,235){\makebox(0,0)[r]{2}}
\thicklines \path(220,357)(240,357)
\thicklines \path(1436,357)(1416,357)
\put(198,357){\makebox(0,0)[r]{4}}
\thicklines \path(220,480)(240,480)
\thicklines \path(1436,480)(1416,480)
\put(198,480){\makebox(0,0)[r]{6}}
\thicklines \path(220,602)(240,602)
\thicklines \path(1436,602)(1416,602)
\put(198,602){\makebox(0,0)[r]{8}}
\thicklines \path(220,724)(240,724)
\thicklines \path(1436,724)(1416,724)
\put(198,724){\makebox(0,0)[r]{10}}
\thicklines \path(220,846)(240,846)
\thicklines \path(1436,846)(1416,846)
\put(198,846){\makebox(0,0)[r]{12}}
\thicklines \path(220,113)(220,133)
\thicklines \path(220,877)(220,857)
\put(220,68){\makebox(0,0){2}}
\thicklines \path(342,113)(342,133)
\thicklines \path(342,877)(342,857)
\put(342,68){\makebox(0,0){2.2}}
\thicklines \path(463,113)(463,133)
\thicklines \path(463,877)(463,857)
\put(463,68){\makebox(0,0){2.4}}
\thicklines \path(585,113)(585,133)
\thicklines \path(585,877)(585,857)
\put(585,68){\makebox(0,0){2.6}}
\thicklines \path(706,113)(706,133)
\thicklines \path(706,877)(706,857)
\put(706,68){\makebox(0,0){2.8}}
\thicklines \path(828,113)(828,133)
\thicklines \path(828,877)(828,857)
\put(828,68){\makebox(0,0){3}}
\thicklines \path(950,113)(950,133)
\thicklines \path(950,877)(950,857)
\put(950,68){\makebox(0,0){3.2}}
\thicklines \path(1071,113)(1071,133)
\thicklines \path(1071,877)(1071,857)
\put(1071,68){\makebox(0,0){3.4}}
\thicklines \path(1193,113)(1193,133)
\thicklines \path(1193,877)(1193,857)
\put(1193,68){\makebox(0,0){3.6}}
\thicklines \path(1314,113)(1314,133)
\thicklines \path(1314,877)(1314,857)
\put(1314,68){\makebox(0,0){3.8}}
\thicklines \path(1436,113)(1436,133)
\thicklines \path(1436,877)(1436,857)
\put(1436,68){\makebox(0,0){4}}
\thicklines \path(220,113)(1436,113)(1436,877)(220,877)(220,113)
\put(45,945){\makebox(0,0)[l]{\shortstack{$R_{\mbox{cr}}/m_{0}^{2}$}}}
\put(828,23){\makebox(0,0){$D$}}
\put(342,755){\makebox(0,0)[l]{Symmetric phase}}
\put(919,235){\makebox(0,0)[l]{Broken phase}}
\thinlines \dottedline{20}(220,113)(220,113)(221,113)(271,143)(321,170)(372,195)(422,219)(473,242)(523,264)(573,285)(624,306)(674,326)(724,346)(775,366)(825,387)(876,408)(926,429)(976,452)(1027,476)(1077,501)(1128,528)(1178,558)(1228,592)(1279,630)(1329,676)(1354,704)(1380,735)(1405,774)(1430,826)(1436,846)
\thinlines \path(222,501)(222,501)(223,502)(226,504)(233,506)(245,512)(271,522)(321,543)(372,562)(423,581)(473,599)(524,616)(575,633)(625,649)(676,664)(727,679)(777,694)(828,708)(879,721)(929,734)(980,747)(1031,759)(1081,771)(1132,783)(1183,794)(1233,805)(1284,816)(1335,826)(1385,837)(1436,846)
\thinlines \dashline[3]{8}(222,115)(222,115)(223,117)(226,121)(233,129)(245,144)(271,174)(321,230)(372,281)(423,327)(473,370)(524,410)(575,447)(625,482)(676,515)(727,546)(777,575)(828,602)(879,628)(929,653)(980,676)(1031,698)(1081,720)(1132,740)(1183,760)(1233,779)(1284,797)(1335,814)(1385,830)(1436,846)
\end{picture}

     \vglue 1ex
     \hspace*{17ex}\mbox{(b) Critical curvature $R_{\mbox{cr}}$ as a function
of dimension $D$.}
     \vglue 1ex
\caption{Phase diagrams in de Sitter space and Einstein
universe.
         The full and the dashed lines
         represent the exact solution in
         de Sitter space and Einstein universe respectively.
         The dotted line represents the critical curvature
obtained
         by the weak curvature approximation.}
\label{fig:crradall}
\end{figure}
It should be noted that Eq. (\ref{eqn:critgap}) reduces to
\begin{equation}
     a_{\mbox{cr}}=\frac{e^{-\gamma}}{2 m_{0}}\, ,
\end{equation}
in two dimensions which reproduces the result obtained in
Refs. \cite{FT} and \cite{KINK}.
Using the relationship (\ref{rl:ein}),
the critical curvature is obtained from the critical radius.
For some special values of $D$ the critical curvature
$R_{\mbox{cr}}$ simplifies to
\begin{equation}
\begin{array}{ll}
     {\displaystyle R_{\mbox{cr}}=0}\,  &; D=2\, ,\\[4mm]
     {\displaystyle R_{\mbox{cr}}=8 m_{0}^{2}}\,  &; D=3\, ,\\[4mm]
     {\displaystyle R_{\mbox{cr}}=12 m_{0}^{2}}
     \,  &; D=4\, .
\end{array}
\label{rc234:ein}
\end{equation}
The critical curvature $R_{\mbox{cr}}$ is plotted in Fig. 18(b) as a
function of
the spacetime dimension $D$.
It is clearly seen in Fig. 18(b) that three lines of the
critical curvature
reach the same value $12 m_{0}^{2}$ at $D\rightarrow 4$.
For $D=2$ the critical curvature obtained by weakly
curvature approximation
is exactly equal to that in Einstein universe.
As is explained in the last section the chiral symmetry
is restored by the effect of an infra-red divergence for any
finite curvature
in two-dimensional weakly curved spacetime.
The critical curvature is thus zero.
In two-dimensional Einstein universe it is obtained that
the critical curvature $R_{\mbox{cr}}=0$.
The situation is, however, different from that in weakly
curved spacetime.
By definition two-dimensional Einstein universe is a flat
spacetime,
$R=0$.
The symmetry restoration is induced by finite size
effects
of the compact and closed space.
The spacetime curvature $R$ is not suitable to represent
the phase structure
in two-dimensional Einstein universe.

Thus we have considered the Gross-Neveu type model as one of the
prototype
models of the dynamical symmetry breaking and investigated the phase
transition induced by the curvature effect without
making any approximation in the spacetime curvature.

In de Sitter, anti-de Sitter space 
and Einstein universe the two-point functions
have been solved exactly.
Then the exact expression of the effective potential
is obtained in such spacetimes.
We calculated the renormalized
effective potential for finite $R$ in the leading
order of the $1/N$ expansion in arbitrary dimensions
$2\leq D < 4$ .
It is found that the broken chiral symmetry is restored at a
certain critical curvature
in de Sitter space and Einstein universe
(i.e., positive curvature spacetimes).
The phase transition from the broken phase to the symmetric
phase is of second order.
In anti-de Sitter space (i.e., negative curvature
spacetime) only the broken phase
is realized irrespective of the value of $\lambda$.

For a negative curvature spacetime the weak curvature
approximation gives the exact phase structure, $R_{\mbox{cr}}=0$,
in the whole range of $D$ considered here: $2 \leq D < 4$.
The critical curvature $R_{\mbox{cr}}$ obtained by weak curvature
approximation
is exactly equal to the one obtained in de Sitter
space and Einstein universe at $D=4$ shown in Fig. 18(b).
In four dimensions ultraviolet divergences appear in terms
independent
of the curvature $R$ and terms linear in $R$ only.
The higher order terms in $R$ of the effective potential
are ultraviolet finite.
Expanding the exact results in de Sitter space and Einstein
universe
asymptotically about $R=0$,
we obtain the $R^{2}$ term of the effective potential
in a compact spacetime with a constant curvature.
\begin{equation}
     V(\sigma)=V_{0}(\sigma)+V_{\sR}(\sigma)+V_{\sR2}(\sigma)
     +O(R^{3})\, ,
\end{equation}
where $V_{0}(\sigma)$ and $V_{\sR}(\sigma)$ are given by
Eqs. (\ref{v:nonren}) and (\ref{vr:nonren:r}) respectively,
the $R^{2}$ term $V_{\sR2}(\sigma)$ reads \cite{WC2}
\begin{equation}
     V_{\sR2}(\sigma)=-\frac{\tr \11}{(4\pi)^{\sD/2}}
     \Gamma\left(1-\frac{D}{2}\right)
     \frac{R^{2}}{5760}\frac{(D-2)(D-3)(2+5D)}{D(D-1)}
     \sigma^{\sD-4}\, .
\label{v:r2}
\end{equation}
At the four-dimensional limit the $R^{2}$ term (\ref{v:r2})
reduces to
\begin{equation}
     \frac{V_{\sR2}^{D=4}(\sigma)}{\mu^{\sD}}
     =\frac{\tr \11}{(4\pi)^{2}}\frac{11 R^{2}}{17280}
     \left(C_{\mbox{div}}-\frac{173}{66}-\ln
\left(\frac{\sigma}{\mu}\right)^{2}\right)\, .
\label{r2:4d}
\end{equation}
The divergent parts in Eq. (\ref{r2:4d}) appear from the
mass singularity
at $\sigma\rightarrow 0$ and the normalization condition
$V(0)=0$.
Only an infrared divergence appears in the $R^{2}$ term
(\ref{r2:4d}).
The infrared divergence does not appear in the de Sitter
space
and Einstein universe since the space components are compact
and closed (i.e. there is an infrared cut-off 
$\Lambda_{IR}=1/(2a)$).
The $R^{2}$ term has no contribution to the symmetry
restoration
in a compact space (de Sitter space and Einstein universe).
Thus the critical curvature is determined by the terms
involving ultraviolet divergences in four dimensions
and the weak curvature approximation gives the exact result.
Therefore the weak curvature approximation seems
to be useful near
four dimensions.\cite{WC2}
If we use the cut-off regularization in four dimensions
we also see the similar situation.
In the case the  weak curvature approximation also gives the
exact result at the limit $\Lambda\rightarrow\infty$.\cite{SACH}

In the other dimensions the critical curvature (\ref{cr:r:w})
is unable to compare with the results obtained
in de Sitter space and Einstein universe directly, because
the global topology of the spacetime may play a crucial
role for the symmetry restoration.
In two-dimensional Einstein universe $R\otimes S$
not the curvature but the size of the space coordinate $a$
determines the minimum of the effective potential.
(See \S 6.)
The critical radius $a_{\mbox{cr}}$ in Einstein universe
is half as many as that in de Sitter space.
It seems that the number of the compactified directions
determine the critical radius.
Thus the finite size effect may play an essential 
role in
two dimensions.\footnote{For $D=2$ the chiral symmetry
may be restored at any value of $a$ in the case of
finite $N$ through the creation of an instanton-antiinstanton
or kink-antikink condensation as is discussed at finite
temperature.\cite{KINK}
In our method we cannot deal with the influence of time
or space dependent configurations.
Instantons and kinks are, however, suppressed at the
large $N$ limit and the phase transition can
take place in two dimensions.
}

\section{Extensions of four-fermion model}

\subsection{Gauged NJL model in curved spacetime}

Gauged NJL model maybe considered as one of the most realistic
four-fermion models.\cite{BLL}
From one side, that is one of the best effective theories to
describe QCD.
From another side, gauged NJL model may well describe
SM (standard model) where composite bound states play the role of
an elementary Higgs fields in the process of dynamical symmetry
breaking.

For a long time the only way to study gauged NJL model
has been related with Schwinger-Dyson equation.
\cite{SD}${}^{,}$\footnote{
It is not known how to formulate consistent Schwinger
Dyson equation even in ladder approximation in general
curved spacetime.\cite{OS}
}
However in Ref. \cite{HKKN} (following the renormalization group
(RG) approach of Ref. \cite{BHL}) it has been shown that some class of
gauge Higgs-Yukawa models in leading order of a modified
$1/N_{c}$ expansion leads to a well-defined, non-trivial
theory.
This theory is equivalent to the gauged NJL model when the
ultraviolet cut-off goes to infinity (using corresponding
compositeness conditions).
Then, it appears the possibility to find the effective
potential and to discuss the phase structure of gauged NJL
model without use of Schwinger-Dyson equations. It has 
been done in flat space in Ref. \cite{HKKN} and in curved space
in Ref. \cite{GO}.
Below, we will discuss the phase structure of gauged
NJL model following closely Ref. \cite{GO}.

\subsubsection{Renormalization group equations for
the class of gauge Higgs-Yukawa models}

In this subsection we review the class of gauge Higgs-Yukawa models in flat 
spacetime we are going to extend and the approximations we are using in 
such an analysis. Thereby we closely follow Ref. \cite{HKKN} where, using the
 RG approach, this model has been considered in detail.

The Lagrangian of the model in flat spacetime is given by (we use the 
notations of Ref. \cite{HKKN})
\begin{eqnarray}
L_m =&-&{\frac{1}{4}}~ G^a_{\mu\nu} G^{a\mu \nu} ~+~ {\frac{1}{2}}~
(\partial_{\mu}\sigma)^2
 ~-~\frac{1}{2}~ m^2\sigma^2~-~{\frac{\lambda}{4}}~\sigma^4 \nonumber\\
&+&{\sum\limits^{N_f}_{i=1}}~ \bar{\psi}_i ~i\gamma^{\mu}D_{\mu}~ \psi_i ~-~ 
{\sum\limits^{n_f}_{i=1}}~y\sigma \bar{\psi}_i\psi_i~.~
\end{eqnarray}
\noindent Here the gauge group $SU(N_c)$ is chosen, with $N_f$ 
fermions $\psi_i (i = 1, 2, ..., N_f)$ belonging to the representation 
$R$ of $SU(N_c)$, $\sigma$ is the scalar field.

Let us now describe the $1/N$ approximation for the perturbative study of 
this theory at high energies through the RG equations:\\
a) The gauge coupling constant is assumed to be small:
\begin{equation}
{\frac{g^2N_c}{4\pi}}\ll 1 ~,
\end{equation}
\noindent and the RG equations are considered only in the first non-trivial
 order on $g^2$.\\
b) The number of fermions should be large enough:
\begin{equation}
N_f\sim N_c~.
\end{equation}
\noindent On the same time it is supposed that only $n_f$ fermions
 ($n_f \ll N_f$) have large Yukawa couplings, whereas the 
remaining fermions have vanishing Yukawa couplings.\\
c) The approach is assumed to be perturbative also in $1/N_c$,
so only the leading order of the $1/N_c$ expansion survives; this means 
that scalar loop contributions should be negligible
\begin{equation}
\left|{\frac{\lambda}{y^2}}\right| \le N_c~.
\end{equation}
\noindent Note that above approximation maybe reasonable
for the minimal SM.

Within the above approximation the RG equations for the coupling constants
 in (5$\cdot$1) are\footnote{
For general discussion of one-loop RG equations in gauge
Higgs-Yukawa models, see Refs. \cite{Chang} and \cite{GW}. 
}
\begin{eqnarray}
&{}& {\frac{dg(t)}{dt}}=-~{\frac{b}{(4\pi)^2}}~ g^3(t)~,\nonumber\\
&{}&{\frac{dy(t)}{dt}}= {\frac{y(t)}{(4\pi)^2}}~[a~y^2(t)-c~g^2(t)]~,\nonumber\\
&{}&{\frac{d\lambda (t)}{dt}} = {\frac{u~y^2(t)}{(4\pi)^2}}~[\lambda (t)-y^2 (t)]~,
\end{eqnarray}
 \noindent where
$ b=(11 N_c-4T(R)N_f)/3$, $c=6~C_2(R)$, $a=u/4=2~n_fN_c$. 
For the fundamental representation we have 
$T(R) = 1/2$, $C_2(R) = (N_c^2 - 1)/(2N_c)$.
Here $t = \ln (\mu / \mu_0)$ and $\mu_0$
is the reference scale to discuss low energy physics.

The RG equation for $g^2$ (respectively $\alpha = g^2/4 \pi$) is solved by
\begin{equation}
\eta (t)\equiv {\frac{g^2(t)}{g^2_0}}\equiv {\frac{\alpha
(t)}{\alpha_0}}=
\left(1+{\frac{b~ \alpha_0}{2\pi}}t\right)^{-1}~.
\end{equation}

To solve the RG equations for the Yukawa and the scalar couplings it is 
convenient to introduce the following RG invariants:
\begin{eqnarray}
&{}& h(t)\equiv -\eta^{-1+c/b}(t)~\left[
1-{\frac{c-b}{a}} ~~
{\frac{g^2(t)}{y^2(t)}}\right],\nonumber\\
&{}& k(t)\equiv-\eta^{-1+2c/b}(t)~\left[1-{\frac{2c-b}{2a}}~~
{\frac{\lambda (t)}{y^2(t)}}~~ {\frac{g^2(t)}{y^2(t)}}\right]~.
\end{eqnarray}
\noindent With their help we obtain
\begin{eqnarray}
&{}& y^2(t)={\frac{c-b}{a}}~ g^2
(t)~\left[1+h_0\eta^{1-c/b}(t)\right]^{-1},\nonumber\\
&{}& \lambda (t) = {\frac{2a}{2c-b}}~~{\frac{y^4(t)}{g^2(t)}}~
\left[1+k_0\eta^{1-2c/b}(t)\right].
\end{eqnarray}
\noindent Of course, these solutions are explicitly known only when the
values of the RG invariants $h$ and $k$ are given. Note that in 
(5$\cdot$8) and
below we use notations $h_0$ and $k_0$ for $h$ and $k$ in order to show
that they are constants (RG invariants).

As it has been shown in detail in Ref. \cite{HKKN} 
for the solutions (5$\cdot$6) $\sim$ (5$\cdot$8)
to be consistent with the above approximation 
(5$\cdot$2) $\sim$ (5$\cdot$4), 
and in order for the theory to be
non-trivial one, we should have
\begin{equation}
c>b~~\mbox {and}~~ 0\le h_0 < \infty ~.
\end{equation}
\noindent That makes the Yukawa coupling to be asymptotically free and 
the theory to be non-trivial. Further analysis shows
that the scalar coupling constant is nontrivial in the above approximation 
only if
\begin{equation}
k_0 = 0~.
\end{equation}

Below we will discuss the solutions (5$\cdot$8) 
with conditions (5$\cdot$9) and (5$\cdot$10) only. 

Let us extend now this model to curved spacetime. 
It is by now well-known (see Ref. \cite{BOS} 
for an introduction) that in order 
to be multiplicatively renormalizable in curved spacetime, the Lagrangian 
of the theory should be:
\begin{equation}
L=L_m+L_{ext} -{\frac{1}{2}} \xi R \sigma^2~,
\end{equation}
\noindent where $L_m$ is given by (5$\cdot$1) with a change of flat (partial)
derivatives to the corresponding covariant 
derivatives: $\partial_\mu \rightarrow \nabla_\mu$, $\xi$ is the 
non-minimal scalar-gravitational coupling constant, and
\begin{equation}
L_{\mbox{ext}} =a_1 R^2+a_2 
C^2_{\mu \nu\alpha \beta} +a_3 G+a_4 \Box
R+\Lambda-{\frac{1}{\kappa}} R~.
\end{equation}
\noindent The Lagrangian $L_{\mbox{ext}}$ 
of the external gravitational field 
is necessarily introduced in order to have the theory to be multiplicatively
renormalizable one; $a_1, ... a_4, \kappa$ are gravitational coupling
constants.

Since the RG equations for those coupling constants which are present in 
flat space do not change in curved spacetime,\cite{BOS} 
all the discussions 
of RG equations for $g, y$ and $\lambda$
(as well as the approximations introduced above) are also valid 
here. In addition, the effective coupling constants corresponding to $\xi, 
a_1, ... , a_4, G$ appear. However, for our purposes, we only need the 
RG equation for the coupling constant $\xi$
which maybe written as following (within the above described 
approximation),
\begin{equation}
{\frac{d\xi (t)}{d t}} = {\frac{1}{(4\pi)^2}}~ 2a y^2 (t)~
\left(\xi (t)- 
{\frac{1}{6}}\right)~.
\end{equation}
\noindent Taking into account that
\begin{equation}
{\frac{d}{dt}} f(t)\equiv {\frac{d}{dt}}\left[ \eta^{c/b} (t) ~~
{\frac{(\xi (t)-{\frac{1}{6}})}{y^2 (t)}} \right]=0\, ,
\end{equation}
\noindent we may use this RG invariant $f(t)$ to find 
the solution of (5.13):
\begin{equation}
\xi (t)={\frac{1}{6}}+y^{2}(t)\eta^{-c/b} (t)~ f_0~,
\end{equation}
\noindent where $f_0 \not= 0$ is the value of the RG invariant. 
Note that in the UV-limit ($t \rightarrow \infty$) we have:
\begin{eqnarray}
&{}& y^2 (t)\sim {\frac{c-b}{a}}~~{\frac{g_0^2}{h_0}}
 ~\eta^{c/b} (t)
\rightarrow +0,~~~~~(h_0\ge 0) \nonumber\\
&{}& \xi (t) \simeq {\frac{1}{6}} + {\frac{c-b}{a}}~ g^2_0~{\frac{f_0}{h_0}}~.
\end{eqnarray}

For $f_0 \approx 0$ the scalar-gravitational coupling constant
$\xi(t) \rightarrow 1/6$, i.e., asymptotic conformal invariance is 
realized.\cite{BO} Hence, the RG invariant $f_0$ characterizes the
deviation from conformal invariance ($\xi = 1/6$). Different types 
of UV-behavior of $\xi (t) $ for different GUT's have been listed
in Ref. \cite{BOS}; the most typical ones are:
\begin{eqnarray}
&\mbox{a)    }& \xi (t)\rightarrow {\frac{1}{6}}~~, 
\nonumber\\
&\mbox{b)    }& |\xi (t)| \rightarrow \infty~~, 
\nonumber\\
&\mbox{c)    }& \xi (t) \rightarrow \xi~~. 
\end{eqnarray}
\noindent As we can see, unlike the analysis in the case of flat 
spacetime, there will not appear any restrictions to the sign
and the value of $f_0$ from the study of the UV-asymptotics of $\xi(t)$.

Let us discuss now the one-loop effective potential for the theory (5.1) 
in the approach where we keep only terms with the accuracy up to linear 
curvature,\cite{BO2,EO} i.e., $\sigma^2 \gg |R|$. 
This approach is actually equivalent to the one described
in \S 3.
Within above approximation we get
\begin{eqnarray}
V&=&{\frac{1}{2}} m^2 \sigma^2 + {\frac{\lambda}{4}} \sigma^4 +
{\frac{1}{2}}
\xi R \sigma^2 - {\frac{a \mu^4_F}{2(4\pi)^2}} \left[ \ln 
{\frac{\mu^2_F}{\mu^2}}-{\frac{3}{2}}\right]\nonumber\\
&{}& -~{\frac{a R \mu^2_F}{12(4\pi)^2}} \left[ \ln
{\frac{\mu^2_F}{\mu^2}}-1\right],
\end{eqnarray}
\noindent where $\mu_F \equiv y \sigma$.

Using such a form for the potential $V$ one may investigate the 
curvature-induced phase transitions which have been discussed in detail in
Refs. \cite{BO2,BOS} and \cite{EO} for different GUT's.

\subsubsection{Gauged NJL model in curved spacetime}

Let us discuss now the gauged NJL model in curved spacetime.

The assumption that not elementary ones but composite bound 
states are mainly relevant in quantum cosmology
motivated the study of the chiral symmetry 
breaking under the influence of the external gravitational field
(see previous sections.)
Such an investigation has been done via explicit calculation of the 
effective potential for the composite field $\langle\bar{\psi}\psi\rangle$.
Our 
purpose here will be to discuss the same question on RG language for
the gauged NJL model using its equivalence with the gauged Higgs-Yukawa 
model (5$\cdot$1) (the explicit loop calculations in such a model are extremely
hard to do).

We start from the gauged NJL model with four-fermion 
coupling constant $G$ 
in curved spacetime
\begin{equation}
L=-{\frac{1}{4}} G^2_{\mu \nu} + {\sum\limits^{N_f}_{i=1}}
\bar{\psi}_{i} i \gamma^{\mu}D_{\mu}
\psi_i + G ~{\sum\limits^{n_f}_{i=1}} (\bar{\psi}_i \psi_i)^2~.
\end{equation}
\noindent As usually one introduces an auxiliary field $\sigma$ to replace
the NJL model by the equivalent Higgs-Yukawa model. As is shown in 
Ref. \cite{BHL}, it is possible within the 
RG approach to use a set of boundary 
conditions for the effective couplings of the gauge Higgs-Yukawa model 
at $t_\Lambda = \ln (\Lambda/\mu_0)$ (where $\Lambda$ is the
UV-cut off
of the gauged NJL model) in order to identify the gauged NJL model with 
the gauge Higgs-Yukawa model. Taking into account Eqs. (5.8) the explicit
expressions for these compositeness conditions (which are specified 
in Ref. \cite{BHL}) 
are given as (see also Eqs. (4$\cdot$7) and (4$\cdot$8) of 
Ref. \cite{HKKN}):
\begin{eqnarray}
&{}& y^2 (t) ={\frac{c-b}{a}}~ g^2 (t)~ \left[
1-\left({\frac{\alpha (t)}
{\alpha (t_\Lambda)}}\right)^{1-c/b} \right]^{-1}
\equiv 
y^2_\Lambda (t),\nonumber\\
&{}& {\frac{\lambda (t)}{y^4(t)}} =
{\frac{2a}{2c-b}}~~{\frac{1}{g^2 (t)}}
\left[ 1- \left( {\frac{\alpha (t)}{\alpha (t_\Lambda)}}\right)^
{1-2c/b} \right] \equiv {\frac{\lambda_\Lambda
(t)}{y^4_\Lambda(t)}}~,
\end{eqnarray}
\noindent where $ t<t_\Lambda$. Hence, the system (5$\cdot$1) with 
cut-off $\Lambda$ is equivalent to the system (5$\cdot$19) with the same 
cut-off when the running coupling constants are given by (5$\cdot$20). 
In this sense the gauged NJL model may be called a renormalizable one.
In addition, as it has been shown in Ref. \cite{HKKN}, 
one should have also a
compositeness condition for the mass, whose most convenient form will 
be (for $b \rightarrow +0$ \cite{HKKN}):
\begin{equation}
m^2 (t) = {\frac{2a}{(4\pi)^2}} \left(
{\frac{\Lambda^2}{\mu^2}}\right)^w
y^2_\Lambda (t)~\mu^2 \left[{\frac{1}{g_4(\Lambda)}} -
{\frac{1}{w}}\right]~,
\end{equation}
\noindent where $G\equiv ((4\pi)^2/a) g_4(\Lambda)/\Lambda^2 ,
 ~g_4 (\Lambda)$ is a dimensionless constant, and $w\equiv
1-\alpha/(2\alpha_C)
,~ \alpha\equiv \alpha_0$ and where $y^2_\Lambda (t)$ 
is given by the first of Eqs. (5$\cdot$20) at $b \rightarrow +0$.

The above compositeness conditions define the gauged NJL model as 
the gauge Higgs-Yukawa model in flat spacetime. Now, since we are 
working in curved spacetime (in linear curvature approximation), 
one has to add the compositeness condition for $\xi(t_\Lambda)$. 
Making the calculations in the same way as in 
Ref. \cite{HS} one obtains 
\begin{equation}
\xi (t_\Lambda) ={\frac{1}{6}}~.
\end{equation}

Obviously, the compositeness condition for $\xi(t)$ is again the 
same one as in the non-gauged NJL model \cite{HS}
(for finite corrections see also Ref. \cite{Reu}). 
Hence, the conformal 
invariance due to the renormalization group \cite{BO} takes place again. 
Analyzing the RG equation (5$\cdot$13) for $ \xi(t)$ we will find that in all 
cases (i.e., for the general situation, where $y(t)$ is given by 
(5$\cdot$20), 
for the case with $b \rightarrow 0$ and with the corresponding $y(t)$, 
and also for the case $\alpha_0 \rightarrow 0$, i.e., the non-gauged 
NJL model) we have
\begin{equation}
\xi (t)={\frac{1}{6}}~.
\end{equation}

Now, one can study the possibility of chiral symmetry breaking in 
the model under discussion using the effective potential language, or
more precisely,
   the RG-improved effective potential (see Refs. \cite{CW}
and \cite{EJ} and
references therein). Because the technique to study the RG-improved 
effective potential is widely known for the flat 
space \cite{CW,EJ} as 
well as for curved spacetime \cite{BO2,EO} we will not give any 
details of its derivation.

Using the fact that the effective potential satisfies the RG equation, 
one can explicitly solve this equation by the method of characteristics,
and we find:
\begin{equation}
V (g,y,\lambda,m^2,\xi,\sigma,\mu)=V(g(t), y(t),\lambda (t),m^2
(t),\xi (t),
\sigma (t), \mu e^t)~,
\end{equation}
\noindent where the effective coupling constants $g(t), ... , \sigma(t)$ 
are defined by the RG equations (5$\cdot$5), (5$\cdot$13) 
(at the scale $\mu e^t$) and 
corresponding RG equations for $m^2 (t), \sigma(t)$ written 
in Ref. \cite{HKKN} for 
the case of the above gauge-Higgs-Yukawa model ($t$ is left unspecified 
for the moment). As boundary condition it is convenient to use the 
one-loop effective potential (5$\cdot$18).

In the case of the gauged NJL model one has to substitute into the 
effective potential (5$\cdot$18) the effective couplings 
fulfilling compositeness 
conditions in accordance with Eq. (5$\cdot$24). 
In this way, one obtains the RG
improved effective potential in the gauged NJL model. In flat spacetime 
such calculation has been already done in Ref. \cite{HKKN} 
for the case of fixed 
gauge coupling ($b \rightarrow +0$), so we may use for the running 
coupling constants (except scalar-gravitational
coupling constant) the results which are known from flat spacetime
calculations.

Taking into account that the condition
\begin{equation}
\mu e^t=\mu_F(t)\, ,
\end{equation}
\noindent serves actually for finding $t$, we get for the case of 
fixed gauge coupling
\begin{equation}
e^t=\left(\frac{\mu_F(\mu)}{\mu}\right)^{1/(2-w)}\, ,
\end{equation}
\noindent where $\mu_F(\mu) = \mu_F(t=0)$. Using (5.20), 
(5$\cdot$23) and (5$\cdot$26) 
in the case $b \rightarrow 0$ one gets (for the case of flat space 
see expression (6$\cdot$13) of Ref. \cite{HKKN}):
\begin{eqnarray}
{\frac{(4\pi)^2}{2a}}~{\frac{V}{\mu^4}}&=&{\frac{1}{2}}~y^2_\Lambda (\mu)
\left({\frac{\Lambda^2}{\mu^2}}\right)^w
\left({\frac{1}{g_4(\Lambda)}}-
{\frac{1}{g^\ast_4}}\right) {\frac{\sigma^2(\mu)}{\mu^2}}\nonumber\\
&{}&
+~{\frac{\alpha_C}{4\alpha}}
 ~\left[\left({\frac{y_\Lambda(\mu)\sigma(\mu)}{\mu}}\right)^{4/(2-w)}
 -
\left({\frac{\mu^2}{\Lambda^2}}
\right)^{\alpha /\alpha_C}
\left({\frac{y_\Lambda(\mu)\sigma (\mu)}{\mu}}\right)^4\right]\nonumber\\
&{}& +~{\frac{3}{8}}\left({\frac{y_\Lambda (\mu)\sigma(\mu)}{\mu}}
\right)^{4/(2-w)} \nonumber\\
&{}&+~ {\frac{R}{24\mu^2}}  
\left({\frac{y_\Lambda(\mu)\sigma(\mu)}{\mu}} 
\right)^{2/(2-w)} 
\left(1+~{\frac{2\alpha_C}{\alpha}}
\right)\nonumber\\
&{}&-~
\frac{R\alpha_C}{12\alpha \mu^2}
\left({\frac{\mu}{\Lambda}}
\right)^{\alpha /\alpha_C}~
{\frac{y^2_\Lambda (\mu)\sigma^2(\mu)}{\mu^2}}~,
\end{eqnarray}
\noindent where $g^\ast_4 \equiv w$. Thus, we have got the RG-improved 
effective potential for the gauged NJL model (with finite cut-off) in
curved spacetime.

Taking the limit $\Lambda \rightarrow \infty$ we will get the renormalized 
effective potential of the gauged NJL model in curved spacetime:
\begin{eqnarray}
{\frac{(4\pi)^2}{2a}}~ {\frac{V}{\mu^4}}&=&{\frac{1}{2}} 
\left( {\frac{1}{
g_{4R}(\mu)}}-{\frac{1}{g^\ast_{4R}}}\right){\frac{y^2_\ast~\sigma^2 (\mu)}
{\mu^2}}\nonumber\\
&{}& +~ {\frac{\alpha_C}{4\alpha}}\left(
1+{\frac{3\alpha}{2\alpha_C}}\right)
\left({\frac{y_\ast~ \sigma(\mu)}{\mu}}\right)^{4/(2-w)}\nonumber\\ 
&{}&+~ {\frac{R}{24\mu^2}} \left(1+{\frac{2\alpha_C}{\alpha}}\right)
\left({\frac{y_\ast ~\sigma (\mu)}{\mu}}\right)^{2/(2-w)}\, ,
\end{eqnarray}
\noindent where 
$y^2_\ast =(4\pi)^2 \alpha /2a\alpha_{C}$ 
and the four-fermion coupling 
renormalization has been done; in particular $g^\ast_{4R}$ is a finite 
constant (for discussion of that renormalization 
see Refs. \cite{HKKN} and \cite{KTY}).

 Note that 
in flat space ($R=0$) the potential (5$\cdot$28) in the same approach has been 
obtained in Ref. 
\cite{KTY} (for non-renormalized potential see also 
Ref. \cite{BL}) using 
 the ladder SD equation. Hence, the RG-improved effective potential
may serve as a very useful tool to study the non-perturbative effects on 
equal footing with the SD equation. It is really surprising that RG
improved effective potential gives the same results as ladder SD equation.

Using the finite effective potential (5$\cdot$28) we are able to discuss the 
chiral symmetry breaking in the gauged NJL model under consideration. 
In flat spacetime, the possibility of chiral symmetry breaking is defined 
by the sign of the first term in (5$\cdot$28). 
That gives the value of the critical 
four-fermion coupling constant $g^\ast_4(\Lambda) = w$. In curved spacetime 
the situation is more complicated as one has additional terms in the
effective potential (5$\cdot$28), even in the linear curvature 
approximation.

Taking into account that
\begin{eqnarray}
&{}&w=1-{\frac{3C_2(R)}{2(2\pi)^2}} g^2\simeq 1-{\frac{3}{4\pi}} 
{\frac{N_cg^2}{4\pi}} \simeq 1,\nonumber\\
&{}& {\frac{\alpha_C}{\alpha}} \simeq {\frac{2\pi}{3}}
{\frac{4\pi}
{N_c g^2}}\gg 1,
\end{eqnarray}
\noindent and introducing $x=y_\ast~\sigma (\mu)/\mu$, one can 
rewrite the quadratic part of the potential (5$\cdot$28) as follows:
\begin{eqnarray}
&{}&{\frac{(4\pi)^2}{2a}}~ {\frac{V^{(2)}}{\mu^4}}\simeq
\left\{ {\frac{1}{2}}\left(
{\frac{1}{g_{4R}(\mu)}} - {\frac{1}{g^\ast_{4R}}}\right) +
{\frac{R}{12\mu^2}}~ {\frac{\alpha_C}{\alpha}}\right\} x^2~.
\end{eqnarray}
\noindent Relation (5$\cdot$30) determines the way to estimate the chiral 
symmetry breaking for the gauged NJL model in curved spacetime.

In particular, in flat spacetime and for (cut-off) dependent effective 
potential we observe that the chiral symmetry is broken for
\begin{equation}
{\frac{1}{g_4(\Lambda)}} - {\frac{1}{g^\ast_4}}<0 ~~\mbox{or}~~
{\frac{1}
{g_{4R}(\mu)}} - {\frac{1}{g^\ast_{4R}}}<0~.
\end{equation}
\noindent Hence, the critical value of the four-fermion coupling 
constant which divides the chiral symmetric and non-symmetric phase 
is $g^\ast_4 = w$.

 In curved spacetime, chiral symmetry is always broken if
\begin{equation}
\left( {\frac{1}{g_{4R}(\mu)}} - {\frac{1}{g^\ast_{4R}}}\right)
+
{\frac{\alpha_C}{6\alpha}}~~ {\frac{R}{\mu^2}}<0~.
\end{equation}
\noindent When this condition (5$\cdot$32) is valid one can easily find the 
curvature-induced dynamical fermion mass.
In a similar way one can find the chiral symmetry breaking
condition for cut-off dependent effective potential,
\begin{eqnarray}
&{}&\left( {\frac{\Lambda^2}{\mu^2}} \right)^w
\left({\frac{1}{g_4(\Lambda)}}-
{\frac{1}{g^\ast_4}} \right) + {\frac{R}{6\mu^2}}
{\frac{\alpha_C}
{\alpha}}\left[ 1-\left(
{\frac{\mu}{\Lambda}}\right)^{\alpha /\alpha_C}\right] < 0~.
\end{eqnarray}

Thus, we have got the condition for chiral symmetry breaking in terms of 
the dimensionless curvature $\tilde{R} \equiv R/{\mu}^2$. From (5$\cdot$32) 
we see that the critical coupling constant depends on the curvature. 
For negative curvature the critical coupling is higher and one has a 
greater chance to find the system in the phase with broken chiral 
symmetry. As an example of spaces being in correspondence with 
(5$\cdot$33) one 
can consider the inflationary universe $S^4$ with small curvature.

Note that the critical value of the curvature at which symmetry 
breaking is absent is defined according to (5$\cdot$32) by
\begin{equation}
{\frac{R_{\mbox{cr}}}{\mu^2}}
={\frac{6\alpha}{\alpha_C g^\ast_{4R}}}~.
\end{equation}
\noindent At all positive curvatures below $R_{\mbox{cr}}$ as well as at 
small negative curvatures the chiral symmetry is broken.

It is quite remarkable that such a simple quantum condition of chiral 
symmetry breaking in gauged NJL model in curved spacetime
is obtained explicitly. Up to now such a simple symmetry breaking 
tree-level condition in curved spacetime has been known only for the 
Higgs sector of (5$\cdot$11):
\begin{equation}
\sigma^2 = -\frac{\xi R +m^2}{\lambda}\, ,
\end{equation}
\noindent where $\xi R + m^2 < 0$.

Thus we discussed the gauged NJL model in curved spacetime using 
quite standard RG language.
The effective potential for composite fermions is found explicitly
and its phase structure is discussed.
Some analytical results are obtained explicitly (in particularly, the
condition of chiral symmetry breaking in the case of fixed gauge
coupling).
There are different possibilities to extend above methods.
First, one can consider other gauged NJL models
(with more scalars) as equivalent to gauge-Higgs-Yukawa
models typical for GUTs.
Second, one can consider other gravitational backgrounds
(say, de Sitter space).
There is no problem also to generalize this approach for the 
situation when external gravity is sufficiently strong 
(see Ref. \cite{GO2}).

\subsection{Higher derivative four-fermion model in curved spacetime}

As it has been mentioned several times, four-fermion models
are considered now as non-renormalizable effective theories
where the presence of an ultraviolet cut-off $\Lambda$
at loop diagrams is a necessary condition.
Then there are different possibilities to extend the NJL
model.
Among such possibilities, a quite interesting one is the 
introduction in the original Lagrangian of higher derivative
terms in the four-fermion interaction \cite{AA}
or the kinetic term.\cite{HK}
It maybe shown \cite{HHJKS} that the physics of such higher
derivative four-fermion model is still equivalent to the 
physics of SM.
Note also that the inclusion of higher derivatives
in effective theories gives the possibility to take
into account the structural effects of the medium and
external fields.
In the present subsection we discuss the effective potential
and phase structure of higher derivative four-fermion model
in curved spacetime.
We mainly follow Ref. \cite{HD}.

We start by presenting the model which we set out to study
(it was introduced in Ref. \cite{AA}).
Its Lagrangian in curved spacetime is given by
\beq
{\cal L} = \bar{\psi} i \gamma ^\mu (x) \nabla _\mu \psi + \frac{1}{4
N_{c} \Lambda ^2} \left\{ \lambda _1 \left( \bar{\psi} \psi \right) ^2 +
3 \lambda _2 \left[ \bar{\psi} \left(1-2 \frac{\not\!\nabla ^2}{\Lambda
^2} \right) \psi \right]^2 \right\} \label{1} \ ,
\eeq
where $N_c$ is the number of fermionic species, 
$\Lambda $ a cut-off parameter, and
$\lambda_1$ and $\lambda_2$ are coupling constants. We work in the
$1/N_c$ expansion scheme.

 By introducing some auxiliary fields $\chi _1$ and $\chi  _2$ we can
give a description of this nonrenormalizable theory by means of the
action
\bea
S &=& \int d^4 x \sqrt{g} \left\{ \bar{\psi} i \gamma ^\mu (x)
\nabla _\mu \psi - N_c \Lambda ^2 \left( \frac{\chi_1 ^2}{\lambda_1} +
\frac{\chi_2 ^2}{\lambda_2} \right) \right.
\nonumber \\
&&\left. - \left[ \chi_1 \bar{\psi} \psi +
\sqrt{3} \chi_2 \bar{\psi} \left( 1- 2 \frac{\not\!\nabla ^2}{\Lambda
^2} \right) \psi \right]
\right\} \ .           \label{2}
\eea

Our purpose is to study the influence of
external gravity on the dynamical
breaking and restoration patterns of the symmetry possessed by the
Lagrangian (\ref{1}), which are given by the transformations
$\psi \rightarrow \gamma _5 \psi  $ and $  \bar{\psi} \rightarrow
\bar{\psi} \gamma _5  \ .$
If we refer to the action (\ref{2}), the symmetry is realized by adding
the following transformations for the auxiliary fields:
$\chi_1 \rightarrow - \chi_1  $ and $  \chi_2 \rightarrow
-\chi_2$.

 The effective potential of this model in the $N_c \rightarrow \infty$
limit is given by
\beq
V = \Lambda ^2 \left( \frac{\chi _2 ^2}{\lambda _1} +
\frac{\chi_2 ^2}{\lambda _2} \right) + V_{1} \ 
\label{veff}
\eeq
with \[ V_{1}=i \left( {\cal V}ol \right) ^{-1} \ln \mbox{Det}
\left( i \not\!\nabla -m + 2 \sqrt{3} \chi_2 \frac{\not\!\nabla
^2}{\Lambda ^2} \right) \ ,\]
where ${\cal V}ol$ is the volume of the spacetime 
and $m=\chi_1 + \sqrt{3} \chi _2 \ .$
 Thus, to leading order in the $1/N_c$ expansion we have the
gap equations, as follows
\bea
\left. \frac{\partial V_{1}}{\partial m}  \right| _{\chi _2} =
-i \left( {\cal V}ol \right) ^{-1} \tr  \frac{1}{i \not\!\nabla -m
+ 2 \sqrt{3} \chi _2 {\not\!\nabla ^2} /\Lambda ^2} \ ,
\label{deriv1}
\\ \left. \frac{\partial V_{1}}{\partial {\chi _2}}  \right| _m =
i \left( {\cal V}ol \right) ^{-1} \tr  \frac{2 \sqrt{3}
{\not\!\nabla ^2}/{\Lambda ^2}}{i
\not\!\nabla
-m + 2 \sqrt{3} \chi _2 {\not\!\nabla ^2}/{\Lambda ^2}} \ .
\label{deriv2}
\eea
First of all, we have to calculate the fermionic propagator, which
satisfies \[ \left( i \not\!\nabla -m + 2 \sqrt{3} \chi _2
\frac{\not\!\nabla ^2}{\Lambda ^2} \right) {\cal G}(x,\,x')=\delta
(x,\,x') \ .\]
Once this propagator is obtained (apart from terms which disappear
under the $\tr $ operation) it will be immediate to produce
Eqs. (\ref{deriv1}) and (\ref{deriv2}) in explicit form and also
the dependence of
the effective potential itself on the fields $\chi _1  $ and $  \chi
_2$. The details of this rather lengthy derivation are written in the
Appendix of Ref. \cite{HD}, where we give details about the calculation
of the effective potential up to terms linear in the curvature.
The technique is actually equivalent to weak curvature expansion
method given in \S 3. (The difference is in the presence of 
higher derivatives in propagator equation).

 The final outcome is shown below in natural variables, which are
obtained by using the following definitions
\beq
r=\frac{{\rm R}}{\Lambda ^2}\,, ~~~x_1=\frac{\chi _1}{\Lambda}\,
,~~~x_2= \frac{\chi_2}{\Lambda}\,,~~~a=x_1 + \sqrt{3} x_2\,
,~~~v=\frac{V}{\Lambda ^4} \ .  \label{varnat}
\eeq
The `dimensionless' effective potential $v$ maybe calculated as \cite{HD}
\bea
v & = & \frac{a^2}{\l_{1}}-\frac{a^2}{8 \pi ^2}+\frac{a^4}{32
\pi^2}-\frac{a^2 r}{96 \pi^2}-2 \sqrt{3} \frac{a x_{2}}{l_{1}}
+\sqrt{3}\frac{a x_{2}}{4 \pi ^2} -\sqrt{3} \frac{a ^3 x_{2}}{2
\pi^2}+3
\frac{x_{2}^2}{l_{1}} \\ \nn
& & + \frac{x_{2}^2}{l_{2}} -\frac{x_{2}^2}{2 \pi ^2} + 9
\frac{a ^2
x_{2} ^2}{4 \pi^2}
+\frac{r x_{2}^2}{16 \pi ^2} - 2 \sqrt{3} \frac{a x_{2} ^3}{\pi
^2}+9 \frac{x_{2} ^4}{4 \pi ^2}
 -\frac{a ^4 \ln  a^2  }{16\pi ^2}- \frac{a^2 r \ln
a^2  }{ 96 \pi^2}  \ , \label{potnat}
\eea
where $l_{1}=\lambda_{1}/\Lambda^{2}$, $l_{2}=\lambda_{2}/\Lambda^{2}$.
Note that in deriving  this expression we have taken into account only
terms up to quartic order on the fields $a$ and $x_2$.
The gap equations ---which one obtains by differentiating $v$ with
respect
to $x_1$ and $x_2$ and equating the results to zero--- are, respectively
\bea
x_{1} \left( 1 - {{8\,{{\pi }^2}}\over {l_{1}}} \right) & = &
  -2\times{3^{{3/2}}}\,{{x_{1}}^2}\,x_{2} -
  18\,x_{1}\,{{x_{2}}^2} - 8\,{\sqrt{3}}\,{{x_{2}}^3} 
 -{a^3}\,\ln a^2 \nonumber \\
 & &- r\,\left( {{a}\over 6} +
     {{a\,\ln a^2}\over {12}} \right)  \ ,   \\
 x_2 \left( 1 - {{8\,\,{{\pi }^2}}\over {l_{2}}} \right) & = &
-2 \,{\sqrt{3}}\,{x_{1}}^3 -18\,{{x_{1}}^2} \,x_{2} -
  8\times{3^{{3/2}}}\,x_{1}\,{{x_{2}}^2} - 24\,{{x_{2}}^3}
  - {\sqrt{3}}\,{a^3}\,\ln  a^2 \nonumber \\
  & & -r\,\left( {{x_{1}}\over {2\,{\sqrt{3}}}} +
     {{a\,\ln  a^2}\over {4\,{\sqrt{3}}}} \right)  \ . \label{diffeq}
\eea

 One can study now the influence of gravity on the symmetry breaking
pattern of the theory. A simple inspection reveals that a positive
curvature tends to protect the symmetry of the vacuum, and also a
negative one may trigger the breaking of the symmetry.

 In flat spacetime the symmetry is broken whenever either $\lambda_1$
or
$\lambda_2$ are greater than $8 \pi ^2 /\Lambda ^2$, that is why
we shall take in the following $k_1$ and $k_2$ to be $l_1 - 8 \pi ^2$ and
$l_2 - 8 \pi^2$ respectively. We may illustrate these remarks by
studying the evolution of the
minimum, given by two coordinates (which we choose to be $a$ and $x_2$)
in several circumstances. Before including gravity, it is worth
discussing in more detail the situation in flat space. In fact, several
cases may be analyzed (see Ref. \cite{AA}). 
To introduce this study it is
convenient to define
\[ \pm \mu_i ^2 \equiv \Lambda^2 \left( 1- \frac{8
\pi ^2}{\lambda_i ^\pm } \right) \ , \]
where $\lambda_i ^{\pm }$ means $\lambda_i > 8 \pi^2$ (for the $+$ sign)
or $\lambda_i < 8 \pi^2$ (for the $-$ sign). One sticks here to the case
when
$\mu_i \ll \Lambda $. We only repeat the two situations given when
$\mu_2^2 > 3 \mu_1^2$ and ($\lambda_1^+,\,\,\lambda_2^-$), or
$\mu_2^2 < 3 \mu_1^2$ and ($\lambda_1^-,\,\,\lambda_2^+$). In both cases
the results may be summarized by
\bea
\chi_1^2 & = & \frac{\mu_1^2}{\ln (\Lambda^2 /m^2)} \ \left|
1-\frac{3 \mu_1^2}{\mu_2^2} \right|^3 \   \left[ 1 +
O \left( \frac{1}{\ln (\Lambda ^2 /m^2)}\right)
\right] \ , \\ \nn
\chi_2 & = & - \sqrt{3} \chi_1 \left( \frac{\mu_1}{\mu_2} \right) ^2
\left[ 1 +
O \left( \frac{1}{\ln (\Lambda ^2 /m^2)}\right)
\right] \ , \\ \nn
m^2 & = & \frac{ \mu_1 ^2 \mu_2^2}{\mid \mu_2^2 - 3 \mu_1 ^2 \mid\,\ln
(\Lambda ^2 /m^2) } \ . \\ \nn
\eea

 As a first example, consider the case in which the coupling constants
are such that the symmetry is not broken in flat space, that is to say,
$\lambda_i < 8 \pi^2$. 
One can see here
the situation is modified as we move from negative to positive
curvature by the numerical analysis.\cite{HD}
We observe that there is a negative value of the
curvature above which the symmetry is restored (phase transition.)

Note that usual four-fermion model
was studied under the influence of a gravitational field in different
situations (see \S 3). 
The conclusion of this section was always that if the
coupling constant
is greater than the critical value in flat space, there is a first
order phase transition at some positive value of the curvature.
In the present case there is no actual contradiction. 
The point is that, till now,
we have
concentrated ourselves in cases where the coupling constants are around
$8 \pi^2$.
We observe that if we explore regions in the space of parameters where
$\lambda_2$ is much smaller, and $\lambda_1$ is greater than $8 \pi^2$
---which would correspond to a limit where our theory approaches the
Gross-Neveu model--- then there is a first order phase transition
for some positive value of the curvature.
Similarly, one can study numerically phase structure
of the theory for other values of parameters and curvature.

 There is an interesting question about this model ---namely whether it
would be possible for some generalization of it to be represented in
renormalizable form. 
For $\lambda_{2}=0$ it is possible to transform theory
to
Yukawa-type model which describes near the critical point
the physics of chiral symmetry breaking. 
Note that the higher
derivative term in (\ref{1}) acts against such generalization at
$\lambda_2 \neq 0$.

Finally, note that in the same way, one can study other higher
derivative four-fermion models in curved spacetime.

\subsection{Supersymmetric NJL model in curved spacetime}

Next we consider the supersymmetric NJL model non-minimally 
interacting with external supergravity. 
It is known that in flat spacetime \cite{snjl}
the fermion and the boson loop effects in this model
are canceled out.
Thus the supersymmetric NJL model does not show the dynamical 
chiral symmetry breaking in flat spacetime.
Hence, could we expect that external (super) gravitational 
field could be the source of phase transitions in SUSY
NJL model?

Below, we will discuss the curvature induced chiral symmetry 
breaking in supersymmetric NJL model following closely 
Ref. \cite{BIO}.

We will start with the action of SUSY NJL model
in an external supergravitational background.
This model can be considered
as local supersymmetry generalization of the model given
in Ref. \cite{snjl}:
\begin{equation}
     S=\displaystyle\int d^{8}z E^{-1}
     \left[\bar{Q}Q+\bar{Q}^{c}Q^{c}
           +\lambda_{0}(\bar{Q}^{c}Q)(\bar{Q}Q^{c})
           +\bar{\xi}_{1}(\bar{Q}Q)
           +\bar{\xi}_{2}(\bar{Q}^{c}Q^{c})\right],
\label{act:org}
\end{equation}
where chiral superfields $Q^{\alpha}$, $Q^{c}_{\alpha}$
carry the color index $\alpha =1, \cdots, N$
and belong to the representations of $SU(N)$,
$E=\mbox{Ber} E^{A}_{M}$, $E^{A}_{M}$ is supertetrade,
$\bar{\xi}_{1}$ and $\bar{\xi}_{2}$ are non-minimal
coupling constants of SUSY NJL model
with external supergravity.
We follow the notation of book \cite{sgra}.
Note also that there exists more general
form of non-minimal interaction above SUSY theory
with supergravity but we consider only a simplest
variant.

After the standard introduction of auxiliary 
superfields in a similar way which is explained in \S 2
and rewriting the action (\ref{act:org}) in component
fields we will limit ourselves to purely gravitational
background.
In the leading order of the $1/N$-expansion
the spinor auxiliary fields are dropped away
since they may contribute only to
next-to-leading order terms in the $1/N$-expansion.
Then the action to start with takes the form:
\begin{eqnarray}
     S&=&\displaystyle\int d^{4}x\sqrt{-g}\biggl[
         -\phi^{\dag}(\nabla^{\mu}\nabla_{\mu}
         +\rho^{2}+\xi_{1} R)\phi
         -{\phi^{c}}^{\dag}(\nabla^{\mu}\nabla_{\mu}
         +\rho^{2}+\xi_{2} R)\phi^{c}\nonumber \\
       &&\displaystyle +\bar{\psi}(i\gamma^{\mu}\nabla_{\mu}
         -\rho)\psi-\frac{1}{\lambda_{0}}\rho^{2}\biggr] ,
\label{s:snjl2}
\end{eqnarray}
where $\rho^{2}=\sigma^{2}+\pi^{2}$ is an auxiliary
scalar as in the original NJL model,
$\psi$ is $N$ component Dirac spinor,
$\xi_{1}=(1+\bar{\xi}_{1})/6$,
$\xi_{2}=(1+\bar{\xi}_{2})/6$.
The minimal interaction with external supergravity 
corresponds to $\xi_{1}=\xi_{2}=1/6$.
The start point action has the chiral symmetry.
If the auxiliary field $\rho$ develops the non-vanishing 
vacuum expectation value, $\langle\rho\rangle =m \neq 0$,
the fermion $\psi$ and the scalar $\phi$ acquire
the dynamical mass $m$ and the chiral symmetry is
eventually broken.%
\footnote{
We started with the locally supersymmetric action 
(\ref{act:org}) in order to justify the consideration 
of the Lagrangian (\ref{s:snjl2}) and clarify its origin.
However, after restriction to pure gravitational 
background the theory with Lagrangian (\ref{s:snjl2}) 
is not supersymmetric since the gravitino field 
( fermionic superpartner for gravitational field ) 
is absent. However, taking into account that we started from 
supersymmetric theory (in flat spacetime) we continue 
to call our model as SUSY NJL model.
}

%
To find the phase structure of the model given by
action (\ref{s:snjl2}) we introduce an effective 
potential.
To evaluate the effective potential ones start with 
generating functional of Green functions. 
\begin{eqnarray}
     Z&=&\displaystyle\int {\cal D}\psi{\cal D}\bar{\psi}
         {\cal D}\rho\ e^{iS}\nonumber \\
      &=&\displaystyle\int {\cal D}\rho\frac{\mbox{Det}
         (i\gamma^{\mu}\nabla_{\mu}-\rho)}
         {\mbox{Det}(\nabla^{\mu}\nabla_{\mu}
         +\rho^{2}+\xi_{1} R)
         (\nabla^{\mu}\nabla_{\mu}
         +\rho^{2}+\xi_{2} R)}
        \exp i\int d^{4}x\sqrt{-g}
         \left(-\frac{1}{\lambda_{0}}\rho^{2}\right)\nonumber \\
      &=&\displaystyle
         \int {\cal D}\rho\exp i\biggl[\int d^{4}x\sqrt{-g}
         \left(-\frac{1}{\lambda_{0}}\rho^{2}\right)
        -i\mbox{ln Det}(i\gamma^{\mu}\nabla_{\mu}-\rho)
                                              \nonumber \\
       &&\displaystyle +i\mbox{ln Det}(\nabla^{\mu}\nabla_{\mu}
         +\rho^{2}+\xi_{1} R)
         +i\mbox{ln Det}(\nabla^{\mu}\nabla_{\mu}
         +\rho^{2}+\xi_{2} R)
         \biggr].
\end{eqnarray}
An internal line of the $\rho$-propagator has no contribution
in the leading order of the $1/N$-expansion.
Assuming that the $\rho$ is slowly variating field and 
applying the $1/N$-expansion method the effective potential 
for $\rho$ is found to be
\begin{eqnarray}
     V(\rho)&=&\frac{1}{\lambda_{0}}\rho^{2}
            -i \mbox{tr} \int^{\rho}_{0}ds\ S(x,x;s)
                                         \nonumber \\
            &&-2i \int^{\rho}_{0}sds\ [G_{1}(x,x;s)
                                    +G_{2}(x,x;s)]
              +O\left(\frac{1}{N}\right),
\label{epot:snjl}
\end{eqnarray}
where $S(x,x;s)$ and $G_{i}(x,x;s)$ are the spinor and scalar
two-point functions respectively.
It should be noted that the effective potential 
(\ref{epot:snjl})
is normalized as $V(0)=0$.

We will evaluate the effective potential taking into
account the terms up to linear curvature and
using local momentum representation of propagators
(see \S 3 above).
Hence the effective potential reads
\begin{eqnarray}
     \displaystyle V(\rho)&=&
     \displaystyle \frac{1}{\lambda_{0}}\rho^{2}
      -4 i \int^{\rho}_{0}sds
       \int \frac{d^{4}p}{(2\pi)^{4}}\left[
       \frac{1}{p^{2}-s^{2}}
      -\frac{1}{12}R\frac{1}{(p^{2}-s^{2})^{2}}
       +\frac{2}{3}R^{\mu\nu}p_{\mu}p_{\nu}
       \frac{1}{(p^{2}-s^{2})^{3}}\right]\nonumber \\
     &&\displaystyle +2 i \int^{\rho}_{0}sds
       \int \frac{d^{4}p}{(2\pi)^{4}}\left[
       \frac{2}{p^{2}-s^{2}}
       -\left(\frac{2}{3}-\xi_{1}-\xi_{2} \right)
       R\frac{1}{(p^{2}-s^{2})^{2}}\right.\nonumber \\
     &&\left.\displaystyle +\frac{4}{3}R^{\mu\nu}p_{\mu}p_{\nu}
       \frac{1}{(p^{2}-s^{2})^{3}}\right]\nonumber \\
     &=&\displaystyle \frac{1}{\lambda_{0}}\rho^{2}
       +2 i R \left(\frac{1}{6}-\frac{2}{3}+\xi_{1}+\xi_{2}\right)
       \int^{\rho}_{0}sds
       \int \frac{d^{4}p}{(2\pi)^{4}}
       \frac{1}{(p^{2}-s^{2})^{2}}
       +O\left(\frac{1}{N}\right).
\label{epot:snjl:wc}
\end{eqnarray}
Thus for $\xi_{1}+\xi_{2}=1/2$ the fermion loop contribution
is cancelled with the boson loop contribution.
To evaluate the integration in Eq. (\ref{epot:snjl:wc}) 
we perform the Wick rotation $p^{0}\rightarrow ip^{0}$
\begin{eqnarray}
     I&=& i  R \int^{\rho}_{0}sds
       \int \frac{d^{4}p}{(2\pi)^{4}}
       \frac{1}{(p^{2}-s^{2})^{2}}\nonumber \\
     &\rightarrow& - R \int^{\rho}_{0}sds
       \int \frac{d^{4}p}{(2\pi)^{4}}
       \frac{1}{(p^{2}+s^{2})^{2}}.
\label{epot:snjl:lhs}
\end{eqnarray}
Applying the Schwinger proper time method \cite{SP} $I$ 
is rewritten as
\begin{eqnarray}
     I&=& - R \int^{\rho}_{0}sds
      \int \frac{d^{4}p}{(2\pi)^{4}}
      \int^{\infty}_{0}tdt\ e^{-t(p^{2}+s^{2})}
      \nonumber \\
      &=&\frac{R}{2(4\pi)^{2}}\int^{\infty}_{0}dt\frac{1}{t^{2}}
      \left(e^{-t\rho^{2}}-1\right).
\end{eqnarray}
Since the integration over $t$ is divergent around $t\sim 0$,
we introduce the proper time cut-off $\Lambda$ and find
\begin{eqnarray}
     I&\rightarrow&\frac{R}{2(4\pi)^{2}}\int^{\infty}_{1/\Lambda^{2}}
       dt\frac{1}{t^{2}}
       \left(e^{-t\rho^{2}}-1\right)\nonumber \\
      &=& \frac{R}{2(4\pi)^{2}}\left[
          \rho^{2}\mbox{Ei}
          \left(-\frac{\rho^{2}}{\Lambda^{2}}\right)
          +\Lambda^{2}
          (e^{-\rho^{2}/\Lambda^{2}}-1)          
          \right],
\end{eqnarray}
where $\mbox{Ei}(-x)$ is the exponential-integral function
which is defined by
\begin{equation}
     \mbox{Ei}(-x)=-\int^{\infty}_{x}dt\frac{e^{-t}}{t}
     =\ln x+\gamma+\sum^{\infty}_{n=1}
        \frac{(-x)^{n}}{n\cdot n!} < 0\ ;
      (x > 0).
\label{pro:ei}
\end{equation}
Hence the effective potential in the leading order of
the $1/N$-expansion for the supersymmetric NJL model
in curved spacetime reads
\begin{equation}
     V(\rho)= \frac{1}{\lambda_{0}}\rho^{2}
     -\frac{R}{(4\pi)^{2}}
          f(\xi_{1},\xi_{2})
          \left[
          \rho^{2}\mbox{Ei}
          \left(-\frac{\rho^{2}}{\Lambda^{2}}\right)
          \frac{}{}+\Lambda^{2}
          (e^{-\rho^{2}/\Lambda^{2}}-1)          
          \right],
\label{epot:snjl:wcfin}
\end{equation}
where $f(\xi_{1},\xi_{2})$ is
\begin{equation}
     f(\xi_{1},\xi_{2})=\frac{1}{2}-\xi_{1}-\xi_{2}.
\label{def:f}
\end{equation}


The ground state of the system corresponds to
the minimum of the effective potential.
To find the ground state we evaluate the effective potential
with varying the $f(\xi_{1},\xi_{2})R\lambda_{0}$. \cite{BIO}

In flat spacetime $R=0$
the effective potential (\ref{epot:snjl:wcfin}) 
takes a very simple form of 
quadratic function. It is evident that the minimum of the effective 
potential will be only at $\rho = 0$ for arbitrary values of the 
nonminimal coupling parameters. 

For $f(\xi_{1},\xi_{2})R\lambda_{0} > 0$ taking into account the property 
(\ref{pro:ei}) of the exponential-integral function we see that 
the effective potenial (\ref{epot:snjl:wcfin}) is 
non-negative and takes minimum at $\rho = 0$. 

\begin{figure}
\setlength{\unitlength}{0.240900pt}
\begin{picture}(1500,900)(0,0)
\tenrm
\thicklines \path(220,113)(240,113)
\thicklines \path(1436,113)(1416,113)
\put(198,113){\makebox(0,0)[r]{0}}
\thicklines \path(220,189)(240,189)
\thicklines \path(1436,189)(1416,189)
\put(198,189){\makebox(0,0)[r]{0.0001}}
\thicklines \path(220,266)(240,266)
\thicklines \path(1436,266)(1416,266)
\put(198,266){\makebox(0,0)[r]{0.0002}}
\thicklines \path(220,342)(240,342)
\thicklines \path(1436,342)(1416,342)
\put(198,342){\makebox(0,0)[r]{0.0003}}
\thicklines \path(220,419)(240,419)
\thicklines \path(1436,419)(1416,419)
\put(198,419){\makebox(0,0)[r]{0.0004}}
\thicklines \path(220,495)(240,495)
\thicklines \path(1436,495)(1416,495)
\put(198,495){\makebox(0,0)[r]{0.0005}}
\thicklines \path(220,571)(240,571)
\thicklines \path(1436,571)(1416,571)
\put(198,571){\makebox(0,0)[r]{0.0006}}
\thicklines \path(220,648)(240,648)
\thicklines \path(1436,648)(1416,648)
\put(198,648){\makebox(0,0)[r]{0.0007}}
\thicklines \path(220,724)(240,724)
\thicklines \path(1436,724)(1416,724)
\put(198,724){\makebox(0,0)[r]{0.0008}}
\thicklines \path(220,801)(240,801)
\thicklines \path(1436,801)(1416,801)
\put(198,801){\makebox(0,0)[r]{0.0009}}
\thicklines \path(220,877)(240,877)
\thicklines \path(1436,877)(1416,877)
\put(198,877){\makebox(0,0)[r]{0.001}}
\thicklines \path(423,113)(423,133)
\thicklines \path(423,877)(423,857)
\put(423,68){\makebox(0,0){$-10$}}
\thicklines \path(929,113)(929,133)
\thicklines \path(929,877)(929,857)
\put(929,68){\makebox(0,0){$-5$}}
\thicklines \path(1436,113)(1436,133)
\thicklines \path(1436,877)(1436,857)
\put(1436,68){\makebox(0,0){0}}
\thicklines \path(220,113)(1436,113)(1436,877)(220,877)(220,113)
\put(45,945){\makebox(0,0)[l]{\shortstack{$m/\Lambda$}}}
\put(828,23){\makebox(0,0){$f(\xi_{1},\xi_{2})R G$}}
\thinlines \path(845,114)(845,114)(806,115)(783,116)(766,117)(752,118)(731,119)(715,122)(689,126)(670,130)(640,138)(617,146)(583,161)(556,177)(515,209)(484,241)(458,273)(436,305)(417,336)(399,368)(383,400)(368,432)(355,464)(342,495)(330,527)(318,559)(307,591)(297,623)(287,654)(278,686)(268,718)(260,750)(251,782)(243,813)(235,845)(227,877)
\end{picture}

\caption{Behaviour of the dynamically generated mass $m$
         with varying the $f(\xi_{1},\xi_{2})R\lambda_{0}$.}
\label{susym}
\end{figure}
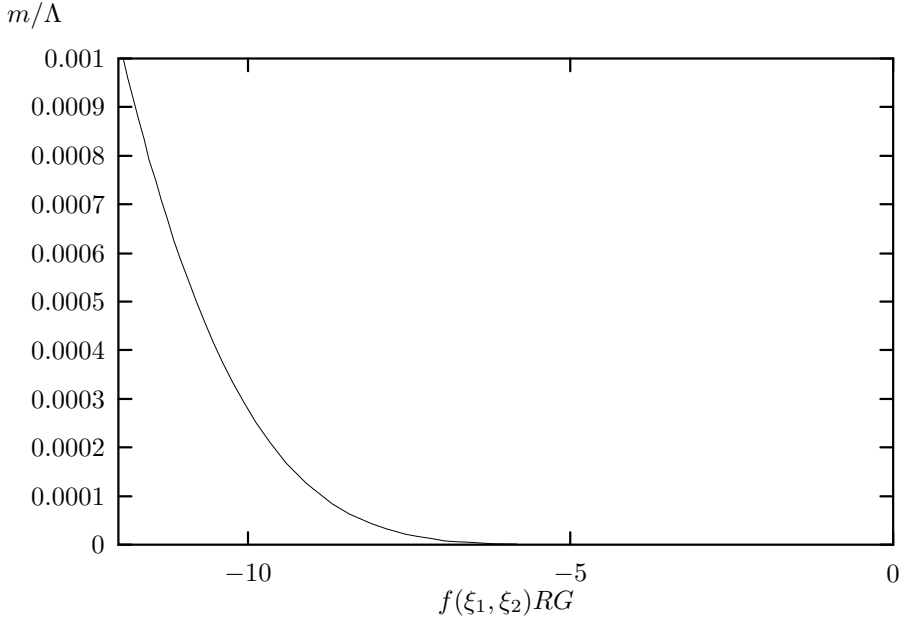

To understand the phase structure of the SUSY NJL model
for a negative $f(\xi_{1},\xi_{2})R\lambda_{0}$
we will analyse the effective potential more precisely.
The dynamically generated mass $m$ of fermion $\psi$ and 
scalar $\phi$ is given by the value of $\rho$ at the
minimum of the effective potential.
Stationary condition for the effective potential
(\ref{epot:snjl:wcfin}) is given by
\begin{equation}
     \frac{\partial V(\rho)}{\partial \rho}=
     \rho\left[\frac{2}{\lambda_{0}}-\frac{2R}{(4\pi)^{2}}
     f(\xi_{1},\xi_{2})
     \mbox{Ei}\left(-\frac{\rho^{2}}{\Lambda^{2}}\right)\right]=0.
\label{gap:snjl}
\end{equation}
Thus the dynamical mass $m=\langle\rho\rangle$ satisfies
\begin{equation}
     \frac{16\pi^{2}}{R\lambda_{0}}=
     f(\xi_{1},\xi_{2})
     \mbox{Ei}\left(-\frac{m^{2}}{\Lambda^{2}}\right) < 0.
\label{mass:snjl}
\end{equation}
This equation has a solution only for a negative 
$f(\xi_{1},\xi_{2})R\lambda_{0}$.
In Fig. \ref{susym} the dynamically generated mass is plotted 
as a function of $f(\xi_{1},\xi_{2})R\lambda_{0}$.
The dynamical mass $m=\langle\rho\rangle$ which 
corresponds to the order parameter smoothly disappears 
as the $f(\xi_{1},\xi_{2})R\lambda_{0}$ increases.

\vglue 2ex

The critical curvature $R_{\mbox{cr}}$ which divides the symmetric
phase and the broken phase is given by the curvature
where the order parameter $m$ disappears.
If we take the limit $m/\Lambda\rightarrow 0$
in Eq. (\ref{mass:snjl}),
we find
\begin{eqnarray}
     R\lambda_{0}&=&\frac{16\pi^{2}}{f(\xi_{1},\xi_{2})}
        \left[\ln\frac{m^{2}}{\Lambda^{2}}+\gamma
        +O\left(\frac{m^{2}}{\Lambda^{2}}\right)
     \right]^{-1}\nonumber \\
     &\rightarrow& 0\, \,  (f(\xi_{1},\xi_{2})\neq 0).
\end{eqnarray}
Due to the stability of the ground state the coupling $\lambda_{0}$
must be real and positive.
Hence the critical curvature is given by $R_{\mbox{cr}}=0$ and
the chiral symmetry is broken down for an
arbitrary negative $f(\xi_{1},\xi_{2})R$
and an arbitrary positive $\lambda_{0}$.

For $m/\Lambda\ll 1$ the dynamical mass $m$ is obtained by
\begin{equation}
     m^{2}=\Lambda^{2}\left[
           \exp\left(\frac{16\pi^{2}}{f(\xi_{1},\xi_{2})R\lambda_{0}}
           -\gamma\right)
           +O\left(\frac{m^{2}}{\Lambda^{2}}\right)
           \right].
\end{equation}
The dynamical mass is suppressed exponentially
for a small negative $f(\xi_{1},\xi_{2})R\lambda_{0}$.

Thus it is found that the chiral symmetry is
broken down for an arbitrary negative $f(\xi_{1},\xi_{2})R$
and an arbitrary positive $\lambda_{0}$ in SUSY NJL model within 
our approximation.
For $\xi_{1}+\xi_{2} < 1/2$
the broken phase is realised in a spacetime with
an arbitrary negative curvature and the symmetric
phase is realized for an arbitrary positive curvature.
On the other hand there is only the broken phase 
in positive curvature spacetime
and the symmetric phase at a negative
curvature for $\xi_{1}+\xi_{2} > 1/2$.
In both cases the critical curvature is given by 
$R_{\mbox{cr}}=0$.
The dynamically generated fermion mass exponentially
disappears at the critical point.
If the dynamically generated fermion mass is
extremely small, the large $R^{2}/m^{2}$ compared
with $R$ may spoil the validity of the linear 
curvature approximation.
The terms of the order $O(R^{2}/m^{2})$
may change the exponential behavior of the dynamical mass
near the critical point.

Note finally that there are few possibilities
to generalize the results of the present consideration.
First of all, it could be interesting to study
phase structure of SUSY NJL model on non-trivial 
supergravitational background (with non-zero
gravitino).
Second, it could be interesting to investigate
the gauged SUSY NJL model where compositeness conditions
a la Bardeen-Hill-Lindner \cite{BHL} maybe implemented.
That could lead to the formulation of compositeness 
condition for $\xi_{1}$ and $\xi_{2}$
which presumably should be given by asymptotically
(super) conformal invariant values. \cite{BOS,BO}

\section{Thermal and curvature induced phase transition}

At the early stage of the universe the combined effects 
of the temperature and curvature to the DSB
may play an important role.
In the present sections we will apply the similar analysis 
as in \S 2 at finite temperature and curvature.
To show the combined effects of the temperature 
and curvature to the DSB,
we fix the coupling constant $\lambda$ above the critical 
one and see whether the broken chiral symmetry is restored
in an environment of the high temperature and/or large
curvature.%
\footnote{
In Ref. \cite{KS} the combined effects of the temperature 
and curvature to the DSB is discussed
in the positive weak curvature limit.
In the paper the variable $k$ is introduced
by using the trace formula:
\begin{equation}
     1=\frac{1}{(4\pi s)^{D/2}}\int\frac{d^{D}k}{(2\pi)^{D}}
     e^{-s k^{2}}\, .
\label{trform}
\end{equation}
Inserting Eq. (\ref{trform}) into Eq. (\ref{v:gn})
the effective potential is rewritten as
\begin{equation}
     V(\sigma ) = \frac{1}{2\lambda_0}\sigma^{2}
                  -i\mbox{tr}\int_{0}^{\sigma}ds 
     \frac{1}{(4\pi s)^{D/2}}\int\frac{d^{D}k}{(2\pi)^{D}}
     e^{-s k^{2}} S(x,x;s)\, .
\label{v:gn:tr}
\end{equation}
The temperature $T$ and chemical potential $\mu$ are introduced by the
replacement
\begin{equation}
     k^{0}\rightarrow \sqrt{(\omega_{n}-i\mu)^{2}}\, ,\, \, 
     \int\frac{dk^{0}}{2\pi}\rightarrow k_{B}T\sum_{n}\, \, ,
\end{equation}
where $k_{B}$ is the Boltzmann constant and
$\omega_{n}=(2n+1)\pi k_{B}T$.
The variable $k$ in Eq. (\ref{v:gn:tr})
does not always correspond to the momentum of the fermion field
in a general curved spacetime.
Thus above method is useful only in flat spacetime
or special spacetime where the variable $k$ is regarded
as the momentum of the fermion field.
So we do not use the trace formula to introduce temperature here.
}
To introduce the temperature in the theory
it is supposed that the system is in equilibrium.
This assumption is not accepted in a general curved
spacetime.
In the spacetime which has no time evolution
the equilibrium state can be defined.
We then restrict ourselves in the positive curvature 
spacetime $R\otimes S^{D-1}$ and the negative curvature 
spacetime $R\otimes H^{D-1}$.
To find the ground state at finite temperature
and curvature we calculate the effective 
potential and analyze its stationary condition
by the gap equation in the leading order of the 
$1/N$ expansion.

In this section we mainly follow Ref. \cite{II}
and discuss the thermal and curvature effects
to the dynamical symmetry breaking.

\subsection{Effective potential at finite temperature}

It is known that 
the broken chiral symmetry is restored for a sufficiently 
high temperature through the second order phase 
transition in flat spacetime.\cite{IKM,FT}
Here we introduce the effect of the finite temperature
to the four-fermion models and give the explicit expression
of the effective potential at finite temperature.

As we have seen in \S 2, the effective potential 
is expressed by the two-point function $S(x,x;s)$ of a massive 
free fermion.
The two-point function at finite temperature is
defined by 
\begin{equation}
     S^{T}(x,x;s)
     =\frac{\sum_{\alpha}{\large e}^{-\beta E_{\alpha}}
      \langle \alpha | \mbox{T}(\psi(x)\bar{\psi}(x))
      | \alpha \rangle}
      {\sum_{\alpha}{\large e}^{-\beta E_{\alpha}}}\, ,
\label{eqn:TPFTemp}
\, 
\end{equation}
where $E_\alpha$ is the energy in the state
specified by quantum number $\alpha$ respectively,
$\beta=1/k_{B}T$ with $k_{B}$ 
the Boltzmann constant and $T$ the temperature.

Following the standard procedure of the dealing with
Matsubara Green function, the two-point function at finite 
temperature is obtained from the one at $T=0$ 
by the Wick rotation and the replacements \cite{Temp}
\begin{equation}
\left\{
\begin{array}{rcl}
\displaystyle
\int_{-\infty}^{\infty}\!\!\frac{dk^0}{2\pi i}
&\rightarrow &
\displaystyle\frac{1}{\beta}\sumomeg,\\[4mm]
k^0 &\rightarrow &
\displaystyle i\omega_{n}\equiv i\frac{2n+1}{\beta}\pi,\\
\gamma^0 &\rightarrow &
\displaystyle i\gamma^0 .\\
\end{array}
\right.
\label{replace}
\end{equation}
In Minkowski space the effective potential at finite temperature
in the leading order of the $1/N$ expansion
reads\cite{FT}
\begin{eqnarray}
     \VT(\sigma) & = &
     \frac{1}{2\lambda_0}\sigma^{2}
     + \mbox{tr}\int_{0}^{\sigma}ds
     \frac{1}{\beta}\sum_{n}\int\frac{d^{D-1}k}{(2\pi)^{D-1}}
     \frac{1}{\diracsh{k} + s}\, .
\label{v:lndet}
\end{eqnarray}
If we perform the integration over $k$, we obtain
\begin{equation}
\VT(\sigma) = \frac{1}{2\lambda_{0}}\sigma^{2}
-\intsig \frac{\tr \11}{(4\pi)^{(D-1)/2}}
\Gamma\left(\frac{3-D}{2}\right)\frac{1}{\beta}
\sumomeg s
\left(s^{2}+\omega_{n}^{2}\right)^{(D-3)/2}.
\label{v:Temp1}
\end{equation}
Performing a summation and integrating over angle variables
and $s$ in Eq. (\ref{v:lndet}), we get
\begin{eqnarray}
\VT(\sigma) &=& \frac{1}{2\lambda_{0}}\sigma^{2}
-\frac{\tr \11}{(4\pi)^{D/2}}\frac{1}{D}
\Gamma\left(1-\frac{D}{2}\right)\sigma^{D}\nonumber \\
&&-\frac{\tr \11}{(4\pi)^{(D-1)/2}}
\frac{1}{\Gamma\left(\frac{D-1}{2}\right)}
\frac{1}{\beta}\int^{\infty}_{0}\!\!\!dt\,
 t^{(D-3)/2}\ln
\frac{1+e^{-\beta\sqrt{t+\sigma^{2}}}}{1+e^{-\beta\sqrt{t}}}.
\label{v:Temp2}
\end{eqnarray}
Comparing Eq. (\ref{v:Temp1}) with Eq. (\ref{v:Temp2}) 
we find the following relation.
\begin{eqnarray}
&&\frac{\tr \11}{(4\pi)^{D/2}}\frac{1}{D}
\Gamma\left(1-\frac{D}{2}\right)\sigma^{2}
+\frac{\tr \11}{(4\pi)^{(D-1)/2}}
\frac{1}{\Gamma\left(\frac{D-1}{2}\right)}\frac{1}{\beta}
\int_{0}^{\infty}\!\!dt\,t^{(D-3)/2}
\ln\frac{1+e^{-\beta\sqrt{t+\sigma^{2}}}}{1+e^{-\beta\sqrt{t}}}
\nonumber \\
&&\hspace{2.5cm} = \intsig\frac{\tr \11}{(4\pi)^{(D-1)/2}}
\Gamma\left(\frac{3-D}{2}\right)\frac{1}{\beta}
\sumomeg s\left(s^{2}+\omega_{n}^{2}\right)^{(D-3)/2}.
\label{eqn:relzero}
\end{eqnarray}
This relation will be used for numerical calculation
of the effective potential at finite temperature in 
curved spacetime.

Starting from the theory with the broken chiral symmetry
at vanishing $T$ and evaluating the effective potential at 
finite temperature, it is known that 
the broken chiral symmetry
is restored at a critical temperature $T_{\mbox{cr}}$
through the second order phase transition.
The critical temperature is given by \cite{FT}
\begin{equation}
        \frac{k_{B}T_{\mbox{cr}}}{m_{0}} = \frac{1}{2\pi}
                \left[
                        \frac{2\Gamma\left({\displaystyle{3-D \over 2}}\right)}
                        {\sqrt{\pi}\Gamma
                                \left({\displaystyle{2-D \over 2}}\right)}
                        (2^{3-D}-1)\zeta(3-D)
                \right]^{1/(2-D)} \, .
\label{cr:Temp}
\end{equation}
Below, we investigate the combined effects of the temperature
and curvature on $R\otimes S^{D-1}$ and $R\otimes H^{D-1}$.

\subsection{Positive curvature space ($R\otimes S^{D-1}$)}

In the positive curvature space, $R\otimes S^{D-1}$,
the effective potential is given by (4$\cdot$60).
According to the definition of the two-point function
at finite temperature (\ref{eqn:TPFTemp}),
we obtain the effective potential in the space
at finite temperature by the replacements (\ref{replace})
\begin{eqnarray}
\VTR(\sigma)&=&\frac{1}{2\lambda_{0}}\sigma^{2}
-\intsig \frac{\tr \11}{(4\pi)^{(D-1)/2}}
\frac{s a^{3-D}}{\beta}\nonumber \\
&&\times
\sumomeg
\frac{\Gamma\left(\frac{D-1}{2}+i\alpha_{n}\right)
      \Gamma\left(\frac{D-1}{2}-i\alpha_{n}\right)}
     {\Gamma\left( 1 +i\alpha_{n}\right)
      \Gamma\left( 1 -i\alpha_{n}\right)}
\Gamma\left(\frac{3-D}{2}\right),
\label{eqn:EfptEinT}
\end{eqnarray}
where $\alpha_{n}$ is defined by
\begin{equation}
\alpha_{n}\equiv a\sqrt{s^{2}+\omega_{n}^{2}}.
\end{equation}
Evaluating the effective potential (\ref{eqn:EfptEinT})
we will find the phase structure of the model at finite
temperature in positive curvature space.

For numerical calculations we need the finite expression 
of the effective potential in summation and integration.
Inserting Eq. (2$\cdot$33) and Eq. (\ref{eqn:relzero})
into Eq.(\ref{eqn:EfptEinT}) the renormalized
effective potential reads
\begin{eqnarray}
\VTR_{R}(\sigma)&=&\frac{1}{2}\left[
\frac{1}{\lambda_{R}}+\frac{\tr \11}{(4\pi)^{D/2}}
\Gamma\left(1-\frac{D}{2}\right)(D-1)\right]\mu^{D-2}\sigma^{2}
\nonumber \\
&&-\frac{\tr \11}{(4\pi)^{D/2}}\frac{1}{D}
\Gamma\left(1-\frac{D}{2}\right)\sigma^{2}\nonumber \\
&&
-\frac{\tr \11}{(4\pi)^{(D-1)/2}}
\frac{1}{\Gamma\left(\frac{D-1}{2}\right)}\frac{1}{\beta}
\int_{0}^{\infty}\!\!dt \,t^{(D-3)/2}
\ln\frac{1+e^{-\beta\sqrt{t+\sigma^{2}}}}{1+e^{-\beta\sqrt{t}}}
 \\
&&+\intsig\frac{\tr \11}{(4\pi)^{(D-1)/2}}
\frac{1}{\beta}\Gamma\left(\frac{3-D}{2}\right)
\nonumber \\
&&\times
\sumomeg s\left[ \left(s^{2}+\omega_{n}^{2}\right)^{(D-3)/2}
-a^{3-D}
\frac{\Gamma\left(\frac{D-1}{2}+i\alpha_{n}\right)
      \Gamma\left(\frac{D-1}{2}-i\alpha_{n}\right)}
     {\Gamma\left( 1 +i\alpha_{n}\right)
      \Gamma\left( 1 -i\alpha_{n}\right)}\right] \nonumber.
\end{eqnarray}
In this representation of the effective potential
the divergence is canceled out in the summation.

The phase structure of the theory is obtained by observing the
minimum of the effective potential.
The necessary condition for the minimum of the effective potential 
is given by the gap equation:
\begin{equation}
     \left. 
     \frac{\partial V^{TR}_{R}(\sigma)}{\partial \sigma}
     \right|_{\sigma = m} =0\, .
\label{eqn:gap}
\end{equation}
If the gap equation has a non-trivial solution which
corresponds to the minimum of the effective potential,
the chiral symmetry is broken down and the dynamical
fermion mass is generated.
The non-trivial solution of the gap equation is given by
\begin{eqnarray}
&&\left(\frac{1}{\lambda_{R}}
-\frac{1}{\lambda_{\mbox{cr}}}\right)\mu^{D-2}
-\frac{\tr \11}{(4\pi)^{D/2}}\Gamma\left(1-\frac{D}{2}\right)m^{D-2}
\nonumber \\
&&+\frac{\tr \11}{(4\pi)^{(D-1)/2}}
\frac{1}{\Gamma\left(\frac{D-1}{2}\right)}
\int_{0}^{\infty}\!\!dt\,t^{(D-3)/2}
\frac{1}{\sqrt{t+m^{2}}}
\frac{e^{-\beta\sqrt{t+m^{2}}}}{1+e^{-\beta\sqrt{t+m^{2}}}}
\label{eqn:GapEinT} \\
&&+\frac{\tr \11}{(4\pi)^{(D-1)/2}}\frac{1}{\beta}
\Gamma\left(\frac{3-D}{2}\right)
\nonumber \\
&&\times
\sumomeg\left[
\left(m^{2}+\omega_{n}^{2}\right)^{(D-3)/2}
-a^{3-D}
\frac{\Gamma\left(\frac{D-1}{2}+i\alpha_{n}\right)
      \Gamma\left(\frac{D-1}{2}-i\alpha_{n}\right)}
     {\Gamma\left( 1 +i\alpha_{n}\right)
      \Gamma\left( 1 -i\alpha_{n}\right)}\right] = 0,
\nonumber
\end{eqnarray}
where $\alpha_{n}=a \sqrt{m^{2}+\omega_{n}^{2}}$,
$\lambda_{\mbox{cr}}$ is defined in Eq. (\ref{cr:l:d})
and $m$ corresponds to the dynamically generated fermion
mass.

In Ref. \cite{II} the typical behaviors of the dynamical
fermion mass $m$ was evaluated numerically
as a function of temperature 
$T$ or curvature $R$.
It was found that the broken chiral 
symmetry is restored for a sufficiently high temperature
and large curvature and only the second order phase transition 
is realized with varying temperature and/or curvature 
for $2\leq D<4$.

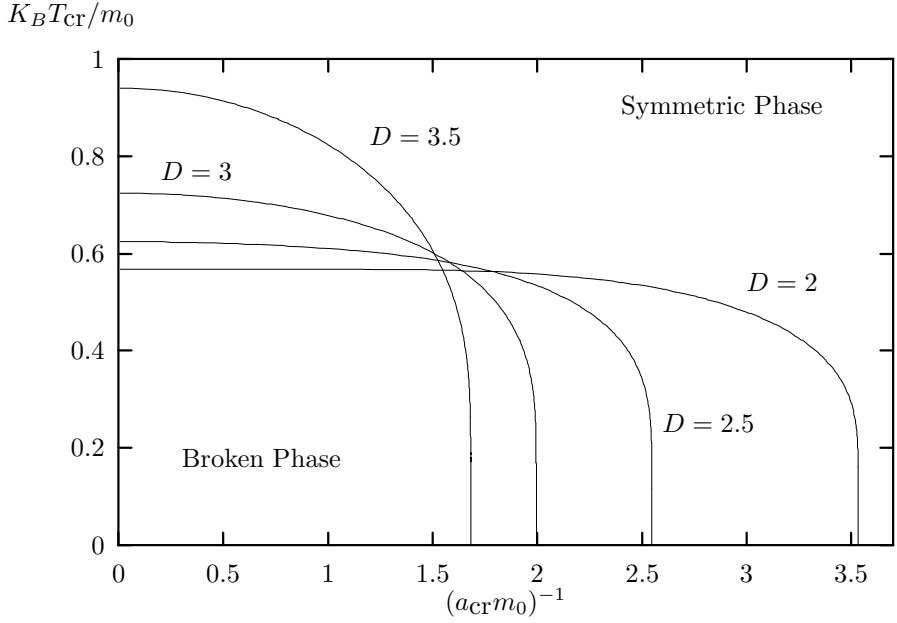
\begin{figure}
                \vglue 1ex
\setlength{\unitlength}{0.240900pt}
\begin{picture}(1500,900)(0,0)
\tenrm
\thicklines \path(220,113)(240,113)
\thicklines \path(1436,113)(1416,113)
\put(198,113){\makebox(0,0)[r]{0}}
\thicklines \path(220,266)(240,266)
\thicklines \path(1436,266)(1416,266)
\put(198,266){\makebox(0,0)[r]{0.2}}
\thicklines \path(220,419)(240,419)
\thicklines \path(1436,419)(1416,419)
\put(198,419){\makebox(0,0)[r]{0.4}}
\thicklines \path(220,571)(240,571)
\thicklines \path(1436,571)(1416,571)
\put(198,571){\makebox(0,0)[r]{0.6}}
\thicklines \path(220,724)(240,724)
\thicklines \path(1436,724)(1416,724)
\put(198,724){\makebox(0,0)[r]{0.8}}
\thicklines \path(220,877)(240,877)
\thicklines \path(1436,877)(1416,877)
\put(198,877){\makebox(0,0)[r]{1}}
\thicklines \path(220,113)(220,133)
\thicklines \path(220,877)(220,857)
\put(220,68){\makebox(0,0){0}}
\thicklines \path(384,113)(384,133)
\thicklines \path(384,877)(384,857)
\put(384,68){\makebox(0,0){0.5}}
\thicklines \path(549,113)(549,133)
\thicklines \path(549,877)(549,857)
\put(549,68){\makebox(0,0){1}}
\thicklines \path(713,113)(713,133)
\thicklines \path(713,877)(713,857)
\put(713,68){\makebox(0,0){1.5}}
\thicklines \path(877,113)(877,133)
\thicklines \path(877,877)(877,857)
\put(877,68){\makebox(0,0){2}}
\thicklines \path(1042,113)(1042,133)
\thicklines \path(1042,877)(1042,857)
\put(1042,68){\makebox(0,0){2.5}}
\thicklines \path(1206,113)(1206,133)
\thicklines \path(1206,877)(1206,857)
\put(1206,68){\makebox(0,0){3}}
\thicklines \path(1370,113)(1370,133)
\thicklines \path(1370,877)(1370,857)
\put(1370,68){\makebox(0,0){3.5}}
\thicklines \path(220,113)(1436,113)(1436,877)(220,877)(220,113)
\put(45,945){\makebox(0,0)[l]{\shortstack{$K_{B}T_{\mbox{cr}}/m_{0}$}}}
\put(828,23){\makebox(0,0){$(a_{\mbox{cr}}m_{0})^{-1}$}}
\put(1009,801){\makebox(0,0)[l]{Symmetric Phase}}
\put(319,251){\makebox(0,0)[l]{Broken Phase}}
\put(614,755){\makebox(0,0)[l]{$D=3.5$}}
\put(286,701){\makebox(0,0)[l]{$D=3$}}
\put(1074,304){\makebox(0,0)[l]{$D=2.5$}}
\put(1206,526){\makebox(0,0)[l]{$D=2$}}
\thinlines \path(223,547)(223,547)(240,547)(256,547)(273,547)(289,547)(305,547)(322,547)(338,547)(355,547)(371,547)(388,547)(404,547)(420,547)(437,547)(453,547)(470,547)(486,547)(503,547)(519,547)(536,547)(552,547)(568,547)(585,547)(601,547)(618,546)(634,546)(651,546)(667,546)(683,546)(700,546)(716,545)(733,545)(749,545)(766,544)(782,544)(798,543)(815,543)(831,542)(835,542)(848,541)(864,540)(881,539)(897,538)(897,538)(913,536)(930,535)(941,534)(946,533)(963,532)(977,530)
\thinlines \path(977,530)(979,530)(996,528)(1007,526)(1012,526)(1028,523)(1033,523)(1045,521)(1056,519)(1061,518)(1076,515)(1078,515)(1094,511)(1095,511)(1111,508)(1112,507)(1127,504)(1128,503)(1143,500)(1144,499)(1157,496)(1160,495)(1169,492)(1176,490)(1181,488)(1192,484)(1193,484)(1203,480)(1209,478)(1213,477)(1223,473)(1226,472)(1231,469)(1240,465)(1242,464)(1248,461)(1256,458)(1259,456)(1263,454)(1270,450)(1275,447)(1276,446)(1282,442)(1288,438)(1291,436)(1294,435)(1299,431)(1304,427)(1308,424)(1309,423)(1314,419)
\thinlines \path(1314,419)(1318,416)(1322,412)(1324,410)(1326,408)(1330,404)(1333,400)(1337,396)(1340,393)(1341,392)(1343,389)(1346,385)(1349,381)(1351,377)(1353,374)(1356,370)(1358,366)(1360,362)(1362,358)(1363,354)(1365,351)(1367,347)(1368,343)(1369,339)(1371,335)(1372,332)(1373,328)(1374,324)(1374,320)(1375,316)(1376,312)(1377,309)(1377,305)(1378,301)(1378,297)(1379,293)(1379,289)(1379,286)(1380,282)(1380,278)(1380,274)(1380,270)(1380,267)(1381,263)(1381,259)(1381,255)(1381,251)(1381,247)(1381,244)(1381,240)(1381,236)
\thinlines \path(1381,236)(1381,232)(1381,228)(1381,225)(1381,221)(1381,217)(1381,213)(1381,209)(1381,205)(1381,202)(1381,198)(1381,194)(1381,190)(1381,186)(1381,183)(1381,175)(1381,167)(1381,144)(1381,114)
\thinlines \path(223,590)(223,590)(240,590)(256,590)(273,590)(289,590)(305,589)(322,589)(338,589)(355,588)(371,588)(388,587)(404,587)(420,586)(437,586)(453,585)(470,584)(478,584)(486,583)(503,582)(519,581)(536,580)(541,580)(552,579)(568,578)(585,576)(591,576)(601,575)(618,574)(631,572)(634,572)(651,570)(665,568)(667,568)(683,566)(694,565)(700,564)(716,561)(719,561)(733,558)(742,557)(749,556)(762,553)(766,552)(780,549)(782,549)(797,545)(798,545)(812,542)(815,541)(826,538)
\thinlines \path(826,538)(831,536)(839,534)(848,531)(851,530)(862,526)(864,526)(873,523)(881,519)(883,519)(892,515)(897,513)(901,511)(909,507)(913,505)(917,503)(924,500)(930,497)(932,496)(938,492)(945,488)(946,487)(951,484)(956,480)(962,477)(963,476)(967,473)(972,469)(977,465)(979,463)(981,461)(986,458)(990,454)(994,450)(996,448)(998,446)(1001,442)(1005,438)(1008,435)(1011,431)(1012,429)(1014,427)(1017,423)(1019,419)(1022,416)(1024,412)(1026,408)(1029,404)(1031,400)(1032,396)(1034,393)
\thinlines \path(1034,393)(1036,389)(1038,385)(1039,381)(1041,377)(1042,374)(1043,370)(1044,366)(1046,362)(1047,358)(1048,354)(1048,351)(1049,347)(1050,343)(1051,339)(1051,335)(1052,332)(1053,328)(1053,324)(1054,320)(1054,316)(1054,312)(1055,309)(1055,305)(1055,301)(1056,297)(1056,293)(1056,289)(1056,286)(1056,282)(1056,278)(1057,274)(1057,270)(1057,267)(1057,263)(1057,259)(1057,255)(1057,251)(1057,247)(1057,244)(1057,240)(1057,236)(1057,232)(1057,228)(1057,225)(1057,221)(1057,217)(1057,213)(1057,209)(1057,205)(1057,202)
\thinlines \path(1057,202)(1057,198)(1057,190)(1057,179)(1057,163)(1057,152)(1057,141)(1057,125)(1057,114)
\thinlines \path(223,666)(223,666)(240,666)(256,666)(273,665)(289,665)(305,664)(317,664)(322,663)(338,662)(355,661)(367,660)(371,659)(388,658)(404,656)(404,656)(420,654)(434,652)(437,652)(453,649)(460,649)(470,647)(483,645)(486,644)(503,641)(504,641)(519,638)(523,637)(536,634)(540,633)(552,630)(556,629)(568,626)(572,626)(585,622)(586,622)(599,618)(601,617)(611,614)(618,612)(623,610)(634,607)(635,607)(645,603)(651,601)(655,599)(665,595)(667,594)(674,591)(683,587)(683,587)
\thinlines \path(683,587)(691,584)(699,580)(700,579)(707,576)(714,572)(716,571)(721,568)(728,565)(733,562)(734,561)(740,557)(746,553)(749,551)(752,549)(757,545)(763,542)(766,539)(768,538)(772,534)(777,530)(782,526)(782,526)(786,523)(790,519)(794,515)(798,511)(798,510)(801,507)(805,503)(808,500)(812,496)(815,492)(815,492)(818,488)(821,484)(824,480)(826,477)(829,473)(831,469)(831,469)(834,465)(836,461)(838,458)(840,454)(842,450)(844,446)(846,442)(848,439)(848,438)(850,435)
\thinlines \path(850,435)(851,431)(853,427)(854,423)(856,419)(857,416)(858,412)(859,408)(861,404)(862,400)(863,396)(864,393)(865,389)(865,385)(866,381)(867,377)(868,374)(868,370)(869,366)(870,362)(870,358)(871,354)(871,351)(872,347)(872,343)(872,339)(873,335)(873,332)(873,328)(874,324)(874,320)(874,316)(874,312)(874,309)(875,305)(875,301)(875,297)(875,293)(875,289)(875,286)(875,282)(875,278)(875,274)(875,270)(875,267)(875,263)(875,259)(875,255)(875,251)(875,247)(875,244)
\thinlines \path(875,244)(876,240)(876,236)(876,232)(876,228)(876,225)(876,221)(876,217)(876,213)(876,209)(876,205)(876,202)(876,198)(876,190)(876,183)(876,186)(876,141)(876,133)(876,114)
\thinlines \path(223,831)(223,831)(240,831)(256,830)(273,829)(289,828)(305,826)(322,824)(338,821)(355,818)(359,817)(371,814)(376,813)(388,810)(391,809)(404,806)(405,805)(417,801)(420,801)(429,798)(437,795)(440,794)(451,790)(453,789)(460,786)(470,783)(470,782)(479,778)(486,776)(487,775)(496,771)(503,768)(504,767)(511,763)(519,759)(526,756)(533,752)(539,748)(546,744)(552,740)(558,736)(564,733)(570,729)(575,725)(581,721)(586,717)(591,714)(596,710)(601,706)(606,702)(610,698)
\thinlines \path(610,698)(615,694)(619,691)(624,687)(628,683)(632,679)(636,675)(640,671)(644,668)(648,664)(651,660)(655,656)(658,652)(662,649)(665,645)(668,641)(671,637)(675,633)(678,629)(681,626)(683,622)(686,618)(689,614)(692,610)(694,607)(697,603)(699,599)(702,595)(704,591)(707,587)(709,584)(711,580)(713,576)(715,572)(717,568)(719,565)(721,561)(723,557)(725,553)(727,549)(729,545)(730,542)(732,538)(734,534)(735,530)(737,526)(738,523)(740,519)(741,515)(743,511)(744,507)
\thinlines \path(744,507)(745,503)(746,500)(748,496)(749,492)(750,488)(751,484)(752,480)(753,477)(754,473)(755,469)(756,465)(757,461)(758,458)(759,454)(759,450)(760,446)(761,442)(762,438)(762,435)(763,431)(763,427)(764,423)(765,419)(765,416)(766,412)(766,408)(767,404)(767,400)(768,396)(768,393)(768,389)(769,385)(769,381)(769,377)(770,374)(770,370)(770,366)(770,362)(771,358)(771,354)(771,351)(771,347)(771,343)(771,339)(772,335)(772,332)(772,328)(772,324)(772,320)(772,316)
\thinlines \path(772,316)(772,312)(772,309)(772,305)(772,301)(772,297)(772,293)(772,289)(772,286)(773,282)(773,278)(773,274)(773,270)(773,267)(773,263)(773,255)(773,259)(773,247)(773,244)(773,251)(773,236)(773,225)(773,221)(773,209)(773,198)(773,190)(773,183)(773,167)(773,148)(773,141)(773,129)(773,118)(773,114)
\end{picture}

    \vglue 1ex
\caption{The phase diagram at $D=2.0, 2.5, 3.0, 3.5$.}
\label{fig:phaseKcrTcr}
\end{figure}

Since the dynamical fermion mass smoothly disappears
at the critical point for second order phase transition,
the critical line on $T$-$R$ plane is given by the massless limit
of Eq. (\ref{eqn:GapEinT}).
To find the equation for the critical line 
in an analytic form we take the limit $m\rightarrow 0$ 
in Eq. (\ref{eqn:GapEinT}) and find 
\begin{eqnarray}
&&\frac{\tr \11}{(4\pi)^{D/2}}\Gamma\left(1-\frac{D}{2}\right)m_{0}^{D-2}
-\frac{\tr \11}{(4\pi)^{(D-1)/2}}\frac{2}{\betacr}
\Gamma\left(\frac{3-D}{2}\right)
 \left(\frac{2\pi}{\betacr}\right)^{D-3}\zeta(3-D,1/2)
\label{eqn:relTcrKcr} \nonumber \\
&&+\frac{\tr \11}{(4\pi)^{(D-1)/2}}\frac{1}{\betacr}
\Gamma\left(\frac{3-D}{2}\right)
\nonumber \\
&&\times
\sumomeg\left[
\left|{\omegacr}_{n}\right|^{D-3}
-a_{cr}^{3-D}
\frac{\Gamma\left(\frac{D-1}{2}+i{\alphacr}_{n}\right)
      \Gamma\left(\frac{D-1}{2}-i{\alphacr}_{n}\right)}
     {\Gamma\left( 1 +i{\alphacr}_{n}\right)
      \Gamma\left( 1 -i{\alphacr}_{n}\right)}\right] = 0,
\nonumber
\end{eqnarray}
where $\zeta(z,a)$ is the generalized zeta function,
${\omegacr}_{n} = (2n+1)\pi/\betacr$,
${\alphacr}_{n} = a_{cr}\left|{\omegacr}_{n}\right|$ and $m_{0}$ is 
the dynamical fermion mass in Minkowski spacetime at $T=0$.
The critical lines are shown in Fig. \ref{fig:phaseKcrTcr}.

\vglue 2ex

The two-dimensional spacetime $R\otimes S^{1}$ is
a flat compact spacetime, $R=0$.
Thus the symmetry restoration which is caused by decreasing $a$
is induced by the finite size effect of the compact
space.
In four dimensions the effective potential at $T=0$
is divergent.
However the thermal effect gives only a finite correction
to the effective potential.
Thus the critical temperature $T_{\mbox{cr}}$ at $R=0$ 
is divergent at the four-dimensional limit.
Near four dimensions the curvature effect
seems to give the
main contribution to the phase transition.
But it results from the non-renormalizability of the
four-fermion model.
In a renormalizable theory the situation must
be changed.

\subsection{Negative curvature space ($R\otimes H^{D-1}$)}

In a negative curvature spacetime the chiral symmetry is always
broken down irrespective of $\lambda$
at $T=0$. Can the thermal effect restore the 
symmetry in the negative curvature spacetime?
Here we consider the model at finite temperature 
in the negative curvature spacetime, 
$R\otimes H^{D-1}$, for both $\lambda >\lambda_{\mbox{cr}}$ 
and $\lambda\leq\lambda_{\mbox{cr}}$.
The manifold $R\otimes H^{D-1}$ is defined by the metric
\begin{equation}
     ds^{2}=dt^{2}+a^{2}(d\theta^{2}+\sinh^{2}\theta\ d\Omega_{D-2})\, .
\end{equation}
It is a constant curvature spacetime with negative  
curvature
\begin{equation}
     R=- (D-1)(D-2)a^{-2}\, ,
\label{curvature}
\end{equation}
respectively ($2 \leq D < 4$).

According to the similar method used in the previous subsection
the effective potential at finite temperature in
$R\otimes H^{D-1}$ reads
\begin{equation}
\VTR(\sigma)=\frac{1}{2\lambda_{0}}\sigma^{2}
-\intsig \frac{\tr \11}{(4\pi)^{(D-1)/2}}
\frac{s a^{3-D}}{\beta}
\sumomeg
\frac{\Gamma\left(\frac{D-1}{2}+\alpha_{n}\right)}
     {\alpha_{n}\Gamma\left(\frac{3-D}{2}+\alpha_{n}\right)}
\Gamma\left(\frac{3-D}{2}\right).
\label{eqn:EfptNegT}
\end{equation}
$R\otimes H^{1}$ is equivalent to the two-dimensional
Minkowski space $R^2$.
At the two-dimensional limit of Eq. (\ref{eqn:EfptNegT})
the effective potential in two-dimensional Minkowski space
is reproduced.
Because of the convenience for numerical calculations
we rewrite the effective potential (\ref{eqn:EfptNegT}) 
in the same form described in the previous subsection.
Inserting Eq. (2$\cdot$33) and Eq. (\ref{eqn:relzero})
into Eq. (\ref{eqn:EfptNegT}) we get
\begin{eqnarray}
\VTR_{R}(\sigma)&=&\frac{1}{2}\left[
\frac{1}{\lambda_{R}}+\frac{\tr \11}{(4\pi)^{D/2}}
\Gamma\left(1-\frac{D}{2}\right)(D-1)\right]\mu^{D-2}\sigma^{2}
-\frac{\tr \11}{(4\pi)^{D/2}}\frac{1}{D}
\Gamma\left(1-\frac{D}{2}\right)\sigma^{2}\nonumber \\
&&-\frac{\tr \11}{(4\pi)^{(D-1)/2}}
\frac{1}{\Gamma\left(\frac{D-1}{2}\right)}\frac{1}{\beta}
\int_{0}^{\infty}\!\!dt \,t^{(D-3)/2}
\ln\frac{1+e^{-\beta\sqrt{t+\sigma^{2}}}}{1+e^{-\beta\sqrt{t}}}
\label{epot:Neg:finite}
\\
&&+\intsig\frac{\tr \11}{(4\pi)^{(D-1)/2}}
\frac{1}{\beta}\Gamma\left(\frac{3-D}{2}\right)
\nonumber \\
&&\times
\sumomeg s\left[ \left(s^{2}+\omega_{n}^{2}\right)^{(D-3)/2}
-a^{3-D}
\frac{\Gamma\left(\frac{D-1}{2}+\alpha_{n}\right)}
     {\alpha_{n}\Gamma\left(\frac{3-D}{2}+\alpha_{n}\right)}
     \right] \nonumber.
\end{eqnarray}

To find the minimum of the effective potential (\ref{epot:Neg:finite})
we analyze the non-trivial solution of the gap equation.
Substituting Eq. (\ref{epot:Neg:finite}) to Eq. (\ref{eqn:gap})
the gap equation reads
\begin{eqnarray}
&&\left(\frac{1}{\lambda_{R}}
-\frac{1}{\lambda_{\mbox{cr}}}\right)\mu^{D-2}
-\frac{\tr \11}{(4\pi)^{D/2}}\Gamma\left(1-\frac{D}{2}\right)m^{D-2}
\nonumber \\
&&+\frac{\tr \11}{(4\pi)^{\frac{D-1}{2}}}
\frac{1}{\Gamma\left(\frac{D-1}{2}\right)}
\int_{0}^{\infty}\!\!dt\,t^{(D-3)/2}
\frac{1}{\sqrt{t+m^{2}}}
\frac{e^{-\beta\sqrt{t+m^{2}}}}{1+e^{-\beta\sqrt{t+m^{2}}}}
\label{eqn:GapNegT} \\
&&+\frac{\tr \11}{(4\pi)^{(D-1)/2}}\frac{1}{\beta}
\Gamma\left(\frac{3-D}{2}\right)
\nonumber \\
&&\times\sumomeg\left[
\left(m^{2}+\omega_{n}^{2}\right)^{(D-3)/2}
-a^{3-D}
\frac{\Gamma\left(\frac{D-1}{2}+\alpha_{n}\right)}
     {\alpha_{n}\Gamma\left(\frac{3-D}{2}+\alpha_{n}\right)}
     \right] = 0.
\nonumber
\end{eqnarray}

Evaluating the gap equation (\ref{eqn:GapNegT})
the dynamical fermion mass $m$ is obtained.\cite{II}
The dynamical fermion mass smoothly disappears as the temperature
increases with the curvature fixed for both 
$\lambda > \lambda_{\mbox{cr}}$ and 
$\lambda\leq\lambda_{\mbox{cr}}$.
Then the broken chiral symmetry is restored for a sufficiently
high temperature. 
On the other hand the dynamical fermion mass becomes 
heavier as the curvature $|R|$ decreases with the temperature 
fixed.
Calculating Eq. (\ref{cr:Temp}) in three dimensions the 
critical temperature for $a\rightarrow\infty$ is given by
\begin{equation}
     k_{B}T_{\mbox{cr}}=\frac{1}{2\ln 2}\, .
\end{equation}
The curvature effects enhance the symmetry breaking
on $R\otimes H^{2}$.
Hence there is only the broken phase for $k_{B}T < 1/(2 \ln 2)$
in the model $\lambda > \lambda_{\mbox{cr}}$.
For $k_{B}T \geq 1/(2 \ln 2)$ or 
$\lambda \leq \lambda_{\mbox{cr}}$ 
the dynamical fermion mass is smoothly generated as the
curvature $|R|$ increases and the chiral symmetry is
broken down by the curvature effect.
After the same analysis done in arbitrary dimensions $2 < D < 4$
we find the same behavior for the dynamical fermion mass.
The thermal effect restores the broken chiral symmetry 
while the negative curvature effect breaks
the chiral symmetry.
Only the second order phase transition occurs with varying
temperature and curvature in $2 < D < 4$.
In two dimensions Eq. (\ref{eqn:GapNegT}) has the same 
behavior in Minkowski space.

\vglue 2ex
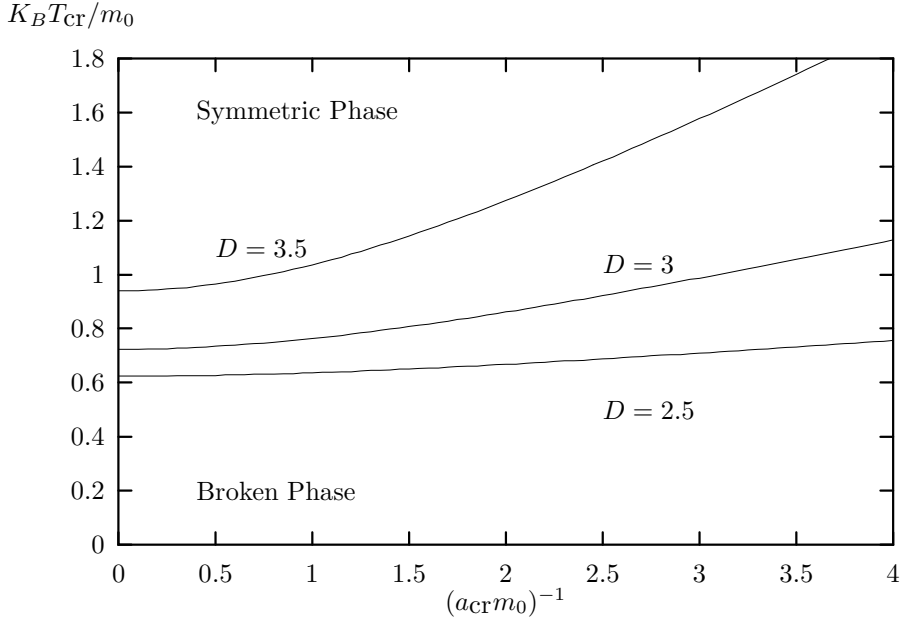
\begin{figure}
\setlength{\unitlength}{0.240900pt}
\begin{picture}(1500,900)(0,0)
\tenrm
\thicklines \path(220,113)(240,113)
\thicklines \path(1436,113)(1416,113)
\put(198,113){\makebox(0,0)[r]{0}}
\thicklines \path(220,198)(240,198)
\thicklines \path(1436,198)(1416,198)
\put(198,198){\makebox(0,0)[r]{0.2}}
\thicklines \path(220,283)(240,283)
\thicklines \path(1436,283)(1416,283)
\put(198,283){\makebox(0,0)[r]{0.4}}
\thicklines \path(220,368)(240,368)
\thicklines \path(1436,368)(1416,368)
\put(198,368){\makebox(0,0)[r]{0.6}}
\thicklines \path(220,453)(240,453)
\thicklines \path(1436,453)(1416,453)
\put(198,453){\makebox(0,0)[r]{0.8}}
\thicklines \path(220,537)(240,537)
\thicklines \path(1436,537)(1416,537)
\put(198,537){\makebox(0,0)[r]{1}}
\thicklines \path(220,622)(240,622)
\thicklines \path(1436,622)(1416,622)
\put(198,622){\makebox(0,0)[r]{1.2}}
\thicklines \path(220,707)(240,707)
\thicklines \path(1436,707)(1416,707)
\put(198,707){\makebox(0,0)[r]{1.4}}
\thicklines \path(220,792)(240,792)
\thicklines \path(1436,792)(1416,792)
\put(198,792){\makebox(0,0)[r]{1.6}}
\thicklines \path(220,877)(240,877)
\thicklines \path(1436,877)(1416,877)
\put(198,877){\makebox(0,0)[r]{1.8}}
\thicklines \path(220,113)(220,133)
\thicklines \path(220,877)(220,857)
\put(220,68){\makebox(0,0){0}}
\thicklines \path(372,113)(372,133)
\thicklines \path(372,877)(372,857)
\put(372,68){\makebox(0,0){0.5}}
\thicklines \path(524,113)(524,133)
\thicklines \path(524,877)(524,857)
\put(524,68){\makebox(0,0){1}}
\thicklines \path(676,113)(676,133)
\thicklines \path(676,877)(676,857)
\put(676,68){\makebox(0,0){1.5}}
\thicklines \path(828,113)(828,133)
\thicklines \path(828,877)(828,857)
\put(828,68){\makebox(0,0){2}}
\thicklines \path(980,113)(980,133)
\thicklines \path(980,877)(980,857)
\put(980,68){\makebox(0,0){2.5}}
\thicklines \path(1132,113)(1132,133)
\thicklines \path(1132,877)(1132,857)
\put(1132,68){\makebox(0,0){3}}
\thicklines \path(1284,113)(1284,133)
\thicklines \path(1284,877)(1284,857)
\put(1284,68){\makebox(0,0){3.5}}
\thicklines \path(1436,113)(1436,133)
\thicklines \path(1436,877)(1436,857)
\put(1436,68){\makebox(0,0){4}}
\thicklines \path(220,113)(1436,113)(1436,877)(220,877)(220,113)
\put(45,945){\makebox(0,0)[l]{\shortstack{$K_{B}T_{\mbox{cr}}/m_{0}$}}}
\put(828,23){\makebox(0,0){$(a_{\mbox{cr}}m_{0})^{-1}$}}
\put(342,792){\makebox(0,0)[l]{Symmetric Phase}}
\put(342,198){\makebox(0,0)[l]{Broken Phase}}
\put(372,580){\makebox(0,0)[l]{$D=3.5$}}
\put(980,554){\makebox(0,0)[l]{$D=3$}}
\put(980,325){\makebox(0,0)[l]{$D=2.5$}}
\thinlines \path(220,378)(220,378)(236,378)(251,378)(266,378)(281,378)(296,378)(312,379)(327,379)(342,379)(357,379)(372,379)(388,380)(403,380)(418,380)(433,381)(448,381)(464,381)(479,382)(494,382)(509,383)(524,383)(540,384)(555,384)(570,385)(585,385)(600,386)(616,387)(631,387)(646,388)(661,389)(676,389)(692,390)(707,391)(722,391)(737,392)(752,393)(768,394)(783,394)(798,395)(813,396)(828,397)(844,397)(859,398)(874,399)(889,400)(904,401)(920,402)(935,402)(950,403)(965,404)
\thinlines \path(965,404)(980,405)(996,406)(1011,407)(1026,408)(1041,409)(1056,410)(1072,411)(1087,412)(1102,412)(1117,413)(1132,414)(1148,415)(1163,416)(1178,417)(1193,418)(1208,419)(1224,420)(1239,421)(1254,422)(1269,423)(1284,424)(1300,425)(1315,426)(1330,427)(1345,428)(1360,429)(1376,430)(1391,431)(1406,432)(1421,433)(1436,434)
\thinlines \path(220,420)(220,420)(236,420)(251,420)(266,421)(281,421)(296,421)(312,422)(327,422)(342,423)(357,424)(372,425)(388,426)(403,427)(418,428)(433,429)(448,430)(464,431)(479,433)(494,434)(509,436)(524,437)(540,439)(555,440)(570,442)(585,444)(600,446)(616,448)(631,450)(646,452)(661,454)(676,456)(692,458)(707,460)(722,462)(737,464)(752,467)(768,469)(783,471)(798,474)(813,476)(828,479)(844,481)(859,484)(874,486)(889,489)(904,491)(920,494)(935,497)(950,499)(965,502)
\thinlines \path(965,502)(980,505)(996,507)(1011,510)(1026,513)(1041,516)(1056,518)(1072,521)(1087,524)(1102,527)(1117,530)(1132,532)(1148,535)(1163,538)(1178,541)(1193,544)(1208,547)(1224,550)(1239,553)(1254,556)(1269,559)(1284,562)(1300,565)(1315,568)(1330,571)(1345,574)(1360,577)(1376,580)(1391,583)(1406,586)(1421,589)(1436,592)
\thinlines \path(220,512)(220,512)(236,512)(251,512)(266,513)(281,514)(296,515)(312,516)(327,517)(342,519)(357,521)(372,523)(388,525)(403,527)(418,530)(433,533)(448,536)(464,539)(479,542)(494,546)(509,549)(524,553)(540,557)(555,561)(570,565)(585,570)(600,574)(616,579)(631,584)(646,588)(661,593)(676,598)(692,604)(707,609)(722,614)(737,620)(752,625)(768,631)(783,636)(798,642)(813,648)(828,654)(844,660)(859,666)(874,672)(889,678)(904,684)(920,691)(935,697)(950,703)(965,710)
\thinlines \path(965,710)(980,716)(996,722)(1011,729)(1026,736)(1041,742)(1056,749)(1072,756)(1087,762)(1102,769)(1117,776)(1132,783)(1148,789)(1163,796)(1178,803)(1193,810)(1208,817)(1224,824)(1239,831)(1254,838)(1269,845)(1284,852)(1300,860)(1315,867)(1330,874)(1337,877)
\end{picture}

                \vglue 1ex
\caption{The phase diagram at D=2.5, 3.0, 3.5 for
$\lambda > \lambda_{cr}$.}
\label{fig:phaseKcrTcrBrknRH}
\end{figure}
\begin{figure}
                \vglue 1ex
\setlength{\unitlength}{0.240900pt}
\begin{picture}(1500,900)(0,0)
\tenrm
\thicklines \path(220,113)(240,113)
\thicklines \path(1436,113)(1416,113)
\put(198,113){\makebox(0,0)[r]{0}}
\thicklines \path(220,198)(240,198)
\thicklines \path(1436,198)(1416,198)
\put(198,198){\makebox(0,0)[r]{0.2}}
\thicklines \path(220,283)(240,283)
\thicklines \path(1436,283)(1416,283)
\put(198,283){\makebox(0,0)[r]{0.4}}
\thicklines \path(220,368)(240,368)
\thicklines \path(1436,368)(1416,368)
\put(198,368){\makebox(0,0)[r]{0.6}}
\thicklines \path(220,453)(240,453)
\thicklines \path(1436,453)(1416,453)
\put(198,453){\makebox(0,0)[r]{0.8}}
\thicklines \path(220,537)(240,537)
\thicklines \path(1436,537)(1416,537)
\put(198,537){\makebox(0,0)[r]{1}}
\thicklines \path(220,622)(240,622)
\thicklines \path(1436,622)(1416,622)
\put(198,622){\makebox(0,0)[r]{1.2}}
\thicklines \path(220,707)(240,707)
\thicklines \path(1436,707)(1416,707)
\put(198,707){\makebox(0,0)[r]{1.4}}
\thicklines \path(220,792)(240,792)
\thicklines \path(1436,792)(1416,792)
\put(198,792){\makebox(0,0)[r]{1.6}}
\thicklines \path(220,877)(240,877)
\thicklines \path(1436,877)(1416,877)
\put(198,877){\makebox(0,0)[r]{1.8}}
\thicklines \path(220,113)(220,133)
\thicklines \path(220,877)(220,857)
\put(220,68){\makebox(0,0){0}}
\thicklines \path(463,113)(463,133)
\thicklines \path(463,877)(463,857)
\put(463,68){\makebox(0,0){2}}
\thicklines \path(706,113)(706,133)
\thicklines \path(706,877)(706,857)
\put(706,68){\makebox(0,0){4}}
\thicklines \path(950,113)(950,133)
\thicklines \path(950,877)(950,857)
\put(950,68){\makebox(0,0){6}}
\thicklines \path(1193,113)(1193,133)
\thicklines \path(1193,877)(1193,857)
\put(1193,68){\makebox(0,0){8}}
\thicklines \path(1436,113)(1436,133)
\thicklines \path(1436,877)(1436,857)
\put(1436,68){\makebox(0,0){10}}
\thicklines \path(220,113)(1436,113)(1436,877)(220,877)(220,113)
\put(45,945){\makebox(0,0)[l]{\shortstack{$K_{B}T_{\mbox{cr}}/m_{0}$}}}
\put(828,23){\makebox(0,0){$(a_{\mbox{cr}}m_{0})^{-1}$}}
\put(317,792){\makebox(0,0)[l]{Symmetric Phase}}
\put(488,580){\makebox(0,0)[l]{$D=3.5$}}
\put(901,444){\makebox(0,0)[l]{$D=3$}}
\put(1168,189){\makebox(0,0)[l]{$D=2.5$}}
\thinlines \path(293,113)(293,113)(305,113)(317,113)(330,113)(342,113)(354,113)(366,114)(378,114)(390,114)(403,114)(415,114)(427,115)(439,115)(451,115)(463,115)(475,116)(488,116)(500,116)(512,117)(524,117)(536,117)(548,118)(561,118)(573,118)(585,119)(597,119)(609,120)(621,120)(634,121)(646,121)(658,121)(670,122)(682,122)(694,123)(707,123)(713,124)(719,124)(725,124)(731,124)(737,125)(743,125)(749,125)(755,125)(761,126)(767,126)(773,126)(779,126)(786,127)(792,127)(798,127)
\thinlines \path(798,127)(804,127)(810,128)(816,128)(822,128)(828,129)(834,129)(840,129)(846,129)(852,130)(859,130)(865,130)(871,131)(877,131)(883,131)(889,131)(895,132)(901,132)(907,132)(913,133)(919,133)(925,133)(931,134)(938,134)(944,134)(950,135)(956,135)(962,135)(968,135)(974,136)(980,136)(986,136)(992,137)(998,137)(1004,137)(1011,138)(1017,138)(1023,138)(1029,139)(1035,139)(1041,139)(1047,140)(1053,140)(1059,140)(1065,141)(1071,141)(1077,141)(1083,142)(1090,142)(1096,142)(1102,143)
\thinlines \path(1102,143)(1108,143)(1114,143)(1120,144)(1126,144)(1132,144)(1138,145)(1144,145)(1150,145)(1156,146)(1163,146)(1169,146)(1175,147)(1181,147)(1187,147)(1193,148)(1199,148)(1205,149)(1211,149)(1217,149)(1223,150)(1229,150)(1235,150)(1242,151)(1248,151)(1254,151)(1260,152)(1266,152)(1272,153)(1278,153)(1284,153)(1290,154)(1296,154)(1302,154)(1308,155)(1315,155)(1321,155)(1327,156)(1333,156)(1339,157)(1345,157)(1351,157)(1357,158)(1363,158)(1369,159)(1375,159)(1381,159)(1387,160)(1394,160)(1400,160)(1406,161)
\thinlines \path(1406,161)(1412,161)(1418,162)(1424,162)(1430,162)(1436,163)
\thinlines \path(299,113)(299,113)(311,114)(323,114)(336,115)(348,117)(360,119)(372,121)(384,123)(396,126)(409,129)(421,132)(433,136)(445,139)(457,143)(463,145)(469,147)(475,150)(482,152)(488,154)(494,156)(500,159)(506,161)(512,163)(518,166)(524,168)(530,171)(536,173)(542,176)(548,178)(555,181)(561,184)(567,186)(573,189)(579,192)(585,194)(591,197)(597,200)(603,203)(609,205)(615,208)(621,211)(627,214)(634,217)(640,220)(646,223)(652,225)(658,228)(664,231)(670,234)(676,237)
\thinlines \path(676,237)(682,240)(688,243)(694,246)(700,249)(707,252)(713,255)(719,258)(725,261)(731,264)(737,267)(743,271)(749,274)(755,277)(761,280)(767,283)(773,286)(779,289)(786,292)(792,295)(798,299)(804,302)(810,305)(816,308)(822,311)(828,314)(834,318)(840,321)(846,324)(852,327)(859,330)(865,333)(871,337)(877,340)(883,343)(889,346)(895,350)(901,353)(907,356)(913,359)(919,362)(925,366)(931,369)(938,372)(944,375)(950,379)(956,382)(962,385)(968,388)(974,392)(980,395)
\thinlines \path(980,395)(986,398)(992,402)(998,405)(1004,408)(1011,411)(1017,415)(1023,418)(1029,421)(1035,425)(1041,428)(1047,431)(1053,434)(1059,438)(1065,441)(1071,444)(1077,448)(1083,451)(1090,454)(1096,458)(1102,461)(1108,464)(1114,468)(1120,471)(1126,474)(1132,478)(1138,481)(1144,484)(1150,488)(1156,491)(1163,494)(1169,498)(1175,501)(1181,504)(1187,508)(1193,511)(1199,514)(1205,518)(1211,521)(1217,524)(1223,528)(1229,531)(1235,534)(1242,538)(1248,541)(1254,544)(1260,548)(1266,551)(1272,554)(1278,558)(1284,561)
\thinlines \path(1284,561)(1290,565)(1296,568)(1302,571)(1308,575)(1315,578)(1321,581)(1327,585)(1333,588)(1339,591)(1345,595)(1351,598)(1357,602)(1363,605)(1369,608)(1375,612)(1381,615)(1387,618)(1394,622)(1400,625)(1406,629)(1412,632)(1418,635)(1424,639)(1430,642)(1436,645)
\thinlines \path(330,113)(330,113)(342,115)(354,118)(366,122)(378,129)(390,139)(403,150)(415,163)(427,178)(433,186)(439,195)(445,203)(451,212)(457,221)(463,230)(469,240)(475,249)(482,259)(488,268)(494,278)(500,287)(506,297)(512,307)(518,316)(524,326)(530,336)(536,346)(542,355)(548,365)(555,375)(561,384)(567,394)(573,404)(579,413)(585,423)(591,432)(597,442)(603,451)(609,461)(615,470)(621,480)(627,489)(634,499)(640,508)(646,518)(652,527)(658,536)(664,546)(670,555)(676,564)
\thinlines \path(676,564)(682,574)(688,583)(694,592)(700,601)(707,611)(713,620)(719,629)(725,638)(731,647)(737,656)(743,666)(749,675)(755,684)(761,693)(767,702)(773,711)(779,720)(786,729)(792,738)(798,747)(804,756)(810,765)(816,774)(822,783)(828,792)(834,801)(840,810)(846,819)(852,828)(859,837)(865,846)(871,854)(877,863)(883,872)(886,877)
\end{picture}

                \vglue 1ex
\caption{The phase diagram at $D=2.5, 3.0, 3.5$
for $\lambda \leq \lambda_{\mbox{cr}}$ .}
\label{fig:phaseKcrTcrSymmRH}
\end{figure}

For the second order phase transition the critical point is
obtained by the massless limit of the gap equation.
Taking the massless limit $m\rightarrow 0$ in 
Eq. (\ref{eqn:GapNegT})
we find the equation that gives the relation between critical
temperature $\beta_{\mbox{cr}}=1/(k_{B}T_{\mbox{cr}})$ 
and critical radius $a_{\mbox{cr}}$:
\begin{eqnarray}
&&
\frac{\tr \11}{(4\pi)^{D/2}}\Gamma\left(1-\frac{D}{2}\right)m_{0}^{D-2}
\nonumber \\
&&
-\frac{\tr \11}{(4\pi)^{(D-1)/2}}\frac{2}{\betacr}
\Gamma\left(\frac{3-D}{2}\right)
 \left(\frac{2\pi}{\betacr}\right)^{D-3}\zeta(3-D,1/2)
\nonumber \\
&&
+\frac{\tr \11}{(4\pi)^{(D-1)/2}}\frac{1}{\betacr}
\Gamma\left(\frac{3-D}{2}\right)
\label{eqn:relTcrKcrNeg}
\\
&&\times
\sumomeg\left[
\left|{\omegacr}_{n}\right|^{D-3}
-a_{cr}^{3-D}
\frac{\Gamma\left(\frac{D-1}{2}+{\alphacr}_{n}\right)}
     {{\alphacr}_{n}\Gamma\left(\frac{3-D}{2}+{\alphacr}_{n}\right)}
     \right] = 0.\nonumber
\end{eqnarray}
Evaluating Eq. (\ref{eqn:relTcrKcrNeg}) numerically
we draw the phase diagram of the four-fermion model
with varying temperature and/or curvature on 
$R\otimes H^{D-1}$ at $D=2.5, 3.0, 3.5$ 
in Figs. \ref{fig:phaseKcrTcrBrknRH}
and \ref{fig:phaseKcrTcrSymmRH}.
In those figures the normalization scale $m_{0}$
is taken to the value defined in Eq. (\ref{def:mass})
for $\lambda > \lambda_{\mbox{cr}}$ and
\begin{equation}
     m_{0}=\mu
   \left[
       -\frac{(4\pi)^{D/2}}{\tr \11 \Gamma(1-D/2)}\frac{1}{\lambda_{R}}
       -D+1
   \right]^{1/(D-2)}\, 
\label{def:mass2}
\end{equation}
for $\lambda\leq\lambda_{\mbox{cr}}$.
At a scale $T\sim a^{-1} \sim m_{0}$ the thermal effect
gives the main contribution to the phase structure
for $D \leq 3$ and the symmetric phase is realized.
At  four-dimensional limit $T_{\mbox{cr}}$ is
divergent for $\lambda > \lambda_{\mbox{cr}}$.
It comes from the divergence which appears
at the four-dimensional limit.
Thus it may be a result caused mainly by the 
non-renormalizability of the theory.

Thus we have investigated the phase structure of the
four-fermion model at finite temperature and curvature
in arbitrary dimensions ($2 < D < 4$).
In positive curvature space the curvature effect
restores the broken chiral symmetry.
On the contrary the curvature effect enhances the 
chiral symmetry breaking in negative curvature space.
At finite temperature the lower limit of the momentum 
for the fermion field ($k^{0}< \pi /\beta$) appears 
from the antiperiodicity and
then the long range effect is suppressed.
Thus the thermal effect restore the broken symmetry
even in a negative curvature spacetime.

In two spacetime dimensions it is impossible to introduce
the combined effect of the temperature and curvature 
within our method because of the assumption of the 
equilibrium.
This assumption is not fulfilled in inflationary
expanding universe, but we may discuss 
the phase transition at early universe 
using our results.

\section{Finite size and topological effects}

In the previous sections we discussed four-fermion models
in curved spacetime.
The main motivation to do such a study was the cosmological
applications.
However, one may expect that the very early universe
had a non-trivial topology
(the introduction of the temperature in the hot universe
maybe classified as the similar type of effects).
This renders interesting to investigate the phenomenon
of dynamical chiral symmetry breaking in curved spacetime
with non-trivial topology. This section will be devoted
to such a study on the background ${\cal M}^{D-1}\otimes S^{1}$
where $S^{1}$ is the one-dimensional sphere and ${\cal M}^{D-1}$
is a $(D-1)$-dimensional arbitrary curved
manifold of trivial topology (see Ref. \cite{TOPO}).

\subsection{Effective potential in ${\cal M}^{D-1}\otimes S^{1}$}

We start again from the action of four-fermion models
in curved spacetime defined in (\ref{ac:gn})
or (\ref{ac:njl}).
Making the explicit calculations in the same way
as in \S 2 we easily get the effective potential
in the leading order of $1/N$ expansion:
$$
     V(\sigma ) = \frac{1}{2\lambda_0}\sigma^{2}
                  -i\mbox{tr} \int_{0}^{\sigma}ds S(x,x;s)
                  +O\left(\frac{1}{N}\right)\, .
$$
Using the known expression for the propagator of a free
Dirac field in a weakly-varying gravitational 
background (see expression (\ref{exp:s})), we will
then write the effective potential with accuracy up to 
linear curvature terms.
In the spacetime ${\cal M}^{D-1}\otimes S^{1}$
there is no constraint for the boundary condition \cite{ISHAM,Dowker} 
along the compactified direction.
Some of the independent spin structures are allowed in our universe.
If we take the antiperiodic boundary condition
along the compactified direction, the field theory
in ${\cal M}^{D-1}\otimes S^{1}$ is equivalent to the
finite temperature field theory (see \S 6).
Here we consider the fermion fields with periodic and 
antiperiodic boundary conditions independently.
Thus, the effective potential is found to be
\begin{eqnarray}
     V(\sigma) & = & \frac{1}{2 \lambda_{0}}\sigma^2
     -i\,{\rm tr}\  \displaystyle \int_{0}^{\sigma} s ds
     \frac{1}{L}
     \sum_{n=-\infty}^{\infty}
     \int \frac{d^{D-1} k}{\left( 2 \pi \right) ^{D-1}} 
\nonumber  \\ 
     && \times \left[
     \frac{1}{k^2-s^2} 
     -\frac{1}{12} R \frac{1}{\left( k^2 - s^2
     \right)^2}
     +\frac{2}{3}
     R_{\mu \nu} k^{\mu} k^{\nu}
     \frac{1}{\left( k^2 - s^2 \right) ^3}\right] ,        
\label{calcul}
\end{eqnarray}
where one should integrate over \( k_0, \cdots , k_{D-2} \) 
and sum over the coordinate
$k_{D-1}$, which is given by: 
$k_{D-1} = \left(2 n + \delta_{p,1}\right) \pi/ L$. 

Evaluating the effective potential (\ref{calcul})
we study the phase structure of the four-fermion
models in weakly curved spacetime with nontrivial
topology below.

\subsection{Flat spacetime with nontrivial topology
($R^{D-1}\otimes S^{1}$)}

Now it is not so clear if the symmetry restoration 
is mainly caused by the curvature effect or finite size 
effect in an arbitrary dimension.
In flat spacetime with nontrivial topology
we clearly see only the finite size effect.
In Refs. \cite{KNSY,Kimetal,IIYF} the dynamical 
symmetry breaking
has been investigated in flat spacetime with nontrivial
topology.
In the present subsection we calculate the effective
potential (\ref{calcul}) in flat spacetime 
$R^{D-1}\otimes S^{1}$ to see only a finite size effect.

In $R^{D-1}\otimes S^{1}$ the effective potential
(\ref{calcul}) reduces to
\begin{equation}
     V(\sigma) = \frac{1}{2 \lambda_{0}}\sigma^2
     -i\,{\rm tr}\  \displaystyle \int_{0}^{\sigma} s ds
     \frac{1}{L}
     \sum_{n=-\infty}^{\infty}
     \int \frac{d^{D-1} k}{(2 \pi )^{D-1}} 
     \frac{1}{k^2-s^2}\, .
\label{calcul:flat}
\end{equation}

If we perform the integration in Eq. (\ref{calcul:flat})
we find that \cite{IKM}
\begin{eqnarray}
     V(\sigma)&=&\frac{\sigma^{2}}{2\lambda_{0}}
     +\frac{\tr\11}{2 (4\pi)^{(D-1)/2}}
     \Gamma\left(\frac{1-D}{2}\right)\nonumber \\
     &&\times\frac{1}{L}\sum^{\infty}_{n=-\infty}
     \left[(k_{D-1}^{2}+\sigma^{2})^{(D-1)/2}
     -{k_{D-1}^{2}}^{(D-1)/2}\right]\, . 
\label{v:flat:topo}
\end{eqnarray}
Thus the effective potential is modified by finite
size effect which is interpreted as a generalized
version of the Casimir effect.\cite{CE}
Analyzing Eq. (\ref{v:flat:topo}) it is found that the
divergences in the effective potential are of the same
form in Minkowski spacetime.
Thus we need not renormalize the parts which include the
finite size effects contributions. 
We apply the same renormalization
condition as that shown in Eq. (\ref{cond:ren})
and obtain the renormalized effective potential as in 
Minkowski spacetime.
The renormalized effective potential is given by
replacing $\lambda_{0}$ in Eq. (\ref{v:flat:topo})
with the renormalized one $\lambda$ in Eq. (\ref{eqn:ren}).

If we apply the expression (\ref{v:flat:topo})
in the gap equation (\ref{gap:w}), we find that
\begin{equation}
     \frac{1}{\lambda}-\frac{1}{\lambda_{\mbox{cr}}}
     -\frac{\tr\11}{(4\pi)^{(D-1)/2}}
     \Gamma\left(\frac{3-D}{2}\right)
     \frac{1}{L\mu}
     \sum_{n=-\infty}^{\infty}
     \left(\frac{k_{D-1}^{2}+m^{2}}{\mu^{2}}\right)^{(D-3)/2}
     =0\, .
\label{gap:flat:topo}
\end{equation}
In Eq. (\ref{gap:flat:topo}) we neglected the trivial
solution $m=0$.

In the case of antiperiodic boundary condition
it is well-known that the field theory in $R^{D-1}\otimes S^{1}$
is equivalent to the finite temperature field theory.
The effective potential (\ref{gap:flat:topo}) is in
agreement with that of the finite temperature four-fermion
theory (see for example Ref. \cite{IKM} and references there in)
with recourse to the relation between the size $L$ of the
compactified direction and the temperature $T$
\begin{equation}
     L\leftrightarrow\frac{1}{k_{B}T}\, 
\end{equation}
with $k_{B}$ the Boltzmann constant.
Evaluating the effective potential (\ref{v:flat:topo})
and the gap equation (\ref{gap:flat:topo})
one can study numerically phase structure of the theory
and find that the second order phase transition takes place 
and the broken chiral symmetry is restored for sufficiently
small $L < L_{\mbox{cr}}$ at the large $N$ limit. 
It agrees with the known result at finite temperature.
Analyzing the gap equation (\ref{gap:flat:topo})
the critical size is given by
\begin{equation}
     L_{cr}m_{0}=2\pi\left[
     \frac{2\Gamma((3-D)/2)}{\sqrt{\pi}\Gamma (1-D/2)}
     (2^{3-D}-1)\zeta(3-D)\right]^{1/(D-2)}\, ,
\label{crl:flat}
\end{equation}
where $m_{0}$ is the dynamical fermion mass in Minkowski
spacetime.
It is equal to the well-known formula of the critical
temperature for the four-fermion model.\cite{IKM}
\begin{figure}
\setlength{\unitlength}{0.240900pt}
\begin{picture}(1500,900)(0,0)
\tenrm
\thicklines \path(220,113)(240,113)
\thicklines \path(1436,113)(1416,113)
\put(198,113){\makebox(0,0)[r]{0}}
\thicklines \path(220,245)(240,245)
\thicklines \path(1436,245)(1416,245)
\put(198,245){\makebox(0,0)[r]{0.05}}
\thicklines \path(220,376)(240,376)
\thicklines \path(1436,376)(1416,376)
\put(198,376){\makebox(0,0)[r]{0.1}}
\thicklines \path(220,508)(240,508)
\thicklines \path(1436,508)(1416,508)
\put(198,508){\makebox(0,0)[r]{0.15}}
\thicklines \path(220,640)(240,640)
\thicklines \path(1436,640)(1416,640)
\put(198,640){\makebox(0,0)[r]{0.2}}
\thicklines \path(220,772)(240,772)
\thicklines \path(1436,772)(1416,772)
\put(198,772){\makebox(0,0)[r]{0.25}}
\thicklines \path(220,113)(220,133)
\thicklines \path(220,877)(220,857)
\put(220,68){\makebox(0,0){2}}
\thicklines \path(342,113)(342,133)
\thicklines \path(342,877)(342,857)
\put(342,68){\makebox(0,0){2.2}}
\thicklines \path(463,113)(463,133)
\thicklines \path(463,877)(463,857)
\put(463,68){\makebox(0,0){2.4}}
\thicklines \path(585,113)(585,133)
\thicklines \path(585,877)(585,857)
\put(585,68){\makebox(0,0){2.6}}
\thicklines \path(706,113)(706,133)
\thicklines \path(706,877)(706,857)
\put(706,68){\makebox(0,0){2.8}}
\thicklines \path(828,113)(828,133)
\thicklines \path(828,877)(828,857)
\put(828,68){\makebox(0,0){3}}
\thicklines \path(950,113)(950,133)
\thicklines \path(950,877)(950,857)
\put(950,68){\makebox(0,0){3.2}}
\thicklines \path(1071,113)(1071,133)
\thicklines \path(1071,877)(1071,857)
\put(1071,68){\makebox(0,0){3.4}}
\thicklines \path(1193,113)(1193,133)
\thicklines \path(1193,877)(1193,857)
\put(1193,68){\makebox(0,0){3.6}}
\thicklines \path(1314,113)(1314,133)
\thicklines \path(1314,877)(1314,857)
\put(1314,68){\makebox(0,0){3.8}}
\thicklines \path(1436,113)(1436,133)
\thicklines \path(1436,877)(1436,857)
\put(1436,68){\makebox(0,0){4}}
\thicklines \path(220,113)(1436,113)(1436,877)(220,877)(220,113)
\put(45,945){\makebox(0,0)[l]{\shortstack{$\displaystyle\frac{L_{\mbox{cr}}m_{0}}{2\pi}$}}}
\put(828,23){\makebox(0,0){$D$}}
\put(1010,772){\makebox(0,0)[l]{Broken phase}}
\put(524,429){\makebox(0,0)[l]{Symmetric phase}}
\thinlines \path(220,853)(220,853)(271,842)(321,832)(372,821)(423,809)(473,797)(524,784)(575,771)(625,758)(676,743)(727,728)(777,711)(828,694)(879,676)(929,656)(980,635)(1031,612)(1081,587)(1132,559)(1183,528)(1233,491)(1284,449)(1335,395)(1360,361)(1385,320)(1398,294)(1411,263)(1417,243)(1423,220)(1427,206)(1430,190)(1431,180)(1433,167)(1434,152)(1436,113)
\end{picture}

\vglue 1ex
\caption{Critical size $L_{\mbox{cr}}$ as the function of $D$
         for $\lambda > \lambda_{\mbox{cr}}$
         in the case of the anti-periodic boundary
condition.}
\vglue 9ex
\setlength{\unitlength}{0.240900pt}
\begin{picture}(1500,900)(0,0)
\tenrm
\thicklines \path(220,113)(240,113)
\thicklines \path(1436,113)(1416,113)
\put(198,113){\makebox(0,0)[r]{0}}
\thicklines \path(220,209)(240,209)
\thicklines \path(1436,209)(1416,209)
\put(198,209){\makebox(0,0)[r]{1}}
\thicklines \path(220,304)(240,304)
\thicklines \path(1436,304)(1416,304)
\put(198,304){\makebox(0,0)[r]{2}}
\thicklines \path(220,400)(240,400)
\thicklines \path(1436,400)(1416,400)
\put(198,400){\makebox(0,0)[r]{3}}
\thicklines \path(220,495)(240,495)
\thicklines \path(1436,495)(1416,495)
\put(198,495){\makebox(0,0)[r]{4}}
\thicklines \path(220,591)(240,591)
\thicklines \path(1436,591)(1416,591)
\put(198,591){\makebox(0,0)[r]{5}}
\thicklines \path(220,686)(240,686)
\thicklines \path(1436,686)(1416,686)
\put(198,686){\makebox(0,0)[r]{6}}
\thicklines \path(220,782)(240,782)
\thicklines \path(1436,782)(1416,782)
\put(198,782){\makebox(0,0)[r]{7}}
\thicklines \path(220,877)(240,877)
\thicklines \path(1436,877)(1416,877)
\put(198,877){\makebox(0,0)[r]{8}}
\thicklines \path(220,113)(220,133)
\thicklines \path(220,877)(220,857)
\put(220,68){\makebox(0,0){3}}
\thicklines \path(463,113)(463,133)
\thicklines \path(463,877)(463,857)
\put(463,68){\makebox(0,0){3.2}}
\thicklines \path(706,113)(706,133)
\thicklines \path(706,877)(706,857)
\put(706,68){\makebox(0,0){3.4}}
\thicklines \path(950,113)(950,133)
\thicklines \path(950,877)(950,857)
\put(950,68){\makebox(0,0){3.6}}
\thicklines \path(1193,113)(1193,133)
\thicklines \path(1193,877)(1193,857)
\put(1193,68){\makebox(0,0){3.8}}
\thicklines \path(1436,113)(1436,133)
\thicklines \path(1436,877)(1436,857)
\put(1436,68){\makebox(0,0){4}}
\thicklines \path(220,113)(1436,113)(1436,877)(220,877)(220,113)
\put(45,945){\makebox(0,0)[l]{\shortstack{$\displaystyle\frac{L_{\mbox{cr}}m_{0}}{2\pi}$}}}
\put(828,23){\makebox(0,0){$D$}}
\put(256,256){\makebox(0,0)[l]{Broken}}
\put(269,180){\makebox(0,0)[l]{phase}}
\put(706,591){\makebox(0,0)[l]{Symmetric phase}}
\thinlines \path(263,877)(264,860)(270,757)(276,678)(282,616)(295,523)(307,457)(320,409)(332,372)(345,342)(358,319)(383,283)(408,257)(433,237)(458,222)(483,210)(533,193)(583,180)(633,171)(684,164)(734,158)(784,153)(834,149)(884,146)(934,143)(985,140)(1035,138)(1085,135)(1135,133)(1185,131)(1235,129)(1286,127)(1336,124)(1386,121)(1398,120)(1411,119)(1417,118)(1423,117)(1427,116)(1430,116)(1431,115)(1433,115)(1434,114)(1436,113)
\end{picture}

\vglue 1ex
\caption{Critical size $L_{\mbox{cr}}$ as the function of $D$
         for $\lambda \leq \lambda_{\mbox{cr}}$
         in the case of the periodic boundary condition.}
\end{figure}
In Fig. 23 the critical size is plotted as a function of
dimension $D$ for $\lambda > \lambda_{\mbox{cr}}$ by 
the use of Eq. (\ref{crl:flat}).

On the other hand finite size $L$ has an opposite
effect for the fermion field with periodic boundary condition.
In this case the first derivative of the
effective potential (\ref{gap:flat:topo})
has a negative value for $D=2$ at the limit $\sigma\rightarrow 0$
\begin{equation}
     \left.\frac{dV}{d\sigma}\right|_{\sigma\rightarrow 0}
     \rightarrow -\frac{\tr\11}{2}\frac{1}{L} < 0\, .
\end{equation}
For $2 < D \leq 3$ the first derivative of $V$ vanishes
but the second derivative of $V$ has a negative value
at the limit $\sigma \rightarrow 0$
\begin{equation}
     \left.\frac{1}{\mu^{D-2}}\frac{\partial^{2}V}{\partial\sigma^{2}}
     \right|_{\sigma\rightarrow 0}
     \rightarrow -\frac{\tr\11}{(4\pi)^{(D-1)/2}}
     \Gamma\left(\frac{3-D}{2}\right)
     \frac{D-2}{L\mu}
     \left(\frac{\mu^{2}}{\sigma^{2}}\right)^{(3-D)/2} < 0\, .
\end{equation}
Thus the symmetric phase cannot be realized
for $2 \leq D \leq 3$ even if the coupling 
constant $\lambda$ is sufficiently small.

For $3< D < 4$ the first derivative of $V$ also vanishes
at the limit $\sigma \rightarrow 0$ but
the sign of the second derivative of $V$ depends on the 
coupling constant $\lambda$.
Evaluating the effective potential (\ref{v:flat:topo})
numerically it is observed that the chiral symmetry is 
broken down for a sufficiently small length $L$ in the case 
$\lambda < \lambda_{\mbox{cr}}$, while only the broken phase is
realized for $\lambda > \lambda_{\mbox{cr}}$.

Analyzing the gap equation (\ref{gap:flat:topo})
the critical value of $L$ is given by
\begin{equation}
     L_{\mbox{cr}}m'_{0}=2\pi\left[
     -\frac{2\Gamma((3-D)/2)}{\sqrt{\pi}\Gamma (1-D/2)}
     \zeta(3-D)\right]^{1/(D-2)}\, ,
\label{crl:flat2}
\end{equation}
where $m_{0}'$ is defined in Eq. (\ref{m0p}).
In Fig. 24 we present the critical size $L_{\mbox{cr}}$ 
as a function of
dimension $D$ by the use of Eq. (\ref{crl:flat2}).
In drawing Fig. 24 we fix the coupling constant 
$\lambda \leq \lambda_{\mbox{cr}}$.
Since the finite size effect gives a finite correction
to the effective potential even at the four-dimensional limit,
the finite size effects disappear at $D\rightarrow 4$
as is shown in Figs. 23 and 24.

Thus the boundary condition of the fermion field has large
contributions to the phase structure.
If the fermion field possesses the antiperiodic boundary condition
the finite size effect raise the vacuum energy 
$V(\langle\sigma\rangle)$. 
Hence the broken chiral symmetry is restored for a sufficiently small $L$.
The fermion field which possesses the periodic boundary condition
has an opposite finite size effect.
The vacuum energy $V(\langle\sigma\rangle)$ 
reduces by the finite size effect.
Thus the chiral symmetry is broken down for a small $L$.

We can interpret above results in the following way.
Under the antiperiodic boundary condition minimum momentum
allowed for the fermion field becomes $p=\pi/L$.
Then the infrared cut-off induced in a finite $L$
spacetime.
Since the lower momentum modes have an essential
role to break the chiral symmetry, the symmetry
restoration occurs for a sufficiently small $L$.
On the other hand the vanishing momentum is allowed under
the periodic boundary condition.
In $R^{D-1}\otimes S^{1}$ a fermion field $\psi(x)$ 
is able to interact with
$\psi(x+nL)$ where $n$ is an arbitrary integer.
Summing up all the correlations, 
$\langle\bar{\psi}(x)\psi(x)\rangle$,
$\langle\bar{\psi}(x)\psi(x+L)\rangle$,
$\langle\bar{\psi}(x)\psi(x+2L)\rangle$,
$\cdots$,
the vacuum expectation value of the composite operator
increases as $L$ decreases.
It is understood as a dimensional reduction.
Compactifying one direction to the size $L$ in $D$-dimensional
space, it looks $(D-1)$-dimensional space for particles
with Compton wavelength much larger than the size $L$.
In the lower dimensional space the influence from 
the lower momentum fermion exceeds. Then  
the finite size effect 
breaks the chiral symmetry under the 
periodic boundary condition.

\subsection{Curved spacetime with nontrivial topology 
(${\cal M}^{3}\otimes S^{1}$)}

Now we are interested in the combined effect of the finite
size and curvature.
The fermion field with an antiperiodic boundary condition
has been already discussed on compact spaces
$S^{D}$ and $R\otimes S^{D-1}$ in \S 4.
Thus we consider the universe where both the independent
spin structures exist below.

For this purpose we evaluate the effective potential 
(\ref{calcul}) in weakly curved spacetime.
Since we are interested in the theory in four dimensions,
we restrict ourselves in four spacetime dimensions
and calculate the effective potential by using the cut-off
regularization.
Notice also that a summation over the two inequivalent spin
structures which are admitted by the spacetime ${\cal M}^{3}
\times S^{1}$ will be performed. Of course, one can
consider also the corrections corresponding to periodic and 
antiperiodic (non-zero temperature) boundary conditions
independently. 

Integration over $s$ in Eq. (\ref{calcul}) is immediate. 
To perform the momentum
integration, one first makes the Wick rotation ($k^0=ik^4$)
and puts then a cut-off
to regularize the resulting expressions. In our case, we
simply restrict
\[\left( k^4\right) ^2+\left( k^1\right) ^2+\left
( k^2\right) ^2 \leq
\Lambda^2  \ ,\]
so that our cut-off is different, when compared with the
cut-off for the case of trivial topology discussed
in \S 3.2.

From now on, we shall call $V_1$ the contribution to the
effective potential which comes from the logarithm of the
determinant
of the operator which appears in Eq.
(\ref{calcul}).
After carrying out the integrations over $s$ and the
momenta, we are
led to the following expression for the contribution to
$V_1$ coming
from the $p=0$ case, namely, purely periodic boundary
conditions
(this corresponds to the contribution to $V(\sigma)$
obtained by
taking only the $p=0$ term in Eq. (\ref{calcul})). Of
course, we
want to study $V_{1}$, which is given by the sum over the
two values of
$p$, but ---as we discuss below--- $V_{1}$ may be written
down
immediately once the $p=0$ contribution has been worked out.
We obtain
\begin{eqnarray}
V_{1}^{p=0} & = & - \frac{1}{L} \displaystyle \sum_{n=-
\infty }^{n = \infty }
\left\{ \frac{1}{3 \pi ^2} \,  \Lambda ^3 \log \left(
1+\frac{\sigma ^2}{\Lambda^2 + 4 \pi ^2 n^2 /L^2}
 \right) \right.   \nonumber \\
&& +  \frac{2}{3\pi^2} \left[ \sigma^2 \Lambda - \left
( \frac{4 \pi^2
n^2}{L^2} + \sigma^2 \right) ^{3/2}
\arctan \left( \frac{\Lambda}{\sqrt{4 \pi^2 n^2 /L^2
+
\sigma^2 }} \right) \right.
\nonumber \\
&&\left.+ \left( \frac{2 \pi n}{L} \right) ^3
\arctan
\left( \frac{\Lambda L}{2 \pi n} \right) \right] \nonumber
\\
&& + \frac{R}{3 \left( 2 \pi \right) ^2 } \left[ \frac{2 \pi
n}{L}
\arctan \left( \frac{\Lambda L}{2 \pi n} \right) - \sqrt
{ \frac{4 \pi^2
n^2}{L^2} +\sigma^2} \ \arctan \left( \frac{\Lambda }{\sqrt
{4
\pi^2 n^2 /L^2 + \sigma^2 }} \right) \right]  \nonumber \\
&& - \left. \frac{R}{2 \left( 3 \pi \right) ^2 } \left
[ \frac{ \left(
2 \pi n/L \right) ^2 \Lambda }{\Lambda^2 + 4
\pi^2
n^2 /L^2 } - \frac{ \left( 4 \pi^2 n^2 /L^2 +
\sigma^2 \right)
\Lambda }{4 \pi^2 n^2 /L^2 + \sigma^2 + \Lambda^2 }
\right] \right\} \ .                       \label{3h}
\end{eqnarray}

    To simplify this expression we use standard techniques
drawn from
complex analysis, such as the expression
\begin{eqnarray}
\displaystyle \sum_{n=-\infty }^{\infty} f \left( \frac{2
\pi n }{L} i \right)
&=&
\frac{L}{2 \pi i} \int_{-i \infty }^{i \infty } dp\,
\frac{1}{2} \left[
f(p)+f(-p) \right] \nonumber \\
&&+ \frac{L}{2 \pi i}
\int_{-i \infty + \epsilon }^{i \infty
+ \epsilon } dp \, \frac{f(p)+f(-p)}{\exp{(Lp)} -1}.
\label{trk}
\end{eqnarray}
One has just to identify the function $f$ for each term in
Eq.
(\ref{3h}) and then perform the integrals in (\ref{trk}).
Another remark is in order here: when computing the term
\begin{eqnarray}
&&- \frac{2}{L} \frac{1}{3 \pi^2}  \sum_{n=\infty }^{\infty }
\left[ \sigma^2 \Lambda - \left( \frac{4 \pi^2 n^2}{L^2}
+ \sigma^2 \right) ^{3/2}
\arctan \left( \frac{\Lambda}{\sqrt{4 \pi^2 n^2 /L^2
+
\sigma^2 }} \right) \right.\nonumber \\
&&\left.+ \left( \frac{2 \pi n}{L} \right) ^3
\arctan
\left( \frac{\Lambda L}{2 \pi n} \right) \right] ,\nonumber
\end{eqnarray}
it is better to rewrite it as
\begin{eqnarray}
- \lim_{\sigma ' \rightarrow 0 } \frac{2}{L} \frac{1}{3
\pi^2}
\sum_{n=\infty }^{\infty }  &&
\left[  \sigma^2 \Lambda - \left( \frac{4 \pi^2 n^2}{L^2}
+ \sigma^2 \right) ^{3/2} \arctan \left( \frac{\Lambda
}{\sqrt{4
\pi^2 n^2 /L^2 + \sigma^2}} \right) \right.  \nonumber \\
&& - \ \ \ \left. \sigma '^2 \Lambda + \left( \frac{4 \pi^2
n^2}{L^2}
+ \sigma '^2 \right) ^{3/2} \arctan \left( \frac{\Lambda
}{\sqrt{4
\pi^2 n^2 /L^2 + \sigma '^2}} \right)  \right] ,  \nonumber
\end{eqnarray}
since now the expression within square brackets satisfies
the properties
which justify the use of Eq. (\ref{trk}).

    As we said before, once $V_{1}^{p=0}$ has been computed,
one may
write immediately
$V_{1}^{p=1}$ and $V_1$, which is given (by definition) by
$V_1 =
V_{1}^{p=0} + V_{1}^{p=1}$, because
\begin{equation}
\frac{1}{L} \sum_{n=-\infty }^{\infty } F \left
( \frac{2n+1}{L}
\right)  =
\frac{1}{L} \sum_{n=-\infty }^{\infty } F \left( \frac{n}{L}
 \right)
-\frac{1}{L} \sum_{n=-\infty }^{\infty } F \left
( \frac{2n}{L} \right)\, ,
\end{equation}
and, then
\begin{equation}
\frac{1}{L} \sum_{n=-\infty }^{\infty } \left[ F \left
( \frac{2n+1}{L}
\right) + F \left( \frac{2n}{L} \right) \right] =
\frac{1}{L}
\sum_{n=-\infty}^{\infty } F \left( \frac{n}{L} \right).
\end{equation}
    Now it is apparent that the physics displayed by the
model in the
case of purely periodic boundary conditions and in the one
where we
consider both spin structures will be essentially the same,
since
\[V_{1,L} = V_{1,L}^{p=0} + V_{1,L}^{p=1} = 2 V_{1,2L}^{p=0}
.\]
Thus we see that both cases are related by a trivial
rescaling of the
length and an overall factor which multiples $V_1$. 

However, the last remark does not apply to the case of
purely
antiperiodic
boundary conditions, where we only take $p=1$ ---which gives
the
thermodynamics of a system in three-dimensional space with
trivial
topology. In fact, from the last expression we get
\[ V_{1,L}^{p=1}=V_{1,L}-V_{1,L}^{p=0}=V_{1,L}-\frac{1}{2}
V_{1,L/2 } .\]
Henceforth, we shall concentrate below only on the analysis
of the case
in which
both spin structures are taken into account, as they appear
in Eq. (\ref{calcul}).

    The first term on the right-hand side
of Eq. (\ref{trk}) may be
computed without difficulty in all cases. The second term
is, in
general, rather
more involved. In order to simplify this contribution as
much as
possible in the cases which come from Eq. (\ref{3h}),
one should
pay careful attention to the determination of the integrand
along the
contour of integration. After some work, the final result is
found to be
\begin{eqnarray}
\frac{V_1}{\Lambda^4}&=&-\frac{2}{3 \pi^2} \left
[ 1-\sqrt{1+x^2}+
\frac{1}{l} \log \left(
\frac {\exp{\left( 2l\sqrt{1+x^2}\right) }-1}{\exp{\left
( 2l\right) }
-1} \right) \right] \nonumber \\
&& + \frac{1}{6 \pi^2} \left[ \sqrt{1+x^2} - 1 - \frac{3}{2}
x^2
\sqrt{1+x^2} + \frac{3}{2} x^4 {\rm arcsinh} \left
( \frac{1}{x}
\right) \right] \nonumber \\
&& + \frac{4}{3 \pi^2} \left[ \displaystyle
\int_{x}^{\sqrt{1+x^2}} d \tau
\frac{ \left( \tau^2 - x^2 \right) ^{3/2}}{\exp {\left( 2 l
\tau
\right) } - 1} - \int_0^1 d \tau \frac{\tau^3 }{\exp{ \left
( 2 l
\tau \right) } - 1} \right]  \nonumber \\
&&+\frac{r}{6 \left( 2 \pi \right) ^2}\left[ 1-\sqrt{1+x^2}
+
x^2 {\rm arcsinh} \left( \frac{1}{x} \right) \right]   \\
&& + \frac{2 r}{ 3 \left( 2 \pi \right) ^2} \left[ \int_0^1
d
\tau \frac{ \tau }{\exp{\left( 2 l \tau \right) } - 1} -
\int_x^{\sqrt{1+x^2}} d\tau \frac{\sqrt{\tau^2 -
x^2}}{\exp{\left(
2 l \tau \right) }-1} \right] \nonumber \\
&& - \frac{r}{2 \left( 3 \pi \right) ^2} \left( 1-
\frac{1}{\sqrt{1+x^2}
} \right) \nonumber \\
&&+ \frac{r}{ \left( 3 \pi \right) ^2} \left
[ \frac{1}{
\sqrt{1+x^2}} \frac{1}{\exp{ \left( 2 l \sqrt{1+x^2} \right)
}-1} -
\frac{1}{\exp{\left( 2 l \right) }-1} \right]   , \nonumber
\label{exact}
\end{eqnarray}
where $x=\sigma /\Lambda $, $l=L \Lambda $ and
$r=R/\Lambda^2 $.

    The value of the field $\sigma $ which satisfies the gap
equation,
$ V'(\sigma)=\partial V(\sigma)/\partial \sigma  = 0 $,
gives a dynamical mass to the fermions. This last equation,
when written
in
terms of the natural variables $x$, $l$, $r$ and $c$ ($c
\equiv \lambda
\Lambda ^2 $), reads
\begin{eqnarray}
0= \frac{V'(x)}{\Lambda ^4}&=& \frac{x^2}{2c} + \frac{5}{6
\pi^2}
\frac{x}{\sqrt{1+x^2}} - \frac{4}{3 \pi^2 }
\frac{x}{\sqrt{1+x^2}}
\frac{1}{1-\exp\left( -2l \sqrt{1+x^2}\right) } \nonumber \\
&&+\frac{x}{\pi^2} \left[ x^2 {\rm arcsinh} \left
( \frac{1}{x}
\right) -\sqrt{1+x^2} \right] + \frac{1}{2 \pi^2}
\frac{x}{\sqrt{1+x^2}}
\nonumber \\
&& +\frac{4}{3 \pi^2} \left[ \frac{x}{\sqrt{1+x^2}}
\frac{1}{\exp\left(2l
\sqrt{1+x^2}\right) -1} - 3x \displaystyle
\int_{x}^{\sqrt{1+x^2}} d \tau
\frac{
\left( \tau ^2 - x^2 \right) ^{1/2}}{\exp(2 l \tau ) -1}
\right]  \nonumber \\
&&+\frac{rx}{3 \left( 2 \pi \right) ^2} \left[ {\rm arcsinh}
\left(
\frac{1}{x} \right) - \frac{1}{\sqrt{1+x^2}} \right] -
\frac{r}{4 \left(
3 \pi \right) ^2 } \frac{x}{\left( 1+x^2 \right) ^{3/2}}
\nonumber  \\
&&-\frac{2rx}{3\left( 2 \pi \right) ^2} \left
[ \frac{1}{\sqrt{1+x^2}}
\frac{1}{\exp\left( 2l\sqrt{1+x^2}\right)
-1}\right.\nonumber \\
&&\left.
-\int_{x}^{\sqrt{1+x^2}} d
\tau \frac{1}{\sqrt{\tau^2 - x^2}}
\frac{1}{\exp(2l\tau )-1} \right] 
\label{6:7} \\
&&-\frac{r}{2\left( 3 \pi \right)^2} \left[ \frac{x}{\left
( 1+x^2
\right) ^{3/2}} \frac{1}{\exp\left( 2l
\sqrt{1+x^2}\right) -1}\right.\nonumber \\
&&\left.-\frac{2l}{1+x^2}
\frac{\exp\left( 2l \sqrt{1+x^2}\right) }{\left( \exp\left(
2l\sqrt{1+x^2}\right) -1 \right)^2} \right]. \nonumber
\end{eqnarray}

Let us discuss some limiting case of above expression (\ref{6:7}).
Assuming $L\Lambda \ll 1$, let us expand $V/\Lambda^4$ in
powers of $l$. Assuming
now that $l \sqrt{1+x^2} \ll 1$, we readily obtain
\begin{eqnarray}
\frac{l V(x)}{\Lambda ^4} & = & \frac{x^2}{2g}-\frac{1}{3
\pi^2}
\log{\left( 1+x^2\right) }+\frac{2}{3 \pi^2} \left( -x^2+x^3
{\rm arcsec}  \sqrt{1+\frac{1}{x^2} } \right) \nonumber \\
&&+\frac{r}{3 \left( 2 \pi \right) ^2}\,x\,{\rm arcsec}
\sqrt{1+\frac{1}{x^2} } \ + \frac{r}{
2 \left( 3 \pi \right) ^2} \left( \frac{1}{1+x^2}-1 \right)
\nonumber \\
&&+O\left( l^2 \right) ,
\label{6:8}
\end{eqnarray}
where $g=\lambda_{0} \Lambda^2/l$.
Using (\ref{6:8}) one can study the phase structure of the model.

\subsubsection{Phase structure in ${\cal M}^{3}\otimes S^{1}$}

Because we are making expansion on the curvature, i.e.,
$r \ll 1$, it is worthwhile studying first the case $r=0$.
Here it
can be proven that it is impossible to have a first order
phase
transition, in fact there is a second order phase transition
at
$g_{\mbox{cr}}=\pi^2/2 $. For $g>g_{\mbox{cr}}$ there is chiral 
symmetry breaking
and the symmetry is restored for $g<g_{\mbox{cr}}$; 
in other words, for a
given value of $\lambda$, the symmetry is restored when $l$
grows
beyond a critical point. In the broken phase the  order
parameter
is given by 
$ x_{\mbox{br}}=\pi \left(2/\pi^2 - 1/g\right) $
or, in terms of the generated mass 
$m_{\mbox{gen}} =2\Lambda /\pi -L \pi /\lambda_{0}$. It
is straightforward to study the behavior of the value of the
effective
potential at $x_{\mbox{br}}$, for varying $g$ or $l$, in this limit
of small
compactification length. One finds \[ V(\sigma
_{\mbox{br}}) = - \Lambda^4 \frac{\pi^2}{6l}
 \left( \frac{2}{\pi^2}-\frac{1}{g} \right)^3 \ ,\]
and for $L\rightarrow 0$ we find
\[ \lim _{L \rightarrow 0} \ L V(\sigma _{\mbox{br}}) =-\Lambda ^3
\frac{4}{3 \pi^4}\, . \]

    Now we shall study the influence that the presence of
curvature has on this behavior. It is immediate to notice
that for negative values of the curvature, the symmetry is
always broken
(the slope of the effective potential is $r/24 \pi$ at the
origin.)
Moreover, for fixed positive values of $r$ (which is kept
always small),
as $l$ grows there is a first-order phase transition: it has
actually
lost
its continuous character. Now the critical value of the
parameter $g$
is \[ g_{\mbox{cr}}
= \frac{\pi ^2}{2} \left( 1 + \frac{2 \pi }{3}
\sqrt{\frac{r}{8}} \right) \ . \]
The approximation is consistent with the small-curvature
limit. The difference between the values
of the order parameter $x$ in the broken and disordered
phases
at the phase transition is given by $x=
r/8 $ (retaining again only the first correction coming from
the
curvature).

    We can now, on the other hand, study the situation when
$g$ is
held fixed and the curvature takes on different values.

   If $g<\pi^2/2 $
there is a continuous phase transition at $r=0$ (the
symmetry is
broken for negative values of $r$ and is restored for
positive
curvature). As $r$ approaches $0$ from below, the order
parameter tends to zero according to the expression \[ x=
\frac{\mid
r \mid }{24 \pi \Delta } \ , \]
or $m_{\mbox{gen}} = \Lambda^{-1} |R|/24 \pi \Delta$, being
$\Delta \equiv 1/g-2/\pi^2$.
In deriving this result we assume that $\mid r \mid \ll 1$
and that
$\mid r \mid \ll \Delta $.

One may also consider the case $\Delta < 0$ or,
equivalently, $g>\pi^2/2$. To be consistent with the
requirement of small curvature near the critical point, we
have to
assume now that $\mid \Delta \mid \ll 1$. In this setting
one
may expect that, as $r$ grows, there will be a phase
transition from the
ordered to the disordered phase  at some positive
value of the curvature. Retaining again the first terms of
the
expansions only, we obtain \[ r_{\mbox{cr}}=\frac{\left( 3 \pi
\right) ^2}
{2} \Delta ^2. \]

    We shall here consider again the phase structure in the
fixed-curvature case but corresponding now to the opposite
limit, when
$L$ is large.

It is easy to see from Eq. (\ref{exact}) that, dropping
exponentially vanishing contributions, one may approximate
it with
\begin{equation}
\frac{V(x)}{\Lambda^4} = \frac{x^2}{2 c} -\frac{1}{4 \pi^2}
\left[
2\left( \sqrt{1+x^2} -1\right) + x^2 \sqrt{1+x^2} - x^4 {\rm
arcsinh}  \frac{1}{x}  \right] ,    \label{largel}
\end{equation}
where $c$ is, as before, $c=\lambda_{0} \Lambda ^2$.

From the analysis of gap equation it is trivial to see that
the symmetry is broken when
\[ \lambda_{0} > \frac{\pi ^2}{\Lambda ^2} \ . \]
Otherwise, the symmetry is respected by the vacuum.

If one takes into account the presence of a background
gravitational
field, one can check that the effective potential
at large $L$ is becoming

\begin{eqnarray}
\frac{V(x)}{\Lambda ^4}&=&\frac{x^2}{2 c} -\frac{1}{4 \pi^2}
\left[
2\left( \sqrt{1+x^2} -1\right)  + x^2 \sqrt{1+x^2} - x^4
{\rm
arcsinh} \frac{1}{x} \right]  \nonumber  \\
 && + \frac{r}{6\left( 2 \pi^2 \right) ^2}
\left( 1-\sqrt{1+x^2} + x^2 {\rm arcsinh}  \frac{1}{x}
 \right) \nonumber \\
&&-  \frac{r}{2\left( 3 \pi \right) ^2}
 \left( 1 - \frac{1}{\sqrt{1+x^2}}  \right)    \ .
\end{eqnarray}
To analyze phase structure let us consider 
$\lambda < \pi^{2}/\lambda_{0}^{2}$.
Then the order parameter maybe found approximately as
\[ x= \left[ \sinh \left( \frac{12}{\mid r \mid }
(\pi^2/c -1 )- \frac{5}{3} \right) \right] ^{-1}\, , \]
(remember that this is valid for negative $r$). The symmetry
is restored
for positive values of $r$.
The numerical analysis for this and some other cases
maybe found in Ref. \cite{TOPO}.

Thus we discussed phase structure of four-fermion model on
${\cal M}^{D-1}\otimes S^{1}$.
In compact flat spacetime $R^{D-1}\otimes S^{1}$
it was found that the boundary condition for the
fermion field drastically changes the phase structure.
In the case of an antiperiodic boundary condition
finite size effect restores the broken chiral symmetry,
while the finite size effect breaks the chiral
symmetry for the fermion field with a periodic boundary 
condition.
In ${\cal M}^{3}\otimes S^{1}$ we found the possibility
of (topology combined with curvature)-induced
phase transitions.

\section{Chiral symmetry breaking in curved spacetime with
magnetic field}

The external electromagnetic (EM) fields play an important
role as a possible probe of quantum field theory
(for a review of quantum field theory in an external
EM fields, see Ref. \cite{b7}). It is quite well-known fact that
an external magnetic field supports chiral symmetry breaking
in four-fermion models. (see Refs. \cite{b4} $\sim$ \cite{b6} 
and references there in).
There was some activity on the study of phase structure
of four-fermion models in an external magnetic field
(for a review, see for example Ref. \cite{b11}).

From another point it maybe possible that strong primordial
magnetic fields should be considered on equal footing with 
the strong curvature in the early universe.
Then, different consequences of the combined effect of the
magnetic and gravitational fields may occur
(for a example, it maybe significant increase in the number
of created particles in the early universe filled with
the constant EM field \cite{b12}).
Hence, for cosmological applications it could be interesting
to discuss the chiral symmetry breaking phenomenon in
four-fermion models under the action of external
gravitational and magnetic fields.
This section will be devoted to such situation where
gravitational field will be taken in linear curvature
approach (see \S 3). We follow Refs. \cite{b13} and \cite{b14}.

\subsection{Three-dimensional four-fermion model with magnetic field}

We start from three-dimensional four-fermion model.
The effective potential maybe easily written as
\begin{equation}\label{2.m}
V(\sigma)=\frac{\sigma^2}{2\lambda_{0}}+i\ln\Det
\left[i\gamma^{\mu}(x)\nabla_{\mu}-
\sigma\right]\; ,
\end{equation}
where at first step there is no magnetic field yet.
It is convenient to work in terms of the derivative from
$V(\sigma)$. Then, using the local momentum representation
of the propagator (see \S 3), we get
\begin{eqnarray}\label{3}
\frac{\partial V(\sigma)}{\partial\sigma}
&=&\frac{\sigma}{\lambda}-i\tr\11\int \frac{d^3
k}{(2\pi)^3}\left[\frac{\sigma}{k^2-\sigma^2}-
\frac{R}{12}\frac{\sigma}{(k^2-\sigma^2)^2}
\right. \nonumber \\
&& \left.
+\frac{2}{3}R_{\mu\nu}k^{\mu}k^{\nu}\frac{\sigma}
{(k^2-\sigma^2)^3}\right]\; .
\end{eqnarray}

Making use of Wick rotation and calculating the trace in
(\ref{3}), we
obtain
\begin{eqnarray}\label{4}
\frac{\partial V(\sigma)}{\partial\sigma}
&=&\frac{\sigma}{\lambda_{0}}+\frac{2\sigma}{\pi^2}\int_{0}^{\infty}
k^2 dk
\left[-\frac{1}{k^2+\sigma^2}-\frac{R}{12}\frac{1}{(k^2+\sigma^2)^2}
+\frac{2}{9}R\frac{k^2}{(k^2+\sigma^2)^3}\right] \nonumber
\\
&=&\frac{\sigma}{\lambda_{0}}+\frac{\sigma}{2\pi^{3/2}}\int_{1/{\Lambda^2}}^{\infty}
ds \exp
(-s\sigma^2)\left(-s^{-3/2}+\frac{R}{12}s^{-1/2}\right)\;.
\end{eqnarray}
In derivation of Eq. (\ref{4}) we have used the expression
for
proper-time representation
\begin{equation}\label{5}
A^{-\nu}=\frac{1}{\Gamma (\nu) }\int_{0}^{\infty}
ds\; s^{\nu-1} e^{-sA}\;,
\end{equation}
and after that, the ultraviolet proper-time cut-off
$\Lambda^2$ has
been introduced in the low limit of proper-time integral.
This cut-off is different from the cut-off used in \S 3.

Performing the integration over $s$ and over $\sigma$, we
get (compare with Ref. \cite{EOS} or \S 3 where 
other regularization has been used):
\begin{eqnarray}\label{6}
V(\sigma)&=&\frac{\sigma^2}{2\lambda_{0}}+\frac{1}{6\pi^{3/2}}
\left\{\Lambda^3\exp\left(-\frac{\sigma^2}{\Lambda^2}\right)-
2\sigma^2\Lambda\exp\left(-\frac{\sigma^2}{\Lambda^2}\right)
+2\sqrt{\pi}\sigma^3{\rm erfc}\left(\frac{\sigma}{\Lambda}
\right)\right.\nonumber \\
&-&\left.\frac{R}{4}\left[\Lambda
\exp\left(-\frac{\sigma^2}{\Lambda^2}\right)-
\sqrt{\pi}\sigma {\rm erfc}\left(\frac{\sigma}{\Lambda}
\right)\right]\right\}\;,
\end{eqnarray}
where $\displaystyle {\rm erfc}(x)= (2/\sqrt{\pi}) \int^\infty_x
e^{-t^2} dt$. Hence,
we obtained the effective potential in three-dimensional 
NJL model in curved
spacetime using proper-time cut-off.

The renormalized effective potential may be found in the
limit $\Lambda\rightarrow \infty$ as following:\footnote{
Three-dimensional four-fermion model is renormalizable in the sense
of $1/N$ expansion.\cite{3DFF}
Thus the theory is defined independent of the regularization
method. If we perform the finite renormalization
\[
     \frac{1}{\lambda}\rightarrow\frac{1}{\lambda}-\frac{2}{\pi}\, ,
\]
and put $\tr\11 = 4$, the effective potential 
(\ref{v:3d:w}) is reproduced.
}
\begin{equation}\label{7}
V(\sigma)=\frac{\sigma^2}{2\lambda}+\frac{|\sigma|^3}{3\pi}+\frac{R|\sigma|}
{24\pi}\;,
\end{equation}
where
\begin{equation}\label{8}
\frac{1}{\lambda}=\frac{1}{\lambda_{0}}-\frac{\Lambda}{\pi^{3/2}}\;.
\end{equation}

Let us analyze now the phase structure of the potential
(\ref{7}). In the
absence of curvature, $R=0$,   for $\lambda > 0$ the
minimum of the
potential (\ref{7}) is given by $\sigma=0$. Hence, there is
no chiral symmetry
breaking. For $\lambda < 0$ the chiral symmetry is broken,
the
dynamically generated fermion mass is given as

\begin{equation}\label{9}
m_{0} \equiv \sigma_{\rm min} = - \frac{\pi}{\lambda}\;.
\end{equation}
When the curvature is not zero we find the following
picture. Let first
$\lambda > 0$. Then the ground state (and dynamically
generated mass)
is defined by
\begin{equation}\label{10}
m \equiv \sigma_{\rm min} = - \frac{\pi}{2 \lambda} +
\frac{1}{2}
\sqrt{\frac{\pi^2}{\lambda^2} - \frac{R}{6}}\;.
\end{equation}
One can see that unlike the case of flat space for positive
$\lambda$ we have the chiral symmetry breaking. 
We also see the possibility of curvature-induced  phase
transitions. The
critical curvature is given by $R_{\mbox{cr}} = 0$ 
(flat space). For negative
curvature, $R< R_{\mbox{cr}} = 0$, we observe the chiral symmetry
breaking, while
for positive curvatures symmetry is not broken.

For negative four-fermion coupling constant $\lambda < 0$ we
get the
following ground state:
\begin{equation}\label{11}
\sigma_{\rm min} = - \frac{\pi}{2 \lambda} - \frac{1}{2}
\sqrt{\frac{\pi^2}{\lambda^2} - \frac{R}{6}}\;.
\end{equation}
The critical curvature is defined by the condition: 
$R_{\mbox{cr}} =9 \pi^2 /2\lambda^2=(9/2)m_{0}^{2}$. 
Between $0<R \leq R_{\mbox{cr}}$ the chiral
symmetry is broken.

Now, let us discuss the situation  when four-fermion model
is considered
in curved spacetime with external magnetic field. That means
that
spinor covariant derivative also contains the
electromagnetic
piece. Treating the magnetic field exactly,\cite{b9} one
can easily find
the effective potential for our model. Considering
now four-fermion model
in curved spacetime with magnetic field, we again work in
linear
curvature approximation as above (but making no
approximations for an
external magnetic field). Moreover, we take into account
only leading
contribution on curvature which does not depend on magnetic
field,
i.e., the contribution discussed above. (One can show that
for not a very
large magnetic field the curvature correction depending
explicitly from
magnetic field is not essential). Then, using Eq. 
(\ref{7}) and the results of the calculation of  
three-dimensional four-fermion effective potential
in an
external magnetic  field,\cite{b4,b6} one can get:
\begin{equation}\label{12}
V=\frac{\sigma^2}{2\lambda}+\frac{|\sigma|^3}{3\pi}+\frac{R|\sigma|}{24\pi
}
+\frac{eH}{4\pi^{3/2}}\int_{0}^{\infty}\frac{ds}{s^{3/2}}e^{-s\sigma^2}
\times \left[\coth(eHs)-\frac{1}{eHs}\right]\;,
\end{equation}
where $H$ is magnetic field and $e$ is electric charge.

The renormalized effective potential (\ref{12}) may be
represented as
following
\begin{equation}\label{13}
V=\frac{\sigma^2}{2\lambda}+\frac{R|\sigma|}{24\pi}+\frac{eH|\sigma|}{2\pi}
-\frac{(2eH)^{3/2}}{2\pi}\zeta\left(-\frac{1}{2},\frac{\sigma^2}{2eH}\right)\;,
\end{equation}
where properties of generalized zeta-function may be found
in Ref. \cite{b10}.

Working in frames of our approximation and considering also
$eH \rightarrow 0$, we find that for positive $\lambda$
\begin{equation}\label{14}
\sigma_{{\rm min}}\simeq\frac{\lambda}{\pi}\left(
\frac{eH}{2}-\frac{R}{24}\right)  \;.
\end{equation}
Hence, for   $R = 0$ chiral symmetry breaking due to
magnetic field
occurs. Negative curvature increases chiral symmetry
breaking. However, positive curvature acts against 
the chiral symmetry breaking. On
the
critical line of phase diagram
\begin{equation}\label{15}
\frac{R}{12} \simeq eH
\end{equation}
the restoration of chiral symmetry breaking occurs. Hence,
in this
example magnetic field and gravitational field act
in the opposite directions with respect to chiral symmetry
breaking.
Gravity tends to restore the chiral symmetry while magnetic
field
tends to break it. On the critical line both effects are
compensated and there is no chiral symmetry breaking.

Similarly, one can analyze other choices for $\lambda$,
$eH$ (see Ref. \cite{b13} for more details).

\subsection{Four-dimensional NJL model in curved spacetime
with magnetic field}

Let us start again from the action for the NJL model
in curved spacetime:
\par$$
S= \int d^4 x \sqrt{-g} \left\{\bar{\psi}i \gamma^{\mu}(x)
\nabla_{\mu} \psi + \frac{\lambda_{0}}{2N} \left[(\bar{\psi}\psi)^2 +
(\bar{\psi}i\gamma_5 \psi)^2 \right]\right\}\, .
\eqno{(8\cdot 15)}$$
Standard calculation gives the effective potential as 
(where as usually it is enough to put pseudo-scalar
$\pi =0$ as is shown in \S 2.)
\par$$
V(\sigma)=\frac{\sigma^2}{2\lambda_{0}}+i\ln\Det\lbrack
i\gamma^{\mu}(x)
\nabla_{\mu}-\sigma\rbrack.
\eqno{(8\cdot 16)}$$
Then calculation similar to the one in previous subsection
gives the effective potential in proper-time cut-off
\par $$
V(\sigma)=\frac{\sigma^2}{2\lambda_{0}}+\frac{1}{8\pi^2}
\int_{1/\Lambda^2}^{\infty}ds
\exp(-s\sigma^2)\left[\frac{1}{s^3}
-\frac{R}{12s^2}\right]
$$
$$
=\frac{\sigma^2}{2\lambda}
+\frac{1}{(4\pi)^2}\left\{(\Lambda^4-\Lambda^2\sigma^2)
\exp\left(-\frac{\sigma^2}{\Lambda^2}\right)\right.
$$
$$
\mbox{\hspace*{6em}}
\left.\quad \quad -\sigma^4 \mbox{Ei}
\left(-\frac{\sigma^2}{\Lambda^2}\right)
-\frac{R}{6}\left[\Lambda^2
\exp\left(-\frac{\sigma^2}{\Lambda^2}\right)
+\sigma^2 \mbox{Ei}
\left(-\frac{\sigma^2}{\Lambda^2}\right)\right]\right\},
\eqno{(8\cdot 17)}$$
where $\mbox{Ei}(-x)$ is defined in Eq.(5$\cdot$54).

Expanding Eq. (8$\cdot$17) and keeping only terms which are not zero
at
$\Lambda\to\infty$ we get
\par$$
V(\sigma)=\frac{\sigma^2}{2\lambda_{0}}
-\frac{1}{(4\pi)^2}\left[2\Lambda^2\sigma^2
+\sigma ^4\left(\ln\frac{\sigma^2}{\Lambda^2}+\gamma
-\frac{3}{2}\right)
+\frac{R\sigma^2}{6}\left(\ln\frac{\sigma^2}{\Lambda^2}+\gamma-1\right)
\right]
$$
$$
+O\left(\frac{\sigma^{2}}{\Lambda^2}\right)\, .
\eqno{(8\cdot 18)}$$
Thus, we have got the effective potential with proper-time
cut-off.

Using Eq. (8$\cdot$18) the gap equation is found as follows
\par$$
\frac{4\pi^2}{\lambda_{0}\Lambda^2}-1=
\frac{\sigma^2}{\Lambda^2}\left(\ln\frac{\sigma^2}{\Lambda^2}
+\gamma-1\right)+\frac{R}{12\Lambda^2}\left(\ln\frac{\sigma^2}{\Lambda^2}
+\gamma\right).
\eqno{(8\cdot 19)}$$

This gap equation defines the possibility of chiral
symmetry breaking in curved spacetime (in linear
curvature approximation).
If our weakly curved spacetime is filled by the magnetic
field
 (the covariant derivative is now
 $\tilde{\nabla_{\mu}}=\nabla_{\mu}-ieA_{\mu},\; A_{\mu}
=-B x_2 \delta_{\mu 1}$) one can get the following
effective potential
\par$$
\mbox{\hspace*{-18em}}V(\sigma)=\frac{\sigma^2}{2\lambda}
+i\ln\Det\left[i\gamma^{\mu}(x)\tilde{\nabla_{\mu}}-\sigma\right]
\quad\quad\quad
$$
$$
=\frac{\sigma^2}{2\lambda_{0}}
+\frac{1}{8\pi^2}\int_{1/\Lambda^2}^{\infty}
{\frac{ds}{s^2}}\exp(-s\sigma^2)\left[|eB|\coth(s|eB|)
+\left(-{\frac{R}{12}}+O(R^2)\right)\right].
\eqno{(8\cdot 20)}$$
In the absence of the gravitational field ($R=0$), the
effective
 potential corresponds to flat space situation where the
magnetic
field is treated exactly.\cite{SP} In the absence of the magnetic
field
 ($B=0$) we are back to the potential (8$\cdot$17). In addition, in
the linear curvature terms the effect of the magnetic field
is not taken into account as in the previous subsection.

Making the calculation of the integrals in Eq. (8$\cdot$20) up to
$O(1/\Lambda^2)$, and taking the derivative with respect to
$\sigma$
 one gets the gap equation,
$\partial V/\partial \sigma=0$,
as follows
\par$$
\frac{4\pi^2}{\lambda_{0}\Lambda^2}-1
=-\frac{\sigma^2}{\Lambda^2}\ln\frac{(\Lambda|eB|^{-1/2})^2}{2}
+\frac{|eB|}{\Lambda^2}\ln\frac{(\sigma|eB|^{-1/2})^2}{4\pi}+
\gamma\frac{\sigma^2}{\Lambda^2}
$$
$$
\mbox{\hspace*{8em}}
+2\frac{|eB|}{\Lambda^2}\ln\Gamma\left(\frac{\sigma^2|eB|^{-1}}{2}
\right)
-\frac{R}{12\Lambda^2}\left(\ln\frac{\Lambda^2}{\sigma^2}-\gamma\right)
+O\left(\frac{1}{\Lambda}\right).
\eqno{(8\cdot 21)}
$$
Using this gap equation one can study the dynamical symmetry
breaking
in different cases. In some cases it can be given
analytically, for
example, for values of the coupling constant $\lambda$ much
below the
critical value, i.e.,
\par$$
\lambda_{0}\ll\frac{4\pi^2}{\Lambda^2} .
\eqno{(8\cdot 22)}$$
One can find (supposing that the second term on the r.h.s.
of (8$\cdot$21)
 is the leading one)
\par$$
\frac{4\pi^2}{\lambda_{0}\Lambda^2}
-1\approx\frac{|eB|}{\Lambda^2}\ln\frac{(\sigma|eB|^{-1/2})^2}{4\pi}
-\frac{R}{12\Lambda^2}\ln\frac{\Lambda^2}{\sigma^2}\, ,
\eqno{(8\cdot 23)}
$$
and finally, the dynamically generated fermionic mass is
given by
\par$$
\sigma^2\approx\left[\left(\frac{|eB|^{-1}}{4\pi}\right)^{-|eB|/\Lambda^2}
\left(\frac{1}{\Lambda ^2}\right)^{-R/(12\Lambda^2)}
\exp\left(\frac{4\pi^2}{\lambda_{0}\Lambda^2}-1\right)
\right]^{1/[|eB|/\Lambda^2+R/( 12\Lambda^2)]}\, .
\eqno{(8\cdot 24)}
$$
In the absence of the magnetic field it gives the analytic
expression for
the dynamically generated fermionic mass due to the
curvature
\par$$
\sigma^2\approx\Lambda^2
\exp\left[\frac{12\Lambda^2}{R}\left(\frac{4\pi^2}{\lambda_{0}\Lambda^2}
-1\right)\right] .
\eqno{(8\cdot 25)}$$
One can see that positive curvature tends to make the first
term
in (8$\cdot$24) less, i.e., it acts against the dynamical symmetry
breaking.
 At the same time, the negative curvature always favors the
dynamical
 chiral symmetry breaking in accordance with the explicit
calculations in as external gravitational field
(see \S 3).

The numerical analysis shows the similar qualitative
behavior (for more details, see Ref. \cite{b13}).
Note finally that in the same way one consider
more complicated situations when both
gravitational and magnetic fields are treated exactly
(see, for example, Ref. \cite{b15} where two dimensional
Gross-Neveu model on sphere with magnetic
monopole has been considered).

\section{Quantum gravity and four-fermion models}

In the present section we give few examples where four-fermion
models maybe relevant to study the dynamics of quantum gravity.
In particular we apply the effective potential of $D=2$ 
four-fermion model to investigate $D=2$ black hole solution
of dilatonic gravity. We discuss conformal factor dynamics
in frame of $1/N$ expansion.
The effective potential for composite gravitino on de Sitter
background is also investigated.

\subsection{Semiclassical approach in the dilatonic gravity
            with Gross-Neveu model}

Motivated by the idea that $2D$ quantum gravity is easier
to study than $4D$ quantum gravity there was recentely much 
of interest in $2D$ quantum gravity and its black holes 
solutions.\cite{W,BH,CG,M,MDS}
We consider now one of such solutions in the $2D$ dilatonic
gravity with Gross-Neveu model following Ref. \cite{MDS}.

The action to start with is
\begin{equation}
     S=\frac{1}{2\pi}\int\! \sqrt{-g} d^{2}x \left[
     e^{-2\phi}(R+\Lambda)+\sum^{N}_{k=1}
     \bar{\psi}_{k}(i\gamma^{\mu}\nabla_{\mu}+\sigma)\psi_{k}
     -\frac{N}{2\lambda_0}\sigma^{2}\right]\, ,
\label{act:dilaton}
\end{equation}
where $\phi$ is the dilaton field and Gross-Neveu action is 
the same as Eq. (2$\cdot$12).

Notice that if we redefine the metric by the relation
\begin{equation}
     g_{\mu\nu}=e^{-2\phi}\hat{g}_{\mu\nu},
\label{def:dilaton}
\end{equation}
the gravitational part of the action (\ref{act:dilaton})
takes the form
\begin{equation}
     S_{1}=\frac{1}{2\pi}\int\! \sqrt{-\hat{g}} d^{2}x 
     e^{-2\phi}\left(\hat{R}+4\hat{g}^{\mu\nu}
     \partial_{\mu}\phi\partial_{\nu}\phi
     +\Lambda e^{-2\phi}\right),
\label{act:dilaton:grav}
\end{equation}
which is similar to the CGHS action \cite{CG} (the cosmological 
constant term is a fixed number in Ref. \cite{CG} while it
is accompanied by the Liouville type potential in Eq.
(\ref{act:dilaton:grav}).)
However, we should emphasize that the theory with action
(\ref{act:dilaton}) is different from the theory with
action (\ref{act:dilaton:grav}) coupled to the Gross-Neveu
fermions (with metric tensor $\hat{g}_{\mu\nu}$)
already at the classical level. The difference becomes
more serious in the semiclassical approach.

Let us start our discussion of the theory described by
Eq. (\ref{act:dilaton}) in the semiclassical
approach ($1/N$ expansion where dilatonic quantum
effects are subleading).
We will assume that background spinors are absent.
In the semiclassical approach we have to add to the action 
(\ref{act:dilaton}) the conformal anomaly term 
\cite{CG,NO}
\begin{equation}
     S_{c}=\frac{1}{2\pi}\int\! \sqrt{-\hat{g}} d^{2}x 
     \left[-\frac{1}{2}(\nabla Z)^2
     +\sqrt{\frac{N}{48}}ZR\right],
\label{act:dilaton:cla}
\end{equation}
where $Z$ is the auxiliary scalar field.
For the semiclassical argument we need the Gross-Neveu effective
potential $V(\sigma,R)$ in curved spacetime which replaces 
classical potential $\sigma^2 /2$ in Eq. (\ref{act:dilaton}).
The dilatonic equation of motion derived from the action
(\ref{act:dilaton})
\begin{equation}
     R+\Lambda =0,
\label{const:cur}
\end{equation}
apparently constrains the curvature to be constant.
Using the Gross-Neveu effective potential in curved spacetime 
with constant curvature in the leading order of the $1/N$ 
expansion, our effective action in the semiclassical
approach reads
\begin{equation}
     S_{\mbox{eff}}=\frac{1}{2\pi}\int\! \sqrt{-g} d^{2}x 
     \left[e^{-2\phi}(R+\Lambda)-V(\sigma,R)\right]
     +S_{c},
\label{def:effact:del}
\end{equation}
where $V(\sigma,R)$ is given by Eq. (3$\cdot$15) with $\sigma$-independent
part to be discared.

Applying the conformal gauge
\begin{equation}
     g_{+-}=g_{-+}=-\frac{1}{2}e^{-2\rho}\, ,\,\, 
     g_{++}=g_{--}=0,
\end{equation}
we derive the following semiclassical equations of motion
from the effective action (\ref{def:effact:del})
(with constant curvature according to 
Eq. (\ref{const:cur}) and without spinors):
\begin{eqnarray}
     T_{++}=e^{-2\phi}
     \left[4\partial_{+}\rho\partial_{+}\phi
     +4(\partial_{+}\phi)^2-2\partial_{+}^2\phi
     \right] \nonumber \\
     +\frac{1}{2}(\partial_{+}Z)^2
     +\sqrt{\frac{N}{48}}(\partial_{+}^2
     -2\partial_{+}\rho\partial_{+})Z=0,
\label{dila:t1}
\end{eqnarray}
\begin{eqnarray}
     T_{--}=e^{-2\phi}
     \left[4\partial_{-}\rho\partial_{-}\phi
     +4(\partial_{-}\phi)^2-2\partial_{-}^2\phi
     \right] \nonumber \\
     +\frac{1}{2}(\partial_{-}Z)^2
     +\sqrt{\frac{N}{48}}(\partial_{-}^2
     -2\partial_{-}\rho\partial_{-})Z=0,
\label{dila:t2}
\end{eqnarray}
\begin{eqnarray}
     T_{+-}=e^{-2\phi}\left[
     2\partial_{+}\partial_{-}\phi
     -4\partial_{+}\phi\partial_{-}\phi
     -\frac{1}{4}\Lambda e^{2\rho}\right]
     -\sqrt{\frac{N}{48}}\partial_{+}\partial_{-}Z
     \nonumber \\
     +\frac{1}{4}e^{2\rho}V(\sigma,-\Lambda)
     +2\partial_{+}\partial_{-}\rho
     \frac{\partial V(\sigma,-\Lambda)}{\partial \Lambda}
     =0,
\label{dila:t3}
\end{eqnarray}
\begin{equation}
     \partial_{+}\partial_{-}\rho
     +\frac{\Lambda}{8}e^{2\rho}=0,
\label{dila:t4}
\end{equation}
\begin{equation}
     \partial_{+}\partial_{-}Z
     -\sqrt{\frac{N}{12}}\partial_{+}\partial_{-}\rho =0
\label{dila:t5}
\end{equation}
and
\begin{equation}
     \frac{\partial V(\sigma,-\Lambda)}{\partial \sigma}=0.
\label{dila:t6}
\end{equation}

The general solution of Eq. (\ref{dila:t4}) is well known:
\begin{equation}
     \rho(x)=\frac{1}{2}\ln
     \left[\frac{F'_{+}(x_{+})F'_{-}(x_{-})}
     {\left(1+\frac{\Lambda}{8}F_{+}(x_{+})F_{-}(x_{-})
     \right)^2}
     \right],
\label{gsol1:dil}
\end{equation}
where $F_{\pm}$ is an arbitrary holomorphic
(anti-holomorphic) function.
The conformal gauge in Eq. (\ref{gsol1:dil})
can be fixed completely by choosing 
$F^{\pm}=x^{\pm}$.

The solution of Eq. (\ref{dila:t5}) is given by
\begin{equation}
     Z=\sqrt{\frac{N}{12}}\rho
     +u_{+}(x^{+})
     +u_{-}(x^{-}),
\label{gsol2:dil}
\end{equation}
where $u_{\pm}$ is an arbitrary holomorphic 
(anti-holomorphic) function.
Using Eqs. (\ref{dila:t4}) and (\ref{dila:t5}) 
in Eq. (\ref{dila:t3}), we get
\begin{equation}
     \partial_{+}\partial_{-}e^{-2\phi}
     +\frac{\Lambda}{4}e^{2\rho}
     (e^{-2\phi}-a)=0,
\label{gsol3:dil}
\end{equation}
where
\begin{equation}
     a=\frac{N}{48}
     +\frac{V(\sigma_{0},-\Lambda)}{\lambda}
     -\frac{\partial V(\sigma_{0},-\Lambda)}{\partial \Lambda},
\label{def:a}
\end{equation}
and $\sigma_{0}$ is a solution of Eq. (\ref{dila:t6}).
Note that the quantity $\Lambda a$ plays the role of the
effective cosmological constant.
It may be interesting to note that in the weak curvature
limit the effective cosmological constant disappears
($a=0$) if
\begin{equation}
     \Lambda=\frac{2\lambda_{0}\mu^2}{N}
     \exp\left(2-\frac{2\pi}{\lambda}\right)
     \left[-\left(1+\frac{\pi}{3}\right)
     \pm \sqrt{\left(1+\frac{\pi}{3}\right)^2+4}\,
     \right].
\label{cond:lambda:dil}
\end{equation}
Accordingly we observe that the quantum effects in the 
Gross-Neveu model provides us with the way to solve the
cosmological constant problem within the model under
investigation.

The general solutiuon of Eq. (\ref{gsol3:dil}) is
\begin{equation}
     e^{-2\phi(x^+,x^-)}=v_+(x^+)
     +\frac{\displaystyle 1-\frac{\Lambda}{8}F_+F_-}
           {\displaystyle 1+\frac{\Lambda}{8}F_+F_-}
     \int^{x^+}\!dy\frac{F'_+(y)}{F_+(y)}v_+(y)+\frac{a}{2}
     +(+\leftrightarrow -).
\label{gsol:gsol3}
\end{equation}
Here the function $u_\pm$ and $v_\pm$
are determined by solving the equations for $T_{++}$
and $T_{--}$.
The equation for $T_{++}$ with the choice $F_\pm =x^\pm$ 
reads
\begin{equation}
     T_{++}=v''_{+}
     +\frac{v'_{+}}{x^+}
     -\frac{v_{+}}{{x^+}^2}
     +\frac{1}{2}{u'_{+}}^2
     +\sqrt{\frac{N}{48}}u''_+ =0,
\end{equation}
and the same for $T_{--}$ with the change of index
$+$ to $-$. 

We rewrite these two equations in the form
\begin{eqnarray}
     v''
     +\frac{v'}{x}
     -\frac{v}{x^2}
     =\frac{N}{24}\{p(x),x\},\nonumber\\
     \frac{1}{2}{u'}^2
     +\sqrt{\frac{N}{48}}u''
     =-\frac{N}{24}\{p(x),x\},
\label{vandu}
\end{eqnarray}
where we omit indices $+$ and $-$, $p(x)$ is an unknown
function and $\{\, ,\, \}$ is the Schwarzian derivative.
By solving Eq. (\ref{vandu}) for $v$ and $u$
and substituting the solutions
into Eq. (\ref{gsol:gsol3}), we finally have
\begin{eqnarray}
     e^{-2\phi(x^+,x^-)}&=&\frac{N}{96}\left[
     C^{+}_{3} x^{+} + \frac{C^{+}_{4}}{x^{+}}
     +x^{+}\int^{x^{+}}\!\{p_{+}(z),z\}dz
     -\frac{1}{x^{+}}\int^{x^{+}}\!z^{2}\{p_{+}(z),z\}dz     
     \right.\nonumber\\
     &&+\frac{\displaystyle 1-\frac{\Lambda}{8}x^{+}x^{-}}
           {\displaystyle 1+\frac{\Lambda}{8}x^{+}x^{-}}
     \int^{x^{+}}\!\frac{dy}{y}\left(
     C^{+}_{3} y + \frac{C^{+}_{4}}{y}
     +y\int^{y}\!\{p_{+}(z),z\}dz
     \right.\nonumber\\
     &&\left. \left. 
     -\frac{1}{y}\int^{y}\!z^{2}\{p_{+}(z),z\}dz\right)
     \right]+\frac{a}{2}+(+\leftrightarrow -),
\end{eqnarray}
where $C^{\pm}_{3}$ and $C^{\pm}_{4}$ are arbitrary constants.
For the simplest case in which $p_{\pm}(x^{\pm})=x^{\pm}$,
the solution for dilaton field is given by
\begin{equation}
     e^{-2\phi(x^+,x^-)}=\frac{N}{48}
     \frac{\displaystyle \left(C^{+}_{3}+
           \frac{\Lambda}{8}C^{-}_{4}\right)x^{+}
           +\left(C^{-}_{3}+
           \frac{\Lambda}{8}C^{+}_{4}\right)x^{-}}
           {\displaystyle 1+\frac{\Lambda}{8}x^{+}x^{-}}
     +a.
\end{equation}
This is the main result of this subsection.
We found the static solution for our model in an explicit
form. In order to compare it with the CGHS-Witten
black hole solution,\cite{CG}
we make the redefinition (\ref{def:dilaton})
of the metric.
The new spacetime interval squared becomes
\begin{eqnarray}
     &&d\hat{s}^{2}=
     -e^{-2\phi(x^+,x^-)}e^{-2\rho(x^+,x^-)}dx^{+}dx^{-}
\nonumber \\
     &&=\frac{-dx^{+}dx^{-}}
           {\displaystyle 
            \left(1+\frac{\Lambda}{8}x^{+}x^{-}\right)
            \left(Ax^{+}+Bx^{-}+a+\frac{\Lambda}{8}ax^{+}x^{-}
            \right)},
\end{eqnarray}
where
\begin{equation}
     A=\frac{N}{48} \left(C^{+}_{3}+
           \frac{\Lambda}{8}C^{-}_{4}\right),
\end{equation}
\begin{equation}
     B=\frac{N}{48} \left(C^{-}_{3}+
           \frac{\Lambda}{8}C^{+}_{4}\right).
\end{equation}
This new solution may be regarded as a modified version of 
the CGHS-Witten black hole.\cite{W,CG}
For the simplest case where all arbitrary constants are
chosen to vanish ($e^{-2\phi}=a$), we get 
\begin{equation}
     d\hat{s}^{2}
     =\frac{-dx^{+}dx^{-}}
           {\displaystyle 
            a \left(1+\frac{\Lambda}{8}x^{+}x^{-}\right)^{2}
            }.
\end{equation}

It is interesting to note that for $\Lambda=0$ in 
Eq. (\ref{act:dilaton}) the semiclassical solution
becomes
\begin{equation}
\begin{array}{rcl}
     \rho&=&0,\\
     \sigma_{0}^{2}&=&\displaystyle \mu^{2}\exp\left(
     2-\frac{2\pi}{\lambda}\right),\\
     e^{-2\phi}&=&\displaystyle\frac{1}{4}V_{0}x^{+}x^{-}-C_{0},
\end{array}
\end{equation}
where $V_{0}=V(\sigma_{0},R=0)$ and $C_{0}$ is an arbitrary
constant. In terms of the new metric (\ref{def:dilaton})
we obtain
\begin{equation}
     d\hat{s}^{2}
     =\frac{dx^{+}dx^{-}}
           {\displaystyle 
            C_{0}- \frac{1}{4}V_{0}x^{+}x^{-}}.
\label{bhs}
\end{equation}
Equation (\ref{bhs}) exactly coincides with the
black hole solution of Refs. \cite{W} and \cite{CG}.

Thus we presented the example of 2D black hole solution 
with constant curvature in dilatonic gravity with Gross-Neveu 
model. It is interesting to remark that one can use the
renormalization group improved effective potential instead 
of one-loop Gross-Neveu effective potential.
Then, one has only change the coupling constant
($\lambda =Ng^{2}$) in the semiclassical effective action
by the running coupling constant
\begin{equation}
     g^{2}(t)=\frac{g^{2}}{1-\lambda t/\pi},
\end{equation}
where $t$ is the renormalization group parameter.
Working in the regime of strong curvature,
$t=\ln R/\mu^{2}$.\cite{NS}
Hence, asymptotic freedom in Gross-Neveu model is
induced by black hole curvature. Stronger
black hole curvature induces stronger asymptotic freedom
as in $d=4$ gauge theories in curved spacetime.\cite{BOS}
In other words, in vicinity of constant curvature black hole,
Gross-Neveu model tends to become free theory.

Note also that one can consider other types of dilatonic 
gravity interacting with Gross-Neveu model and investigate
the constant curvature black holes solutions there.\cite{NS}
However, to study other types of black hole solutions one 
should calculate the effective potential of Gross-Neveu model
in an arbitrary curved spacetime.

\subsection{Conformal factor dynamics in 1$/N$ expansion}

The conformal factor dynamics describes the conformally-flat
solution of the gravitational theories at classical 
level. At the quantum level the dynamics of conformal factor
(induced by the conformal anomaly) was suggested as the tool 
for the description of quantum gravity 
in the infrared phase.\cite{A} 
Conformal factor dynamics gives rise to the 
effective potential for the conformal factor which seems
to be quite relevant in quantum gravity.\cite{E1}
In particular it maybe responsible for infrared phase 
transitions. It is very interesting to note that
composite bound states which are typical for four-fermion
theories maybe also the origin of the conformal factor 
dynamics. We follow Ref. \cite{E2} in discussion of this 
question.

Our starting point is the two-dimensional theory with action
\beq
S = \int d^2x \, \sqrt{-g} \left[ \overline{\psi} \left( i
\gamma^\mu (x) \nabla_\mu -m \right) \psi + R - \frac{
\Lambda}{2} \right], \label{qg:1}
\eeq
where the massive $N$-component spinor $\psi$ is considered to be
a quantum field. The gravitational field,  on the other hand,
may be either classical or quantum. We also consider the
conformal parametrization of the metric
\beq
g_{\mu\nu} = \rho^2 \eta_{\mu\nu}, \label{qg:2}
\eeq
where $\rho$ is the conformal factor (in general it is $\rho =
\rho (x)$) and $\eta_{\mu\nu}$ is  the flat fiducial metric.
In the QG case, the choice (\ref{qg:2}) corresponds to the gauge
fixing. Substituting (\ref{qg:2}) into (\ref{qg:1}) one gets, at the
classical level,
\beq
S = \int d^2x \left[ \overline{\chi} \left( i \gamma^\mu \partial_\mu
- m \rho \right) \chi - \frac{\Lambda}{2} \, \rho^2 \right],
\label{qg:3}
\eeq
where $\chi = \rho^{1/2} \psi$. Rescaling $\rho \rightarrow
\Lambda^{1/2} \rho$, we get
\beq
S= \int d^2x \, \left[ \overline{\chi} \left( i \gamma^\mu \partial_\mu
- h \rho \right) \chi - \frac{1}{2} \rho^2 \right],  \label{qg:4}
\eeq
where $h= m \Lambda^{-1/2}$. As one can see, action (\ref{qg:4}) has the
form that is typical for the Gross-Neveu
model ($\rho \sim \overline{\chi}
\chi$, see Eq. (2$\cdot$12)). 
The dynamics of this model are quite well known:
asymptotic freedom in the UV limit
\beq
h^2(t) = \frac{h^2}{1 + h^2 N t /\pi},  \label{qg:5}
\eeq
where $t$ is the RG parameter.
However, since (\ref{qg:4}) describes also the dynamics of the conformal
factor, the interpretation of the function $h^2(t)$ is now completely
different from the original interpretation. The $h^2(t)$ here
 is a combination of  the fermionic mass in (\ref{qg:1})
 and of the two-dimensional cosmological
constant $\Lambda$. Using the anomalous scaling dimension one gets
the running composite field
\beq
\rho (x,t) = \rho (x) \left( 1 + h^2 N\, t/\pi \right)^{-1/2}.
\label{qg:6}
\eeq
The conformal factor has acquired the classical dimension after
the rescaling $\rho \rightarrow \Lambda^{1/2} \rho$. Hence, one may
argue that the $t$ dependence is due completely to that of the
cosmological (dimensional) constant, i.e., $\Lambda (t) \sim
\Lambda \ (1 + h^2 Nt /\pi )^{-1}$. Therefore, there appears to
be a screening of the cosmological constant in the $1/N$
expansion in the UV regime
(quantum gravitational corrections are subleading in $1/N$
expantion).

One can also investigate specific features of
 the effective potential for the conformal factor,
which coincides with the GN effective potential.\cite{GN}
In particular, the appearance of a minimum, i.e., a  non-zero
vacuum expectation value (v.e.v.),  for the conformal factor
\beq
\rho = \rho_0 \exp \left( 1 - \frac{\pi}{h^2N} \right)  \label{qg:7}
\eeq
is interesting. After having shown the possibility to study the dynamics
of the conformal factor in two dimensions as the dynamics of the GN
model, we now turn to the four-dimensional theory, which is
physically
more interesting.

We start from the multiplicatively renormalizable theory \cite{BOS}
with  action
\beq
S = \int d^4x \, \sqrt{-g} \, \left[ \overline{\psi}
\left( i \gamma^\mu (x) \nabla_\mu - m \right) \psi - \frac{
\Lambda}{\kappa^2} -\frac{R}{\kappa^2} + \frac{W}{\lambda_1}
-\frac{UR^2}{3\lambda_1} \right], \label{qg:8}
\eeq
where $\psi$ is an $N$-component spinor and $W$ the square
of the Weyl tensor. The gravitational field may be chosen to be
classical or quantum, and the theory remains multiplicatively
renormalizable in both cases.

We  shall work again with the conformal parametrization (\ref{qg:2})
for the four-dimensional metric. In the case of four-dimensional
QG this does not fix the gauge, contrary to what happens in two
dimensions, but it can
still be considered as a convenient background.\cite{E1}
Rewriting action (\ref{qg:8}), we get
\bea
S&=& \int d^4x \, \left\{ \overline{\chi} \left( i \gamma^\mu
\partial_\mu - m\rho \right) \chi -\frac{\Lambda}{\kappa^2}
\rho^4 - \frac{6}{\kappa^2} (\partial \rho )^2 \right. \nn \\
&& \hspace{2cm} \left. +
\frac{12U}{\lambda_1} \left[ \sigma \Box^{2} \sigma + 2 (\partial
\sigma )^2 \Box \sigma + (\partial \sigma)^2 (\partial
\sigma)^2 \right] \right\},  \label{qg:9}
\eea
where $\chi =\rho^{3/2}\psi$ and $\sigma = \ln \rho$.
In this way we have got the classical theory for the conformal
factor. At the quantum level, the theory (\ref{qg:9}) may be
considered as an effective theory for QG (see also Ref. \cite{A}).
If we drop the $\rho$-terms with derivatives from action
(\ref{qg:9}), we obtain a model that is reminiscent of the
NJL model (where, of course, owing to the absence of the
$M^2\rho^2$-term, it is $\rho \sim (\overline{\chi} \chi
)^{1/3}$).

Now we are going to study the theory (\ref{qg:9}) in the large-$N$
limit, while concentrating our attention
on the RG and low-derivative terms
in (\ref{qg:9}). The higher-derivative terms are actually of
lesser importance, moreover, they simply disappear in the
subsequent analysis of the effective potential for the conformal
factor.
First of all, we rescale $\rho \rightarrow \sqrt{12} \,
\rho /\kappa$ and denote $h=m\kappa /\sqrt{12}$ and $\lambda =
\Lambda \kappa^2 /6$. Thus,
\beq
S= \int d^4x \, \left[ \overline{\chi} \left( i\gamma^\mu
\partial_\mu - h\rho \right) \chi -\frac{\lambda}{24}
\rho^4 - \frac{1}{2} (\partial \rho )^2 \right].
\label{qg:10}
\eeq
By integrating over the fermionic field, we get the effective potential
for the conformal factor at large $N$:
\beq
V(\rho ) = \frac{\lambda \rho^4}{24} + iN \mbox{Tr} \ln 
\left( i \gamma^\mu
\partial_\mu - h\rho \right),  \label{qg:11}
\eeq
where $\rho$ is  constant. Supposing that, as usually, for large $N$
$\lambda$ scales as $\lambda \sim \wt{\lambda} N$, where $\wt{\lambda
}$ does not depend on $N$, and using a finite cut-off $\mu$ (see
also \S 3) we get (notice that $N$ has been factored out)
\beq
V(\rho )= \frac{\wt{\lambda} \rho^4}{24} -\frac{1}{(4\pi)^2} \left[
\rho^2 \mu^2 + \mu^4 \ln \left( 1 + \frac{\rho^2}{\mu^2} \right) -
\rho^4 \ln \left( 1 + \frac{\mu^2}{\rho^2} \right)\right].
\label{qg:12}
\eeq

The v.e.v. of the conformal factor can be found from (\ref{qg:12}) as
the solution of the equation
\beq
\frac{\partial V(\rho)}{\partial \rho} = \frac{\wt{\lambda}}{6}
\, \rho^2
- \frac{1}{(2\pi)^2} \left[ \mu^2 -\rho^2 \ln \left( 1+
\frac{\mu^2}{\rho^2} \right) \right] = 0.
\label{qg:13}
\eeq
One can present a sample of numerical values of the solution
$\rho^2 / \mu^2$ (which lie between 0 and 1 for $0.04644 \leq
\wt{\lambda} \leq 1$).
Notice that from the point of view of the original theory, a non-zero
v.e.v. for $\rho$ is more acceptable physically, because for $\rho =0$
the conformal parametrization  (\ref{qg:2}) becomes degenerate.

Now we turn to the study of the renormalization structure of theory
(\ref{qg:10}),
which is rather non-trivial. There are a few different ways to 
renormalize this theory, the actual problem being the fact that 
$\rho$ is dimensionless.

We may consider a theory (\ref{qg:10})
---as it stands--- which will correspond eventually to some
QG phase. After renormalization of $\rho$ (taking into account the
negative sign for $(\partial \rho )^2$), we obtain
\beq
\beta_h = -\frac{4Nh^4}{(4\pi)^2}, \ \ \ \  \beta_\lambda =
\frac{3\lambda^2 -8N\lambda h^2 - 48 N h^4}{(4\pi)^2}.  \label{qg:16}
\eeq
The theory has now a UV stable fixed point ($t \rightarrow +\infty$),
 where the
behavior of $h^2$ and $\lambda$ is
\beq
h^2(t) \sim \frac{4\pi^2}{Nt}, \ \ \ \ \ \  \lambda (t) \sim -\frac{48
\pi^2}{Nt}.   \label{qg:17}
\eeq
Hence, we now obtain a decrease of the cosmological constant 
in the UV limit.
Notice that in comparing this theory with the standard scalar 
self-interacting
theory, we do not have here physical restrictions on the sign 
of $\lambda$, and a negative sign is perfectly acceptable.
That indicates to the possibility of solution of cosmological
constant problem in frames of conformal factor dynamics
based on four-fermion like theory.

\subsection{Effective action in $N$=1 supergravity}

In \S 5.3 we already discussed supersymmetric extention of 
NJL model. There are also other possibilities to introduce 
supersymmetry into the theory.
In particular, in present subsection we will discuss $N=1$
supergravity on de Sitter background.
We calculate the effective action for non-zero gravitino
condensate 
$\sigma\sim\langle\bar{\psi}_{a}\Gamma^{ab}\psi_{b}\rangle$.
Hence, such a model maybe considered also as kind of 
four-fermion (four-gravitino model).
Note, however, that unlike the above discussion we work in 
one-loop approximation (not in $1/N$-expansion).
We follow Ref. \cite{OBO} below.

The Lagrangian for the theory of $N=1$ supergravity is
\begin{equation}
     {\cal L}=-\frac{1}{k^2}R(e,\psi)+\frac{1}{2}
     \epsilon^{\mu\nu\rho\sigma}
     \bar{\psi}_{\mu}\gamma_{5}\gamma_{\nu}\nabla_{\rho}
     \psi_{\sigma}+\frac{1}{3}(A_{\mu}^2-S^2-P^2)\, ,
\label{def:SG}
\end{equation}
where $\Gamma^{\mu\nu}=\frac{1}{4}[\gamma^{\mu},\gamma^{\nu}],
R(e,\psi)=R(e)+\frac{11}{4}k^{4}(\bar{\psi}_{\mu}
\Gamma^{\mu\nu}\psi_{\nu})^{2}+\cdots$
(in this expression the terms corresponding to the interaction
of the gravitino with the gravitational field and four-fermion
terms with insertions containing the $\gamma_{5}$ matrix are not
written in the explicit form), 
$\nabla_{\lambda}\equiv\nabla_{\lambda}(e), A_{\mu},
S, P$ is the minimal set of auxiliary fields.
Let us represent the Lagrangian (\ref{def:SG}) in the 
form:\cite{OBO,JA}
\begin{eqnarray}
     {\cal L}=-\frac{1}{k^2}R(e,\psi)+\frac{1}{2}
     \epsilon^{\mu\nu\rho\sigma}
     \bar{\psi}_{\mu}\gamma_{5}\gamma_{\nu}\nabla_{\rho}
     \psi_{\sigma}+\sigma^{2}
     \nonumber \\
     -\sqrt{11}k\sigma(\bar{\psi}_{\mu}\Gamma^{\mu\nu}\psi_{\nu})
     +\frac{1}{3}(A_{\mu}^2-S^2-P^2)\, ,
\label{def:SG:lag}
\end{eqnarray}
where $\sigma$ is the auxiliary scalar field.
Let us calculate the effective action in the theory with the
Lagrangian (\ref{def:SG:lag}) in de Sitter spacetime
($R=4\Lambda$) with constant scalar field $\sigma$.
If the effective equations
\begin{equation}
     \frac{\delta\Gamma}{\delta\sigma}=0,
\label{eqmo:1}
\end{equation}
\begin{equation}
     \frac{\delta\Gamma}{\delta\Lambda}=0,
\label{eqmo:2}
\end{equation}
have the real solutions $\sigma\neq 0$,
then these solutions indicate dynamical supersymmetry
breaking.\cite{JA}
Equation (\ref{eqmo:2}) represents the effective equation 
for the gravitational field which depends on the parameter
$\Lambda$ only.

Let us write $g_{\mu\nu}\rightarrow g_{\mu\nu}+h_{\mu\nu},
\sigma\rightarrow \sigma +\sigma_{q}$, where $g_{\mu\nu}$
is the de Sitter space metric, $\sigma=$const, and
$h_{\mu\nu}, \sigma_{q}$ are quantum fields. The background
values of other fields of the theory we put equal to zero.
One can obtain that the sector of quantum fields 
$(h_{\mu\nu}, \sigma_{q})$ does not interact with the 
gravitino sector.

At first one can integrate over the field $\sigma_{q}$
in the functional integral. Then the additional term
$\triangle {\cal L}_{\sigma_{q}}=-\sigma^{2}h^{2}/4$
appears in the expansion of the classical gravitational
Lagrangian on the quantum field $h_{\mu\nu}$.

In order to calculate the effective action, one needs
to add a gauge-fixing term to the Lagrangian.
We shall choose the one-parameter gauge:
\begin{equation}
     {\cal L}_{\mbox{gauge fix}}=\frac{\gamma}{2k^{2}}\left(
     \nabla^{\mu}h_{\mu\nu}-\frac{1}{2}\nabla_{\nu}h
     \right)^{2},
\label{gfix}
\end{equation}
where $\gamma$ is the gauge parameter, $\gamma=1$
corresponds to the De Donder gauge, 
$\gamma\rightarrow\infty$ corresponds to the 
Landau-DeWitt gauge. Let us write the bilinear
part of the Einstein Lagrangian on the $S^{4}$
background \cite{Frad} with account of 
$\triangle {\cal L}_{\sigma_{q}}$
and ${\cal L}_{\mbox{gauge fix}}$ (\ref{gfix}).
\begin{eqnarray}
     {\cal L}_{2}&=&\frac{1}{2k^{2}}\left[
     \frac{1}{2}\bar{h}^{\bot}\triangle_{2}
     \left(\frac{8}{3}\Lambda -2\Lambda_{0}
     \right)\bar{h}^{\bot}
     +2(\Lambda-\Lambda_{0})
     \xi^{\bot}\triangle_{1}(-\Lambda)\xi^{\bot}
     \right.\nonumber \\
     &&-\frac{3}{16}h_{1}\Box\left(\Box +\frac{4}{3}\Lambda
     \right)\left(-\Box +4\Lambda_{0}-4\Lambda+3\gamma
     \Box+4\gamma\Lambda\right)h_{1}\nonumber \\
     &&\left.
     +2(1-\gamma)h_{1}\Box(\Box+\frac{4}{3}\Lambda)h
     +h\left(-\Box-\frac{4}{3}\Lambda_{0}
     +\frac{1}{3}\gamma\Box\right)h\right]
     -\frac{1}{4}\sigma^{2}h^{2}.
\end{eqnarray}
Here 
\begin{equation}
\Lambda_{0}=\frac{1}{2}k^{2}\sigma^{2},
\end{equation}
\begin{equation}
\bar{h}_{\mu\nu}=\bar{h}_{\mu\nu}^{\bot}+
2\nabla_{(\mu}\xi^{\bot}_{\nu)}+\nabla_{\mu}\nabla_{\nu}h_{1}
-\frac{1}{4}g_{\mu\nu}\Box h_{1},
\end{equation}
\begin{equation}
h_{\mu\nu}=\bar{h}_{\mu\nu}+\frac{1}{4}g_{\mu\nu}h,
\end{equation}
\begin{equation}
g^{\mu\nu}\bar{h}_{\mu\nu}=0,
\end{equation}
\begin{equation}
\nabla^{\mu}\bar{h}_{\mu\nu}^{\bot}=0,
\end{equation}
\begin{equation}
\nabla_{\mu}\xi^{\bot\mu}=0,
\end{equation}
\begin{equation}
\triangle_{i}(X)=\11_{i}(-\Box+X), i=0,1,2,
\end{equation}
where $\11$ is the unit in the space of the corresponding
fields.

Integrating over quantum fields $\bar{h}_{\mu\nu}^{\bot},
\xi^{\bot\mu},h_{1},h,$ taking account of the corresponding
Jacobians and ghost contributions, we have in the limit
$\gamma\rightarrow \infty$
(we omit the details of this straightforward
calculation)
\begin{eqnarray}
     \Gamma_{h_{\mu\nu}}=\frac{1}{2}\mbox{Sp}\ln
     \triangle_{2}\left(\frac{8}{3}\Lambda-k^{2}
     \sigma^{2}\right)-\frac{1}{2}\mbox{Sp}\ln
     \triangle_{1}(-\Lambda)
     -\mbox{Sp}\ln
     \triangle_{0}(-2\Lambda)
     \nonumber \\
     +\frac{1}{2}\mbox{Sp}\ln
     \triangle_{0}\left[
     \frac{5}{2}\sigma^{2}k^{2}-\Lambda
     +\left(\Lambda^{2}+k^{2}\sigma^{2}\Lambda
     +\frac{25}{4}k^{2}\sigma^{4}\right)^{1/2}
     \right]
     \nonumber \\
     +\frac{1}{2}\mbox{Sp}\ln
     \triangle_{0}\left[
     \frac{5}{2}\sigma^{2}k^{2}-\Lambda
     -\left(\Lambda^{2}+k^{2}\sigma^{2}\Lambda
     +\frac{25}{4}k^{2}\sigma^{4}\right)^{1/2}
     \right].
\label{849}
\end{eqnarray}
The explicit expression for $\frac{1}{2}\mbox{Sp}\ln
\triangle_{i}(X), i=0,1,2$ is obtained in Ref. \cite{Frad}
in frames of $\zeta$-regularization \cite{b10}
\begin{equation}
     \frac{1}{2}\mbox{Sp}\ln\triangle_{i}(X)
     =\frac{1}{2}B^{i}_{4}
     \ln\left(\frac{\Lambda}{3\mu^{2}}\right)
     -\frac{1}{6}(2i+1)F'_{i}(b_{i}),
\label{850}
\end{equation}
where $b_{i}=9/4+i-3X/\Lambda$, $B^{i}_{4}$ is the well-known
De Witt $B_{4}$ coefficient for the corresponding operator,
$F'_{i}(b_{i})$ is given by Eq. (2$\cdot$36) in Ref. \cite{Frad}.
Using (\ref{850}) in (\ref{849}), one can easily obtain the
explicit form for $\Gamma_{h_{\mu\nu}}$.
The final expression is very complicated, so we do not write
it here. It is enough for our purposes to have 
$\Gamma_{h_{\mu\nu}}$ in the form (\ref{849}).

Let us now calculate the gravitino contribution to the
effective action. We choose the gauge:
\begin{equation}
     {\cal L}_{\psi}=\frac{1}{2}\gamma_{1}k\bar{\chi}\chi,
\end{equation}
where $\chi =\nabla^{\mu}\psi_{\mu}$. Then the Faddeev-Popov determinant 
is $M=\11\Box=\hat{\nabla}\hat{\nabla}+\Lambda$.
Represent \cite{Frad}
$\psi_{\mu}=\phi_{\mu}+\gamma_{\mu}\psi /4$,
$\gamma^{\mu}\phi_{\mu}=0$,
$\phi_{\mu}=\phi_{\mu}^{\bot}+(\nabla_{\mu}
-\gamma_{\mu}\hat{\nabla}/4)\zeta$,
$\nabla^{\mu}\phi_{\mu}^{\bot}=0$,
$\hat{\nabla}\equiv\gamma^{\mu}\nabla_{\mu}$
in the bilinear part of the gravitino Lagrangian (\ref{def:SG:lag})
taking account of ${\cal L}_{\psi}$.
Integrating over the fields $\phi_{\mu}^{\bot},\zeta,\psi$
taking account of the corresponding Jacobians and ghost contribution
we get in the limit $\gamma_{1}\rightarrow\infty$:
\begin{eqnarray}
     \Gamma_{\psi_{\mu}}
     =-\frac{1}{4}\mbox{Sp}\ln\triangle_{3/2}(m^{2})
      -\frac{1}{2}\mbox{Sp}\ln\left[
      \left(-\frac{1}{3}\hat{\nabla}+2m\right)
      \triangle_{1/2}\left(-\frac{4}{3}\Lambda\right)
      \right. \nonumber \\
      \left. -\frac{1}{9}\triangle_{1/2}(0)(\hat{\nabla}+2m)
      \right]+\mbox{Sp}\ln\triangle_{1/2}(-\Lambda),
\label{852}
\end{eqnarray}
where $m=-\sqrt{11}k\sigma$.
One can write the explicit expression for $\triangle_{1/2}(X),
\triangle_{3/2}(X)$ using relations such as (\ref{850}) 
(see Ref. \cite{Frad}). We do not write this complicated 
expression here.

A natural question now is about the presence of the negative
modes in the effective action 
$\Gamma=\Gamma_{h_{\mu\nu}}+\Gamma_{\psi_{\mu}}$.
The spectra of all operators in (\ref{849}) and
(\ref{852}) are given in Ref. \cite{Frad}.
Using the result of Ref. \cite{Frad} for $\Gamma_{h_{\mu\nu}}$
(\ref{849}) one can show that the operator 
$\triangle_{2}(8\Lambda/3-k^{2}\sigma^{2})$
does not include the negative modes if 
$1\leq k^{2}\sigma^{2}/2\Lambda\leq 8/3$.
The three last terms in (\ref{849}) contain the negative
modes for arbitrary values $\sigma$ and $\Lambda$
($\Lambda >0$). Therefore, $\Gamma_{h_{\mu\nu}}$
contains an imaginary part for all $\sigma$ and $\Lambda$.
Then $\Gamma$ also contains an imaginary part.
(Remember that the structure of the fermion
sector does not influence the bosonic sector.)
Thus the effective equations also contain an imaginary 
part. It indicates a vacuum instability of the considered
background and the absence of dynamical supersymmetry
breaking at the 1-loop level.

It is interesting that one can easily obtain the effective action 
in $N=1$ supergravity in flat spacetime
using the effective action (\ref{849}) and (\ref{852}).
This effective action also contains an imaginary part for
all $\sigma$.

Thus, we found the effective action for composite gravitino
in $N=1$ supergravity and discussed its properties.

In summary, in the present section the use of four-fermion
type models in the theories of quantum gravity has been
discussed.
Few particular examples presented here indicate that four-fermion
models could find wider applications than it is usually believed.

\section{Summary and outlook}

In the present paper we discussed four-fermion models
in an external gravitational field.
As is well-known these models are extremely useful for
analytical study of dynamical symmetry breaking.
Our main purpose was to investigate the influence
of the external gravitational field to dynamical
symmetry breaking.

Using $1/N$-expansion and working in the leading
order of this expansion we calculated the effective
potential in four-fermion models in $D$-dimensional
spacetime ($2\leq D\leq 4$) for a variety of backgrounds.
In particular, we considered the class of spacetimes where
derivative expansion maybe applied.
Here the effective potential for the composite operator
$\bar{\psi}\psi$ has been calculated taking into
account terms linear on curvature.
We also evaluated the effective potential on the background
with constant curvature; Minkowski space, 
de Sitter background, anti-de Sitter
background and Einstein universe.
In all these cases an external curvature has been taken
into account exactly.

Phase structure of the effective potential has been
carefully studied using analytical and numerical
methods.
The possibility of chiral symmetry breaking due to
curvature effects and curvature-induced phase transitions
has been discussed in all detail.
The phase diagrams for dynamically generated fermion mass
have been constructed.
For example, for negative curvature chiral symmetry
is always broken down.

Similar questions have been addressed for the extensions
of four-fermion models; higher derivative four-fermion 
model and gauged NJL model.
In particular, for gauged NJL model (which may play
the role of SM or GUT without elementary scalars)
the technique of renormalization group improvement
has been applied.
The effective potential equivalent to Schwinger-Dyson
approach has been found analytically.
Chiral symmetry breaking has been estimated.

We also discuss the combined influence of two effects
(non-trivial topology and external curvature,
non-vanishing magnetic field and external curvature)
to phase structure of four-fermion model.
The effective potential has again been calculated for
few specific backgrounds.
Phase structure has been analyzed.
In particular we found that in some case two external
effects (for example, magnetic field and gravitational
field) may compensate each other.
That leads to the restoration of chiral symmetry,
which was originally broken.

We presented the number of examples where an analytical
study of composite states is possible even in the
presence of non-trivial gravitational background.
Having in mind the cosmological applications, it is
straitforward now to use the effective potential
obtained above for the construction of inflationary
universe where the role of inflaton will be played
by a composite state of fermion fields.\cite{Inf}

In our calculation we limited ourselves to case of weakly
curved or constant curvature spacetimes.
In this case the effective action is reduced to
the effective potential except for the volume factor.
However, for more general backgrounds one has to calculate
the effective action in four-fermion models.
It is expected that instabilities (due to creation
of instanton-antiinstanton condensation) may appear
in such situation.
That may completely change the phase structure of the theory.
Such an extremely hard task as the calculation of effective
action on general curved background is quite natural
extension of the presented results.

Another very interesting line of research is related with
the study of realistic model of elementary particles
in the early universe.
One can start from some GUT scenarios which maybe
reformulated as gauged four-fermion model
in the same way as the model of \S 5.1.
Then, the role of gravitational background to phase
transitions maybe discussed in all detail,
even in non-perturbative regime.

Note also that methods developed in the present paper
maybe applied to other theories.
For example, in supergravities one can study the effective
potential of composite gravitino (see \S 9.3), 
for gauge theories
one can study the effective potential for composite
vectors \cite{BOS} and so on.

It is interesting to note also that above approach maybe 
well extended to supersymmetric theories. For example,
in \S 5.3 we discussed supersymmetric NJL 
model in curved spacetime and found the possibility of 
dynamical chiral symmetry breaking due to gravitational 
effects.

It would be also extremely interesting to analyze the
next-to-leading corrections to effective potential
of four-fermion model on curved background.
Such multi-loop calculations may clarify the phase
structure in many cases when the creation of multi-critical
points is possible.

Finally, it is interesting to note that in realistic
situation (like early universe) the combination of few 
external effects maybe quite natural.
Hence, it could be interesting to study phase structure
of four-fermion model in curved spacetime with
finite temperature,\cite{II} magnetic field, chemical 
potential and non-trivial topology.

\section*{Acknowledgements}

We would like to thank our friends 
I.~L.~Buchbinder, E.~Elizalde, K.~Fukazawa,
B.~Geyer, D.~M.~Gitman, L.~N.~Granda,
K.~Ishikawa,
T.~Kouno, S.~Leseduarte, S.~Mukaigawa,
H.~T.~Sato, Yu.~I.~Shilnov and K.~Yamamoto
for stimulating discussions and cooperations
in the various
occasions of studies
of four-fermion models in curved spacetime.
T.~M. would like to thank the financial support provided by
the Monbusho Grant-in-Aid for Scientific Research (C) with
the contract No. 08640377.
S.~D.~O. would like to thank
COLCIENCIAS (Colombia) and JSPS (Japan,
Contract No. RC39626102)
for partial support of this work.

\end{document}